\documentclass[aps,prd,preprint,superscriptaddress,amsmath,amssymb, nofootinbib]{revtex4-1}
\usepackage{graphicx}
\usepackage{makecell}
\usepackage{hyperref}

\hypersetup{
    colorlinks=false, 
    linktoc=all,     
}

\usepackage{footmisc}

\usepackage[usenames,dvipsnames]{xcolor}

\def\simgt{\mathrel{\lower2.5pt\vbox{\lineskip=0pt\baselineskip=0pt
           \hbox{$>$}\hbox{$\sim$}}}}
\def\simlt{\mathrel{\lower2.5pt\vbox{\lineskip=0pt\baselineskip=0pt
           \hbox{$<$}\hbox{$\sim$}}}}
           
\begin{document}

\title{Optimal Impedance-Matching and Quantum Limits of Electromagnetic Axion and Hidden-Photon Dark Matter Searches}

\author{Saptarshi Chaudhuri}
\affiliation{Department of Physics, Stanford University, Stanford, CA 94305}

\author{Kent D. Irwin}
\affiliation{Department of Physics, Stanford University, Stanford, CA 94305}
\affiliation{Kavli Institute for Particle Astrophysics and Cosmology, Stanford University, Stanford, CA 94305}
\affiliation{SLAC National Accelerator Laboratory, Menlo Park, CA 94025}

\author{Peter W. Graham}
\affiliation{Department of Physics, Stanford University, Stanford, CA 94305}
\affiliation{Kavli Institute for Particle Astrophysics and Cosmology, Stanford University, Stanford, CA 94305}
\affiliation{Stanford Institute for Theoretical Physics, Department of Physics, Stanford University, Stanford, CA 94305}

\author{Jeremy Mardon}
\affiliation{Department of Physics, Stanford University, Stanford, CA 94305}
\affiliation{Stanford Institute for Theoretical Physics, Department of Physics, Stanford University, Stanford, CA 94305}

\date{\today}

\begin{abstract}
For the first time, we determine the properties of the optimal single-moded, linear, passive search for electromagnetic coupling to axion- and hidden-photon dark matter, subject to the Standard Quantum Limit (SQL) on phase-insensitive amplification. We begin by posing the question of why dark-matter detection through electromagnetic coupling is difficult, even though the dark-matter field possesses enough energy flux per square meter to power a light bulb. We thus introduce the concept of impedance-matching to dark matter, a critical component in optimization. We establish the set of parameters that must be considered to determine the optimal search: the impedance match of a receiver to dark matter; the set of possible receiver frequency-response functions, which may be tuned periodically in an arbitrary way; irreducible noise sources such as thermal and quantum noise; and prior information on the properties of the dark-matter signal. Using complex-power flow equations to characterize the excitation of an electromagnetic receiver, we identify the two categories of couplings to the dark-matter signal: radiative couplings and reactive couplings. We illustrate a primary limitation in extracting power from (or equivalently, impedance-matching to) the dark-matter field, which is the self-impedance of photons acting on electromagnetic charges in the receiver. We motivate a focus on searches using reactive couplings, as receivers using solely radiative couplings are limited by the mismatch between the free-space impedance and the effective source impedance of dark matter. 

Focusing thereafter on single-moded, reactively coupled receivers, we develop a framework to optimize dark matter searches using prior information about the dark-matter signal. Priors can arise, for example, from cosmological or astrophysical constraints, constraints from previous direct-detection searches, or preferred search ranges. We define integrated sensitivity as a figure of merit in comparing searches over a wide frequency range and show that the Bode-Fano criterion sets a limit on integrated sensitivity in a reactively coupled receiver. We examine single-pole resonators, which are a broadly used form of reactive coupling in axion and hidden-photon dark matter searches, and show that when resonator thermal noise dominates amplifier noise, substantial sensitivity is available away from the resonator bandwidth. The optimization of this sensitivity is found to be closely related to noise mismatch with the phase-insensitive amplifier and the concept of measurement backaction. 
We show that not all receivers that optimize integrated power transfer for the dark-matter signal necessarily optimize integrated sensitivity. Nevertheless, \emph{the Bode-Fano constraint establishes the single-pole resonator as a near-ideal (but not precisely ideal) method for single-moded dark-matter detection.} 
Furthermore, we show that for single-moded, linear, passive receivers subject to the SQL, the optimized resonator is superior, in signal-to-noise ratio of an integrated scan, to the optimized reactive broadband receiver at all frequencies at which a resonator may practically be made. 

Owing to the near-ideal integrated sensitivity, we thereafter focus on single-pole resonators. We optimize time allocation in a scanned tunable resonator search using priors and derive quantum limits on resonant search sensitivity. At low frequencies, the application of our optimization may enhance scan rates by a few orders of magnitude. We show that, in contrast to some previous work, resonant searches benefit from quality factors above one million, which corresponds to the characteristic quality factor (inverse of fractional bandwidth) of the dark-matter signal. We discuss our optimization results in the context of practical tradeoffs that may be made in the course of an experimental design and implementation. Finally, we discuss prospects for evading the quantum limit on scan sensitivity using backaction evasion, photon counting, squeezing, entanglement, and other nonclassical approaches, in the context of directions for further investigation. While our results broadly inform laboratory searches for light fields, they are the basis for DMRadio, a DOE-funded program in axion and hidden-photon dark matter detection.
\end{abstract}

\maketitle

\tableofcontents

\newpage

\section{Introduction}
\label{sec:intro}

A significant body of astrophysical and cosmological evidence points to the existence of cold dark matter, which comprises 27\% of the mass-energy in the universe.\cite{ade2014planck} Cold dark matter is a window into physics beyond the Standard Model and plays a significant role not only in particle physics, but also in the formation of galaxies and large-scale structure. Because of dark matter's broad significance, there has been a decades-long effort to directly detect dark matter and determine its properties, with extensive theoretical work in developing candidate models for particle dark matter and experimental probes to search for these candidates. 

A number of candidates may be probed through their feeble coupling to the Standard Model photon. Astrophysical measurements indicate that the local dark-matter density is $\rho_{\rm DM} \sim 0.45$ GeV/cm$^{3}$\cite{tanabashi2018review}, with a virial velocity of $\sim 10^{-3}$c, resulting in an energy flux of $\sim$10 Watts per square meter. In each square meter of flux, there is then enough power to turn on a household LED lamp! This naturally begs the questions: for dark-matter candidates which couple to the Standard Model through the electromagnetic interaction, why are dark-matter searches difficult? What exactly are the physical mechanisms that prevent us from harnessing the entire energy flux of dark matter, for example, as an alternative energy source? Equivalently, why is it impractical to impedance match to dark matter? Over the following two sections of this paper, we will answer these questions, in the context of the principal purpose of this work, which is to conduct a broad optimization of electromagnetic searches for axion and hidden-photon dark matter. Below, we overview axions and hidden photons as light-field dark-matter candidates and motivate the need for a first-principles optimization, based on present dark-matter searches. We then lay out our optimization strategy, in which the analysis of impedance-matching to dark matter plays a critical role.


The class of ultralight bosons, with mass below $\sim$1 eV, has gained much attention in recent years as potential dark-matter candidates. \cite{horns2013searching,graham2015experimental,sikivie2021invisible} The extremely low mass of ultralight-boson dark matter, combined with the observed dark-matter density, implies a large number density. As a result, these bosons are most appropriately described not as individual particles, but as classical fields oscillating at a frequency slightly greater than their rest frequency, $\nu_{\rm DM}^{0} = m_{\rm DM}c^{2}/h$ ($m_{\rm DM}$ is the rest mass of the dark matter, $c$ is the speed of light, and $h$ is Planck's constant); the actual oscillation frequency is slightly higher than $\nu_{\rm DM}^{0}$ as a result of the small kinetic energy. Two prominent candidates in the class of ultralight bosons are axions and hidden photons.

The ``QCD axion'' is a spin-0 pseudoscalar originally motivated as a solution to the strong CP problem\cite{peccei1977cp}, which can also be dark matter\cite{Preskill:1982cy}. However, spin-0 pseudoscalar dark matter may exist even with parameters that do not solve the strong CP problem. Such particles are sometimes referred to as ``axion-like particles.'' In this work, we refer to both QCD axions and axion-like particles as ``axions.'' Axions may be produced nonthermally (as would be required for a sub-eV particle to be cold, nonrelativistic dark matter) through the misalignment mechanism or through inflationary mechanisms\cite{Dine:1982ah,Preskill:1982cy,abbott1983cosmological,graham2018stochastic,takahashi2018qcd}. One may search for axions via their coupling to the strong force \cite{Budker:2013hfa} or their coupling to electromagnetism \cite{Sik83,Sik85}. The latter interaction, described by the Lagrangian
\begin{equation}\label{eq:L_axion} 
\mathcal{L}_{\textrm{int}, a} = g_{a\gamma\gamma}a F\tilde{F},
\end{equation}
is discussed further in this paper. This interaction dictates that in the presence of a background electromagnetic field, the axion converts to a photon. 


The hidden photon is a spin-1 vector. Such particles emerge generically from models for physics beyond the Standard Model, often from theories with new U(1) symmetries and light hidden sectors \cite{holdom1986searching}. The hidden photon was initially described as a dark-matter candidate in \cite{nelson:2011sf} and is further investigated in \cite{arias2012wispy}. Like axions, hidden photons may be produced through the misalignment mechanism. They may also be produced during cosmic inflation. In fact, a vector particle in the $\sim$10 $\mu$eV- 10 meV mass range produced from quantum fluctuations during inflation would naturally have the proper abundance to be a dominant component of the dark matter \cite{Graham:2015rva}. One may search for hidden-photon dark matter via its coupling to electromagnetism \cite{Chaudhuri:2014dla}, which arises from kinetic mixing:
\begin{equation}\label{eq:L_HP}
\mathcal{L}_{\textrm{int}, \gamma'}=\varepsilon F F'.
\end{equation}

A traditional particle detector registers energy deposition from the scattering of a single dark-matter particle with a nucleus or electron. For an ultralight boson, such a measurement scheme is not appropriate, as the energy deposition would be too small to measure. Instead, it is possible to search for the weak collective interactions of the dark-matter field. For example, as a result of their coupling to electromagnetism, the effect of the axion or hidden-photon field may be modeled as an effective electromagnetic current density modifying Maxwell's equations. These current densities produce observable electromagnetic fields oscillating at frequency slightly greater than $\nu_{\rm DM}^{0}$, which couple to a receiver and may be read out with a sensitive amplifier or photon detector. 

In ADMX \cite{Sik83,Sik85,asztalos2010squid,du2018search}, HAYSTAC \cite{brubaker2017first,zhong2018results}, and Dark Matter Radio (DMRadio) \cite{Chaudhuri:2014dla,Silva-Feaver:2016qhh,phipps2020exclusion}, the receiver takes the form of a tunable high-Q resonator. If dark matter exists at a frequency near the resonance frequency, the electromagnetic fields produced by the dark matter are resonantly enhanced in the receiver. 
By tuning the resonator across a wide frequency range, one may obtain strong limits on light-field dark matter. In this manner, a resonant search for axion- or hidden-photon dark matter operates much like an AM radio, tuning into a radio station at a particular frequency. Using these sensitive methods, ADMX has recently established the first constraints on the benchmark DFSZ QCD axion model\cite{du2018search}.

On the other hand, a number of broadband search strategies have also been proposed, including ABRACADABRA\cite{kahn2016broadband,ouellet2019first,salemi2021search} and antenna-based searches such as BRASS\cite{horns2013searching}. They utilize information at the receiver output over the entire range of search frequencies, simultaneously probing many octaves instead of a narrow band of frequencies. Such searches do not require tuning, but also do not benefit from resonant enhancement of signal power. The advent of these searches begs a number of questions regarding the characteristics of the optimal, single-moded receiver. All of these questions must be answered to determine the properties of the fundamentally optimal, single-moded search and to determine whether that search is resonant, broadband, or some other type of receiver. 

First, as alluded to at the beginning of this paper, one must quantify the electromagnetic power flow from light-field dark matter into an arbitrary receiver. For the vast majority of electromagnetic sources, one can design a broadband impedance match to absorb an order-one fraction of the source power. For instance, to detect free-space electromagnetic waves governed by (unmodified) Maxwell's equations, phased-array antennas are routinely constructed to absorb 50\% of the incident-wave power, independent of the wave frequency.\cite{stahl2005cryogenic,hadley1947reflection} For a meter-scale dark-matter receiver, an analogous impedance match to the free-space electromagnetic fields sourced by axion or hidden-photon dark matter could absorb $\sim$5 Watts of power, independent of rest-mass frequency! This is far more power than that expected in widely-used cavity searches\cite{du2018search}, for which the signal power from QCD axions reaches $\sim 10^{-22}$ Watts and only if the axion signal is near resonance. As such, a receiver with a broadband match would yield far more efficient searches than present resonant searches. To determine the optimal search, it is thus important to investigate limitations on impedance-matching to dark matter and the set of possible frequency-response functions of a receiver to a dark-matter excitation. Because a receiver can be periodically varied in time (examples being the tuning of a cavity's resonance frequency or the spacing between dielectrics in a dielectric haloscope\cite{caldwell2017dielectric}), we must allow for the frequency-response function to be periodically varied as well.

Of course, one must remember that the question of detection is not simply one of signal power, but one of signal-to-noise ratio, so a determination of the optimal search must account for irreducible noise sources in a receiver, such as thermal noise and (if a phase-insensitive amplifier is used, which amplifies both signal quadratures equally) quantum noise, including quantum backaction. Additionally, the rest-mass frequency of the dark-matter signal is a priori unknown, so any optimized scan over the model parameter space must also incorporate priors, defined by the experimentalist, and must consider how such priors affect receiver architecture. For instance, the optimal receiver for a wideband, many-octave search governed by uninformative priors may differ from the optimal receiver for a search rescanning candidate signals from previous probes. In fact, as we will show, the optimal receiver in the two situations is indeed different, and the optimization of receiver frequency-response in a wideband search at low frequencies is strongly dependent on the physical temperature (which governs the thermal noise level) as well as the amplifier noise level. Consequently, appropriately tailoring the search to the irreducible noise sources and priors can dramatically reduce search times and make a search more efficient.

All of these factors---the impedance-match to dark matter, the set of possible frequency response functions, periodically-varied receiver parameters, irreducible noise sources, and priors---must be considered simultaneously to determine the properties of the optimal single-moded search. (As we will discuss, no factor may be optimized without the others!)

\emph{The principal purpose of this work is to determine the properties of the fundamentally optimal single-moded receiver with linear, passive matching to search for the electromagnetic coupling of axion- and hidden-photon dark matter, subject to the Standard Quantum Limit (SQL) on phase-insensitive amplification. The receiver may have an arbitrarily complex structure, which can be periodically varied in an arbitrary way. We thus establish a quantum limit on detection of axion and hidden-photon dark matter, including the effects of priors and time-variable parameters in the architecture. While our work provides a baseline search strategy for broad classes of axion and hidden-photon dark matter receivers, it is the basis for DMRadio, a new DOE-funded program in axion and hidden-photon dark matter detection.}

\subsection{Optimization Strategy}
\label{ssec:optimization_strategy}

The analysis for determining the properties of the optimal single-moded, linear, passive quantum-limited receiver utilizes the Maxwellian approach to electromagnetism, which primarily consists of direct manipulations of the partial differential equations governing axion and hidden-photon electrodynamics, as well as the equivalent-circuit approach to electromagnetism, which maps equivalent circuits onto these manipulations in order to model the flow of signal and noise fields. We further require fundamental statements regarding impedance matching, quantum fluctuations, and noise in electromagnetic systems. To perform the optimization, we must then lay a broad foundation for combining insights derived from these various frameworks.

\begin{figure}[htp] 
\includegraphics[width=\textwidth]{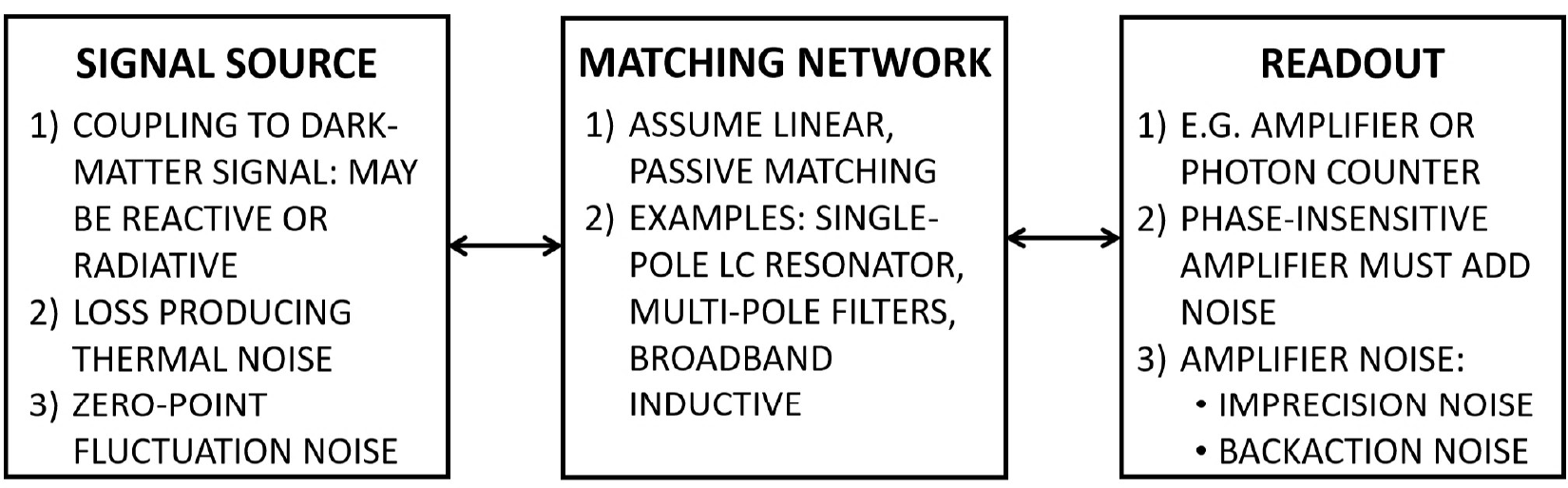}
\caption{Diagram showing the elements of a dark-matter receiver: the signal source (including loss in the coupling element), the matching network, and the readout element, which may be, for example, an amplifier or photon counter. The double arrows signify that signals can travel in both directions through the receiver. For instance, the dark-matter signal propagates to the readout, but the backaction noise of the amplifier propagates to the dark-matter coupling element in the signal source. \label{fig:DetectorBlockDiagram}}
\end{figure}

The development of such a foundation is the subject of this section, which defines the interfaces for optimization. We introduce here the basic structure of a dark-matter receiver, irreducible noise sources in such receivers, and the role of impedance matching and amplifier noise-matching. The elements of a dark-matter receiver are shown as a schematic block diagram in Fig. \ref{fig:DetectorBlockDiagram}. This paper is organized around optimizing each of these blocks, and globally optimizing the blocks and their interactions across a full scan. See the accompanying letter \cite{chaudhuri2019optimal} for a summary of the main results. The reader may prefer to skim this section and refer back to it while working through the optimization.

As investigated in Sections \ref{sec:power_flow}-\ref{sec:DM_circuits}, considerations of the impedance match to axions and hidden photons are critical in optimizing the receiver element that couples to the dark-matter field, represented as item (1) in the ``Signal Source'' block of Fig. \ref{fig:DetectorBlockDiagram}. In Section \ref{ssec:DM_Maxwell}, from the underlying equations of axion and hidden-photon electrodynamics, we write the expressions governing impedance-matching and complex-power flow between the dark-matter field and a general electromagnetic receiver. The expressions demonstrate the two categories of receiver-coupling to the dark-matter electromagnetic signal: reactive coupling and radiative coupling. Reactive coupling describes power coupled from the dark-matter field into energy-storing elements, e.g. wire-wound inductors and parallel-plate capacitors, free-space cavities, and dielectric resonators. Radiative coupling describes power coupled from the dark-matter field into radiating elements, as governed by the elements' free-space-radiation receiving pattern. Radiative coupling can be used to describe, for example, power flow into free-space antennas of broadband real impedance. Reactive and radiative coupling represent the two classes of methods for impedance-matching to dark matter, which must be compared to determine which is advantageous.

Because broadband radiative coupling allows for an efficient impedance match to free-space electromagnetic plane waves (see above), it is natural to expect that it could be equally efficient for the absorption of power from free-space electromagnetic fields induced by axion and hidden-photon dark matter, absorbing $\sim$5 Watts of power independent of rest-mass frequency. However, we will see that, for impedance-matching to dark matter and, consequently, scan sensitivity, radiative couplings possess a general disadvantage, relative to reactive couplings. In Section \ref{ssec:ResvBroad}, we conduct a toy, apples-to-apples comparison of a dark-matter receiver using broadband radiative coupling to a receiver using a narrowband, tunable cavity (reactive) coupling. We find that, when both receivers utilize readout with a phase-insensitive amplifier, the latter is superior in integrated scan sensitivity for a wideband search.


Through a calculation of the source impedance of axion and hidden-photon dark matter, Section \ref{sec:DM_circuits} generalizes the findings of the toy model. We begin by using the arguments of Schwinger to motivate the use of the equivalent-circuit approach to electromagnetism, as opposed to the Maxwellian approach, to calculate the dark-matter source impedance and to carry out complex calculations on arbitrary receivers, including receivers that are not physically circuits. In Section \ref{ssec:toy_circuits}, by following the principles of Dicke\cite{montgomery1987principles} and parametrizing the complex-power flow equations in Section \ref{ssec:DM_Maxwell}, we develop equivalent circuits of the toy receivers analyzed in Section \ref{ssec:ResvBroad}. We calculate that, for a virialized DFSZ axion\cite{dine1981simple,zhitnitskij1980possible} in a 10 Tesla magnetic field, the effective source impedance is $\sim 10^{-28} Z_{fs}$, where $Z_{fs}=\sqrt{\mu_{0}/\epsilon_{0}} \approx 377\ \Omega$ is the free-space impedance. The effective source impedance is the same for a virialized hidden photon with kinetic mixing angle $\sim 3 \times 10^{-16}$. Our equivalent-circuit calculations illustrate a primary limitation in extracting power from the dark-matter field, which is that the self-impedance of photons produced by electromagnetic charges in the coupling element of the receiver is much larger than the effective dark-matter source impedance. \emph{As a result, it is impractical to obtain an efficient impedance match to dark matter.} In particular, using a modal analysis\cite{zmuidzinas2003thermal} to describe free-space receiver radiation, we demonstrate that the sensitivity of a linear, passive receiver using only radiative coupling to the dark-matter electromagnetic signal and using readout with a phase-insensitive amplifier is generally limited by the mismatch between the free-space impedance $Z_{fs}$ and the dark-matter source impedance. Reactive coupling then generally outperforms radiative coupling. 


Thereafter, we focus on single-moded reactive couplings (couplings that can be represented as occurring through a single inductor or capacitor in the receiver's equivalent-circuit representation) for the global receiver optimization across all three blocks of Fig. \ref{fig:DetectorBlockDiagram}; single-moded reactive couplings are used in a large majority of the experiments proposed, currently under construction, or running. We argue that, without loss of generality, we may focus on single-moded inductive, rather than capacitive, couplings. Receivers that may be modeled as single-moded inductive couplings include, for example, a single mode of a free-space cavity, such as in ADMX and HAYSTAC (see Appendix \ref{sec:cav_circuit}), or a physically-lumped-element pickup inductor, such as in ABRACADABRA or DMRadio. In Section \ref{ssec:reactive_circuit}, we quantify the dark-matter excitation of a reactive coupling element as an equivalent-circuit voltage. We briefly discuss practical aspects of maximizing the voltage, which is dependent on receiver volume, the level of coupling-element alignment with dark-matter drive fields, and (for axions) background electromagnetic field; a more detailed treatment is left to Section \ref{ssec:FOM}, which describes practical tradeoffs for a receiver. 

We then lay the foundations for the final optimization analysis, discussing additional factors that greatly impact sensitivity. We explain the manner in which a receiver optimization must consider not only dark-matter signal, but also all noise sources. There are two fundamental noise sources associated with the ''Signal Source.'' The reactive coupling element possesses some loss, i.e. equivalent-circuit resistance, which produces thermal noise, item (2) in ''Signal Source.'' If the circuit is cold and $h\nu_{\rm DM}^{0} \gg kT$, where $k$ is Boltzmann's constant and $T$ is the physical temperature, one observes the effects of the zero-point fluctuations in the receiver (item (3)). For the purposes of the global optimization that follows Section \ref{sec:DM_circuits}, we hold fixed the signal-source properties (equivalent-circuit inductance and resistance, volume, the level of coupling-element alignment with dark-matter drive fields, temperature, and for axions, background electromagnetic field strength). Additionally, we explain why a complete optimization of a single-moded, reactively coupled receiver must consider not only the signal source element, but also the elements on its output. In theory, the follow-on elements (e.g. the elements at the output of a cavity or following a lumped-element pickup inductor) may be used to dramatically enhance the impedance match to dark matter, yielding much better performance than resonant searches. In this context, we explain the practical importance of linear, passive impedance-matching networks. 

The impedance-matching network, represented by the ``Matching Network'' box in Fig. \ref{fig:DetectorBlockDiagram}, is the second element of every receiver. The impedance-matching network transfers the excitation between the signal source and the readout and sets the frequency-response function of the reactively coupled receiver. It may be used to improve both the impedance match to dark matter, as well as the impedance match to the readout. As the set of matching networks/frequency-response functions is infinitely broad, much of the challenge in determining the optimal single-moded receiver lies in constraining this infinitely-broad set. Here, we give a few examples from present searches to guide the reader. A single-pole\footnote{We consider poles of the Laplace transform of the equivalent-circuit response. We restrict the domain to the upper-left quadrant of the complex plane, in which the imaginary part of the pole is nonnegative. Under this convention, an RLC circuit has a single pole. The real part of the pole must be negative, as required for stability.\label{poleconvention}} equivalent-RLC resonator (e.g. DM Radio\cite{Chaudhuri:2014dla}) is an example of a matching network. It uses an equivalent capacitance or network of equivalent capacitors to transform an equivalent inductance to a real impedance on resonance, as seen by the readout. In cavity detectors (e.g. ADMX and HAYSTAC), each mode can be modeled as an RLC circuit in which the the dark-matter excitation is coupled to the equivalent inductance and the equivalent capacitance serves as the matching network.\cite{pozar2012microwave} Another matching network is a multi-pole filter, which, for instance, could have many LC poles at the same frequency. One may also use broadband inductive coupling (e.g. ABRACADABRA\cite{kahn2016broadband}), where a wire-wound pickup coil is wired directly to the input of a SQUID. We restrict our attention to linear, passive matching networks and do not consider the use of active elements or active feedback.\cite{rybka2014improving}

The third and final element is an amplifier or photon counter to read out the signal passed through the impedance matching network (the ``Readout'' element in Fig. \ref{fig:DetectorBlockDiagram}). Photon-counting and quantum-squeezing techniques using phase-sensitive amplifiers are being developed for light field dark-matter detection \cite{lehnert2016quantumaxion}. Here, we focus on readout with phase-insensitive amplifiers, which is used in the majority of presently operating searches and is baselined for numerous planned searches. In a readout with a phase-insensitive amplifier, both quadratures of the incoming signal are coherently amplified with the same gain and analyzed for a dark-matter signal.

In addition to the thermal noise in the signal source, every receiver with a phase-insensitive amplifier readout is also limited by quantum noise and possibly by excess noise in amplifier and data acquisition chains. All of these readout noise sources can be broken down into two components: imprecision noise and backaction noise. The imprecision noise effectively adds some uncertainty to the output of the amplifier, independent of the input. The backaction injects noise into the input (formed by the Signal Source and Matching Network). This noise, having been filtered according to the impedance of the receiver, is added to the amplifier input and appears as additional noise on the output. In contrast to imprecision noise, backaction noise, referred to the amplifier input, is then inherently dependent on the matching network. When the noise temperature is minimized with respect to the impedance of the input circuit, e.g. by varying the matching network, it is said that the input circuit is \emph{noise matched} to the amplifier. The impedance at which the noise temperature is minimized is known as the noise impedance. \cite{clerk2010introduction, wedge1991computer} We will demonstrate that, in a resonant search, optimization of the noise matching can increase scan rate by orders of magnitude at low search frequencies, at which the thermal occupation number of the receiver is much greater than unity.


In Section \ref{ssec:amps}, we discuss the two categories of amplifier measurements in the context of axion and hidden-photon dark-matter searches. The first category is scattering-mode amplifier measurements, in which forward-scattering power is measured. Such measurements can be performed, for example, using a Josephson parametric amplifier.\cite{brubaker2017first,castellanos2008amplification} The second category is op-amp mode amplifier measurements, in which input voltage or current is measured, e.g. as in a SQUID amplifier.\cite{clarke2006squid} The distinctions between scattering mode and op-amp mode necessitate two different formalisms for amplifier noise and the SQL.\cite{clerk2010introduction} In the main text, we focus on scattering-mode measurements, leaving the optimization of op-amp mode measurements for Appendices \ref{sec:FtoV_Amps} and \ref{sec:FtoV_opt}. Though the description is somewhat more complicated in the case of op-amp mode/flux-to-voltage measurements, the primary conclusions are identical to those in the scattering-mode case.


Because the amplifier noise temperature is dependent on the matching network, the second and third blocks in Fig. \ref{fig:DetectorBlockDiagram} must be optimized simultaneously. Sections \ref{sec:resonator_SNR}-\ref{sec:resonator_sensitivity} discuss these blocks and complete the global optimization of the receiver. We start this portion of the optimization by considering the signal-to-noise ratio (SNR) of the receiver.

Because of the need to analyze an arbitrary single-moded, reactively coupled receiver, our theoretical analysis of SNR must be far more general than previous work, which universally considers specific receiver implementations. This is the subject of Section \ref{sec:resonator_SNR}. Building on the standard Dicke radiometer formula, we discuss the signal processing steps for a dark-matter receiver. Based on this discussion, in Sections \ref{ssec:resonator_scattering}-\ref{ssec:reactive_scattering}, we develop a scattering-matrix representation of the receiver, which provides a framework for evaluating signal and noise power transfer between the inductive signal source and the amplifier through the impedance matching network. The SNR is then calculated in Section \ref{ssec:SNR_search}, including the effects of optimal filtering (Section \ref{sssec:SNR_one}) and periodically-varied frequency response (Section \ref{sssec:SNR_scan}). Particular attention is paid to SNR for a quantum-limited scattering-mode amplifier (Section \ref{sssec:SNR_QLA}), which represents a fundamental noise floor and is thus the basis for further optimization.


As we will demonstrate in the course of this paper, tailoring a search to priors can dramatically reduce scan times. Careful consideration of priors is required to rigorously define a value function for the matching network and to make accurate conclusions in the comparison of single-pole resonant and reactive broadband searches. (See Appendix \ref{sec:ResvBroad_SQUID}.) Thus, a receiver optimization must also take into account prior probabilities on the dark matter signal. Priors may, for example, take the form of astrophysical and direct detection constraints or well-motivated regimes of parameter space, such as the parameter space corresponding to QCD axion models. Building on the SNR analysis in Section \ref{sec:resonator_SNR}, we present in Section \ref{sec:scan_opt} a priors-driven optimization of the scan. 

The first part of the priors-driven optimization leads to the conclusion that a single-pole resonator is a near-optimal single-moded, linear, passive method for detecting the axion or hidden-photon dark-matter field. For a fixed scan step, we optimize the impedance-matching network (Section \ref{ssec:opt_impedance}). Using the results from Section \ref{sssec:SNR_QLA}, we determine a priors-based value function for evaluating the merits of a matching network. We optimize a ``log-uniform'' search, to be defined in Section \ref{sssec:neutral_search_opt}, which makes the natural assumption of a logarithmically uniform probability for the mass and electromagnetic coupling strength of dark matter within the search band. For this search, under which the dark-matter rest-mass frequency is \emph{a priori unknown}, the value function simplifies to a measure of frequency-integrated receiver sensitivity. Ideally, the value function would be maximized by noise matching the signal source to the quantum-limited amplifier at all frequencies. However, we show that such a match is not possible for a passive impedance-matching network. In Section \ref{sssec:Bode_Fano_opt}, we establish an upper bound on the integrated sensitivity using the Bode-Fano criterion, which constrains the match between the equivalent-LR (inductance-resistance) signal source, of complex-valued impedance, and the quantum-limited amplifier, of real-valued noise impedance. Note, importantly, that while the Bode-Fano criterion is typically used to constrain frequency-integrated signal transfer, we use it to constrain frequency-integrated signal-to-noise. We show that the Bode-Fano bound is approached by a multipole LC Chebyshev matching network. Such networks are narrowband and difficult to tune, so we consider the integrated sensitivity of the much simpler (and easier to implement) single-pole resonator. 

In Sections \ref{sssec:res_match_opt} and \ref{sssec:matching_opt_discussion}, we show that the single-pole resonator, when its coupling to the quantum-limited amplifier is optimized---in particular, when the resonator is optimally noise-mismatched---, possesses an integrated sensitivity that is approximately 75\% of the fundamental Bode-Fano limit. \emph{The Bode-Fano limit demonstrates that the single-pole equivalent-RLC resonator is a near-ideal single-moded, linear, passive technique to detect the axion or hidden-photon dark matter field.}  One notable consequence is that, subject to the SQL, single-pole resonant searches are superior to reactive broadband searches, such as that used in ABRACADABRA\cite{kahn2016broadband}, at all frequencies at which a resonator may practically be constructed. A detailed discussion is included in Appendix \ref{sec:ResvBroad_SQUID}. An analogous Bode-Fano bound can be derived for receivers capacitively coupled to the dark-matter-induced electromagnetic fields, with the same conclusion that a single-pole RLC resonator is 75\% of the limit. The optimization of integrated sensitivity in a single-pole resonator is related to the sensitivity available outside of the resonator bandwidth and the concepts of noise matching and measurement backaction. None of these have been analyzed in detail in previous work. In particular, some previous work only accounts for the information available within the resonator bandwidth and therefore dramatically underestimates the sensitivity of resonant searches at low frequencies.\cite{Chaudhuri:2014dla, kahn2016broadband, Sik14}


Because the single-pole resonator is close to the Bode-Fano limit, thereafter we further develop the sensitivity analysis of single-pole resonators in the remaining text. In the second part of the priors-driven optimization, given a fixed total search time, we determine the optimal allocation of time across resonant scan steps for the log-uniform search. This part is presented in Section \ref{ssec:opt_time} and builds upon the results of Sections \ref{sssec:SNR_scan} and \ref{sssec:SNR_QLA}. We introduce the notion of a dense scan, where each search frequency is probed by multiple resonance frequencies. We also briefly consider other possible value functions for time allocation based on different prior assumptions about the probability distribution of dark matter.


Combining the results in Sections \ref{sec:power_flow}-\ref{sec:scan_opt} yields a fundamental limit on the performance of axion and hidden-photon dark-matter searches read out by a phase-insensitive amplifier subject to the SQL. We calculate this limit in Section \ref{ssec:QLA_sensitivity}. Owing to the identical results provided in the scan optimization, the limit is the same for scattering-mode and op-amp mode/flux-to-voltage readouts. We discuss the parametric dependence of the fundamental limit on quality factor and contrast it with previous works.  \cite{Chaudhuri:2014dla,Sik83,Sik85} In particular, we find that the sensitivity of a resonant scan increases as the Q is increased above $10^{6}$, the characteristic quality factor of the dark matter signal defined by its bandwidth. That is, the loss in the receiver should be made as low as possible. 
We show that use of the optimized scan strategy can increase the sensitivity to dark-matter coupling strength by as much as 1.25 orders of magnitude at low frequencies. \emph{This corresponds to an increase in scan rate of five orders of magnitude.} In Section \ref{ssec:FOM}, we develop a figure-of-merit for searches and discuss our fundamental limit in the context of practical tradeoffs that may be made in the course of an experiment.

We conclude in Section \ref{sec:conclude}, where we provide directions for further investigation. By establishing a clear Standard Quantum Limit for integrated scan sensitivity (rather than simply integrated power transfer), the results in this paper provide strong motivation for the use of quantum measurement techniques that can evade the Standard Quantum Limit of the dark-matter measurement. We briefly discuss these techniques and prospects for implementing backaction evasion, squeezing, entanglement, photon counting, and other nonclassical approaches in axion- and hidden-photon dark matter searches.
\section{Electromagnetic Power Flow From The Dark-Matter Field}
\label{sec:power_flow}


\subsection{Complex Power Flow Statements for Electromagnetic Axion and Hidden-Photon Dark-Matter Receivers}
\label{ssec:DM_Maxwell}

To quantify impedance matching and complex-power flow from dark matter into the receiver, we work from the underlying modified-Maxwell equations of axion and hidden-photon electrodynamics. The electromagnetic coupling of axion and hidden-photon dark matter can be modeled as effective current densities. For the axion, the effective current density is\cite{Sik83,Sik85,Sik14}
\begin{equation}\label{eq:J_axion_0}
\vec{J}_{a}^{\rm eff}(\vec{x},t)= -\frac{\kappa_{a}}{\mu_{0} c} \left( \vec{B}_{b}(\vec{x},t) \partial_{t}a(\vec{x},t) - \vec{E}_{b}(\vec{x},t) \times \vec{\nabla}a(\vec{x},t) \right),
\end{equation}
where $a(x,t)$ is the pseudoscalar axion potential and $\vec{E}_{b}, \vec{B}_{b}$ are the background electric and magnetic fields required for axion-to-photon conversion. The coupling $\kappa_{a}$ is related to the axion-photon coupling $g_{a\gamma\gamma}$ by
\begin{equation}\label{eq:kappa_def}
    \kappa_{a}=g_{a\gamma\gamma}\sqrt{\hbar c \epsilon_{0}}
\end{equation}
For virialized axions and for background fields of equal energy density, a background electric field gives rise to an effective current density (and thus, an oscillating electromagnetic field) that is smaller than that from a background magnetic field by a factor of the dark matter velocity $v/c \sim 10^{-3}$; in other words, the effect of a background electric field is relatively small.\cite{caldwell2017dielectric} In practice, DC magnetic fields produced in the lab can be orders of magnitude larger than their AC counterparts. As such, throughout the paper, we assume zero background electric field and a background DC magnetic field. Nevertheless, we stress that the primary results of our paper--namely, that reactive couplings outperform radiative couplings and that single-pole resonators, when optimally noise-mismatched, are the near-ideal single-moded linear, passive receiver--are the same in the DC and AC background field cases. We address this idea further throughout the paper and in discussing practical tradeoffs in Section \ref{ssec:FOM}. Limiting ourselves to the case of a background DC magnetic field also allows us to treat the optimization of axion and hidden-photon dark-matter searches simultaneously. Under the assumption of zero background electric field and a background DC magnetic field,
\begin{equation}
\vec{E}_{b}=0,\ \vec{B}_{b}(\vec{x},t)=\vec{B}_{b}(\vec{x}),
\end{equation}
the effective current density then becomes, from eq. (\ref{eq:J_axion_0}),
\begin{equation}\label{eq:J_axion}
\vec{J}_{a}^{\rm eff}(\vec{x},t)= -\frac{\kappa_{a}}{\mu_{0} c} \vec{B}_{b}(\vec{x}) \partial_{t}a(\vec{x},t).
\end{equation}

For the hidden photon, the effective current density is
\begin{equation}\label{eq:J_HP}
\vec{J}_{hp}^{\rm eff}(\vec{x},t)= -\varepsilon \epsilon_{0} \left( \frac{m_{\rm DM}c^{2}}{\hbar} \right)^{2} \vec{A}'(\vec{x},t),
\end{equation} 
where $\vec{A}'(\vec{x},t)$ is the hidden-photon three-vector potential in the interaction basis (omitting the scalar component of the four-potential).\cite{Chaudhuri:2014dla} One may note that, for both the axion and hidden photon, the effective current density is accompanied by an effective charge density\cite{millar2017dielectric,Chaudhuri:2014dla}, given by the continuity equation
\begin{equation}\label{eq:rho_eff}
    \partial_{t}\rho_{\rm DM}^{\rm eff}(\vec{x},t)=-\vec{\nabla} \cdot \vec{J}_{\rm DM}^{\rm eff}(\vec{x},t)
\end{equation}
We have changed the subscript in the effective current, from ``axion" ($a$) or ``hidden photon" ($hp$) to ``DM'', for generality. For virialized dark matter, the magnitude of the charge density is suppressed relative to that of the current density by a factor of the velocity $v/c \sim 10^{-3}$. Thus, we will primarily ignore the effects of the former.

The effective current/charge density produces electromagnetic fields governed by the Maxwell equations
\begin{align}
    \vec{\nabla} \cdot \vec{E}_{\rm DM} = \rho_{\rm DM}^{\rm eff}/\epsilon_{0},\ & \vec{\nabla} \times \vec{E}_{\rm DM} = -\partial_{t} \vec{B}_{\rm DM},\ \nonumber \\
    \vec{\nabla} \cdot \vec{B}_{\rm DM} = 0,\ & \vec{\nabla} \times \vec{B}_{\rm DM} = \mu_{0}\vec{J}_{\rm DM}^{\rm eff} + \mu_{0}\epsilon_{0}\partial_{t} \vec{E}_{\rm DM} \label{eq:DM_fields}
\end{align}
These fields excite electromagnetic charges and currents in the receiver, which we denote $\rho_{\rm rec}(\vec{x},t)$ and $\vec{J}_{\rm rec}(\vec{x},t)$. These charges and currents produce their own electromagnetic fields, governed by
\begin{align}
    \vec{\nabla} \cdot \vec{E}_{\rm rec} = \rho_{\rm rec},\ & \vec{\nabla} \times \vec{E}_{\rm rec} = -\partial_{t} \vec{B}_{\rm rec},\ \nonumber \\
    \vec{\nabla} \cdot \vec{B}_{\rm rec} = 0,\ & \vec{\nabla} \times \vec{B}_{\rm rec} = \mu_{0}\vec{J}_{\rm rec} + \mu_{0}\epsilon_{0}\partial_{t} \vec{E}_{\rm rec} \label{eq:rec_fields}
\end{align}
The total electric and magnetic fields are the sum of those in eqs. (\ref{eq:DM_fields}) and (\ref{eq:rec_fields}), i.e.
\begin{equation}
    \vec{E}_{\rm tot}(\vec{x},t)=\vec{E}_{\rm DM}(\vec{x},t) + \vec{E}_{\rm rec}(\vec{x},t)
\end{equation}
and similarly for the magnetic field.

Assume that the receiver response to the dark-matter excitation is in steady state. Upon Fourier-transforming, the effective dark-matter current density can be represented as
\begin{equation}\label{eq:JDM_FFT}
    \vec{J}_{\rm DM}^{\rm eff}(\vec{x},t)= Re \left( \int \frac{d\omega}{2\pi}\ \vec{J}_{\rm DM}^{\rm eff}(\vec{x},\omega,\omega_{\rm DM}^{0}) \exp(+i\omega t) \right),
\end{equation}
where $\omega=2\pi\nu$ (in particular, $\omega_{\rm DM}^{0}=2\pi\nu_{\rm DM}^{0}$) and similarly for electric and magnetic fields. 
The limits of integration are determined by the bandwidth of the dark-matter signal. Virialization and Earth's motion in the galactic rest frame (the frame in which the bulk motion of dark matter is zero) give dark matter a velocity of $v/c \sim 10^{-3}$ in the receiver rest frame. The $10^{-3}$ velocity yields a $10^{-6}$ dispersion in kinetic energy and therefore, a signal bandwidth
\begin{equation} \label{eq:delta_nuDM}
\Delta \omega_{\rm DM}(\omega_{\rm DM}^{0}) \sim 10^{-6} \omega_{\rm DM}^{0}.
\end{equation}
The dark-matter current density and sourced electromagnetic fields thus span the frequency range $\omega_{\rm DM}^{0} \leq \omega \lesssim \omega_{\rm DM}^{0} + \Delta \omega_{\rm DM}(\omega_{\rm DM}^{0})$. 

We fix the dark-matter search frequency $\omega_{\rm DM}^{0}$ and consider the rate of complex work performed by a single frequency-component $\omega$ of the dark-matter electromagnetic signal on the electromagnetic charges in the receiver. Let $V$ be a volume surrounding the charges (which are assumed to occupy a finite volume). Suppose that the volume is sufficiently large that the receiver fields on its surface represent far-field electromagnetic radiation.\cite{Jackson:1998nia} The rate of complex work is
\begin{equation}\label{eq:CPF_DM}
    \tilde{P}_{in}(\omega,\omega_{\rm DM}^{0})=\frac{1}{2} \int_{V} d^{3}\vec{x}\ \vec{J}_{\rm rec}^{*}(\vec{x},\omega,\omega_{\rm DM}^{0}) \cdot \vec{E}_{\rm DM}(\vec{x},\omega,\omega_{\rm DM}^{0})
\end{equation}

By manipulating this equation, we write expressions governing both how the receiver responds to a dark-matter excitation, as well as the ways in which the receiver may couple power from the dark matter. Since the total electric field is the sum of the dark-matter-induced electric field and receiver electric field, we may write
\begin{align}
    \tilde{P}_{in} &(\omega,\omega_{\rm DM}^{0})= \frac{1}{2} \int_{V} d^{3}\vec{x}\  \vec{J}_{\rm rec}^{*}(\vec{x},\omega,\omega_{\rm DM}^{0}) \cdot \vec{E}_{\rm tot}(\vec{x},\omega,\omega_{\rm DM}^{0}) - \frac{1}{2} \int_{V} d^{3}\vec{x}\ \vec{J}_{\rm rec}^{*}(\vec{x},\omega,\omega_{\rm DM}^{0}) \cdot \vec{E}_{\rm rec}(\vec{x},\omega,\omega_{\rm DM}^{0}) \nonumber \\
    &= \frac{1}{2} \int_{V} d^{3}\vec{x}\  \vec{J}_{\rm rec}^{*}(\vec{x},\omega,\omega_{\rm DM}^{0}) \cdot \vec{E}_{\rm tot}(\vec{x},\omega,\omega_{\rm DM}^{0}) + \frac{1}{2\mu_{0}} \int_{\partial V} (\vec{E}_{\rm rec}(\vec{x},\omega,\omega_{\rm DM}^{0}) \times \vec{B}_{\rm rec}^{*}(\vec{x},\omega,\omega_{\rm DM}^{0}))\cdot \vec{da} \nonumber \\
    &+\frac{i\omega}{2\mu_{0}} \int_{V} d^{3}\vec{x}\ |\vec{B}_{\rm rec}(\vec{x},\omega,\omega_{\rm DM}^{0})|^{2} -\frac{i\omega \epsilon_{0}}{2} \int_{V} d^{3}\vec{x}\ |\vec{E}_{\rm rec}(\vec{x},\omega,\omega_{\rm DM}^{0})|^{2} \label{eq:workout}
\end{align}
where, in the second equality, we have applied complex Poynting's theorem to the receiver fields. 

Equation (\ref{eq:workout}) describes how the receiver responds to an excitation from dark matter. The first term on the right-hand side of eq. (\ref{eq:workout}) represents the complex-power flow into the electromagnetic charges. Specifically, we may write the total current $J_{\rm rec}(\vec{x},\omega,\omega_{\rm DM}^{0})$ as a sum of free and bound currents\cite{Jackson:1998nia}:
\begin{equation}\label{eq:free_bound_currents}
    \vec{J}_{\rm rec}(\vec{x},\omega,\omega_{\rm DM}^{0}) = \vec{J}_{\rm f}(\vec{x},\omega,\omega_{\rm DM}^{0}) + i\omega \vec{\mathcal{P}}(\vec{x},\omega,\omega_{\rm DM}^{0}) + \vec{\nabla} \times \vec{\mathcal{M}}(\vec{x},\omega,\omega_{\rm DM}^{0})
\end{equation}
where $\vec{J}_{f}$ is the free current, $\vec{\mathcal{P}}$ is the polarization vector, generated by any electric dipoles in the receiver (e.g. as in a dielectric), and $\vec{\mathcal{M}}$ is the magnetization vector, generated by any magnetic dipoles in the receiver (e.g. as in a high-permeability material). The complex power-flow into electromagnetic charges is then
\begin{align}
    \tilde{P}_{elec}&(\omega,\omega_{\rm DM}^{0})= \frac{1}{2} \int_{V} d^{3}\vec{x}\  \vec{J}_{\rm rec}^{*}(\vec{x},\omega,\omega_{\rm DM}^{0}) \cdot \vec{E}_{\rm tot}(\vec{x},\omega,\omega_{\rm DM}^{0}) \nonumber \\
    &= \frac{1}{2} \int_{V} d^{3}\vec{x}\  \vec{J}_{\rm f}^{*}(\vec{x},\omega,\omega_{\rm DM}^{0}) \cdot \vec{E}_{\rm tot}(\vec{x},\omega,\omega_{\rm DM}^{0}) -\frac{i\omega}{2} \int_{V} d^{3}\vec{x}\  \vec{\mathcal{P}}^{*}(\vec{x},\omega,\omega_{\rm DM}^{0}) \cdot \vec{E}_{\rm tot}(\vec{x},\omega,\omega_{\rm DM}^{0}) \nonumber \\
    &- \frac{i\omega}{2}\int_{V} d^{3}\vec{x}\  \vec{\mathcal{M}}^{*}(\vec{x},\omega,\omega_{\rm DM}^{0}) \cdot \vec{B}_{\rm tot}(\vec{x},\omega,\omega_{\rm DM}^{0})  \label{eq:Pelec}
\end{align}
The first term represents complex-power flow into free electromagnetic charges/currents, while the second and third terms represent, respectively, complex-power flow into bound electric and magnetic dipoles. These complex-power flow expressions may have both real and imaginary parts, representing dissipation and energy storage in charges/currents, as revealed by relating the current and total field by linear constitutive equations. For example, at a given point in space, the free current may be related to the total electric field by a complex-valued conductivity tensor. The real part of the complex-power flow into free charges/currents represents ohmic dissipation. Similarly, the polarization may be related to the total electric field by a complex-valued susceptibility tensor. The real part of the complex-power flow into the electric dipoles represents loss, e.g. dielectric dissipation, whereas the imaginary part represents a change in polarization energy. One may write similar statements for the magnetization. While we focus our attention on linear, passive matching networks, we note that eq. (\ref{eq:Pelec}) also accounts for complex-power flow into active-matching elements, e.g. negative inductors\cite{clarke2006squid} and capacitors\cite{thorne1980gravitational} and cavities implementing negative dispersion\cite{wicht1997white,pati2007demonstration}.

The second term on the right-hand side of eq. (\ref{eq:workout}) represents complex-power flow through the surface $\partial V$ in the form of electromagnetic radiation emitted by the receiver. The third and fourth terms on the right-hand side of eq. (\ref{eq:workout}) represents changes in magnetic-field energy and electric-field energy produced by the receiver within the volume $V$.

The divergence theorem allows us to rewrite eq. (\ref{eq:CPF_DM}) as
\begin{align}
    \tilde{P}_{in}(\omega,\omega_{\rm DM}^{0}) &= -\frac{1}{2\mu_{0}} \int_{\partial V} (\vec{E}_{\rm DM}(\vec{x},\omega,\omega_{\rm DM}^{0}) \times \vec{B}_{\rm rec}^{*}(\vec{x},\omega,\omega_{\rm DM}^{0}))\cdot \vec{da} \nonumber \\
    &+ \frac{i\omega\epsilon_{0}}{2} \int_{V} d^{3}\vec{x}\ \vec{E}_{\rm DM}(\vec{x},\omega,\omega_{\rm DM}^{0}) \cdot \vec{E}_{\rm rec}^{*}(\vec{x},\omega,\omega_{\rm DM}^{0}) \nonumber \\
    &- \frac{i\omega}{2\mu_{0}} \int_{V} d^{3}\vec{x}\ \vec{B}_{\rm DM}(\vec{x},\omega,\omega_{\rm DM}^{0}) \cdot \vec{B}_{\rm rec}^{*}(\vec{x},\omega,\omega_{\rm DM}^{0}) \label{eq:workin}
\end{align}
The three terms on the right-hand side are indicative of the three ways that dark-matter power may couple into the receiver. The first term on the right-hand side is indicative of power coupled into the receiver through the free-space radiation modes of the receiver. We refer to this form of coupling as \emph{radiative coupling}. The second and third terms in (\ref{eq:workin}) are indicative of power coupled into electric and magnetic-energy-storing elements of the receiver. We refer to these forms of coupling, respectively, as \emph{capacitive} and \emph{inductive} couplings. We will refer to these latter two couplings collectively as \emph{reactive} couplings. Equating (\ref{eq:workout}) and (\ref{eq:workin}) and using (\ref{eq:Pelec}) yields
\begin{align}
    \tilde{P}_{in}(\omega,\omega_{\rm DM}^{0}) &=-\frac{1}{2\mu_{0}} \int_{\partial V} (\vec{E}_{\rm DM}(\vec{x},\omega,\omega_{\rm DM}^{0}) \times \vec{B}_{\rm rec}^{*}(\vec{x},\omega,\omega_{\rm DM}^{0}))\cdot \vec{da} \nonumber \\
    &+ \frac{i\omega\epsilon_{0}}{2} \int_{V} d^{3}\vec{x}\ \vec{E}_{\rm DM}(\vec{x},\omega,\omega_{\rm DM}^{0}) \cdot \vec{E}_{\rm rec}^{*}(\vec{x},\omega,\omega_{\rm DM}^{0}) \nonumber \\
    &- \frac{i\omega}{2\mu_{0}} \int_{V} d^{3}\vec{x}\ \vec{B}_{\rm DM}(\vec{x},\omega,\omega_{\rm DM}^{0}) \cdot \vec{B}_{\rm rec}^{*}(\vec{x},\omega,\omega_{\rm DM}^{0}) \nonumber \\
    &= \frac{1}{2} \int_{V} d^{3}\vec{x}\  \vec{J}_{\rm f}^{*}(\vec{x},\omega,\omega_{\rm DM}^{0}) \cdot \vec{E}_{\rm tot}(\vec{x},\omega,\omega_{\rm DM}^{0}) \nonumber \\
    &-\frac{i\omega}{2} \int_{V} d^{3}\vec{x}\  \vec{\mathcal{P}}^{*}(\vec{x},\omega,\omega_{\rm DM}^{0}) \cdot \vec{E}_{\rm tot}(\vec{x},\omega,\omega_{\rm DM}^{0}) \nonumber \\
    &- \frac{i\omega}{2}\int_{V} d^{3}\vec{x}\  \vec{\mathcal{M}}^{*}(\vec{x},\omega,\omega_{\rm DM}^{0}) \cdot \vec{B}_{\rm tot}(\vec{x},\omega,\omega_{\rm DM}^{0}) \nonumber \\
    & + \frac{1}{2\mu_{0}} \int_{\partial V} (\vec{E}_{\rm rec}(\vec{x},\omega,\omega_{\rm DM}^{0}) \times \vec{B}_{\rm rec}^{*}(\vec{x},\omega,\omega_{\rm DM}^{0}))\cdot \vec{da} \nonumber \\
    &+\frac{i\omega}{2\mu_{0}} \int_{V} d^{3}\vec{x}\ |\vec{B}_{\rm rec}(\vec{x},\omega,\omega_{\rm DM}^{0})|^{2} -\frac{i\omega \epsilon_{0}}{2} \int_{V} d^{3}\vec{x}\ |\vec{E}_{\rm rec}(\vec{x},\omega,\omega_{\rm DM}^{0})|^{2} \label{eq:DM_Poynting} 
\end{align}

Eq. (\ref{eq:DM_Poynting}) is an important result that guides our analysis of signal-source optimization over Sections \ref{sec:power_flow} and \ref{sec:DM_circuits}, in particular in formulating equivalent circuits for arbitrary receivers. By virtue of its derivation from Maxwell's equations, it describes power-flow dynamics in an arbitrary hidden-photon receiver or axion receiver embedded in a background DC magnetic field. A similar power-flow statement can be written for background AC fields, with a similar identification of radiative and reactive couplings to the dark-matter power.

Radiative and reactive couplings represent the two methods for coupling power from dark matter, and both have found use in experimental proposals for axion and hidden-photon dark-matter detection. For instance, broadband dish-antenna searches\cite{horns2013searching} and wire array experiments\cite{sikivie2021invisible} rely on radiative coupling to couple power from the dark-matter, while cavity haloscopes\cite{Sik85} and lumped-element inductive pickup searches\cite{Chaudhuri:2014dla,kahn2016broadband} rely on reactive coupling to couple power from the dark-matter. Radiative coupling seems particularly attractive because, as described in the introduction, it provides an efficient impedance match in other areas of electromagnetic measurement and can enable broadband performance. On the other hand, it is widely known that reactive, resonant couplings enhance the received dark-matter power, albeit in a narrow bandwidth around resonance. We now compare the sensitivity of radiative and reactive couplings using a toy model.


\subsection{A Toy Comparison of Broadband Radiative and Narrowband Resonant Searches}
\label{ssec:ResvBroad}

We perform an apples-to-apples comparison of the sensitivity of two toy receivers in quasi-one dimension (1D) searching for dark matter over a wide frequency range: a broadband resistive antenna and a tunable cavity resonator. We first describe the receiver setups in our quasi-1D toy model and then calculate the size of the signal power. We use this calculation to show explicitly that the excitation of the toy antenna and the toy cavity correspond to the integral terms indicative of radiative and reactive coupling, respectively, as introduced in the previous section. This is followed by a brief SNR analysis, assuming both receivers are read out by phase-insensitive amplifiers. We compare the SNRs at each dark-matter rest-mass frequency in our search band and show that the SNR for the cavity is always larger, scaling with the square root of quality factor. We contrast our calculation with previous results in the literature for axion and hidden-photon searches. 

\begin{figure}[htp] 
\includegraphics[width=\textwidth]{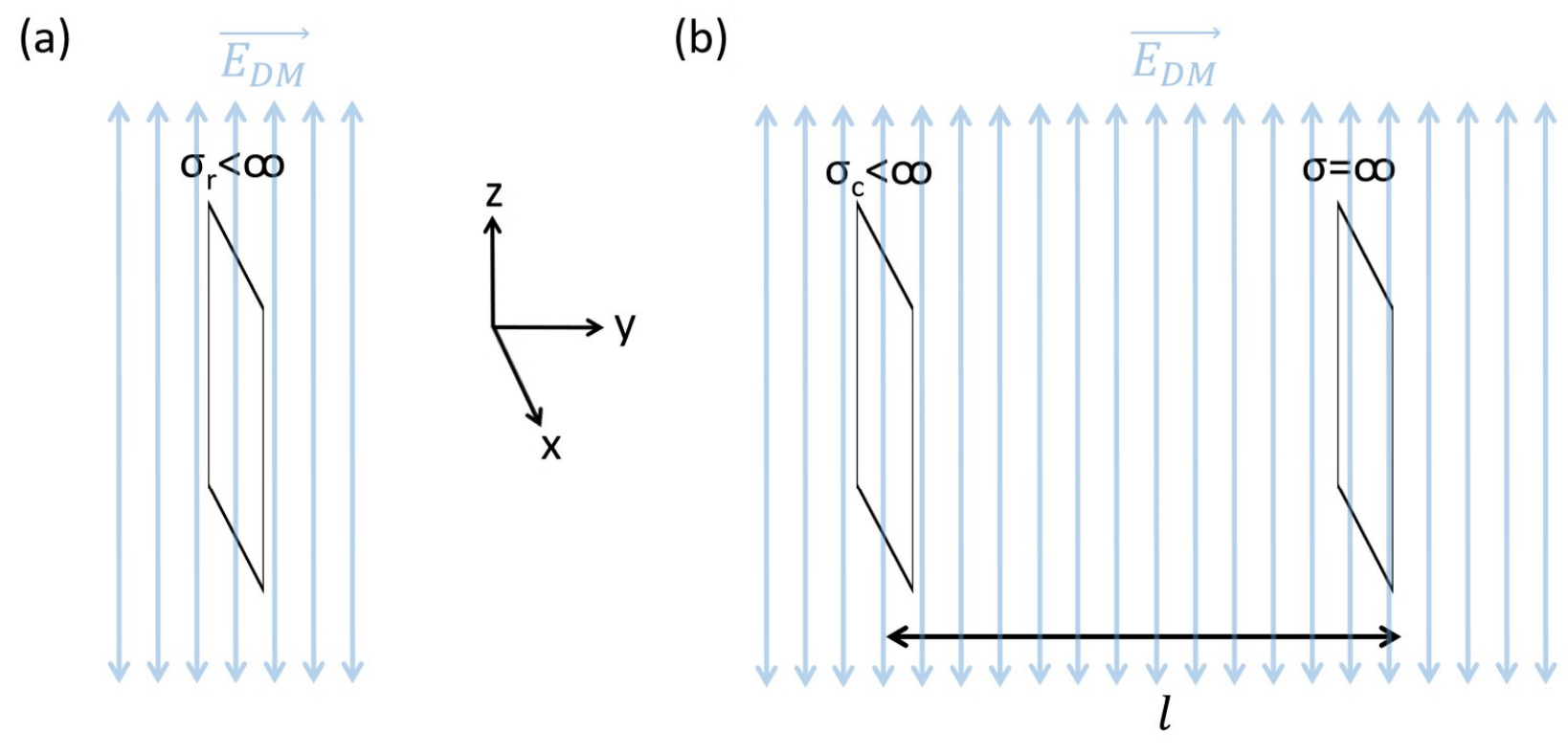}
\caption{Setup for comparison of resistive sheet antenna and tunable cavity as dark-matter detectors. (a) Resistive sheet. (b) Half-wave cavity, of tunable length $\ell$, bounded by a sheet of finite conductivity and a sheet of perfect conductivity. The dark matter produces a free-space electric field, which we assume is tangent to the surface of the sheets. \label{fig:SheetvCavity}}
\end{figure}

See Fig. \ref{fig:SheetvCavity}. In Section \ref{ssec:DM_Maxwell}, we showed that the dark-matter-induced electromagnetic fields can be modeled as sourced from effective current densities. For the hidden photon, the current density intrinsically fills all space. In this toy model, we assume that the axion effective current density also fills all space, as obtained with a uniform DC magnetic field of infinite spatial extent. Then, as can be derived from eq. (\ref{eq:DM_fields}) \cite{horns2013searching,millar2017dielectric}, the current densities source electric and magnetic fields filling all space. Relative to the dark-matter-induced free-space electric field, the dark-matter-induced free-space magnetic field is suppressed, in amplitude, by the dark-matter velocity $v/c \sim 10^{-3}$. The dominant observable is then an electric field, which we denote as $\vec{E}_{\rm DM}$ in the figure. 

In Fig.~\ref{fig:SheetvCavity}a, we show a sheet of real-valued, frequency-independent conductivity $\sigma_{r}$. We set the sheet thickness to be $h$ and the area to be $A$. We assume that, for all frequencies $\omega_{\rm DM}^{0}$ in our search band, the thickness is much less than both the skin depth and the Compton wavelength. We define a quantity
\begin{equation}\label{eq:Z_sheet}
    Z_{r} \equiv 1/(\sigma_{r}h)
\end{equation}
which we call the sheet impedance. The sheet lies in the x-z plane and is sensitive to electric fields tangent to its surface. These electric fields, as dictated by Ohm's Law, dissipate power in the sheet and also cause the sheet to radiate electromagnetic fields into free space. The power dissipation can be measured using a phase-insensitive amplifier (e.g. one located at the feedpoint of phased array elements).

The sheet is a simple model for a broadband free-space antenna. It can represent both a physical resistor as well as a phased array of identical antenna elements. In the latter case, the radiation pattern of each antenna element is a spherical wavefront, but due to interference between the elements, engineered using delays in the electrical network, the radiation pattern of the phased array is in the form of plane waves and identical to that of a large sheet resistor.\cite{hansen2009phased} The equivalent sheet impedance of the phased array is set by the feedpoint impedance of each antenna element. 

In Fig.~\ref{fig:SheetvCavity}b, we show two sheets separated by length $l$: one of frequency-independent, real-valued conductivity $\sigma_{c}$ and the other possessing perfect conductivity. We set the sheets to have the same thickness $h$ and area $A$ as the broadband detector. We assume, that for the finite-conductivity sheet, the thickness is much less than the skin depth and Compton wavelength. The sheet impedance is
\begin{equation} \label{eq:Z_cavity}
    Z_{c} \equiv 1/(\sigma_{c}h)
\end{equation}

This is a quasi-1D toy model of a resonant cavity. We show below that if the frequency $\omega$ of the dark-matter-induced electric field $\vec{E}_{\rm DM}$ satisfies (or nearly satisfies) the resonance condition
\begin{equation}\label{eq:halfwave_res}
\textrm{exp}(i\omega l/c)=-1,\ \textrm{exp}(2 i\omega l/c)=1,
\end{equation}
then the drive is resonantly enhanced. Electric-field and magnetic-field energy is stored in between the sheets, and the resonantly-enhanced fields dissipate power in the left-side sheet, which may be measured with an amplifier. We assume that only a single detector mode is used: the half-wavelength fundamental resonance at $\omega_{r}=\pi c/l$. Although one might use multiple cavity modes or wider band information across the search range (the latter of which is discussed in detail later in this paper), in this section we only consider signal sensitivity within the bandwidth of this single mode. The cavity resonance frequency may be tuned by changing length $l$. 

We assume that the cavity separation $l$ is much smaller than the de Broglie wavelength of the dark matter, which sets the coherence length of $\vec{E}_{\rm DM}$. (See Appendix \ref{ssec:DM_coh}.) This is an appropriate assumption, given that for dark-matter signals near the fundamental resonance frequency, the length is comparable to the Compton wavelength $\lambda_{\rm DM}^{0}=2\pi c/ \omega_{\rm DM}^{0}$, and is thus much less than the coherence length $\sim 1000 \lambda_{\rm DM}^{0}$.\cite{Chaudhuri:2014dla} In both Fig. \ref{fig:SheetvCavity}a and Fig. \ref{fig:SheetvCavity}b, we assume that the lateral extent of the sheets is much smaller than the coherence length, so that we may treat $\vec{E}_{\rm DM}$ as spatially uniform. Furthermore, we assume that the lateral extent of the sheets is much larger than the Compton wavelength, so that we may ignore fringe-field effects and effectively reduce the three-dimensional problem to a one-dimensional problem. 


The dark-matter-induced electric field $\vec{E}_{\rm DM}$ possesses a bandwidth given by eq. (\ref{eq:delta_nuDM}). Nevertheless, in this toy calculation, we assume that the cavity linewidth is larger than this bandwidth, so that for the purpose of calculating signal power we may treat the dark-matter signal as monochromatic. We assume that the dark-matter electric field lies in the $\hat{z}$ direction. For the axion, this may be arranged by applying a uniform DC magnetic field in the $\hat{z}$ direction:
\begin{equation}
    \vec{B}_{b}(\vec{x})=B_{0}\hat{z}
\end{equation}
where $B_{0}$ is the magnitude of the applied magnetic field. For the hidden photon, the experimentalist does not control the direction of the electric field. However, this is not of consequence for the sensitivity comparison. Misalignment with the detector results in the same multiplicative reduction in sensitivity for both the broadband resistive absorber and the cavity. 

Under these assumptions, using the Maxwell equations (\ref{eq:DM_fields}), the dark-matter electric field can be solved and approximated as
\begin{equation}\label{eq:Efield_DM}
\vec{E}_{\rm DM}(t)=Re(\tilde{E}_{\rm DM} \exp( i\omega_{\rm DM}^{0}t)) \hat{z}.
\end{equation}
We relate the complex amplitude of this electric field to the complex amplitude of the z-component of the hidden-photon three-vector potential, denoted $\tilde{A}'$ \cite{Graham:2014sha,Chaudhuri:2014dla,horns2013searching}:
\begin{equation}\label{eq:Efield_hidden}
\tilde{E}_{\rm DM}= -i\omega_{\rm DM}^{0} \varepsilon \tilde{A}'.
\end{equation}
One may write a similar relationship for the axion pseudoscalar potential, whose complex amplitude is $\tilde{a}$ \cite{millar2017dielectric}:
\begin{equation}\label{eq:Efield_axion}
\tilde{E}_{\rm DM}=\kappa_{a}cB_{0} \tilde{a}.
\end{equation}

The signal size is determined by the steady-state power dissipated in the finite conductivity sheet in each of the two experiments. 
Ohm's Law dictates that the volumetric current density in each sheet is related to the electric field by
\begin{equation}\label{eq:OhmsBulk}
\vec{J}_{r,c}(\vec{x},t)=\sigma_{r,c} \vec{E}^{\rm tot}_{r,c}(\vec{x},t),
\end{equation}
where the subscript indicates whether the sheet belongs to the broadband detector (``r") or the cavity (``c") and $\vec{E}^{\rm tot}_{r,c}(\vec{x},t)$ is the total electric field in the sheet. There is also a current in the perfect conductor (the right-hand sheet in the cavity) to screen electric fields from its interior and maintain the boundary condition that the electric field parallel to the surface vanish. We denote this current density as $\vec{J}_{\infty}(\vec{x},t)$. The current in each of the three sheets arises both from the ``incident'' dark-matter electric field and the electric field radiated in response to this drive.


Since there is no free charge density, Maxwell's equations (\ref{eq:rec_fields}) yield a relationship between the volume current density and the resulting electric field at any point in space:
\begin{equation}\label{eq:Maxwell_EJ}
(\partial_{t}^{2}-c^{2}\vec{\nabla}^{2})\vec{E}_{\rm rec}(\vec{x},t)= -\mu_{0}c^{2} \partial_{t} \vec{J}_{\rm rec}(\vec{x},t).
\end{equation}
For the broadband receiver, $\vec{J}_{\rm rec}(\vec{x},t)$ represents the current density for the sheet, while for the cavity, $\vec{J}_{\rm rec}(\vec{x},t)$ includes both the current densities on the finite-conductivity and the perfect-conductivity sheet: $\vec{J}_{\rm rec}(\vec{x},t)=\vec{J}_{c}(\vec{x},t)+ \vec{J}_{\infty}(\vec{x},t)$. 

In general, the current density and the electric field are position-dependent in a sheet. However, because of assumptions on dimensions which enable us to ignore fringing effects, the current density and the electric field are approximately uniform in each sheet. We may turn the volumetric current density into an effective surface current density $\vec{K}_{r,c}(t)$, so that, from eqs. (\ref{eq:Z_sheet}), (\ref{eq:Z_cavity}), and (\ref{eq:OhmsBulk}),
\begin{equation}\label{eq:OhmsLaw_sheet}
\vec{K}_{r,c}(t)=\vec{E}^{\rm tot}_{r,c}(t)/Z_{r,c},
\end{equation}
where $\vec{E}^{\rm tot}_{r,c}(t)$ is now most readily interpreted as the surface field. We may also define an effective surface current density for the perfect conductor, $\vec{K}_{\infty}(t)$. 

We may solve for the current in each sheet, and consequently the produced fields and power dissipation, using superposition. Because we ignore fringe-field effects, the electric field is only a function of $y$, the coordinate normal to the sheets.

\paragraph{Power Dissipation in Resistive Broadband Detector} \mbox{}\\ 

All quantities oscillate at frequency $\omega_{\rm DM}^{0}$ and the fields and currents point in the $\hat{z}$ direction, so we may write
\begin{equation}\label{eq:CurrentPhasor}
\vec{K}_{r}(t)=Re(\tilde{K}_{r} \textrm{exp}( i\omega_{\rm DM}^{0}t)) \hat{z}.
\end{equation}
Setting the position of the sheet to be $y=0$, the current density in eq. (\ref{eq:Maxwell_EJ}) is $\vec{J}_{\rm rec}(\vec{x},t) \approx \vec{K}_{r}(t) \delta(y)$ where $\delta(y)$ is a Dirac delta function. The electric field produced by the sheet is then the plane wave
\begin{equation}\label{eq:Efield_sheet}
\vec{E}_{r}(y,t)=Re(\tilde{E}_{r} \textrm{exp}( i\omega_{\rm DM}^{0} (t- |y|/c)) \hat{z},
\end{equation}
where 
\begin{equation}\label{eq:Erec_sheet}
\tilde{E}_{r}=-Z_{fs}\tilde{K}_{r}/2.
\end{equation}
Equation (\ref{eq:OhmsLaw_sheet}) gives
\begin{equation}
\tilde{K}_{r}=\frac{1}{Z_{r}}(\tilde{E}_{r} + \tilde{E}_{DM}) = \frac{1}{Z_{r}} \left( \tilde{E}_{\rm DM} -\frac{Z_{fs}}{2} \tilde{K}_{r} \right),
\end{equation}
or, rearranging,
\begin{equation}\label{eq:current_field_sheet}
\tilde{K}_{r}=\frac{2\tilde{E}_{\rm DM}}{2Z_{r}+ Z_{fs} }.
\end{equation}
The power dissipated by per unit area is then
\begin{equation}\label{eq:Pdiss_sheet}
\frac{P_{r}}{A}= \frac{Z_{r}}{2} |\tilde{K}_{r}|^{2} = \frac{Z_{r}}{2} \frac{|\tilde{E}_{\rm DM}|^{2}}{(Z_{r} + Z_{fs}/2)^{2} } = \frac{|\tilde{E}_{\rm DM}|^{2}}{Z_{fs}} \frac{Z_{fs}/2Z_{r}}{(1+Z_{fs}/2Z_{r})^{2}}.
\end{equation}
The power dissipation is maximized for sheet impedance $Z_{r}=Z_{fs}/2$, at which
\begin{equation}\label{eq:Pdiss_sheet_max}
\frac{P_{r}}{A} \Bigg|_{\rm max} = \frac{1}{4} \frac{|\tilde{E}_{\rm DM}|^{2}}{Z_{fs}}.
\end{equation}
We see that the power dissipation in the resistive sheet is limited by the free-space self-impedance of the photons, and not by the impedance of the dark-matter source. In fact, this structure is a very poor impedance match to dark matter, absorbing a very small fraction of the available power. (See also the discussion surrounding eq. (\ref{eq:Pavail}) below.)
We explain the above result further in Section \ref{sec:DM_circuits} in terms of impedance mismatch with the dark-matter source.

To further demonstrate that the excitation of the antenna corresponds to a radiative coupling, we show that the surface integral on the left hand side of (\ref{eq:DM_Poynting}) is non-zero and that the input power flow is the sum of the receiver radiation and dissipation, as would be expected from (\ref{eq:DM_Poynting}). Consider a rectangular-prism volume $V_{r}$ which surrounds the sheet and is infinitesimally larger than the sheet volume $Ah$. To evaluate the integrals, we make the identifications, valid over $V_{r}$,
\begin{equation}\label{eq:EDM_id_rad}
    \vec{E}_{DM}(\vec{x},\omega_{\rm DM}^{0}) \rightarrow \tilde{E}_{DM} \hat{z}
\end{equation}
\begin{equation}\label{eq:Erec_id_rad}
    \vec{E}_{\rm rec}(\vec{x},\omega_{\rm DM}^{0}) \rightarrow \tilde{E}_{r} \hat{z}.
\end{equation}
We have omitted the second argument $\omega$ of the three arguments used in parametrizing the field dependence (see previous section) because the fields are monochromatic. Because the radiation emitted from the receiver is in the form of electromagnetic plane waves, the receiver magnetic-field vector at the surface $\partial V_{r}$, identified with $\vec{B}_{\rm rec}(\vec{x}, \omega_{\rm DM}^{0})$, is orthogonal to both the electric-field vector and the surface normal (the direction of propagation). We then evaluate the power-flow integrals using eqs. (\ref{eq:OhmsLaw_sheet}) and (\ref{eq:Erec_sheet}). The input power flow from the dark-matter field is
\begin{equation}\label{eq:sheet_Pin}
-\frac{1}{2\mu_{0}} \int_{\partial V_{r}} (\vec{E}_{\rm DM}(\vec{x},\omega_{\rm DM}^{0}) \times \vec{B}_{\rm rec}^{*}(\vec{x},\omega_{\rm DM}^{0}))\cdot \vec{da}= -\frac{1}{Z_{fs}} A \tilde{E}_{\rm DM} \tilde{E}_{r}^{*}= \frac{1}{2} A \tilde{E}_{\rm DM} \tilde{K}_{r}^{*},
\end{equation}
the electronic power dissipation is
\begin{equation}\label{eq:sheet_Pdiss}
\frac{1}{2} \int_{V_{r}} d^{3}\vec{x}\  \vec{J}_{\rm rec}^{*}(\vec{x},\omega_{\rm DM}^{0}) \cdot \vec{E}_{\rm tot}(\vec{x},\omega_{\rm DM}^{0}) = \frac{1}{2} A |\tilde{K}_{r}|^{2} Z_{r},
\end{equation}
and the radiated power is
\begin{equation}\label{eq:sheet_Prad}
\frac{1}{2\mu_{0}} \int_{\partial V_{r}} (\vec{E}_{\rm rec}(\vec{x},\omega_{\rm DM}^{0}) \times \vec{B}_{\rm rec}^{*}(\vec{x},\omega_{\rm DM}^{0}))\cdot \vec{da} = \frac{1}{2Z_{fs}} \int_{\partial V} |\vec{E}_{\rm rec}(\vec{x},\omega_{\rm DM}^{0})|^{2}\ da = \frac{1}{4} A |\tilde{K}_{r}|^{2} Z_{fs}.
\end{equation}
where we have used, at each point on $\partial V_{r}$,
\begin{equation}
    \vec{B}_{rec}= \hat{n} \times \vec{E}_{rec}/c
\end{equation}
where $\hat{n}$ is the surface normal. Plugging in eq. (\ref{eq:current_field_sheet}), relating the effective surface current $\tilde{K}_{r}$ and dark-matter-induced electric field $\tilde{E}_{\rm DM}$, shows that the input power is equal to the power dissipated plus the power radiated.

\paragraph{Power Dissipation in Cavity Detector} \mbox{}\\ 

Similar to equations (\ref{eq:CurrentPhasor}) and (\ref{eq:Efield_sheet}), we may write the electric field produced by the sheet of impedance $Z_{c}$ as
\begin{equation}\label{eq:Ec_field}
\vec{E}_{c}(y,t)=Re(\tilde{E}_{c} \textrm{exp}(i\omega_{\rm DM}^{0} (t- |y|/c)) \hat{z},
\end{equation}
where 
\begin{equation}\label{eq:cav_eq_1}
\tilde{E}_{c}=-Z_{fs}\tilde{K}_{c}/2.
\end{equation}
As before, we have set the position of this sheet to $y=0$. We may write the electric field produced by the perfectly conducting sheet as
\begin{equation}\label{eq:Einf_field}
\vec{E}_{\infty}(y,t)=Re(\tilde{E}_{\infty} \textrm{exp}( i\omega_{\rm DM}^{0} (t- |y-l|/c)) \hat{z},
\end{equation}
where 
\begin{equation}\label{eq:cav_eq_2}
\tilde{E}_{\infty}=-Z_{fs}\tilde{K}_{\infty}/2.
\end{equation}
Since the electric field must vanish at the perfect conductor, we have
\begin{equation} \label{eq:cav_eq_3}
\tilde{E}_{\infty}+\tilde{E}_{\rm DM} + \tilde{E}_{c} \textrm{exp}(- i \omega_{\rm DM}^{0} l/c)=0.
\end{equation}
Ohm's Law gives
\begin{equation} \label{eq:cav_eq_4}
\tilde{K}_{c}=\frac{1}{Z_{c}} \left( \tilde{E}_{c} + \tilde{E}_{\infty} \textrm{exp}(-i \omega_{\rm DM}^{0} l/c) + \tilde{E}_{\rm DM} \right).
\end{equation}
We have four equations (\ref{eq:cav_eq_1}), (\ref{eq:cav_eq_2}), (\ref{eq:cav_eq_3})-(\ref{eq:cav_eq_4}), and four unknowns, $\tilde{E}_{c}$, $\tilde{E}_{\infty}$, $\tilde{K}_{c}$, and $\tilde{K}_{\infty}$. 
Solving the system gives the complex current amplitude on the sheet at $y=0$,
\begin{equation}\label{eq:I_cavity}
\tilde{K}_{c}= \frac{2 (1-\textrm{exp}(-i\omega_{\rm DM}^{0}l/c)) \tilde{E}_{\rm DM} }{2Z_{c} + Z_{fs}(1-\textrm{exp}(-2 i\omega_{\rm DM}^{0}l/c))} ,
\end{equation}
and a power dissipation per unit area of
\begin{equation}\label{eq:Pdiss_cavity}
\frac{P_{c}}{A}= \frac{Z_{c}}{2}|\tilde{K}_{c}|^{2} = \frac{|\tilde{E}_{\rm DM}|^{2}}{Z_{fs}} \frac{(Z_{fs}/2Z_{c})|1-\textrm{exp}(-i\omega_{\rm DM}^{0}l/c)|^{2}}{|1+(Z_{fs}/2Z_{c})(1-\textrm{exp}(-2i\omega_{\rm DM}^{0}l/c))|^{2}}.
\end{equation}
For dark-matter resonance frequencies close to resonance, $\omega_{\rm DM}^{0} \approx \omega_{r}=c/2l$,  this equation may be expanded as
\begin{equation}\label{eq:Pdiss_cavity_Lorentzian}
\frac{P_{c}}{A} \approx \frac{|\tilde{E}_{\rm DM}|^{2}}{Z_{fs}} \frac{4(Z_{fs}/2Z_{c})}{|1+2\pi i (Z_{fs}/2Z_{c}) (\omega_{\rm DM}^{0}-\omega_{r})/\omega_{r}|^{2}} = \frac{4}{\pi} \frac{|\tilde{E}_{\rm DM}|^{2}}{Z_{fs}} \frac{Q}{|1+2iQ (\omega_{\rm DM}^{0}-\omega_{r})/\omega_{r}|^{2}}.
\end{equation}
The response is a Lorentzian, as would be expected, and the quality factor of the cavity, as determined by the full width at half maximum, is $Q=\pi Z_{fs}/2Z_{c}$. The result can also be derived using the traditional cavity-overlap formalism; see Appendix \ref{sec:cav_circuit}. For an on-resonance dark-matter signal, the power dissipation is maximized,
\begin{equation}\label{eq:Pdiss_cavity_mode_max}
\frac{P_{c}}{A} \Bigg|_{\rm max} =\frac{2|\tilde{E}_{\rm DM}|^{2}}{Z_{c}} = \frac{4}{\pi} \frac{|\tilde{E}_{\rm DM}|^{2}}{Z_{fs}} Q,
\end{equation}
and the power dissipation increases linearly in $Q$. Note, from the first equality of (\ref{eq:Pdiss_cavity_mode_max}), that the power dissipation is independent of the free-space impedance, an observation that is explored in far greater detail in Section \ref{ssec:toy_circuits}. Taking the limit $Q \rightarrow \infty$ (sheet impedance to zero), it seems that, subject to the approximations in this model, arbitrarily large amounts of power can be dissipated. Of course, this is unphysical. Two assumptions that we have made break down. First, for high quality factors giving cavity linewidths narrower than the dark-matter linewidth ($Q \gtrsim 10^{6}$), not all parts of the dark-matter spectrum are fully resonantly-enhanced. A more complicated calculation, in which the shape of the dark-matter spectrum is convolved with the resonator response, is needed to calculate the power dissipation. Such a calculation is carried out in Sec. \ref{sec:resonator_SNR}.

More fundamentally, power conservation dictates that at some quality factor, we must begin to backreact on the dark-matter source, producing enough dark matter through the electromagnetic interaction to locally change the value of the dark-matter density and effectively modify the value of $\tilde{E}_{\rm DM}$. We show here that backreaction is negligible in practice. Specifically, we calculate parametrically the range of Q factors for which backreaction on the dark-matter field can be ignored and conclude that practical cavity quality factors are well within this range. As stated earlier, the power available from the dark matter field is determined by the flux density through the detector, roughly equal to the product of the dark-matter energy density and the velocity and expressed as
\begin{equation} \label{eq:Pavail}
\frac{P_{\rm avail}^{\rm DM}}{A} \sim \begin{cases} \frac{1}{2} \frac{(\omega_{\rm DM}^{0})^{2} |\tilde{A}'|^{2}}{Z_{fs}} \frac{v}{c}, \hspace{1 cm} \textrm{hidden photons} \\
\frac{1}{2} \frac{( \omega_{\rm DM}^{0}/c)^{2} |\tilde{a}|^{2}}{Z_{fs}} \frac{v}{c}, \hspace{1 cm} \textrm{axions}
\end{cases}
\end{equation}
Backreaction can be ignored as long as the power dissipated in the cavity is much smaller than the available power, $P_{c} \ll P_{\rm avail}^{\rm DM}$. Combining equations (\ref{eq:Efield_DM}), (\ref{eq:Efield_hidden}), and (\ref{eq:Efield_axion}) with (\ref{eq:Pdiss_cavity_Lorentzian}) and (\ref{eq:Pavail}), we find the condition on quality factor for neglecting backreaction:
\begin{equation} \label{eq:Q_backaction}
Q \ll \begin{cases} \varepsilon^{-2} \frac{v}{c} \hspace{1 cm} \textrm{hidden photons} \\
\left( \frac{\kappa_{a} c B_{0}}{m_{\rm DM}c/\hbar}  \right)^{-2} \frac{v}{c}, \hspace{1 cm} \textrm{axions}
\end{cases}
\end{equation}
The Q factor below which we can ignore backaction varies as $\varepsilon^{-2}$ ($\kappa_{a}^{-2}$) because the production of hidden photon and axion fields is an order $\varepsilon^{2}$ ($\kappa_{a}^{2}$) effect. One order of $\varepsilon$ ($\kappa_{a}$) comes from the dark matter fields driving currents on the conductors, while another order of $\varepsilon$ ($\kappa_{a}$) comes from those currents producing hidden photons and axions. For a virialized hidden photon with $v/c \sim 10^{-3}$, possessing mixing angle $\varepsilon \sim 10^{-10}$ (slightly below existing constraints on the parameter space \cite{Chaudhuri:2014dla}), eq. (\ref{eq:Q_backaction}) demonstrates that backaction can be ignored as long as $Q \ll 10^{17}$. For the virialized axion, the range of Q values for which backaction can be ignored decreases as the DC magnetic field strength increases because the axion conversion into photons is increased. Suppose that the magnetic field strength is 10 Tesla. For the KSVZ axion\cite{kim1979weak,shifman1980can}, a benchmark QCD axion model, we then find that backaction can be ignored as long as $Q \ll 10^{28}$. Practical cavity Q factors are well within these regimes for both axions and hidden photons, even for low-loss superconducting cavities. As such, backreaction can be ignored, and the dark matter field presents as a stiff electromagnetic source. We also note that, for the broadband resistive antenna, $P_{r} \ll P_{\rm avail}^{\rm DM}$ for any sheet resistance. It is thus appropriate to ignore backreaction in that case as well. In Section \ref{sec:DM_circuits}, we will use eq. (\ref{eq:Q_backaction}) to derive the scale of the source impedance of dark-matter.

Similar to the calculation we performed for the resistive broadband antenna, we may calculate power-flow integrals from Section \ref{ssec:DM_Maxwell} and show that the signal-coupling to the fundamental mode of the cavity is a reactive coupling. Specifically, over a rectangular-prism volume $V_{c}$ which surrounds the cavity and is infinitesimally larger than the cavity volume $Al$, we show that the second integral on the right-hand side of (\ref{eq:DM_Poynting}) is nonzero on-resonance and that it is equal to the power dissipated in the cavity, represented by the first integral on the right-hand side of (\ref{eq:DM_Poynting}). Note that we need not consider any of the other integrals in (\ref{eq:DM_Poynting}). Unlike the antenna and as we discuss further in Section \ref{ssec:toy_circuits}, radiation effects for the half-wave cavity mode are suppressed by the high-conductivity sheets, so we need not consider the surface integral terms. Moreover, in a cavity on-resonance, the volume integrals of electric field energy and magnetic field energy are equal, so they cancel on the right-hand side of (\ref{eq:DM_Poynting}). For a high-Q cavity on-resonance, $\omega_{DM}^{0} l/c =\pi$ and the electric fields produced by the sheets, $\tilde{E}_{c}$ and $\tilde{E}_{\infty}$ are much larger in magnitude than the drive field $\tilde{E}_{\rm DM}$, so eqs. (\ref{eq:cav_eq_1}), (\ref{eq:cav_eq_2}), and (\ref{eq:cav_eq_3}) yield $\tilde{K}_{c} = \tilde{K}_{\infty}$. As with the antenna receiver, we make the identification (\ref{eq:EDM_id_rad}) and identify $\vec{E}_{\rm rec}(\vec{x},\omega_{\rm DM}^{0})$ with the sum of $\vec{E}_{c}(y,t)$ in (\ref{eq:Ec_field}) and $\vec{E}_{\infty}(y,t)$ in (\ref{eq:Einf_field}).
Using eqs. (\ref{eq:cav_eq_1})-(\ref{eq:cav_eq_4}), one then obtains 
\begin{align}
    \frac{i\omega_{\rm DM}^{0}\epsilon_{0}}{2} & \int_{V_{c}} d^{3}\vec{x}\ \vec{E}_{\rm DM}(\vec{x},\omega_{\rm DM}^{0}) \cdot \vec{E}_{\rm rec}^{*}(\vec{x},\omega_{\rm DM}^{0}) \nonumber \\
    &\approx  -\frac{i(2\pi\omega_{\rm DM}^{0})}{4 c} A \tilde{E}_{\rm DM} \tilde{K}_{c}^{*}  \int_{0}^{l} dy\ (\exp(+i\omega_{\rm DM}^{0}y/c) + \exp(-i\omega_{\rm DM}^{0}(y-l)/c)) \nonumber \\
    &= A \tilde{E}_{\rm DM} \tilde{K}_{c}^{*} \label{eq:cavity_drive}
\end{align}
\begin{equation}
\frac{1}{2} \int_{V_{c}} d^{3}\vec{x}\  \vec{J}_{\rm rec}^{*}(\vec{x},\omega_{\rm DM}^{0}) \cdot \vec{E}_{\rm tot}(\vec{x},\omega_{\rm DM}^{0}) = \frac{1}{2} A |\tilde{K}_{c}|^{2} Z_{c} \label{eq:cavity_diss}
\end{equation}
Eq. (\ref{eq:I_cavity}) gives $\tilde{K}_{c} = 2 \tilde{E}_{\rm DM}/Z_{c}$, so the two integrals are equal.

To perform the sensitivity comparison between the resistive broadband and cavity detectors, we set a search band between frequencies $\omega_{l}$ and $\omega_{h}$ and determine the SNR from our power dissipation calculations. While the broadband detector does not change during the search, always being set to the optimal sheet impedance $Z_{r}=Z_{fs}/2$, the cavity resonance frequency is stepped between $\omega_{l}$ and $\omega_{h}$.

We assume that the physical temperature of the two detectors--each at thermal equilibrium with its environment-- is the same. We assume that the noise temperature of the amplifier in the broadband search is the same as the noise temperature of the amplifier in the cavity search at the resonance frequency. Assuming then that the total noise is dominated by the sum of the thermal and amplifier noise, we find that the system temperature $T_{S}$ \cite{asztalos2010squid,brubaker2017first} of the two receivers is the same. Therefore, the noise power $P_{n}=kT_{S}\Delta \omega_{\rm DM}(\nu_{\rm DM}^{0})/(2\pi)$ is also the same.

Additionally, we set the total experiment time, for each experiment, at $T_{\rm tot}$. We assume that the total experiment time is large enough such that the cavity can ring up at every tuning step and so that the dark-matter signal can be resolved in a Fourier spectrum; the former requires time $\sim 2\pi Q/\omega_{r}$ while the latter requires time $\sim 2\pi \cdot 10^{6}/\omega_{\rm DM}^{0}$. We also assume that the tuning time is negligible compared to the integration time at each step.

The Dicke radiometer equation \cite{zmuidzinas2003thermal,dicke1946measurement} gives the SNR in power (as opposed to amplitude) for each experiment
\begin{equation}
SNR(\omega_{\rm DM}^{0}) \approx \frac{P_{\rm diss}}{P_{n}}\sqrt{ \frac{\Delta\omega_{\rm DM}(\nu_{\rm DM}^{0})T(\omega_{\rm DM}^{0})}{2\pi}},
\end{equation}
where $P_{diss}$ (= $P_{r}$ or $P_{c}$) is the power dissipated in the detector and $T(\omega_{\rm DM}^{0})$ is the integration time at dark-matter frequency $\omega_{\rm DM}^{0}$. For the broadband antenna search, this integration time is simply the total experiment time $T_{\rm tot}$, while for the cavity detector, it can be taken as the amount of time during which the dark-matter frequency $\omega_{\rm DM}^{0}$ is within the bandwidth of the resonator, $\Delta t_0$.
The ratio of the SNRs for the cavity and resistive broadband detectors is then
\begin{equation}\label{eq:SNR_ratio_sheet}
\frac{SNR_{c}(\omega_{\rm DM}^{0})}{SNR_{r}(\omega_{\rm DM}^{0})}\approx \frac{P_{c}}{P_{r}} \sqrt{ \frac{\Delta t_0}{T_{\rm tot}}} = \frac{16Q}{\pi} \sqrt{\frac{\Delta t_0}{T_{\rm tot}}},
\end{equation}
where, in the second equality, we have used eqs. (\ref{eq:Pdiss_sheet_max}) and (\ref{eq:Pdiss_cavity}). 

If, in the cavity search, we spend an equal time $\Delta t_0$ at each frequency and step at one part in $Q$, the total integration time can be taken as the $\Delta t_0$ times the number of frequency steps between $\omega_{l}$ and $\omega_{h}$. For broad scans ($\omega_{h} \gtrsim 2\omega_{l}$), the number of frequency steps can be approximated as an integral, so

\begin{equation}\label{eq:IntegrationTimeIntegral}
\frac{T_{\rm tot}}{\Delta t_0} \approx \int_{\omega_{l}}^{\omega_{h}} \frac{Q}{\omega} d\omega = Q  \ln{\frac{\omega_{h}}{\omega_{l}}} \sim  Q.
\end{equation}

The SNR ratio in (\ref{eq:SNR_ratio_sheet}) then scales with $Q$ as
\begin{equation}\label{eq:SNR_ratio_sheet_param}
\frac{SNR_{c}(\omega_{\rm DM}^{0})}{SNR_{r}(\omega_{\rm DM}^{0})} \sim \sqrt{Q}.
\end{equation}
This demonstrates that the high-Q cavity search is superior to the resistive broadband search for dark matter. The advantage of high-Q searches can be even larger if information outside of the resonator bandwidth is used (which has been ignored here), as is shown later. The result may have been intuited from the form of the Dicke radiometer equation: an experiment gains sensitivity much faster with higher signal power (linear relationship) than with longer integration time (square root relationship). Note that, for the scan sensitivity of the cavity, the SNR in power varies as square root of quality factor. This, in turn, implies that the minimum dark-matter coupling ($g_{a\gamma\gamma}$ for the axion and $\varepsilon$ for the hidden photon) to which the cavity is sensitive scales as $Q^{-1/4}$.  Interestingly, as we show in Sec. \ref{ssec:QLA_sensitivity}, the scaling even applies when the resonator quality factor is larger than the characteristic $\sim 10^{6}$ quality factor of the dark-matter signal, for which resonator response and dark-matter spectrum must be convolved.

We note that our analysis and the conclusion that the high-Q cavity is superior to the resistive broadband antenna depends substantially on the assumption of readout with a phase-insensitive amplifier subject to the Standard Quantum Limit. The result may change for readout techniques beyond the Standard Quantum Limit, such as photon counting, quantum-squeezing, or backaction evasion. With these quantum readout techniques, one may need to consider other noise sources in the sensitivity comparison of the two toy receivers. For example, for photon counting, one must take into account shot noise from the electric field signal. As the signal power increases, the shot noise power also increases, so that, in contrast to our above analysis, the noise power cannot be held equal for the two setups.\cite{lehnert2016quantumaxion}

It is useful to contrast our results here with previous related results in the literature. In ref. \cite{horns2013searching}, the authors compare a cavity of volume $\sim (\lambda_{\rm DM}^{0})^{3}$ to an antenna of receiving size $A$, finding that the ratio of signal powers is (assuming that the cavity is on-resonance)
\begin{equation}\label{eq:Pratio_horns}
    \frac{P_{c}}{P_{r}} \sim Q \frac{(\lambda_{\rm DM}^{0})^{2}}{A}
\end{equation}
Equation (\ref{eq:Pratio_horns}) differs substantially from our comparison which yields a ratio $P_{c}/P_{r} \sim Q$. As a result, from (\ref{eq:Pratio_horns}), one would conclude that for sufficiently high frequencies and short Compton wavelengths for which $\frac{(\lambda_{\rm DM}^{0})^{2}}{A} \lesssim 1/Q$, the antenna signal power is larger than the cavity signal power, and thus, that the broadband antenna actually outperforms the cavity. The difference can be reconciled by observing that the comparison in ref. \cite{horns2013searching} does not constitute an apples-to-apples comparison between an antenna and a cavity. While typical, practical cavity volumes are limited to $\sim (\lambda_{\rm DM}^{0})^{3}$, the volume limitation is not fundamental and can be circumvented with proper engineering of the cavity mode. Moreover, one can use a receiver scheme in which $N>1$ identical cavities, each of volume $V_{c}$, are driven by the dark-matter and the output of each cavity amplified. By coherently summing the amplifier timestreams, one then obtains the same search sensitivity as a single cavity of volume $NV_{c}$. Such a scheme is central to the idea of cavity arrays for axion searches at several GHz.\cite{Sho14} (See also Section \ref{sssec:BF_evade}.) We then see that an appropriate apples-to-apples comparison for understanding fundamental optimization is one in which the antenna receiving area is comparable to the cross-sectional area of the cavity; such is the nature of the comparison that we have conducted in this section.

The calculations in this section have demonstrated two important ideas. First, in our toy model, the reactively coupled cavity receiver is superior, on an apples-to-apples basis, to the radiatively coupled broadband receiver when using phase-insensitive amplifier readout. Second, both receivers are a poor impedance match to dark-matter, absorbing a minuscule fraction of the available power. One might say that a poor impedance to dark matter is practically inevitable because dark-matter is ``feebly coupled.'' While the kinetic mixing angle and the axion-photon coupling are ``small'' numbers, such a statement is inherently ill-defined. One must rigorously quantify what it means for an electromagnetic source to be feebly coupled. This is established in the next section where we calculate the scale of the effective source impedance of dark matter. Our calculation enables us to generalize the disadvantage of radiatively coupled receivers relative to reactively coupled receivers, owing to the mismatch of the dark-matter source impedance and the free space impedance. This justifies a focus on reactive coupling for the global optimization in the remainder of the paper.

\section{Equivalent-Circuit Representations of Light-Field Dark-Matter Receivers}
\label{sec:DM_circuits}


So far, we have analyzed axion and hidden-photon dark-matter detection and impedance matching using the Maxwellian approach to electromagnetism. In our quasi-1D toy comparison, we directly manipulated the governing partial differential equations and constitutive relations (Ohm's Law) to solve for dark-matter-induced fields and the receiver response-fields and currents. This yielded a power dissipation per unit area on the conducting sheets, which was combined with a simplified noise analysis to obtain receiver sensitivity. However, when it comes to more realistic receiver configurations, especially those that are employed in light-field dark-matter searches and cannot be modeled as quasi-1D, the Maxwellian approach of directly solving the differential equations to obtain receiver sensitivity becomes much more challenging to use. Analytic techniques for calculating the signal fields and currents are limited, and computationally-intensive numerical approaches, tailored to the specific geometry of the receiver, must often be employed. Moreover, it is unclear how to rigorously incorporate into a Maxwellian framework a comprehensive noise analysis, which requires study of noise matching and the frequency-dependent flow of noise fields and currents throughout the receiver arising from measurement backaction. By itself, the Maxwellian approach to electromagnetism is then of limited utility in conducting a broad, systematic receiver optimization.

However, to rigorously determine the properties of the optimal linear, passive, single-moded, quantum-limited receiver, we do not require a full solution of the field equations. Rather, we need only focus on particular receiver parameters. Such is the driving philosophy behind a second approach to electromagnetism, known as the equivalent-circuit approach. Indeed, Julian Schwinger, remarking on the utility of the equivalent-circuit approach to perform complex electromagnetic analyses, states, ``Most of the information in Maxwell's equations is really superfluous...A limited number of quantities that can be measured or calculated tell you exactly...what the system is doing...The only role of Maxwell's equations is to calculate the few parameters, the effective lumped constants that characterize the equivalent circuits.''\cite{schwinger1969julian,milton2007appreciation} The equivalence of and relationship between the Maxwellian and equivalent-circuit approaches is described in detail by Dicke, Purcell, and others in ref. \cite{montgomery1987principles}. For more information regarding the equivalent-circuit approach and the construction of equivalent circuits, we refer the reader to refs. \cite{Jackson:1998nia,fano1963electromagnetic,fano1968electromagnetic,montgomery1987principles,pozar2012microwave,milton2007appreciation,schwinger1969julian}. 

The use of the equivalent-circuit approach to elucidate and optimize receiver impedance-matching to dark matter is the primary subject of Section \ref{sec:DM_circuits}. We first overview the application of the equivalent-circuit approach to dark-matter detection, as the technique plays a central role for the remainder of the paper. Based on this discussion, in Section \ref{ssec:toy_circuits}, we develop equivalent circuits of the toy receivers of Section \ref{ssec:ResvBroad}, which enables a calculation of the dark-matter source impedance, and, when combined with a modal analysis of radiation, generalizes the advantage of reactive couplings over radiative couplings. In Section \ref{ssec:reactive_circuit}, we narrow our focus to studying impedance-matching in receivers with single-moded reactive couplings, which describes the majority of present axion and hidden-photon searches. We also lay the foundations for following global optimization, which considers noise, as well as the matching network and readout (second and third boxes in Fig. \ref{fig:DetectorBlockDiagram}). 


Equivalent circuits are appropriately understood as parametrizing complex-power flow within a receiver. As such, similar to ref. \cite{montgomery1987principles}, we explain the equivalent-circuit construction as a mapping onto statements of the form (\ref{eq:DM_Poynting}). 

In an equivalent circuit, there are three types of elements: inductors, capacitors, and resistors, which capture the dynamics represented on the right-hand side of (\ref{eq:DM_Poynting}). Inductors represent receiver elements which store magnetic field energy, including in the form of magnetization. As we discussed in Section \ref{ssec:DM_Maxwell}, such magnetic energy storage corresponds to the fifth term on the right-hand side of (\ref{eq:DM_Poynting}), as well as the imaginary part of the third term. Similarly, capacitors represent receiver elements which store electric field energy, including in the form of electric polarization. Such electric energy storage corresponds to the sixth term on the right-hand side of (\ref{eq:DM_Poynting}), as well as the second term. Inductors and capacitors can also be used to model the reactive portion of the power flow in free electrons (first integral), e.g. as in the kinetic inductance of a normal metal or superconductor.\cite{tinkham2004introduction} Resistors represent receiver elements through which power leaves the receiver. They can represent electronic dissipation, corresponding to the real parts of the first, second, and third integrals on the right-hand side of (\ref{eq:DM_Poynting}), as well as electromagnetic radiation, corresponding to the fourth term on the right-hand side of (\ref{eq:DM_Poynting}). 


The three classes of dark-matter signal coupling into a receiver, introduced in Section \ref{ssec:DM_Maxwell}, can be represented by equivalent-circuit voltage sources. Inductive coupling can be represented as a voltage source in series with an inductor, while capacitive coupling can be represented as a voltage source in series with a capacitor. Radiative coupling can be represented as a voltage source in series with a radiation resistor. The voltage sources drive equivalent-circuit currents through the inductors, capacitors, and resistors, representing receiver energy storage, power dissipation, and radiation in response to the dark-matter drive. The circuit currents can represent physical flow of electrons, as well as displacement currents.\cite{montgomery1987principles}

The voltage sources, currents, and inductors, capacitors, and resistors are quantified by parameter values, which may be related by Kirchhoff's Laws. One can produce a mapping between the equivalent-circuit power-flow statements implied by Kirchhoff's Laws and Maxwellian power-flow statements of the form (\ref{eq:DM_Poynting}). As we show below for the equivalent circuits of our toy receivers, one may in fact compute the circuit parameter values through the mapping. In this manner, as Schwinger states, Maxwell's equations are used to calculate the parameter values that characterize the equivalent circuit. The amplitude of a voltage source is determined by the overlap of the dark-matter-induced electromagnetic drive pattern with the field pattern of the receiver coupling elements. See left-hand side of (\ref{eq:DM_Poynting}). The inductances, capacitances, and resistances, relate the circuit-element currents to energy storage and power dissipation/radiation. See right-hand side of (\ref{eq:DM_Poynting}). Radiation resistances are intimately related to the free-space impedance $Z_{fs}=\sqrt{\mu_{0}/\epsilon_{0}}\approx 377\ \Omega$. 

By utilizing these methods, one can produce equivalent circuits of an arbitrary dark-matter receiver, including receivers that are not physically lumped-element circuits. In particular, as Dicke describes in ref. \cite{montgomery1987principles}, lumped-element equivalent-circuit methods can be used rigorously even when the size of the receiver elements is a wavelength or larger. To demonstrate these ideas---in particular, the equivalence of the Maxwellian and equivalent-circuit approaches for analyzing axion and hidden-photon detection and the mapping of Kirchhoff's Laws to Maxwellian power-flow statements---and the utility of the equivalent-circuit approach for understanding impedance-matching to dark matter, we now proceed to produce equivalent circuits of the toy receivers of Section \ref{ssec:ResvBroad}.

\subsection{Comparing Radiative and Reactive Couplings Using Equivalent Circuits}
\label{ssec:toy_circuits}

\begin{figure}[htp] 
\includegraphics[width=\textwidth]{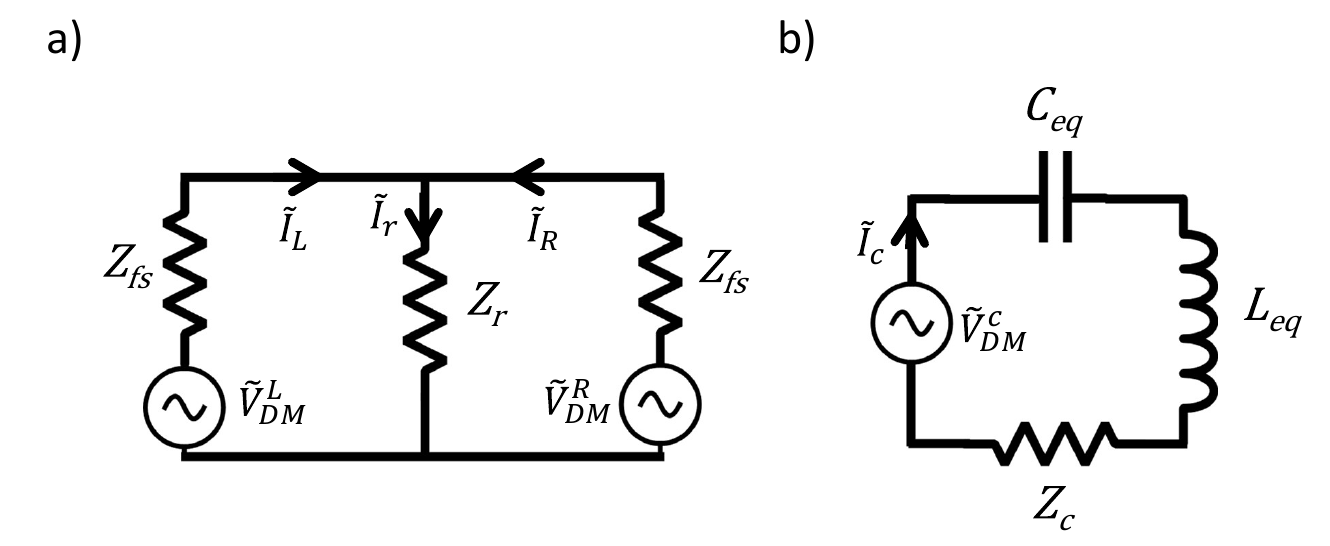}
\caption{Equivalent circuits of the (a) resistive sheet antenna and (b) half-wave cavity mode from Fig. \ref{fig:SheetvCavity}. The dark-matter source impedance, which presents as an equivalent resistor in series with the voltage drives, is not shown. \label{fig:SheetvCavity_CircuitModels}}
\end{figure}

See Fig. \ref{fig:SheetvCavity_CircuitModels}a for the resistive-antenna equivalent circuit. The sheet is represented by a resistor with value equal to its sheet impedance $Z_{r}$. As in Section \ref{ssec:ResvBroad}, the sheet, upon excitation by the dark-matter-induced electromagnetic field $\vec{E}_{DM}(t)$, radiates into two $\hat{z}$-polarized modes propagating in the $+\hat{y}$ and $-\hat{y}$ directions. The two radiation modes are represented by the resistors of value equal to the free-space impedance $Z_{fs}$, which is the characteristic impedance of radiation. The receiver also receives through these modes (eqs. (\ref{eq:sheet_Pin})-(\ref{eq:sheet_Prad})), so we model the receiver excitation as a voltage source in series with each radiation resistor. Each of the voltages and currents shown in the figure is a complex phasor, i.e. the time-domain voltage $V_{\rm DM}^{L,R}(t)$ is related to the voltage amplitude $\tilde{V}_{\rm DM}^{L,R}$ by
\begin{equation}
    V_{\rm DM}^{L,R}(t)=Re(\tilde{V}_{\rm DM}^{L,R} \exp(+i \omega_{\rm DM}^{0} t)).
\end{equation}
As we will show with Kirchhoff's Laws, the voltage sources $\tilde{V}_{\rm DM}^{L,R}$ map on to the dark-matter-induced electric field $\tilde{E}_{\rm DM}$ incident on the antenna. The voltage sources drive currents in the circuit. The current $\tilde{I}_{r}$ represents electronic current flow on the sheet, mapping to $\tilde{K}_{r}$ in (\ref{eq:CurrentPhasor}), while $\tilde{I}_{L}$ and $\tilde{I}_{R}$ represent the flow of displacement currents, i.e. the electromagnetic radiation from the sheet. 

Using symmetry and Kirchhoff's Current Law,
\begin{equation}\label{eq:KCL_sheet}
    \tilde{V}_{DM}^{L}=\tilde{V}_{DM}^{R},\ \tilde{I}_{L}=\tilde{I}_{R}=I_{r}/2,
\end{equation}
i.e. the sheet receives equally and radiates equally in the $\pm \hat{y}$ directions.  Then, from Kirchoff's Voltage Law, we find
\begin{equation}\label{eq:KVL_sheet}
    \tilde{V}_{\rm DM}^{L,R}=\tilde{I}_{r}Z_{r}+\frac{\tilde{I}_{r}}{2}Z_{fs} 
\end{equation}
for each of the voltage sources. The power dissipated in the sheet is then
\begin{equation}\label{eq:Pdiss_sheet_circuit}
    P_{r}=\frac{1}{2} |\tilde{I}_{r}|^{2} Z_{r}= \frac{Z_{r}}{2} \frac{|\tilde{V}_{\rm DM}|^{2}}{(Z_{r}+Z_{fs}/2)^{2}} = \frac{|\tilde{V}_{\rm DM}^{L,R}|^{2}}{Z_{fs}} \frac{Z_{fs}/2Z_{r}}{(1+Z_{fs}/2Z_{r})^{2}},
\end{equation}
which is maximized over sheet impedance when $Z_{r}=Z_{fs}/2$, for which
\begin{equation}\label{eq:Pdiss_sheet_max_circuit}
P_{r} \Big|_{\rm max} = \frac{1}{4} \frac{|\tilde{V}_{\rm DM}^{L,R}|^{2}}{Z_{fs}}.
\end{equation}

We may produce a mapping between the Kirchhoff's Law statements for the circuit and complex-power flow statements of the form (\ref{eq:DM_Poynting}), linking the equivalent-circuit voltages and currents to physical drive fields and currents. Combining the Kirchoff's Law statements, eqs. (\ref{eq:KCL_sheet}) and (\ref{eq:KVL_sheet}), we find
\begin{equation}\label{eq:KL_sheet_mapping}
    \frac{1}{2}\tilde{I}_{L}^{*}\tilde{V}_{DM}^{L} + \frac{1}{2}\tilde{I}_{R}^{*}\tilde{V}_{DM}^{R} = \frac{1}{2} |\tilde{I}_{L}|^{2}Z_{fs} + \frac{1}{2} |\tilde{I}_{R}|^{2}Z_{fs} + \frac{1}{2} |\tilde{I}_{r}|^{2}Z_{r}.
\end{equation}
This is the statement of power flow in the equivalent circuit: the power delivered by the dark-matter-induced voltage sources (left-hand side) is equal to the sum of the power dissipated in the radiation resistors $Z_{fs}$ and the power dissipated in the ohmic resistor $Z_{r}$ (right-hand side). It is mathematically equivalent to the complex-power flow statement from the Maxwellian approach,
\begin{align}
-\frac{1}{2\mu_{0}} \int_{\partial V_{r}} (\vec{E}_{\rm DM}(\vec{x},\omega_{\rm DM}^{0}) \times \vec{B}_{\rm rec}^{*}(\vec{x},\omega_{\rm DM}^{0}))\cdot \vec{da} &= \frac{1}{2} \int_{V_{r}} d^{3}\vec{x}\  \vec{J}_{\rm rec}^{*}(\vec{x},\omega_{\rm DM}^{0}) \cdot \vec{E}_{\rm tot}(\vec{x},\omega_{\rm DM}^{0}) \label{eq:MW_sheet_mapping}\\
&+\frac{1}{2\mu_{0}} \int_{\partial V_{r}} (\vec{E}_{\rm rec}(\vec{x},\omega_{\rm DM}^{0}) \times \vec{B}_{\rm rec}^{*}(\vec{x},\omega_{\rm DM}^{0}))\cdot \vec{da} \nonumber
\end{align}
under the mapping
\begin{equation}\label{eq:sheet_mapping}
    \tilde{V}_{\rm DM}^{L,R} \rightarrow \tilde{E}_{\rm DM}\sqrt{A},\ \tilde{I}_{r} \rightarrow \tilde{K}_{r}\sqrt{A}, 
\end{equation}
where the evaluation of the integrals can be found in eqs. (\ref{eq:sheet_Pin})-(\ref{eq:sheet_Prad}). Note also that the power dissipation expressions (\ref{eq:KVL_sheet}) and (\ref{eq:Pdiss_sheet_circuit}) match the analogous results from the Maxwellian approach, (\ref{eq:current_field_sheet}) and (\ref{eq:Pdiss_sheet}), under this mapping.
We may interpret $\frac{1}{2}|\tilde{I}_{L}|^{2}Z_{fs}$ as the radiation emitted into one mode (e.g. the $z$-polarized mode propagating in the $-\hat{y}$-direction), and $\frac{1}{2}|\tilde{I}_{R}|^{2}Z_{fs}$ as the radiation emitted into the other mode (e.g. the $z$-polarized mode propagating in the $+\hat{y}$-direction). In sum, these two equivalent-circuit power expressions map to the surface integral term, representing radiation, in (\ref{eq:MW_sheet_mapping}). We will make further use of the modal equivalent-circuit analysis in developing a general treatment of radiative coupling to the dark-matter signal.


The equivalent-circuit for the half-wave cavity mode is shown in Fig. \ref{fig:SheetvCavity_CircuitModels}b. The cavity mode is modeled as an equivalent, series-RLC circuit.\cite{montgomery1987principles} As in the antenna circuit, the ohmic resistor represents the sheet impedance $Z_{c}$. The inductor $L_{eq}$ and capacitor $C_{eq}$ represent the magnetic-field and electric-field energy storage between the sheets. The dark-matter drive, which we characterized in eq. (\ref{eq:cavity_drive}) as an effective capacitive coupling, is represented as a voltage $\tilde{V}_{DM}^{c}$ in series with the capacitor. The voltage drives a current $\tilde{I}_{c}$, which represents an electron current in the sheet resistor as well as displacement currents in the space between sheets. 

The resonance frequency and quality factor of the RLC circuit are
\begin{equation}\label{eq:toy_cavity_circuit_fr_Q}
    \omega_{r}=\frac{1}{\sqrt{L^{eq}C^{eq}}},\ Q=\frac{\omega_{r} L^{eq}}{Z_{c}}.
\end{equation} 
Kirchhoff's Laws applied to the equivalent circuit yields, for dark-matter frequencies close to resonance,
\begin{equation}\label{eq:KVL_cavity}
    \tilde{V}_{\rm DM}^{c}=\tilde{I}_{c} \left( Z_{c} + i \omega_{\rm DM}^{0} L^{eq} + \frac{1}{i\omega_{\rm DM}^{0} C^{eq}} \right) \approx \tilde{I}_{c} Z_{c} \left( 1 + 2iQ \frac{\omega_{\rm DM}^{0}-\omega_{r}}{\omega_{r}} \right),
\end{equation}
and an ohmic power dissipation of
\begin{equation}\label{eq:Pdiss_cavity_Lorentzian_circuit}
    P_{c} = \frac{1}{2} |\tilde{I}_{c}|^{2}Z_{c} = \frac{|V_{DM}^{c}|^{2}}{2Z_{c}} \frac{1}{|1+2iQ (\omega_{\rm DM}^{0}-\omega_{r})/\omega_{r}|^{2}}.
\end{equation}
The power dissipation is maximized on resonance,
\begin{equation}\label{eq:Pdiss_cavity_max_circuit}
    P_{c} \Big|_{\rm max}= \frac{|\tilde{V}_{\rm DM}^{c}|^{2}}{2Z_{c}}.
\end{equation}

Similar to the toy antenna, we may produce a mapping between the Kirchoff's Law statements for the cavity and complex-power flow statements from the Maxwellian approach, linking the equivalent-circuit parameters to field and current quantities. We demonstrate this mapping on resonance, $\omega_{DM}^{0}=\omega_{r}=c/2l$. The expression for complex-power flow in the equivalent-circuit can be found by multiplying (\ref{eq:KVL_cavity}) through by $\tilde{I}^{*}_{c}/2$,
\begin{equation}\label{eq:KL_cavity_mapping}
    \frac{1}{2}\tilde{I}^{*}_{c}\tilde{V}_{\rm DM}^{c}= \frac{1}{2}|\tilde{I}_{c}|^{2} \left( Z_{c} + i \omega_{\rm DM}^{0} L^{eq} + \frac{1}{i\omega_{\rm DM}^{0} C^{eq}} \right),
\end{equation}
i.e. complex-power delivered by the voltage source is equal to the complex-power flow into the resistor, inductor, and capacitor. Note that the electric-field energy and magnetic-field energy terms in (\ref{eq:DM_Poynting}), evaluated over the cavity volume $V_{c}$ of Section \ref{ssec:ResvBroad}, are
\begin{align}
    \frac{i\omega_{\rm DM}^{0}}{2\mu_{0}} & \int_{V_{c}} d^{3}\vec{x}\ |\vec{B}_{\rm rec}(\vec{x},\omega_{\rm DM}^{0})|^{2} \nonumber \\
    &\approx  \frac{i\omega_{r}}{2\mu_{0}} A |\tilde{K}_{c}|^{2} \frac{Z_{fs}^{2}}{4} \frac{1}{c^{2}} \int_{0}^{l} dy\ |\exp(-i \omega_{r} y/c) - \exp(+i\omega_{r}(y-l)/c))|^{2} \nonumber \\
    &= \frac{1}{2}A|\tilde{K}_{c}|^{2} i\omega_{r} \frac{\mu_{0} l}{2} \label{eq:cavity_B}
\end{align}
\begin{align}
    -\frac{i\omega_{\rm DM}^{0}\epsilon_{0}}{2} & \int_{V_{c}} d^{3}\vec{x}\ |\vec{E}_{\rm rec}(\vec{x},\omega_{\rm DM}^{0})|^{2} \nonumber \\
    &\approx  -\frac{i\omega_{\rm DM}^{0}\epsilon_{0}}{2} A  |\tilde{K}_{c}|^{2} \frac{Z_{fs}^{2}}{4} \int_{0}^{l} dy\ |\exp(-i\omega_{r} y/c) + \exp(+i\omega_{r}(y-l)/c))|^{2} \nonumber \\
    &= -\frac{i\omega_{r}\epsilon_{0}}{2} A  |\tilde{K}_{c}|^{2} \frac{Z_{fs}^{2}}{2} l= \frac{1}{2}A|\tilde{K}_{c}|^{2} \frac{1}{i\omega_{r}} \frac{\pi^{2}}{2\epsilon_{0}l} \label{eq:cavity_E}
\end{align}
Combining these equations with (\ref{eq:cavity_drive}) and (\ref{eq:cavity_diss}) shows that (\ref{eq:KVL_cavity}) maps onto the complex-power flow statement
\begin{align}
    \frac{i\omega_{\rm DM}^{0}\epsilon_{0}}{2} & \int_{V_{c}} d^{3}\vec{x}\ \vec{E}_{\rm DM}(\vec{x},\omega_{\rm DM}^{0}) \cdot \vec{E}_{\rm rec}^{*}(\vec{x},\omega_{\rm DM}^{0})= \frac{1}{2} \int_{V_{c}} d^{3}\vec{x}\  \vec{J}_{\rm rec}^{*}(\vec{x},\omega_{\rm DM}^{0}) \cdot \vec{E}_{\rm tot}(\vec{x},\omega_{\rm DM}^{0}) \nonumber \\
    &+ \frac{i\omega_{\rm DM}^{0}}{2\mu_{0}} \int_{V_{c}} d^{3}\vec{x}\ |\vec{B}_{\rm rec}(\vec{x},\omega_{\rm DM}^{0})|^{2} -\frac{i\omega_{\rm DM}^{0}\epsilon_{0}}{2} \int_{V_{c}} d^{3}\vec{x}\ |\vec{E}_{\rm rec}(\vec{x},\omega_{\rm DM}^{0})|^{2}
\end{align}
under the mapping
\begin{equation}\label{eq:cavity_mapping}
    \tilde{V}_{\rm DM}^{c} \rightarrow 2\tilde{E}_{\rm DM}\sqrt{A},\ \tilde{I}_{c} \rightarrow \tilde{K}_{c}\sqrt{A},\ L_{eq} \rightarrow \frac{\mu_{0}l}{2},\ C_{eq} \rightarrow \frac{2\epsilon_{0}l}{\pi^2}
\end{equation}
Note the factor of 2 in the mapping for the voltage source, relative to (\ref{eq:sheet_mapping}). This is due to the fact that the dark-matter-induced electric field coherently drives two sheet conductors in the toy cavity, as opposed to a single sheet conductor in the toy antenna. As with the antenna, under the mapping of equivalent-circuit voltage and current, the power dissipation expressions (\ref{eq:Pdiss_cavity_Lorentzian_circuit}) and (\ref{eq:Pdiss_cavity_max_circuit}) match the analogous results from the Maxwellian approach, (\ref{eq:Pdiss_cavity_Lorentzian}) and (\ref{eq:Pdiss_cavity_mode_max}). One may also have derived the equivalent self-inductance and self-capacitance by matching the resonance frequency and quality factor derived in Section \ref{ssec:ResvBroad}, $\omega_{r}=\pi c/l$ and $Q=\pi Z_{fs}/2Z_{c}$, to the equivalent-circuit resonance frequency and quality factor in (\ref{eq:toy_cavity_circuit_fr_Q}). As expected, the inductance and capacitance are proportional to the vacuum permeability and permittivity, respectively, which describe the ability to generate magnetic and electric fields in free space. A general treatment of the cavity equivalent-circuit, deriving circuit parameters from Maxwellian complex-power flow for an arbitrary resonant cavity mode (applying beyond the toy cavity), may be found in Appendix \ref{sec:cav_circuit}.

From eq. (\ref{eq:Pdiss_cavity_max_circuit}), we see that the on-resonance power dissipation for the half-wave cavity mode is independent of the free-space impedance. The $\pi$ phase shift between the sheet resistor and the backshort yields destructive interference between their radiation patterns, suppressing the radiation into free-space and the participation of the free-space impedance. As one reduces the sheet impedance $Z_{c}$, the power dissipation increases. However, the power available in the dark-matter source is finite, and the power dissipation cannot be increased arbitrarily, as power conservation must be obeyed. Once the power dissipation becomes an order-one fraction of the available power, backreaction processes become significant. Photons, produced by the cavity, convert into axions or hidden photons, representing an additional loss channel (in addition to conductive loss) for the cavity which limits the dark-matter power dissipated in the receiver to no more than the power available.


This backreaction on the dark-matter field can be modeled with a source impedance $Z_{\rm DM}$ in series with the voltage sources of Fig. \ref{fig:SheetvCavity_CircuitModels}. The effect on eq. (\ref{eq:KVL_cavity}) is to substitute $Z_{c} \rightarrow Z_{c}+Z_{DM}$. When the resonance frequency is tuned to the dark-matter signal and $Z_{c}$ is reduced below $Z_{\rm DM}$, the current in the circuit is dictated by $Z_{\rm DM}$ instead of $Z_{c}$, i.e. dictated by radiation of dark-matter instead of conductive loss. It is in this regime in which backreaction is significant. From (\ref{eq:Q_backaction}), we know the cavity quality factor, and therefore, the equivalent circuit resistance $Z_{c}$, at which backreaction on the dark-matter field becomes important. Eq. (\ref{eq:Q_backaction}) thus gives a scale for the effective source impedance of dark matter:
\begin{equation}\label{eq:ZDM}
Z_{\rm DM} \sim \begin{cases} \varepsilon^{2} Z_{fs} \frac{c}{v}, \hspace{1 cm} \textrm{hidden photons} \\
\left( \frac{\kappa_{a} c B_{0}}{m_{\rm DM}c/\hbar}  \right)^{2} Z_{fs} \frac{c}{v}, \hspace{1 cm} \textrm{axions}
\end{cases}
\end{equation}
See ref. \cite{chaudhuri2019dark} for a calculation of the axion and hidden-photon source impedance of a single plane-wave mode, directly from the governing equations of electrodynamics. The source impedance varies as the square of the coupling strength ($\varepsilon$ for hidden photons and $\kappa_{a}$ for axions). One power of coupling strength comes from the dark matter producing photons, and the other power of coupling strength comes from those photons driving the receiver, which in turn produces additional photons that convert to dark matter. For a virialized DFSZ axion\cite{dine1981simple,zhitnitskij1980possible} in a 10 Tesla magnetic field, the effective source impedance is $\sim 10^{-28} Z_{fs}$. The effective source impedance is the same for a virialized hidden photon with kinetic mixing angle $\sim 3 \times 10^{-16}$. Owing to the fact that dark matter is ``feebly coupled'' to electrons through electromagnetism (quantified by $\kappa_{a}$ for axions and $\varepsilon$ for hidden photons), the source impedance is ``small.'' 

Comparing the source impedance (\ref{eq:ZDM}) with the scale of the self-impedances in the equivalent circuits now reveals a primary limitation in impedance-matching to dark matter. In order to couple to the weak dark-matter source, either radiatively or reactively, the charges in the receiver must produce their own fields. The self-impedance exerted by these fields is much larger than the source impedance, producing an impedance mismatch which limits power dissipation in the receiver. For example, in the antenna, this self-impedance is the free-space impedance, representing receiver radiation and characterized by the real impedance of value $Z_{fs}$ in series with the voltage source. The mismatch between the source impedance and free-space impedance limits absorption at all possible dark-matter rest-mass frequencies, as represented mathematically by the presence of the free-space impedance in the denominators of eqs. (\ref{eq:Pdiss_sheet_max}) and (\ref{eq:Pdiss_sheet_max_circuit}). In the cavity, the self-impedance takes the form of a self-inductance and self-capacitance. (There is no radiation impedance. The cavity mode decouples from free-space, as described above.) Each of these impedances is imaginary, with magnitude near resonance on the order of the free-space impedance. On-resonance, the two self-impedances cancel. One may then absorb an order-one fraction of the available power in the dark-matter field with a sheet resistance comparable to the dark-matter source impedance. However, such low loss (extremely high quality factor) is impractical. For a dark-matter signal sufficiently far off-resonance, the two self-impedances do not cancel and in sum, are much larger than the source impedance, again limiting power dissipation in the receiver.

\emph{It is impractical to obtain an efficient impedance match to axion and hidden-photon dark matter. The impedance match is limited by the fact that the self-impedance of photons, produced by receiver charges coupling to the dark-matter excitation, is much larger than the source impedance. In particular, for a radiative coupling, the power dissipated in the receiver is limited by the mismatch between the free-space impedance and the dark-matter source impedance.}

To explicitly demonstrate the limitation in a generic radiatively coupled receiver (generalizing beyond the toy antenna), and consequent sensitivity disadvantage relative to reactive coupling, we use a modal analysis similar to that in ref. \cite{zmuidzinas2003thermal}. See also the discussion of waveguides in ref. \cite{pozar2012microwave}. We write the complex-power flow equation for the radiatively coupled receiver. By mapping an equivalent circuit to this power flow equation, we determine the maximum amount of power that may be dissipated in a radiatively coupled receiver, showing that it is limited by the free-space impedance. As before, we make the assumptions of a monochromatic dark-matter field, and (in the case of axions) a uniform, background DC magnetic field of infinite spatial extent, so that the dark-matter-induced electric field is uniform and given by eqs. (\ref{eq:Efield_hidden}) and (\ref{eq:Efield_axion}). We discuss at the end how one may relax these assumptions, with no significant effect on the conclusions. 

A radiatively coupled receiver may be described as possessing a set of transverse electromagnetic modes, defined over a surface $S$, through which the receiver is driven. We denote the modes, indexed by frequency, with complex electric-field and magnetic-field vectors
\begin{equation}
\{ \vec{E}_{k}(\vec{x},\omega_{\rm DM}^{0}), \vec{B}_{k}(\vec{x},\omega_{\rm DM}^{0}) \}_{k=1}^{N}.
\end{equation}
For each mode, the electric-field mode function, magnetic-field mode function, and surface normal vector are mutually orthogonal. Furthermore, the modes obey at each point on the surface
\begin{equation}
    \vec{B}_{k}= (\hat{n} \times \vec{E}_{k})/c,
\end{equation}
where $\hat{n}$ is the outwards surface normal. The modes also obey an orthogonality relation
\begin{equation}
    \int_{S} (\vec{E}_{p} \times \vec{B}^{*}_{q}) \cdot \vec{da} = \int_{S} \vec{E}_{p} \cdot \vec{E}^{*}_{q}\ da = \int_{S} \vec{B}_{p} \cdot \vec{B}^{*}_{q}\ da=0
\end{equation}
for $p \ne q$. As an example, the modes could be plane-wave modes, as in the toy antenna of Section \ref{ssec:ResvBroad}, with prescribed polarization and propagation direction; the modes could also represent the far-field of a vertically-polarized omnidirectional antenna, for which the modes, orthogonal over a sphere, are constructed from spherical Hankel functions and Legendre polynomials.\cite{chu1948physical}

The complex-power flow from the dark matter into the $k$th mode is, analogous to (\ref{eq:sheet_Pin}),
\begin{align}
    \tilde{P}_{in}^{k}(\omega_{\rm DM}^{0}) &= -\frac{1}{2\mu_{0}} \int_{S} (\vec{E}_{DM}(\vec{x},\omega_{\rm DM}^{0}) \times \vec{B}^{*}_{k}(\vec{x},\omega_{\rm DM}^{0})) \cdot \vec{da}\nonumber \\
    &= -\frac{1}{2Z_{fs}} \int_{S} (\vec{E}_{DM}(\vec{x},\omega_{\rm DM}^{0}) \cdot \vec{E}^{*}_{k}(\vec{x},\omega_{\rm DM}^{0}))\ da \label{eq:radcpl_Pin}
\end{align}
and the receiver-power radiated through the $k$th mode is, analogous to (\ref{eq:sheet_Prad}), 
\begin{equation}\label{eq:radcpl_Prad}
    P_{rad}^{k}(\omega_{\rm DM}^{0}) = \frac{1}{2\mu_{0}} \int_{S} (\vec{E}_{k}(\vec{x},\omega_{\rm DM}^{0}) \times \vec{B}^{*}_{k}(\vec{x},\omega_{\rm DM}^{0})) \cdot \vec{da} = \frac{1}{2Z_{fs}} \int_{S} |\vec{E}_{k}(\vec{x},\omega_{\rm DM}^{0})|^{2}\ da
\end{equation}
Thus, the statement of complex-power flow for the generic radiatively coupled receiver is, analogous to (\ref{eq:MW_sheet_mapping}),
\begin{equation}\label{eq:Poynting_radcpl}
    \sum_{k=1}^{N} \tilde{P}_{in}^{k}(\omega_{\rm DM}^{0}) = \sum_{k=1}^{N} P_{rad}^{k}(\omega_{\rm DM}^{0}) + P_{diss} (\omega_{\rm DM}^{0})+ j\mathcal{Q}(\omega_{\rm DM}^{0})
\end{equation}
where $P_{diss}(\omega_{\rm DM}^{0})$ represents receiver electronic dissipation and $\mathcal{Q}(\omega_{\rm DM}^{0})$ represents system reactances. For our quasi-1D toy antenna, the reactance term vanished, as there were no energy-storing elements in the receiver. Note that it is possible for an antenna to radiate, but not receive, through a mode, in which case $\tilde{P}_{in}^{k}(\omega_{\rm DM}^{0})=0$ but $P_{rad}^{k}(\omega_{\rm DM}^{0}) >0$.

\begin{figure}[htp] 
\includegraphics[width=\textwidth]{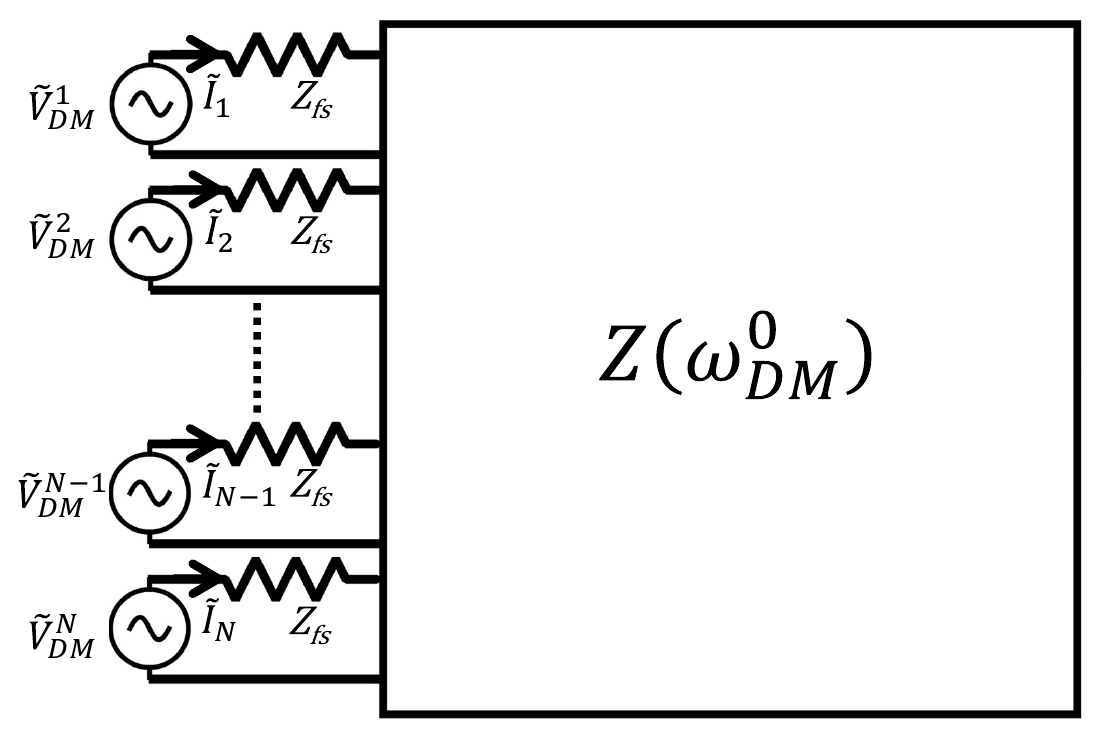}
\caption{Equivalent circuit for generic radiatively coupled receiver with $N$ modes. We have omitted the dependence of the voltage and current quantities on frequency, for brevity. \label{fig:RadiationCircuit}}
\end{figure}

The equivalent circuit that maps to this complex-power flow statement is shown in Fig. \ref{fig:RadiationCircuit}. Each of the modes is represented by a port, with dark-matter-induced voltage source $\tilde{V}_{\rm DM}^{k} (\omega_{\rm DM}^{0})$ driving current phasor $\tilde{I}_{k} (\omega_{\rm DM}^{0})$. The port possesses impedance $Z_{fs}$, the characteristic impedance of radiation, representing the receiver-power radiated through the mode. The receiver electronic dissipation and reactance is represented by an $N \times N$ impedance matrix, denoted $\mathbf{Z}(\omega_{\rm DM}^{0})$.\cite{pozar2012microwave} The impedance matrix relates the voltage sources to the currents,
\begin{equation}\label{eq:Z_matrix}
    \mathbf{\tilde{V}}_{\rm DM}(\omega_{\rm DM}^{0})=(Z_{fs}\mathbf{I}_{N} + \mathbf{Z}(\omega_{\rm DM}^{0})) \mathbf{\tilde{I}}(\omega_{\rm DM}^{0}),
\end{equation}
where $\mathbf{\tilde{I}}(\omega_{\rm DM}^{0})$ and $\mathbf{\tilde{V}}_{\rm DM}(\omega_{\rm DM}^{0})$ are column vectors of the port currents and port voltages, respectively, and $\mathbf{I}_{N}$ is the $N \times N$ identity matrix. To put Fig. \ref{fig:RadiationCircuit} in the context of our earlier work, note that the toy antenna corresponds to an equivalent circuit with $N=2$ modes and impedance matrix representing a shunt resistor of value $Z_{r}$.

Multiplying both sides of (\ref{eq:Z_matrix}) by one-half of the Hermitian conjugate of the port-current vector yields
\begin{equation}\label{eq:KL_radcpl_mapping}
    \frac{1}{2}\mathbf{\tilde{I}}^{\dag}(\omega_{\rm DM}^{0}) \mathbf{\tilde{V}}_{\rm DM}(\omega_{\rm DM}^{0})=\frac{1}{2}Z_{fs}\mathbf{\tilde{I}}^{\dag}(\omega_{\rm DM}^{0})\mathbf{\tilde{I}}(\omega_{\rm DM}^{0}) + \frac{1}{2}\mathbf{\tilde{I}}^{\dag}(\omega_{\rm DM}^{0}) \mathbf{Z}(\omega_{\rm DM}^{0}) \mathbf{\tilde{I}}(\omega_{\rm DM}^{0}).
\end{equation}
Analogous to (\ref{eq:KL_sheet_mapping}), this is the statement of complex-power flow in the equivalent circuit. It maps onto the Maxwellian statement of complex-power flow given in (\ref{eq:radcpl_Pin})-(\ref{eq:Poynting_radcpl}), via
\begin{equation}
    \frac{1}{2}\tilde{I}^{*}_{k}(\omega_{\rm DM}^{0})\tilde{V}_{\rm DM}^{k}(\omega_{\rm DM}^{0}) \rightarrow \tilde{P}^{k}_{in}(\omega_{\rm DM}^{0})
\end{equation}
\begin{equation}
    \frac{1}{2}|\tilde{I}_{k}(\omega_{\rm DM}^{0})|^{2} Z_{fs} \rightarrow \tilde{P}^{k}_{rad}(\omega_{\rm DM}^{0})
\end{equation}
\begin{equation}
    \frac{1}{2}\mathbf{\tilde{I}}^{\dag}(\omega_{\rm DM}^{0}) \mathbf{Z}(\omega_{\rm DM}^{0}) \mathbf{\tilde{I}}(\omega_{\rm DM}^{0}) \rightarrow P_{diss} (\omega_{\rm DM}^{0})+ j\mathcal{Q}(\omega_{\rm DM}^{0})
\end{equation}
In particular, combining the first two equations gives us an expression for the magnitude of the voltage source at each port, in terms of field quantities:
\begin{equation}\label{eq:radcpl_voltage}
    |\tilde{V}^{k}_{\rm DM}(\omega_{\rm DM}^{0})|^{2}= \frac{ |\int_{S} (\vec{E}_{DM}(\vec{x},\omega_{\rm DM}^{0}) \cdot \vec{E}^{*}_{k}(\vec{x},\omega_{\rm DM}^{0}))\ da|^{2}}{\int_{S} |\vec{E}_{k}(\vec{x},\omega_{\rm DM}^{0})|^{2}\ da}.
\end{equation}
The magnitude of the voltage excitation at the $k$th port is thus determined by the projection of the dark-matter-induced electric field onto the mode function for the $k$th mode. The third equation gives us an expression for power dissipation,
\begin{equation}
    P_{diss}(\omega_{\rm DM}^{0}) = Re \left( \frac{1}{2}\mathbf{\tilde{I}}^{\dag}(\omega_{\rm DM}^{0}) \mathbf{Z}(\omega_{\rm DM}^{0}) \mathbf{\tilde{I}}(\omega_{\rm DM}^{0}) \right),
\end{equation}
which we maximize in terms of the dark-matter-induced electric field. Re-expressing in terms of the scattering matrix\cite{pozar2012microwave},
\begin{equation}
    \mathbf{S}(\omega_{\rm DM}^{0})= (\mathbf{Z}(\omega_{\rm DM}^{0}) + Z_{fs}\mathbf{I}_{N})^{-1} (\mathbf{Z}(\omega_{\rm DM}^{0}) - Z_{fs}\mathbf{I}_{N}),
\end{equation}
the power dissipated in the receiver is
\begin{align}
    P_{diss}(\omega_{\rm DM}^{0}) &= \frac{1}{8Z_{fs}} Re \left( \mathbf{V}_{\rm DM}^{\dag} (\mathbf{I}_{N} - \mathbf{S}^{\dag} \mathbf{S}) \mathbf{V}_{\rm DM} \right) \nonumber \\
    & \leq \frac{1}{8Z_{fs}} \sum_{k=1}^{N} |\tilde{V}^{k}_{\rm DM}(\omega_{\rm DM}^{0})|^{2} \nonumber \\
    & \leq \frac{\int_{S} |\vec{E}_{\rm DM}(\vec{x},\omega_{\rm DM}^{0})|^{2}\ da}{8Z_{fs}} \label{eq:Pdiss_max_radcpl} 
\end{align}
where in the third line, we have used eq. (\ref{eq:radcpl_voltage}) as well as the fact that the electric-field mode functions are orthogonal. Plugging in the uniform dark-matter-induced electric fields of (\ref{eq:Efield_hidden}) and (\ref{eq:Efield_axion}) yields
\begin{equation} \label{eq:Pdiss_max_radcpl_uniform}
    P_{diss}(\omega_{\rm DM}^{0}) \leq  \frac{1}{8Z_{fs}} |\tilde{E}_{\rm DM}|^{2} |S|,
\end{equation}
where $|S|$ denotes the area of the receiving surface $S$. Note the presence of $Z_{fs}$ in the denominator of the maximum-power expression. Like eq. (\ref{eq:Pdiss_sheet_max}), it is indicative of the mismatch between the dark-matter source impedance and the free-space impedance, which limits the power received.\footnote{With some additional work, one may calculate the maximum power dissipation directly from (\ref{eq:Poynting_radcpl}), instead of using the equivalent circuit. However, relative to the equivalent-circuit approach, it is somewhat more cumbersome and yields little information as to why power absorption through radiative couplings is limited.} 

To illustrate the disadvantage that the impedance mismatch induces in sensitivity, consider a reactive, free-space cavity-mode coupling, of quality factor $Q$, driven by the dark-matter-induced electric field $\tilde{E}_{\rm DM}$ over an area comparable to that of $S$. (See discussion surrounding (\ref{eq:Pratio_horns}).) For an on-resonance dark-matter signal, the cavity absorbs $\sim Q$ times more power than the radiative coupling which saturates the bound (\ref{eq:Pdiss_max_radcpl_uniform}). The penalty in integration time at a possible dark-matter rest-mass frequency from needing to tune the cavity resonator is at most $\sim Q$, which means that the scan-integrated SNR advantage for the cavity is $\sim \sqrt{Q}$, similar to eq. (\ref{eq:SNR_ratio_sheet_param}). Thus, we find that radiatively coupled receivers are disadvantaged relative to reactively coupled receivers due to the impedance mismatch with the dark matter.

Note that, to derive the limitation on radiatively coupled receivers, we assumed a monochromatic dark-matter signal, and for axions, a uniform, background DC magnetic field of infinite spatial extent. Our statements regarding the advantage of reactive couplings generally still hold when these assumptions are relaxed. Our derivation of the key result (\ref{eq:Pdiss_max_radcpl}) was essentially agnostic to the bandwidth of the dark-matter signal; further discussion of the effect of the nonzero linewidth on the receiver impedance match to dark matter is included in Section \ref{ssec:reactive_circuit}. Following the same steps, one can derive an analogous result for background AC electromagnetic fields, showing that the power coupled to the receiver is limited by the mismatch between the axion source impedance and the free-space impedance. For finite-extent, non-uniform background fields, yielding non-uniform axion-induced electric fields, there is the possibility of constructive/destructive interference of axion-induced electric fields generated by different spatial points within the effective current source (\ref{eq:J_axion_0}). This effect is inherently frequency-dependent, resulting in enhanced field magnitudes on the receiving surface at some spatial points and possible rest-mass frequencies and reduced field magnitudes at other spatial points and frequencies. Such enhancements are the the foundation for certain experimental concepts based on collinear conversion and wire arrays \cite{sikivie2021invisible} in which engineering the periodicity of the background-field pattern allows one to achieve a multiplicative power enhancement comparable to the length scale of the magnet, expressed in terms of the number of Compton wavelengths, over a narrow band of possible dark-matter rest-mass frequencies. However, for virialized axions, the amount of possible power enhancement is limited by the coherence length ($\sim$1000 Compton wavelengths). More importantly for our discussion, eq. (\ref{eq:Pdiss_max_radcpl}) still applies, and the power coupled is ultimately limited by mismatch with the free-space impedance. For sufficiently high quality factor, a reactive cavity coupling achieves a superior scan-integrated sensitivity over a radiatively coupled receiver utilizing enhancement from periodicity engineering. One should also note that, practically, experiments based on periodicity engineering of the background field possess calibration challenges as misalignments of the receiver geometry or imperfections in the periodicity can destroy enhancement. Our above discussion regarding periodicity engineering of the background electromagnetic field does not apply to periodic dielectric haloscopes \cite{caldwell2017dielectric,millar2017dielectric,baryakhtar2018axion}, in which enhancements in coupled power are the result of reactive couplings to dielectric polarization. For our analysis, we thus distinguish between coupled-power enhancements resulting from a uniform background field interacting with a periodic structure of electrons and enhancements resulting from periodicity embedded in the background field itself. 


\emph{We thus conclude that, for scan-integrated sensitivity, receivers utilizing purely-radiative couplings and phase-insensitive amplifier readout are generally disadvantaged relative to reactive couplings on an apples-to-apples basis, due to the mismatch between the dark-matter source impedance and the free-space impedance. This motivates focusing on receivers with reactive couplings for the remainder of the optimization in this paper.}

\subsection{Equivalent-Circuits for Single-Moded Reactively Coupled Receivers}
\label{ssec:reactive_circuit}

\begin{figure}[htp] 
\includegraphics[width=\textwidth]{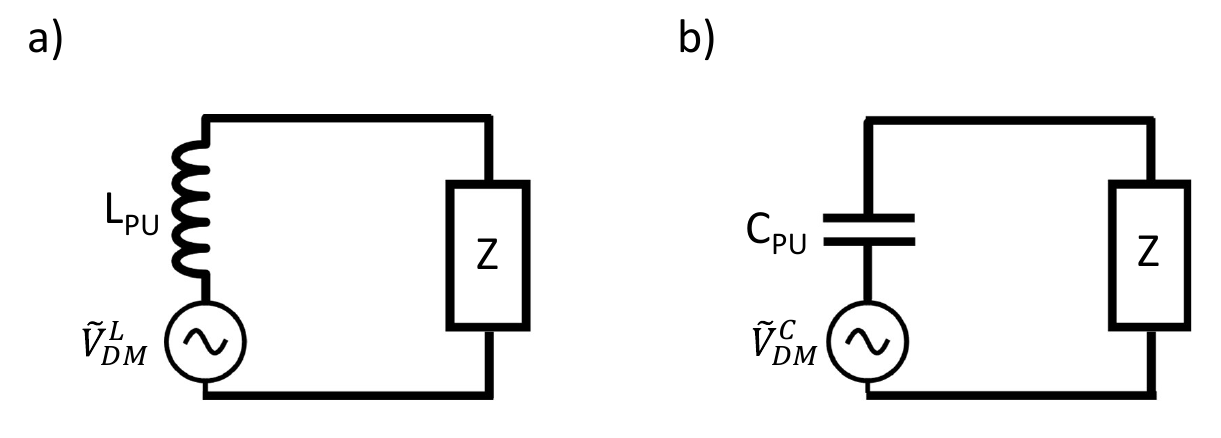}
\caption{Equivalent circuit for a) single-moded inductively coupled receiver and b) single-moded capacitively coupled receivers. The dark-matter-induced voltage, coupled through the pickup reactance, delivers power to the load impedance $Z$. The dependence of the voltage and load impedance $Z$ on drive frequency has been suppressed for brevity.  \label{fig:ReactiveCircuit}}
\end{figure}

In this section, we further narrow our focus to single-moded reactively coupled receivers, i.e. receivers which can be modeled in their equivalent circuit as being driven through a single inductor or a single capacitor. See Fig. \ref{fig:ReactiveCircuit}. While this is a considerable simplification, relative to the larger class of receivers reactively coupled to the dark-matter signal (which may generically be modeled as driving one or more equivalent inductances or capacitances and may also possess radiative couplings depositing power into the receiver), single-moded reactive couplings describe the vast majority of axion and hidden-photon searches. A single-moded reactive coupling describes searches in the quasi-static regime, such as DMRadio and ABRACADABRA, in which the receiver is driven through an equivalent inductance. It also describes cavity resonators, which are widely used in axion searches and can be modeled as single-pole, equivalent-RLC circuits driven through the inductor or capacitor. (See equivalent-circuit of toy cavity, as well as Appendix \ref{sec:cav_circuit}.) It is also important to remember that, practically, receivers using multi-moded couplings often possess significant calibration challenges associated with coherent interference from the excitations. We discuss how our optimization framework may be applied to multi-moded couplings in Section \ref{sssec:BF_evade}.

We compare impedance-matching to dark matter with inductive and capacitive coupling and then quantify the excitation from the dark matter as an equivalent voltage source in series with the reactance. This completes the optimization of the coupling element (first element in the first box of Fig. \ref{fig:DetectorBlockDiagram}). We then explain the importance of the matching network and noise in a global receiver optimization, motivating the calculations of signal-to-noise ratio produced in Section \ref{sec:resonator_SNR}.

To determine whether it is better to couple to the dark-matter signal inductively or capacitively (i.e. better to couple through magnetic-field energy-storing elements or electric-field energy-storing elements), one must compare the size of the receiver to the Compton wavelength of the dark matter. When the size of the receiver is much less the Compton wavelength of the dark matter, the effective dark-matter current source (\ref{eq:JDM_FFT}) can be analyzed quasi-statically, e.g. using Biot-Savart.\cite{Chaudhuri:2014dla,kahn2016broadband} The energy in the dark-matter-induced magnetic field is then much larger than the energy in the dark-matter-induced electric field, and one should couple to the dark-matter signal inductively. Any parasitic excitation of receiver capacitance is generally negligible. When the size of the receiver is comparable to or larger than the Compton wavelength of the dark matter, the dark-matter-induced electric and magnetic fields are comparable in strength. One may then couple to the signal either inductively or capacitively; there is no advantage of one over the other in terms of excitation strength. In the particular case of a resonant cavity mode (discussed further in Appendix \ref{sec:cav_circuit}), the receiver may physically possess both inductive and capacitive couplings, i.e. it may couple to both dark-matter-induced magnetic energy and electric energy. However, mathematically, for the equivalent circuit, the resonant cavity may be modeled as an equivalent-RLC circuit with solely an inductive coupling or solely a capacitive coupling, i.e. with the voltage source in series with inductor or in series with the capacitor. As we will see in optimizing the matching network and readout, the constraints on inductively coupled receivers and capacitively coupled receivers are similar, yielding analogous limits on search sensitivity. Thus, without loss of generality, we may consider receivers coupled to the dark-matter signal via an equivalent-circuit inductance for our global optimization.   

The excitation of a receiver driven through an equivalent-circuit inductance may be modeled with a voltage in series with the inductor $L_{\rm PU}$, denoted with phasor $\tilde{V}_{DM}^{L}$ in Fig. \ref{fig:ReactiveCircuit}a. We quantify the voltage for a monochromatic dark-matter signal and (for axions) for a background DC magnetic field $\vec{B}_{b}(\vec{x})$ of finite spatial extent. We discuss at the end of this section how we may generalize for a non-monochromatic signal, e.g. for a virialized dark-matter candiate. We may parametrize the square-modulus of the voltage phasor as
\begin{equation}\label{eq:VDM_mono}
    |\tilde{V}_{DM}^{L}(\omega_{\rm DM}^{0})|^{2}=4(\omega_{\rm DM}^{0})^{2} L_{\rm PU}U_{DM}^{L}(\omega_{\rm DM}^{0})
\end{equation}
where 
\begin{equation}\label{eq:UDM_mono}
    U_{DM}^{L}(\omega_{\rm DM}^{0})= \begin{cases}
     k(\omega_{\rm DM}^{0})^{2} \left( \frac{\kappa_{a} c^{2}}{\omega_{\rm DM}^{0}}  \right)^{2} \rho_{\rm DM} \int d^{3}\vec{x}\ |\vec{B}_{b}(\vec{x})|^{2}, \hspace{1 cm} \textrm{axions}\\
    k(\omega_{\rm DM}^{0})^{2}\varepsilon^{2} \rho_{\rm DM} V_{s}, \hspace{1 cm} \textrm{hidden photons}
    \end{cases} 
\end{equation}
will be referred to as coupled energy. (The quantity possesses units of energy.) $k(\omega_{\rm DM}^{0})$ is a non-negative geometric factor, relating to the shape of the dark-matter-induced drive fields, as well as the alignment of the pickup element with these drive fields. It may be calculated, as Schwinger states, by linking the equivalent-circuit quantities to Maxwellian field and current quantities. For example, in a resonant cavity, $k(\omega_{\rm DM}^{0})$ is directly related to the cavity overlap factor. (See Appendix \ref{sec:cav_circuit}.) For a given physically-lumped-element pickup inductor, it is maximized by aligning the field pattern of the inductor with the dark-matter-induced magnetic field. In practice, the geometric factor is determined by calibration and detector modeling. For axions, the integral over magnetic-field square-magnitude is performed over all space. For hidden photons, $V_{s}$ is the volume of the shield surrounding the apparatus, required to mitigate electromagnetic interference. Though the hidden-photon current source fills all space, as discussed in Section \ref{ssec:DM_Maxwell}, the receiver cannot couple dark-matter power from outside of the shield.\cite{Chaudhuri:2014dla} (This assumes that the shield is high-conductivity or superconducting and many skin depths or penetration depths thick.) The shield volume $V_{s}$ thus sets the size of the voltage excitation for the inductively coupled hidden-photon receiver. We return to the parametrization of the dark-matter excitation voltage (\ref{eq:VDM_mono}) in Section \ref{ssec:FOM}, where we discuss practical tradeoffs in optimizing an experimental design.

Having demonstrated the disadvantage of radiative couplings and quantified the excitation of the single-moded inductively coupled receiver as an equivalent-circuit voltage, we have completed the optimization of the coupling element of the dark-matter receiver, represented as element (1) in Fig. \ref{fig:DetectorBlockDiagram}. We now proceed to the remainder of the optimization. While single-pole resonators are broadly used in axion and hidden-photon searches (DMRadio in the quasistatic limit, ADMX and HAYSTAC in the cavity limit), they clearly do not constitute the optimal matching structure. In the context of the equivalent circuit shown in Fig. \ref{fig:ReactiveCircuit}a, for a single-pole resonator, the impedance $Z=Z(\omega_{\rm DM}^{0})$ consists of a series capacitor and resistor. The self-impedance of the coupling element, which limits the impedance match to dark matter, is nulled only at the resonance frequency, resulting in enhanced power absorption at that frequency. If the series capacitor is replaced with an equivalent negative inductor, then one can null the self-impedance over a broad range of possible dark-matter rest-mass frequencies, achieving a much better frequency-integrated impedance match to dark matter and a much better search sensitivity. Such reactance cancellation schemes are possible for any frequency range, e.g. both in the quasi-static limit and beyond the quasi-static limit.\cite{clarke2006squid,thorne1980gravitational,sussman2009non,wicht1997white} See also the discussion of negative dispersion and white-light cavities in ref. \cite{wicht1997white,salit2010enhancement}. This higlights the importance of considering in the optimization not only the receiver element that couples to the dark-matter signal, but also the elements that sit at the output of the coupling element. As described in the introduction, the elements at the output constitute the matching network, which can provide both an improved impedance match to the dark matter as well as an improved impedance match to the readout. 

At the same time, it is important to remember that not all matching networks are practically realizable. Negative inductors, as described above, are active-matching elements, requiring an external energy source to implement. The realization of active matching to enhance search sensitivity is challenging due to the tendency of active elements to generate excess noise and introduce instabilities. Matching networks that are readily realizable are generally linear and passive, so we restrict our attention to linear, passive matching networks and determine, under this assumption, the optimal matching network (single-pole resonant, reactive broadband as in ABRACADABRA, or otherwise) for an inductively coupled receiver. 

Because we require prohibitively low loss to absorb an order-one fraction of the $\sim10$ Watts of available power in the dark-matter field, the signal power transferred to a receiver is extremely small, below $\sim 10^{-22}$ Watts for a state-of-the-art cavity haloscope searching for DFSZ axion dark matter.\cite{brubaker2017first} Since the signal is small, it is necessary to also consider noise in the search optimization. One irreducible source of noise is thermal noise produced by loss in the receiver. The coupling element inevitably possesses loss, e.g. in the form of normal-metal ohmic loss, dielectric loss in wires and bulk dielectrics, and quasiparticle loss in superconductors. This loss may be represented as an equivalent-circuit resistance in series with the equivalent inductive coupling element. If one is using a phase-insensitive amplifier for readout, then the sensitivity is subject to quantum noise, including quantum backaction whose effect is dependent upon the matching network.\cite{caves1982quantum,clerk2010introduction} Thus far, we have considered power-matching to the dark-matter source in our optimization. Now we must simultaneously consider power-matching and noise-matching to the amplifier. Ideally, one would achieve an efficient power match and noise match at all frequencies in the search range, but as we will show, this is severely constrained by the Bode-Fano criterion. The next step in our optimization is to produce a sensitivity calculation for a light-field dark-matter receiver. Because we are analyzing an arbitrary single-moded inductively coupled receiver (not just single-pole resonators or reactive broadband structures), our calculation of SNR must be far more general than those that appear in the literature and universally consider specific receiver implementations.

An important aspect of our SNR analysis is taking into account the nonzero linewidth of the dark matter signal. Thus far, we have considered a monochromatic signal, but axion and hidden photon dark matter are most often considered as virialized, possessing bandwidth given in (\ref{eq:delta_nuDM}), $\Delta \omega_{\rm DM}(\omega_{\rm DM}^{0}) \sim 10^{-6} \omega_{\rm DM}^{0}$. As shown in Appendix \ref{ssec:DM_coh} and ref. \cite{Chaudhuri:2014dla}, the dark matter field can be treated as a superposition of oscillators spanning the bandwidth. Any two distinct frequency components beat against each other, causing the frequency of the signal to wander in the time domain. Consider such a field interacting with a resonant cavity, with mean frequency equal to the resonant frequency. If the range of the frequency wander is larger than the resonator linewidth, then the signal energy in the resonator is significantly reduced, relative to a monochromatic, on-resonance signal of equal drive strength. Analogous effects are possible in other receivers (not single-pole resonators) if they possess features in frequency space narrower than the dark-matter linewidth. To capture the physics of a non-monochromatic dark-matter signal in our sensitivity analysis, we must convolve the frequency response of the receiver with the dark-matter signal spectrum.

To facilitate further analysis, we thus generalize eq. (\ref{eq:VDM_mono}) to non-monochromatic dark-matter signals. We also create a unified language for discussing axions and hidden photons simultaneously. In particular, we define an effective, dimensionless coupling constant, denoted below as $g_{\rm DM}$ and based on the energy coupled to the receiver, that applies to both axions and hidden photons. In the axion case,$g_{\rm DM}$ is also a function of the magnetic field. For a virialized dark matter signal, the voltage is not described by a single phasor, but by a spectral density, which we denote $\mathcal{S}_{VV}^{\rm DM}$. The spectral density is proportional to the distribution of dark-matter energy over frequency denoted by $\frac{d\rho_{\rm DM}}{d\nu}(\nu,\nu_{\rm DM}^{0})$. $d\rho_{\rm DM}/d\nu$ has units of energy per-unit-volume per-unit-bandwidth and must satisfy
\begin{equation}
\int d\nu \frac{d\rho_{\rm DM}}{d\nu}(\nu,\nu_{\rm DM}^{0})= \rho_{\rm DM},
\end{equation}
where $\rho_{\rm DM} \sim$ 0.45 GeV/cm$^{3}$ is the local energy density. $d\rho_{\rm DM}/d\nu$ can be calculated from the energy density distribution over velocity; as an example, in Appendix \ref{sec:SH_Model}, we calculate $d\rho_{\rm DM}/d\nu$ explicitly for the standard halo model. 
From (\ref{eq:VDM_mono}), the voltage spectral density may then be written as
\begin{align}
\mathcal{S}_{VV}^{\rm DM}\left( \nu, g_{\rm DM}, \frac{d\rho_{\rm DM}}{d\nu}(\nu,\nu_{\rm DM}^{0}) \right) &= 2(2\pi\nu)^{2}L_{\rm PU} U_{\rm DM} \left( \nu, g_{\rm DM}, \frac{d\rho_{\rm DM}}{d\nu}(\nu,\nu_{\rm DM}^{0}) \right), \label{eq:SVV_DM}
\end{align}
where
\begin{equation}\label{eq:UDM_def}
    U_{\rm DM} \left( \nu, g_{\rm DM}, \frac{d\rho_{\rm DM}}{d\nu}(\nu,\nu_{\rm DM}^{0}) \right) = g_{\rm DM}^{2} \frac{d\rho_{\rm DM}}{d\nu} V_{\rm PU}
\end{equation}
is the spectral density describing coupled energy (possessing units of energy per unit bandwidth) and 
\begin{equation}\label{eq:gDM_def}
    g_{\rm DM}^{2}V_{\rm PU} \equiv \begin{cases}
     k(\nu_{\rm DM}^{0})^{2}  \left(\frac{\kappa_{a}} {m_{DM}/\hbar} \right)^{2} \int d^{3}\vec{x}\ |\vec{B}_{b}(\vec{x})|^{2}, \hspace{1 cm} \textrm{axions}\\
    k(\nu_{\rm DM}^{0})^{2} \varepsilon^{2} V_{s}, \hspace{1 cm}. \textrm{hidden photons}
    \end{cases} 
\end{equation}
Note that we have substituted frequency $\nu_{\rm DM}^{0}$ for angular frequency $\omega_{\rm DM}^{0}$. This is convenient for adopting the notation of ref. \cite{zmuidzinas2003thermal}, from which techniques are used for the sensitivity calculation. $\kappa_{a}$ is related to $g_{a\gamma\gamma}$ in eq. (\ref{eq:kappa_def}). $V_{\rm PU}$ is a fiducial ``pickup" volume chosen by the experimentalist. It can be conveniently set to, e.g., the volume of the magnet bore for axions or the shield volume for hidden photons. For convenience, we define a fiducial DC magnetic field strength $B_{0}$
\begin{equation}\label{eq:DC_energy_param}
    B_{0}^{2} \equiv \frac{1}{V_{\rm PU}} \int d^{3}\vec{x}\ |\vec{B}_{b}(\vec{x})|^{2} 
\end{equation}
$g_{\rm DM}$ is dependent on rest-mass frequency, i.e. $g_{\rm DM}=g_{\rm DM}(\nu_{\rm DM}^{0})$, but we omit the explicity dependence for brevity. Additionally, while the geometric factor for axions tends to be approximately independent of frequency $\nu$ within the dark-matter distribution, the geometric factor is generally dependent on frequency $\nu$ for hidden photons, i.e. we should write $k(\nu,\nu_{\rm DM}^{0})$ instead of $k(\nu_{\rm DM}^{0})$. While the direction of the dark-matter-induced electromagnetic fields driving the receiver is chosen by the experimentalist probing axions (by choosing the background DC magnetic field profile), it is unknown for the hidden photon. The direction may be different for different Fourier components within the dark-matter distribution, resulting in a different geometric factor for each component. For the consideration of signal to noise in the main text, we assume that the direction is the same for all Fourier components, so that the geometric factor is independent of $\nu$. In Appendix \ref{ssec:direction_variation}, we show how detector misalignment with the hidden-photon field may be mitigated by using three identical receivers oriented in mutually orthogonal directions, with a degradation in SNR no greater than a factor of $\sqrt{3}$. In the main text, we assume that the dark-matter density, (for hidden photons) direction, and velocity distribution are fixed during a search. Discussion of how to modify scan strategy to account for changes in these quantities can be found in Appendix \ref{sec:variation}.

\section{Sensitivity Calculation of a Search for Light-Field Dark Matter}
\label{sec:resonator_SNR}

For our sensitivity analysis, we first qualitatively discuss the processing of a signal through the receiver chain. The dark-matter signal has a finite coherence time, set by the inverse of the dark-matter bandwidth, (see Appendix \ref{ssec:DM_coh})
\begin{equation} \label{eq:DM_tcoh}
t_{\rm coh} \sim (\Delta\nu_{\rm DM})^{-1} \sim 10^{6}/\nu_{\rm DM}^{0}.
\end{equation}
On timescales much shorter than the dark-matter coherence time, the dark-matter signal behaves as a monochromatic wave with frequency $\nu_{\rm DM}^{0}$. One may determine if a timestream contains a signal by using a time-domain Wiener optimal filter \cite{wiener1949extrapolation}. This filter gives an estimate for the amplitude of the dark-matter signal if one exists in the timestream. If this amplitude is much greater than the uncertainty, set by the noise, then a candidate signal has been detected.

\begin{figure}[htp] 
\includegraphics[width=\textwidth]{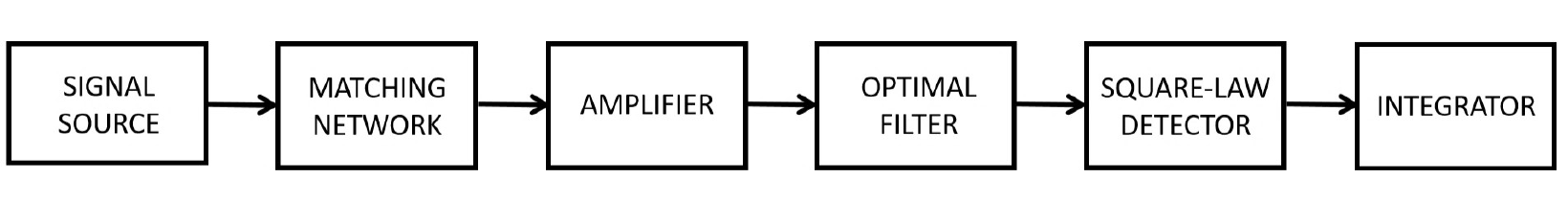}
\caption{Diagram showing the conceptual components of a Dicke radiometer. \label{fig:DickeDiagram}}
\end{figure}

However, stronger constraints on coupling parameter space require integration longer than the coherence time. Such integrations are the limit of interest in our paper.
For integration times much longer than the coherence time, the dark-matter signal no longer behaves as a phase-coherent amplitude signal. The complex phasor describing the oscillation at frequency $\nu_{\rm DM}^{0}$  (e.g. the integral expression in eq. (\ref{eq:axion_fft_slow})) has drifted significantly since the start of integration, and any averaging of the phasor over these timescales vanishes. The dark-matter signal in this regime behaves as an incoherent power signal. Conceptually, an incoherent power signal can be processed using a Dicke radiometer, first described in \cite{dicke1946measurement}. A schematic diagram of a Dicke radiometer, as it pertains to the dark-matter receiver, is shown in Fig. \ref{fig:DickeDiagram}. 

As indicated in Fig. \ref{fig:DetectorBlockDiagram}, the signals from the source, which in this case are the dark-matter signal, thermal noise, and zero-point fluctuation noise, are fed into a matching network and read out with an amplifier. The output signal of the amplifier, which includes both imprecision and backaction noise, is sent to an optimal filter to maximize SNR and then to a square-law detector, whose output voltage is proportional to the input power. This voltage is sent to an integrator. The square-law detector and integrator average down fluctuations in the noise power (i.e. thermal, zero-point fluctuation, and amplifier noise power), which set the variance (uncertainty) in the measurement scheme. When power is observed in excess of the mean noise power, then a candidate dark-matter signal has been detected, and follow-up is required to validate or reject this signal. In a typical modern experiment, the output of the amplifier is fed into a computer, and the optimal filtering, square-law detection, and integration steps are executed in software after digitization. 

The well-known expression for the SNR of a Dicke radiometer is given by
\begin{equation}\label{eq:Dicke}
SNR \approx \frac{P_{sig}}{P_{n}} \sqrt{\Delta \nu \cdot t},
\end{equation}
where $\Delta \nu$ is the signal bandwidth, $t$ is the integration time, and $P_{sig}$ and $P_{n}$ are, respectively, the signal and mean noise power within that bandwidth. 

Our analysis of signal processing does not directly use this formula. Rather, it uses signal manipulation parallel to that implemented in a Dicke radiometer in order to provide a rigorous framework for understanding sensitivity in dark-matter receivers. Importantly, our framework applies to any single-moded, reactively coupled receiver (single-pole resonant, broadband, or otherwise), which is necessary for optimization. Our framework also allows us to derive the optimal filter and to consider the effects on SNR due to combining information from multiple receiver configurations, e.g. from multiple frequency steps in a tunable resonator. The method is easily extended to the nonclassical detection schemes (e.g. squeezing and photon counting) motivated in Section \ref{sec:conclude}. Nevertheless, we use (\ref{eq:Dicke}) to build intuition about our results.

In Section \ref{ssec:amps}, we describe phase-insensitve amplifier measurement schemes and the Standard Quantum Limit on amplification. An accurate description of amplifier noise is central to our sensitivity calculations. In Sec. \ref{ssec:resonator_scattering}, we produce a scattering-matrix representation of a single-pole resonator, including the effect of the dark-matter signal quantified in Sec. \ref{ssec:reactive_circuit}, to model signal and noise transfer. In Sec. \ref{ssec:reactive_scattering}, we then generalize the scattering-matrix representation to any receiver with equivalent inductive-coupling to the dark matter. Sections \ref{ssec:resonator_scattering}-\ref{ssec:reactive_scattering} cover the first three blocks of Fig. \ref{fig:DickeDiagram}. Section \ref{ssec:SNR_search} covers the last three blocks and yields the expressions for SNR needed for search optimization.

\subsection{Amplifier Measurement Schemes and the Standard Quantum Limit}
\label{ssec:amps}

In a phase-insensitive amplifier, both quadratures of the signal are amplified equally. Ref. \cite{clerk2010introduction} illustrates that there are two forms of phase-insensitive amplification. The first is scattering-mode amplification, in which the amplifier measures a forward-scattering power wave. One example of a scattering-mode amplifier is a Josephson parametric amplifier, used in cavity haloscopes such as ADMX and HAYSTAC.\cite{castellanos2008amplification} The second is op-amp-mode amplification, in which the amplifier measures an input current or voltage. An example of an op-amp-mode amplifier is a SQUID (Superconducting Quantum Interference Device), which, given a current transduced to an input flux, produces an amplified voltage. The SQUID may be a dc SQUID, or if lower receiver loss is desired, a dissipationless rf SQUID.\cite{clarke2006squid,mates2008demonstration} SQUIDs are used in lumped-element searches such as DMRadio and ABRACADABRA.

For both scattering-mode amplifiers and op-amp mode amplifiers, the Heisenberg uncertainty principle dictates that the experimental sensitivity is subject to noise added by the phase-insensitive amplifier, characterized by a noise temperature $T_{N}(\nu)$.\footnote{The noise temperature of the amplifier $T_{N}(\nu)$ is defined as follows. The resistor at temperature $T$ within the input circuit (in our case, within the signal source specifically), produces a thermal noise spectral density proportional to the Bose-Einstein thermal occupation number $n(\nu)=(\textrm{exp}(h\nu/kT)-1)^{-1}$; see equations (\ref{eq:a1n}), (\ref{eq:th_occ}), and (\ref{eq:SVV_FD}). We may define an added noise number $N_{A}(\nu)$ as the increase in thermal-occupation number such that the increase in thermal noise equals the amplifier noise. The noise temperature is then given by $T_{N}(\nu) \equiv N_{A}(\nu)h\nu/k$. This definition is equivalent to that used in \cite{clerk2010introduction}, but differs from that in \cite{caves1982quantum}. We stress that this difference is simply one of convention, rather than one originating in physical principles. \label{TNnote}} 
The limitation set by Heisenberg is referred to as the Standard Quantum Limit (SQL)\cite{caves1982quantum} and in the high-gain regime, corresponds to a minimum amplifier noise temperature of one-half photon, $kT_{N}(\nu) = h\nu/2$. The quantum-limited amplifier noise constitutes one half of the quantum noise associated with the SQL, with the other half being provided by the zero-point fluctuation noise. In sum, the total signal-source-referred noise (amplifier noise + zero-point fluctuation noise + thermal noise) is at least one photon in excess of the thermal noise.

As described in Section \ref{sec:intro}, the amplifier noise can be decomposed into two effective noise sources: imprecision noise and backaction noise.   Under the definition of Standard Quantum Limit (see Appendices \ref{sec:SM_QL} and \ref{ssec:FtoV_QL} for more information as well as refs. \cite{clerk2010introduction,caves1980quantum,caves1981quantum,jaekel1990quantum}), the imprecision and backaction noise of a quantum-limited amplifier are uncorrelated, so the noise impedance (the impedance seen by the amplifier when noise-matched and thus the impedance at which the noise temperature of $kT_{N}(\nu)=h\nu/2$ is achieved) is real-valued.

Scattering-mode amplifiers and op-amp-mode amplifiers require two distinct formalisms to describe imprecision and backaction noise and the SQL. As such, we must treat them separately in our sensitivity and optimization analysis. The treatment of scattering-mode amplifiers is carried out in the main text, while the treatment for op-amp mode amplifiers is carried out in Appendices \ref{sec:FtoV_Amps} and \ref{sec:FtoV_opt}. There, we focus on the particular case of a flux-to-voltage amplifier, such as a SQUID. The results for optimization are identical for scattering-mode and op-amp-mode amplifiers.

\subsection{Scattering Representation of A Single-Pole Resonator}
\label{ssec:resonator_scattering}

\begin{figure}[htp] 
\includegraphics[width=\textwidth]{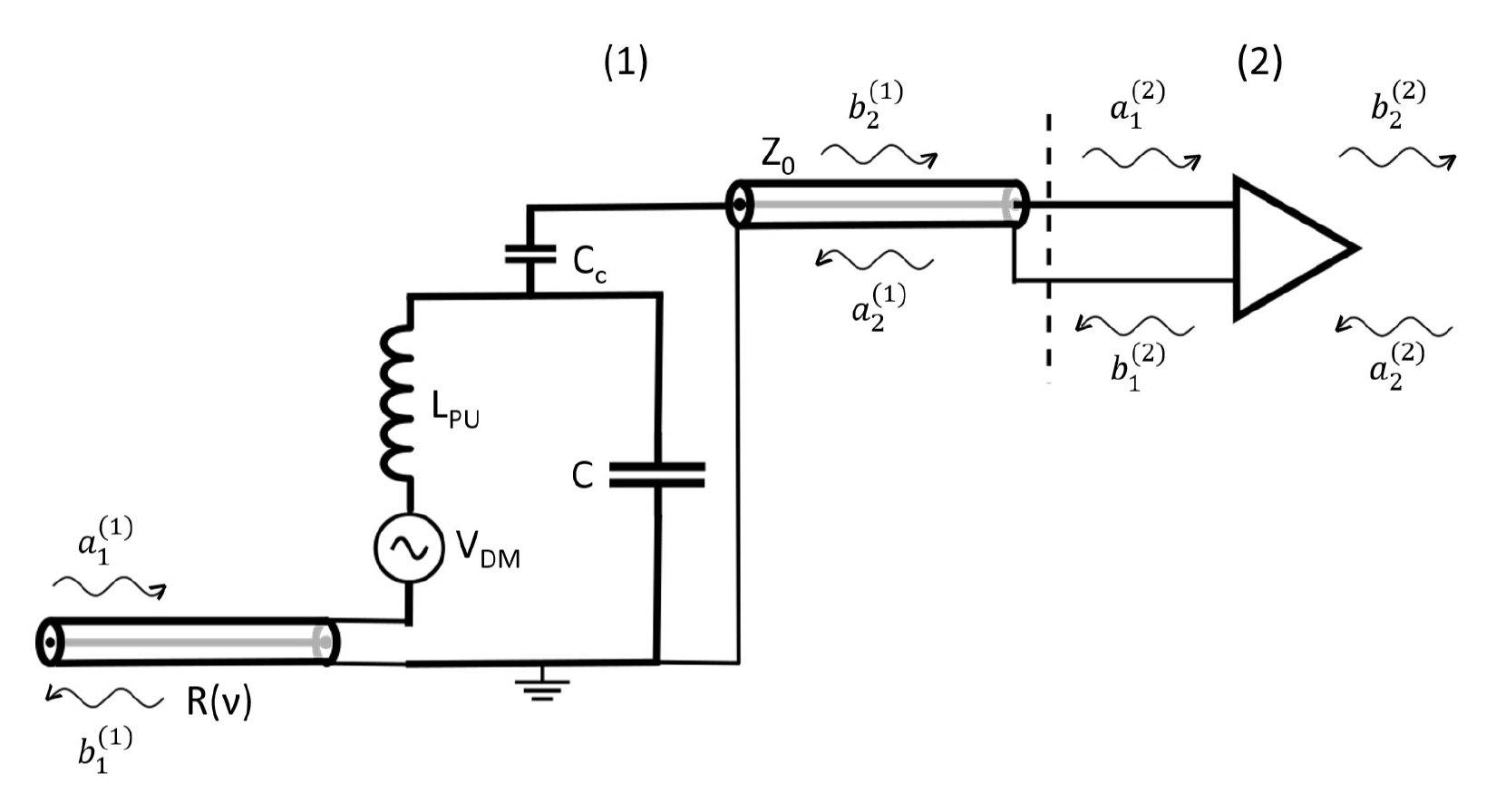}
\caption{Scattering representation of single-pole resonant receiver as a cascade of two-port circuits. Circuit (1), to the left of the dashed line, represents the resonator, while circuit (2), to the right of the dashed line, represents the amplifier. \label{fig:ResonatorScattering}}
\end{figure}

See Fig. \ref{fig:ResonatorScattering}, which displays a single-pole resonant receiver, capacitively coupled to a scattering-mode amplifier through a transmission line of characteristic impedance $Z_{0}$. In a typical cavity haloscope, the transmission line impedance is $Z_{0}=50$ Ohms. As described in Sections \ref{sec:intro} and \ref{ssec:reactive_circuit}, within the signal source, the single-pole resonator is driven by three equivalent-circuit voltage sources: a thermal noise voltage, a zero-point-fluctuation noise voltage, and a dark-matter signal voltage. The two noise voltages are sourced by the equivalent-circuit resistor, while the signal voltage is sourced by the equivalent-circuit inductor. The resistance $R(\nu)$ is represented by a semi-infinite, lossless transmission line of equal characteristic impedance; on this line, power is carried away from the system--equivalent to dissipation-- and thermal noise and zero-point-fluctuation noise are injected into the system. This is a typical representation in the quantum optics literature\cite{yurke1984quantum}. The dark-matter signal voltage, denoted in frequency space as $V_{\rm DM}=V_{\rm DM} \left( \nu,g_{\rm DM}, \frac{d\rho_{\rm DM}}{d\nu} \right)$ and characterized by spectral density (\ref{eq:SVV_DM}), is displayed in series with the inductance $L_{\rm PU}$, as in Fig. \ref{fig:ReactiveCircuit}.
Each of the voltage drives creates a forward-scattering wave on the transmission line that is amplified and appears at the output of the amplifier, along with the amplifier noise. To analyze the sensitivity, we must then describe the transfer function between the LR signal source and the amplifier output. The necessary description is obtained by expressing the resonant detection scheme in a scattering-matrix representation. We refer the reader to ref. \cite{pozar2012microwave} for more information on scattering matrix formalism.

As shown in Fig. \ref{fig:ResonatorScattering}, the single-pole resonant receiver may be represented as a cascade of two separate two-port circuits. The first of the two equivalent circuits describes the matching network, whose scattering parameters may be quantified in terms of the resonance frequency and quality factor. The receiver has resonance frequency $\nu_{r}=1/(2\pi\sqrt{L_{\rm PU}(C+C_{c})})$. The quality factor $Q$ of the resonator is determined by two sources of loss. First, resonator energy is lost by dissipation in the transmission line of impedance $R(\nu)$. This frequency-dependent resistance represents, for example, internal losses of the resonator due to loss in metals or loss in the quasiparticle system (if superconductors are used), loss in wire insulation (in the case of physically-lumped-element inductor coils), loss by coupling to parasitic electromagnetic modes, and loss in dielectrics. Second, resonator energy is lost by power coupling into the transmission line. The resonator quality factor is thus
\begin{equation}\label{eq:Q_def}
Q=(Q_{\rm int}^{-1} + Q_{\rm cpl}^{-1})^{-1},
\end{equation}
where the internal quality factor is
\begin{equation}\label{eq:Qint_def}
Q_{\rm int}=2\pi \nu_{r}L_{\rm PU}/R(\nu_{r}),
\end{equation}
and the coupled quality factor is
\begin{equation}\label{eq:Qc_def}
Q_{\rm cpl}=\frac{1}{(2\pi\nu_{r}C_{c})^{2}Z_{0}(2\pi\nu_{r}L_{\rm PU})}.
\end{equation}

The incoming and outgoing waves at each port in Fig. \ref{fig:ResonatorScattering} are defined in the frequency domain by the complex phasors (in units ${\rm \sqrt{Watts}/Hz}$)
\begin{equation} \label{eq:a_11}
a_{1}^{(1)}(\nu)= \frac{V_{1}(\nu)+I_{1}(\nu)R(\nu)}{2\sqrt{R(\nu)}}
\end{equation}
\begin{equation} \label{eq:a_21}
a_{2}^{(1)}(\nu)= \frac{V_{2}(\nu)+I_{2}(\nu)Z_{0}}{2\sqrt{Z_{0}}}
\end{equation}
\begin{equation} \label{eq:b_11}
b_{1}^{(1)}(\nu)=\frac{V_{1}(\nu)-I_{1}(\nu)R(\nu)}{2\sqrt{R(\nu)}}
\end{equation}
\begin{equation} \label{eq:b_21}
b_{2}^{(1)}(\nu)=\frac{V_{2}(\nu)-I_{2}(\nu)Z_{0}}{2\sqrt{Z_{0}}}
\end{equation}
where $V_{1,2}(\nu)$ and $I_{1,2}(\nu)$ are, respectively, input and output voltages and currents at the far ends of the two transmission lines. (Subscript 1 corresponds to the transmission line of impedance $R(\nu)$, and subscript 2 corresponds to the transmission line of impedance $Z_{0}$.) These wave amplitudes are related by
\begin{align} 
\left[ \begin{array}{c} b_{1}^{(1)}(\nu) \\ 
b_{2}^{(1)}(\nu) \end{array} \right] &= \left[ \begin{array}{cc} S_{11}^{(1)}(\nu,\nu_{r}) & S_{12}^{(1)}(\nu,\nu_{r}) \\ S_{21}^{(1)}(\nu,\nu_{r}) & S_{22}^{(1)}(\nu,\nu_{r}) \end{array} \right]
\left[ \begin{array}{c} a_{1}^{(1)}(\nu) \\ 
a_{2}^{(1)}(\nu) \end{array} \right] \nonumber \\
&+ \left[ \begin{array}{c} -\frac{S_{12}^{(1)}(\nu,\nu_{r})}{2\sqrt{R(\nu)}} V_{\rm DM}(\nu, g_{\rm DM}, \frac{d\rho_{\rm DM}}{d\nu}(\nu,\nu_{\rm DM}^{0})) \sqrt{\frac{Q_{\rm cpl}}{Q_{\rm int}}} \exp \left( +i \frac{2\pi \nu}{\bar{c}} l \right) \\ \frac{S_{21}^{(1)}(\nu,\nu_{r})}{2\sqrt{R(\nu)}} V_{\rm DM}(\nu, g_{\rm DM}, \frac{d\rho_{\rm DM}}{d\nu}(\nu,\nu_{\rm DM}^{0})) \end{array} \right]  \label{eq:S_res}
\end{align}
where
\begin{equation}\label{eq:S11}
S_{11}^{(1)}(\nu,\nu_{r})= \frac{ Q_{\rm int}-Q_{\rm cpl} +2i Q_{\rm int}Q_{\rm cpl} \left( \frac{\nu}{\nu_{r}} -1 \right)} {Q_{\rm int} + Q_{\rm cpl} + 2iQ_{\rm int}Q_{\rm cpl} \left( \frac{\nu}{\nu_{r}} -1 \right)}
\end{equation}
\begin{equation}\label{eq:S21}
S_{21}^{(1)}(\nu,\nu_{r})=S_{12}^{(1)}(\nu,\nu_{r}) = \frac{ 2\sqrt{Q_{\rm int}Q_{\rm cpl}}} {Q_{\rm int} + Q_{\rm cpl} + 2iQ_{\rm int}Q_{\rm cpl} \left( \frac{\nu}{\nu_{r}} -1 \right)} \exp \left( -i \frac{2\pi \nu}{\bar{c}} l \right)
\end{equation}
\begin{equation}\label{eq:S22}
S_{22}^{(1)}(\nu,\nu_{r})= \frac{ Q_{\rm cpl}-Q_{\rm int} +2i Q_{\rm int}Q_{\rm cpl} \left( \frac{\nu}{\nu_{r}} -1 \right)} {Q_{\rm int} + Q_{\rm cpl} + 2iQ_{\rm int}Q_{\rm cpl} \left( \frac{\nu}{\nu_{r}} -1 \right)} \exp \left(-i \frac{4\pi\nu}{\bar{c}} l \right)
\end{equation}
are the scattering parameters for the equivalent circuit. $l$ and $\bar{c}$ are the length and phase velocity of the $Z_{0}$ transmission line. In deriving eqs. (\ref{eq:S11})-(\ref{eq:S22}), we assume $|\nu-\nu_{r}| \ll \nu_{r}$; we also assume $Q, Q_{\rm int} \gg 1$, and $2\pi \nu_{r} Z_{0}C_{c} \ll 1$, which are typical design parameters for resonators.

An amplifier is represented in the second of the two two-port circuits. We assume that the amplifier has perfect input and output match to the transmission line, i.e. its input impedance is $Z_{0}$. Input and output match prevent the emergence of cavity modes on the transmission line, which complicate readout of the dark-matter signal. We also assume that the amplifier possesses reverse isolation, so that waves incoming on the right-hand port do not transmit to the resonant detector. Reverse isolation is an experimentally desired attribute for amplifiers, as it prevents heating of the resonator and interference due to spurious signals on the feedline. The amplifier is described by the following scattering relation for a signal at frequency $\nu$:
\begin{equation} \label{eq:S_amp}
\left[ \begin{array}{c} b_{1}^{(2)}(\nu) \\
b_{2}^{(2)}(\nu) \end{array} \right] = \left[ \begin{array}{cc} 0 & 0 \\ \sqrt{G(\nu)} & 0 \end{array} \right]
\left[ \begin{array}{c} a_{1}^{(2)}(\nu) \\
a_{2}^{(2)}(\nu) \end{array} \right] 
+ \left[ \begin{array}{c} c_{1}^{(2)}(\nu) \\
c_{2}^{(2)}(\nu) \end{array} \right] ,
\end{equation}
where $|G(\nu)|$ is the power gain of the amplifier. $c_{1}^{(2)}(\nu)$ and $c_{2}^{(2)}(\nu)$ are, respectively, the backward and forward-traveling noise waves generated by the amplifier. As dictated by the Heisenberg uncertainty principle and the Standard Quantum Limit, a phase-insensitive amplifier must necessarily add such noise waves in the detection scheme.\cite{caves1982quantum} The backward-traveling noise wave may be thought of as the backaction noise of the amplifier, while the forward-traveling noise wave is the imprecision noise. We assume that the amplifier has high power gain $|G(\nu)| \gg 1$ over the entire search band such that we may ignore the noise introduced by follow-on electronics (e.g. further analog amplification and filtering, digitization). This condition may be realized by using different high power gain amplifiers in the different frequency regimes. A possible implementation of this is discussed in \cite{Chaudhuri:2014dla}. 

Combining equations (\ref{eq:S_res}) and (\ref{eq:S_amp}) with the connection relations $a_{1}^{(2)}(\nu)=b_{2}^{(1)}(\nu)$ and $a_{2}^{(1)}(\nu)=b_{1}^{(2)}(\nu)$, we obtain the scattering relations for the cascaded circuit:
\begin{align}
\left[ \begin{array}{c} b_{1}^{(1)}(\nu) \\ b_{2}^{(2)}(\nu) \end{array} \right] &= \left[ \begin{array}{cc} S_{11}^{(1)}(\nu, \nu_{r}) & 0 \\ \sqrt{G(\nu)} S_{21}^{(1)}(\nu, \nu_{r}) & 0 \end{array} \right] 
\left[ \begin{array}{c} a_{1}^{(1)}(\nu) \\ a_{2}^{(2)}(\nu) \end{array} \right] \nonumber \\
&\hspace{0.5 cm} + \left[ \begin{array}{c} S_{12}^{(1)}(\nu,\nu_{r})c_{1}^{(2)}(\nu) \\ \sqrt{G(\nu)}S_{22}^{(1)}(\nu,\nu_{r})c_{1}^{(2)}(\nu) + c_{2}^{(2)}(\nu) \end{array} \right] \nonumber \\
&\hspace{0.5 cm} + \left[ \begin{array}{c} -\frac{S_{12}^{(1)}(\nu,\nu_{r})}{2\sqrt{R(\nu)}} V_{\rm DM} \left( \nu, g_{\rm DM}, \frac{d\rho_{\rm DM}}{d\nu}(\nu,\nu_{\rm DM}^{0}) \right) \sqrt{\frac{Q_{\rm cpl}}{Q_{\rm int}}} \exp \left( +i \frac{2\pi \nu}{\bar{c}} l \right) \\ \sqrt{G(\nu)} \frac{S_{21}^{(1)}(\nu,\nu_{r})}{2\sqrt{R(\nu)}} V_{\rm DM} \left( \nu, g_{\rm DM}, \frac{d\rho_{\rm DM}}{d\nu}(\nu,\nu_{\rm DM}^{0}) \right) \end{array} \right] .  \label{eq:S_total}
\end{align}

The left-hand side of equation (\ref{eq:S_total}) is a single vector. The top term of this vector $b_{1}^{(1)}(\nu)$ represents the wave that is dissipated in the resonator. The bottom term $b_{2}^{(2)}(\nu)$ represents the wave that is amplified and read out with further electronics. The signal and noise content of 
$b_{2}^{(2)}(\nu)$ determines the sensitivity of the detector. 

On the right-hand side of equation (\ref{eq:S_total}), there are three terms. The first term represents the response of the receiver to waves injected at the ports. Note that the response is independent of the incoming wave at port 2 of circuit (2); this is a result of output match and reverse isolation. For our calculation, $a_{1}^{(1)}(\nu)$ represents the thermal- and zero-point-fluctuation noise waves injected into the system. Assuming that the detector system is at temperature $T$, the noise correlation for this wave is
\begin{equation}\label{eq:a1n}
<a_{1}^{(1)}(\nu) (a_{1}^{(1)}(\nu'))^{*}>=h\nu (n(\nu,T) + 1/2) \delta(\nu-\nu'),
\end{equation}
where 
\begin{equation}\label{eq:th_occ}
n(\nu,T)=\frac{1}{\exp(h\nu/k_{B}T)-1}
\end{equation}
is the thermal occupation number of the resonator. We suppress the dependence on temperature $T$ where convenient. We have lumped together the thermal and zero-point noise; in (\ref{eq:a1n}), the ``$n(\nu,T)$'' term represents the thermal noise, while the ``$1/2$'' term represents the zero-point fluctuations. 

The second vector in the equation represents the response of the receiver to amplifier noise. The top term is the backaction noise that is dissipated in the resonator. The bottom term is the sum of the imprecision noise and the backaction noise which is reflected at the resonator and transmitted through the amplifier. The amplifier noise is typically quantified by use of a 2$\times$2 noise correlation matrix $C(\nu)$ with values given by
\begin{equation}\label{eq:Cn_amp}
<c_{i}^{(2)}(\nu)c_{j}^{(2)}(\nu')^{*}>=C_{ij}(\nu) \delta (\nu-\nu') ,
\end{equation}
where the values $i$ and $j$ are either 1 or 2. Note that $C_{12}(\nu)=C_{21}(\nu)^{*}$.

The third vector is the response of the receiver to the dark-matter signal, eqs. (\ref{eq:SVV_DM}) and (\ref{eq:UDM_def}). The top term is the portion of the signal that is dissipated in the resonator. The bottom term is the portion of the signal that is transmitted and amplified.

\subsection{Scattering Representation of A Single-Moded, Inductively Coupled Receiver}
\label{ssec:reactive_scattering}

\begin{figure}[htp] 
\includegraphics[width=\textwidth]{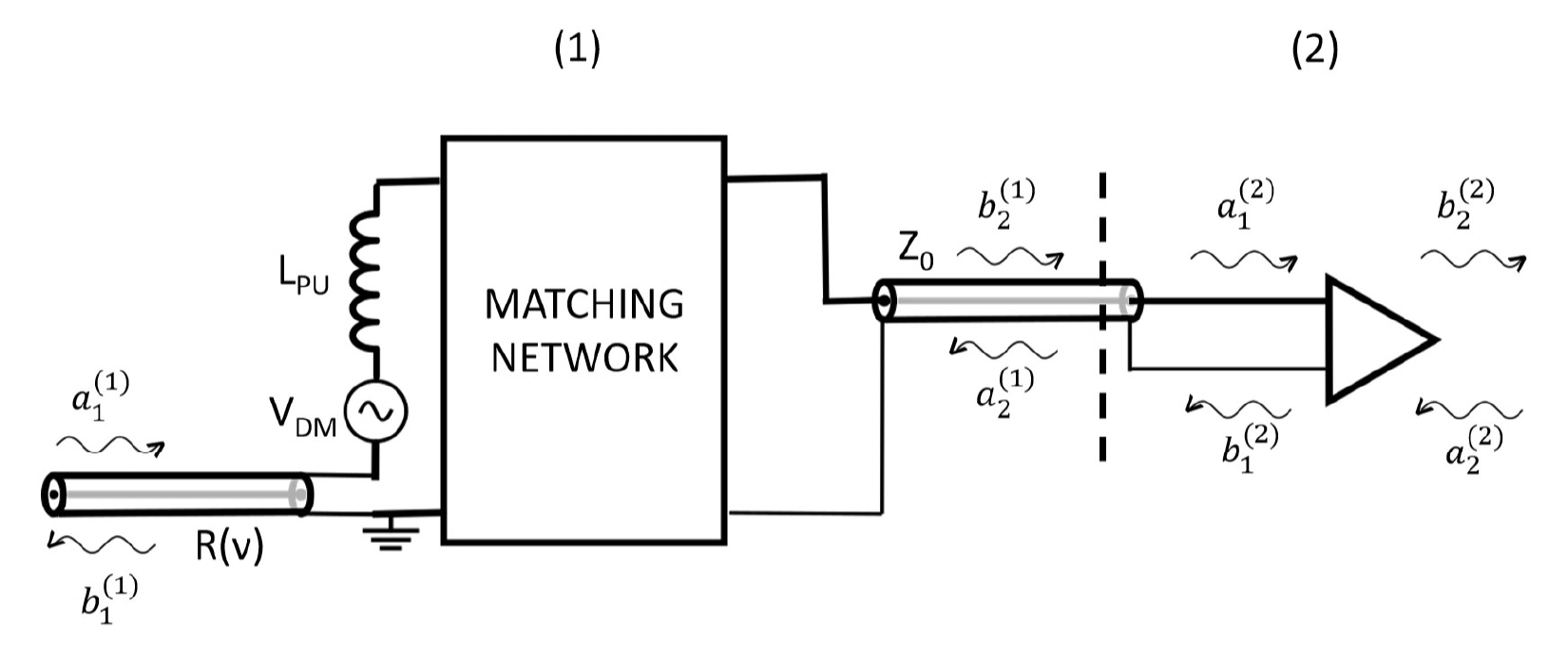}
\caption{Scattering representation of a generic inductively coupled detector as a cascade of two-port circuits. Equivalent circuit (1) represents the signal source and matching network, while equivalent circuit (2) represents the amplifier. \label{fig:ReactiveScattering}}
\end{figure}

Fig. \ref{fig:ReactiveScattering} illustrates the generalization of our resonator treatment to a single-moded detector coupled through an equivalent inductance to the dark matter signal and read out by a scattering-mode amplifier. 

As in the case of the single-pole resonator, the dark-matter signal presents as a voltage in series with the pickup inductor $L_{\rm PU}$, with spectrum described by (\ref{eq:SVV_DM}) and (\ref{eq:UDM_def}). The inductor is inevitably accompanied by some loss, denoted as the frequency-dependent quantity $R(\nu)$. The resistance is replaced by a transmission line. The signal and noise from the LR source are fed through the matching network to an amplifier.

The signal source and matching network are represented by one $2\times 2$ scattering matrix, denoted here as $S^{(1)}(\nu)$:
\begin{equation}\label{eq:S1_def}
S^{(1)}(\nu)= \left[ \begin{array}{cc} S_{11}^{(1)}(\nu) & S_{12}^{(1)}(\nu) \\ S_{21}^{(1)}(\nu) & S_{22}^{(1)}(\nu) \end{array} \right].
\end{equation}
For a resonator, the values of this scattering matrix are given by (\ref{eq:S11}), (\ref{eq:S21}) and (\ref{eq:S22}). As stated in the introduction, we assume that the matching network is linear and contains only passive components. We further assume that all loss is contained within $R(\nu)$, so that the matching network itself is lossless. (We extend to lossy matching networks in Section \ref{sssec:BF_evade}. Unsurprisingly, a lossy matching network is suboptimal.) Then, $S^{(1)}(\nu)$ is unitary at each frequency $\nu$:
\begin{equation}\label{eq:S1_unitary}
S^{(1)}(\nu) [S^{(1)}(\nu)]^{\dag} =I,
\end{equation}
where $I$ is the $2\times 2$ identity matrix.

The amplifier is described by (\ref{eq:S_amp}) and produces outgoing noise waves at both its input and output. In analogy with (\ref{eq:S_total}), it is straightforward to write an expression for the wave amplitude at the output of the amplifier:
\begin{align}
b_{2}^{(2)}(\nu) & = \sqrt{G(\nu)} S_{21}^{(1)}(\nu) a_{1}^{(1)}(\nu) + \sqrt{G(\nu)}S_{22}^{(1)}(\nu) c_{1}^{(2)}(\nu) + c_{2}^{(2)}(\nu) \nonumber \\
& \hspace{1 cm} + \sqrt{G(\nu)} \frac{S_{21}^{(1)}(\nu)}{2\sqrt{R(\nu)}} V_{\rm DM} \left( \nu, g_{\rm DM}, \frac{d\rho_{\rm DM}}{d\nu}(\nu,\nu_{\rm DM}^{0}) \right) \label{eq:b_out_amp},
\end{align}
where $a_{1}^{(1)}(\nu)$ obeys (\ref{eq:a1n}) and the amplifier noise waves obey (\ref{eq:Cn_amp}).

Armed with our results for the amplifier output, we now calculate the SNR that results from the optimal filtering, square-law detection, and integration steps, the last three blocks of the Dicke radiometer schematic in Fig. \ref{fig:DickeDiagram}. Our treatment focuses on statistical moments of the frequency components $b_{2}^{(2)}(\nu)$ at the output of the amplifier. See refs. \cite{zmuidzinas2003thermal} and \cite{zmuidzinas2015use} for a similar analysis of thermal noise, photon noise and amplifier noise in astronomical receivers. We assume that the amplifier gain is sufficiently large that the signal and noise can be treated as classical complex amplitudes, rather than introducing quantum-mechanical creation and annihilation operators. Sufficiently high gain implies that shot noise from the electromagnetic field in the receiver can be ignored in the square-law detection step. 

\subsection{Signal-to-Noise Ratio of Search}
\label{ssec:SNR_search}

We calculate the SNR at a single dark-matter search frequency in two steps. In the first step, we calculate the SNR from a single receiver configuration. For example, for a tunable single-pole resonator, a single receiver configuration consists of a single resonance frequency, which may or may not align with the dark-matter search frequency. In the second step, we discuss how information from multiple receiver configurations may be combined. We thus answer the question of how to calculate SNR from the combined datastream of multiple resonance frequencies, including multiple modes, or two entirely different detection setups. The third and final part of this section is dedicated to specific results for quantum-limited amplifiers, which represent a fundamental noise floor and are therefore used extensively in the search optimization of Section \ref{sec:scan_opt}. We conclude with some implications of the SNR analysis for a scan, which naturally introduce the value functions for the matching network optimization.

\subsubsection{SNR for single receiver configuration}
\label{sssec:SNR_one}

Suppose we have a single receiver configuration and we wish to test whether there is dark matter at frequency $\nu_{\rm DM}^{0}$. At the output of the amplifier (including further electronics, e.g. room-temperature amplification and mixing), we receive a timestream of values $b(t)$, which is related to the frequency-domain function $b_{2}^{(2)}(\nu)$ by
\begin{equation}\label{eq:bt_def}
b(t)= \int_{0}^{\infty} d\nu\ b_{2}^{(2)}(\nu) \exp(+i2\pi\nu t).
\end{equation}
In the remainder of this section, we drop the subscript and superscript in $b_{2}^{(2)}(\nu)$ and simply write $b(\nu)$, unless otherwise stated (since the wave $b_{1}^{(1)}(\nu)$ does not affect the SNR and is not frequently considered). We write $b^n(\nu)$ to indicate solely the noise terms in $b(\nu)$--the terms present even in the absence of a dark-matter signal. We also drop the limits on the frequency integrals; $(0,+\infty)$ is implied. 

The length $\tau$ of this timestream is assumed to be much longer than the dark-matter coherence time $t_{\rm coh} \sim 10^{6}/\nu_{\rm DM}^{0}$, as well as any characteristic time associated with the slowest pole in the receiver. We denote this characteristic time as $t_{\rm pole}$; it sets the time required for the receiver to reach steady state given a voltage excitation in the LR signal source. For a single-pole resonator, the characteristic time is the resonator ring-up time $\sim Q/\nu_{r}$. The assumption on the integration time implies that we need not consider transients and that Fourier transforms and time-harmonic representations of the amplifier output are appropriate mathematical tools. The assumption also implies that we can resolve the dark-matter linewidth, so that the measurement is not degraded by noise power outside of the signal bandwidth. 
We use this assumption frequently, so we define a timescale
\begin{equation}\label{eq:tstar_def}
t_{\rm rec}^{*}(\nu_{\rm DM}^{0}) \equiv \textrm{max} (10^{6}/\nu_{\rm DM}^{0}, t_{\rm pole}).
\end{equation}
The timescale is inherently dependent on both the dark-matter rest mass and the receiver parameters.

The timestream from the output of the amplifier is fed through a convolution filter described by the function $f(t)$, resulting in an output $(f * b)(t)$. The convolution filter is usually implemented in software, and the SNR can be maximized by deriving an optimal convolution filter.\cite{wiener1949extrapolation} In the frequency-domain, this convolution filter implements a multiplicative weighting on $b(\nu)$ and their complex conjugates $b^{*}(\nu)$.
\begin{equation}
b(\nu) \rightarrow f(\nu)b(\nu)
\end{equation}
\begin{equation}
b^{*}(\nu) \rightarrow f^{*}(\nu)b^{*}(\nu)
\end{equation}
Here, $f(\nu)$ are the Fourier components of $f(t)$ and $f^{*}(\nu)$ their complex conjugates. 

The filtered timestream is then fed through a square-law detector and integrator (also typically implemented in software). The output of the integrator is the time-averaged power at the output of the convolution filter, quantified by
\begin{equation} \label{eq:d_def}
d= \frac{1}{\tau}\int_{0}^{\tau} dt\ |(f \ast b)(t)|^{2}.
\end{equation}
The statistics of $d$ gives the signal-to-noise ratio. The dark-matter power in $d$ sets the signal, while the standard deviation of $d$, as produced by thermal, zero-point, and amplifier fluctuations, sets the noise. If, in an experiment, the total power in $d$ is in excess of the mean noise power in $d$ by many standard deviations, then a candidate dark-matter signal has been detected, and follow-up characterization is required.

We optimize SNR with respect to the filter $f$. We show below that the optimum filter depends on the shape of the dark-matter spectrum, but not on the coupling of dark matter to electromagnetism, quantified by $g_{\rm DM}$. We may therefore write
\begin{equation} \label{eq:f_full_param}
f(\nu)=f_{\rm rec} \left( \nu,\nu_{\rm DM}^{0}, \frac{d\rho_{\rm DM}}{d\nu} (\nu,\nu_{\rm DM}^{0}) \right).
\end{equation}
The optimal filter is also dependent on the receiver parameters, e.g. S-parameters. Indeed, we find that this is the case for all of the quantities calculated in this section. As such, for brevity, we skip full parametrization where convenient and write $f(\nu)$ with the other dependencies implicitly understood.

We examine the noise first. A useful quantity is the total noise correlator for the output Fourier components. The noise in the frequency domain is, from (\ref{eq:b_out_amp}),
\begin{equation} \label{eq:bn_nu_def}
b^{n}(\nu)=\sqrt{G(\nu)}S_{21}^{(1)}(\nu) a_{1}^{(1)}(\nu) + \sqrt{G(\nu)}S_{22}^{(1)}(\nu)c_{1}^{(2)}(\nu) + c_{2}^{(2)}(\nu).
\end{equation}
The correlator is defined by
\begin{equation} \label{eq:B_nu_def}
<b^{n}(\nu)b^{n}(\nu')^{*}> \equiv B(\nu) \delta(\nu-\nu').
\end{equation}
Plugging (\ref{eq:a1n}), (\ref{eq:Cn_amp}), (\ref{eq:bn_nu_def}) into (\ref{eq:B_nu_def}) yields
\begin{align} 
B(\nu) &= |G(\nu)| |S_{21}^{(1)}(\nu)|^{2} h\nu (n(\nu)+1/2) \label{eq:B_nu_eval} \\
&+ |G(\nu)||S_{22}^{(1)}(\nu)|^{2} C_{11}(\nu)+ C_{22}(\nu) + 2Re(\sqrt{G(\nu)}S_{22}(\nu)C_{12}(\nu)). \nonumber 
\end{align}
The first term on the right-hand side of the expression represents the thermal and zero-point noise. The second through fourth terms represent the amplifier noise contributions from, respectively, the backaction, the imprecision, and their correlation.

The mean noise power (the contribution of the noise modes to the mean of the time-averaged power $d$) is
\begin{align} 
<d^{n}> &=\frac{1}{\tau} \int_{0}^{\tau} dt \int_{0}^{\infty} d\nu \int_{0}^{\infty} d\nu'\ f(\nu) f(\nu)^{*} <b^{n}(\nu) b^{n}(\nu')^{*}> e^{+i2\pi(\nu-\nu')t} \nonumber \\
&= \frac{1}{\tau} \int_{0}^{\tau} dt \int d\nu \int d\nu' f(\nu) f(\nu')^{*} B(\nu) \delta(\nu-\nu') e^{+i2\pi(\nu-\nu')t} \nonumber \\
&= \int_{0}^{\infty} d\nu |f(\nu)|^{2} B(\nu). \label{eq:dn_mean}
\end{align}
The contribution of the noise modes to the second moment of $d$ is
\begin{align}
<(d^{n})^{2}> &= \frac{1}{\tau^{2}} \int_{0}^{\tau} dt \int_{0}^{\tau} dt' \int d\nu_{1} \int d\nu_{2} \int d\nu_{3} \int d\nu_{4}\ f(\nu_{1}) f^{*}(\nu_{2}) f(\nu_{3}) f^{*}(\nu_{4}) \nonumber \\
& \hspace{2 cm} <b^{n}(\nu_{1}) (b^{n}(\nu_{2}))^{*} b^{n}(\nu_{3}) (b^{n}(\nu_{4}))^{*}> e^{+i2\pi(\nu_{1}-\nu_{2})t} e^{+i2\pi(\nu_{3}-\nu_{4})t'}. \label{eq:d_soM}
\end{align}
Assuming that the noise modes are Gaussian (as one would have with a quantum-limited amplifier and thermal noise in the signal source), the four-point correlation may be evaluated by decomposing it into sums of products of two-point correlations. In particular,
\begin{align}
<b^{n}(\nu_{1}) (b^{n}(\nu_{2}))^{*} b^{n}(\nu_{3}) (b^{n}(\nu_{4}))^{*}> &= <b^{n}(\nu_{1}) (b^{n}(\nu_{2}))^{*}> <b^{n}(\nu_{3}) (b^{n}(\nu_{4}))^{*}> \nonumber \\
& \hspace{0.5 cm} <b^{n}(\nu_{1}) (b^{n}(\nu_{4}))^{*}> <b^{n}(\nu_{2}) (b^{n}(\nu_{3}))^{*}>.
\end{align}
Using equations (\ref{eq:B_nu_def}) and (\ref{eq:dn_mean}) gives
\begin{align}
<(d^{n})^{2}> &= <d^{n}>^{2} \\
& \hspace{0.5 cm} + \frac{1}{\tau^{2}} \int_{0}^{\tau} dt \int_{0}^{\tau} dt' \int d\nu_{1} \int d\nu_{3}\ |f(\nu_{1})|^{2}|f(\nu_{3})|^{2}B(\nu_{1}) B(\nu_{3}) e^{+i2\pi(\nu_{1}-\nu_{3})(t-t')}. \nonumber
\end{align}
When $|\nu_{1}-\nu_{3}|\tau \gg 1$,
\begin{equation}
\frac{1}{\tau^{2}} \int_{0}^{\tau} dt \int_{0}^{\tau} dt'\ e^{+i2\pi(\nu_{1}-\nu_{3})(t-t')} \approx \frac{1}{\tau}\delta(\nu_{1}-\nu_{3}). \label{eq:d_delta_func_approx}
\end{equation}
We may use this approximation because the integration time is much longer than $t_{\rm rec}^{*}(\nu_{\rm DM}^{0})$, set by the two characteristic times in the detection system--the dark-matter coherence time and the longest timescale associated with receiver poles.\cite{zmuidzinas2003thermal} The contribution of the noise modes to the variance in $d$ is thus
\begin{align} 
\sigma_{d} & \left( \tau, f_{\rm rec} \left( \nu,\nu_{\rm DM}^{0}, \frac{d\rho_{\rm DM}}{d\nu} (\nu,\nu_{\rm DM}^{0}) \right) \right)^{2} = <(d^{n})^{2}> - <d^{n}>^{2} \nonumber \\
&= \frac{1}{\tau} \int d\nu\ \left| f_{\rm rec} \left( \nu,\nu_{\rm DM}^{0}, \frac{d\rho_{\rm DM}}{d\nu} (\nu,\nu_{\rm DM}^{0}) \right) \right|^{4} B(\nu)^{2} \label{eq:noise_1}.
\end{align} 
This variance sets the uncertainty in power at the output of the integrator.

We now add the dark-matter signal to our analysis of the statistics of $d$. Using (\ref{eq:SVV_DM}), (\ref{eq:UDM_def}), (\ref{eq:b_out_amp}), and (\ref{eq:d_def}), we may write down the contribution of the signal to the integrated power $d$:
\begin{align} 
d_{\rm DM} & \left( \nu_{\rm DM}^{0}, g_{\rm DM}, \frac{d\rho_{\rm DM}}{d\nu} (\nu,\nu_{\rm DM}^{0}),  f_{\rm rec} \left( \nu,\nu_{\rm DM}^{0}, \frac{d\rho_{\rm DM}}{d\nu} (\nu,\nu_{\rm DM}^{0}) \right) \right)  \nonumber \\
& =\frac{1}{\tau} \int_{0}^{\tau} dt \int d\nu \int d\nu'\ f(\nu) f^{*}(\nu') e^{+i2\pi(\nu-\nu')t} \nonumber \\ 
&\hspace{1 cm} \sqrt{G(\nu)G(\nu')^{*}} S_{21}^{(1)}(\nu,\nu_{r}) S_{21}^{(1)}(\nu',\nu_{r})^{*} \frac{V_{\rm DM} \left( \nu, g_{\rm DM}, \frac{d\rho_{\rm DM}}{d\nu}(\nu,\nu_{\rm DM}^{0}) \right) V_{\rm DM}^{*} \left( \nu' , g_{\rm DM}, \frac{d\rho_{\rm DM}}{d\nu}(\nu',\nu_{\rm DM}^{0}) \right)}{4\sqrt{R(\nu)}\sqrt{R(\nu')}} \nonumber \\
&= \frac{1}{\tau} \int_{0}^{\tau} \int d\nu \int d\nu'\  f(\nu) f^{*}(\nu') \delta(\nu-\nu') e^{+i2\pi(\nu-\nu')t} \nonumber \\ 
&\hspace{1 cm} \sqrt{G(\nu)G(\nu')^{*}} S_{21}^{(1)}(\nu) S_{21}^{(1)}(\nu')^{*} \frac{\mathcal{S}_{VV}^{\rm DM}\left(\nu, g_{\rm DM}, \frac{d\rho_{\rm DM}}{d\nu}(\nu,\nu_{\rm DM}^{0}) \right)}{4\sqrt{R(\nu)}\sqrt{R(\nu')}} \nonumber \\
&= \int d\nu \left| f_{\rm rec} \left( \nu,\nu_{\rm DM}^{0}, \frac{d\rho_{\rm DM}}{d\nu} (\nu,\nu_{\rm DM}^{0}) \right) \right|^{2} P_{\rm DM} \left( \nu, g_{\rm DM}, \frac{d\rho_{\rm DM}}{d\nu} (\nu,\nu_{\rm DM}^{0}) \right), \label{eq:sig_1}
\end{align}
Implicit in our use of the delta-function normalization and the derivation of eq. (\ref{eq:sig_1}) are the assumptions that we can resolve the dark matter frequency spectrum into multiple small bins at the output of the amplifier and that the system is approximately in steady-state; these assumptions require that the integration time $\tau$ be much longer than $t^{*}_{\rm rec}(\nu_{\rm DM}^{0})$. $P_{\rm DM} \left( \nu, g_{\rm DM}, \frac{d\rho_{\rm DM}}{d\nu} (\nu,\nu_{\rm DM}^{0}) \right)$ is the power density of the signal at the output of the amplifier, given from (\ref{eq:SVV_DM}) and (\ref{eq:UDM_def}) by
\begin{align} \label{eq:PDM_sig}
P_{\rm DM} \left( \nu, g_{\rm DM}, \frac{d\rho_{\rm DM}}{d\nu} (\nu,\nu_{\rm DM}^{0}) \right) &= |G(\nu)| |S_{21}^{(1)}(\nu)|^{2} \frac{\mathcal{S}_{VV}^{\rm DM}\left(\nu, g_{\rm DM}, \frac{d\rho_{\rm DM}}{d\nu}(\nu,\nu_{\rm DM}^{0}) \right)}{4R(\nu)} \\
&= |G(\nu)| |S_{21}^{(1)}(\nu)|^{2} \frac{2(\pi\nu)^{2} L_{\rm PU}}{R(\nu)} U_{\rm DM} \left( \nu, g_{\rm DM}, \frac{d\rho_{\rm DM}}{d\nu} (\nu,\nu_{\rm DM}^{0}) \right). \nonumber
\end{align}
Note that the dark-matter signal also contributes to the variance of $d$, through cross-terms with the noise of the form $B(\nu)P_{\rm DM}(\nu)$. This is most readily observed by substituting $b^{n}(\nu) \rightarrow b(\nu)=b^{n}(\nu) + \sqrt{G(\nu)}S_{21}^{(1)}(\nu,\nu_{r}) \frac{V_{\rm DM} \left( \nu, g_{\rm DM}, \frac{d\rho_{\rm DM}}{d\nu}(\nu,\nu_{\rm DM}^{0}) \right)}{2\sqrt{R(\nu)}}$ into (\ref{eq:d_soM}) as required for a calculation of the total second-moment. We assume such cross-terms contribute negligibly to the variance in $d$. If they do not, then the dark matter signal power is comparable to or larger than the mean noise power and we can resolve the signal within a time shorter than or comparable to $t_{\rm rec}^{*}(\nu_{\rm DM}^{0})$. 

The SNR for the single measurement of length $\tau$ is then given by the dark-matter power (\ref{eq:sig_1}) (which is the power in excess of the mean noise power) divided by the standard deviation in noise power (\ref{eq:noise_1}):
\begin{align}
SNR & \left[ \nu_{\rm DM}^{0}, g_{\rm DM}, \frac{d\rho_{\rm DM}}{d\nu} (\nu,\nu_{\rm DM}^{0}), \tau, f_{\rm rec} \left( \nu,\nu_{\rm DM}^{0}, \frac{d\rho_{\rm DM}}{d\nu} (\nu,\nu_{\rm DM}^{0}) \right) \right] \nonumber \\
&= \frac{d_{\rm DM} \left( \nu_{\rm DM}^{0}, g_{\rm DM}, \frac{d\rho_{\rm DM}}{d\nu} (\nu,\nu_{\rm DM}^{0}),  f_{\rm rec} \left( \nu,\nu_{\rm DM}^{0}, \frac{d\rho_{\rm DM}}{d\nu} (\nu,\nu_{\rm DM}^{0}) \right) \right)}{\sigma_{d} \left(\tau, f_{\rm rec} \left( \nu,\nu_{\rm DM}^{0}, \frac{d\rho_{\rm DM}}{d\nu} (\nu,\nu_{\rm DM}^{0}) \right) \right)} \nonumber \\
&= \frac{\int d\nu \left| f_{\rm rec} \left( \nu,\nu_{\rm DM}^{0}, \frac{d\rho_{\rm DM}}{d\nu} (\nu,\nu_{\rm DM}^{0}) \right) \right|^{2} P_{\rm DM} \left( \nu, g_{\rm DM}, \frac{d\rho_{\rm DM}}{d\nu} (\nu,\nu_{\rm DM}^{0}) \right)}{  \left( \tau^{-1} \int d\nu |f_{\rm rec} \left( \nu,\nu_{\rm DM}^{0}, \frac{d\rho_{\rm DM}}{d\nu} (\nu,\nu_{\rm DM}^{0}) \right)|^{4} B(\nu)^{2} \right)^{1/2}}. \label{eq:SNR_func_f}
\end{align}  

The optimal filter, which maximizes SNR, is given by
\begin{equation}\label{eq:f_opt}
\left| f_{\rm rec}^{\rm opt}\left( \nu,\nu_{\rm DM}^{0}, \frac{d\rho_{\rm DM}}{d\nu} (\nu,\nu_{\rm DM}^{0}) \right) \right|^{2} = \frac{ P_{\rm DM} \left( \nu, g_{\rm DM}, \frac{d\rho_{\rm DM}}{d\nu} (\nu,\nu_{\rm DM}^{0}) \right)}{ g_{\rm DM}^{2} B(\nu)^{2}}.
\end{equation}
From equation (\ref{eq:SNR_func_f}), observe that multiplication of this filter by any constant preserves the value of the SNR. In particular, because $g_{\rm DM}$ is a constant, the optimal filter can be scaled to be independent of coupling. We show this explicitly by omitting $g_{\rm DM}$ from the arguments on the left-hand side of $(\ref{eq:f_opt})$ and by adding a normalization $1/g_{\rm DM}^{2}$ to cancel the $g_{\rm DM}^{2}$ factor in $P_{\rm DM}$, as can be seen from equations (\ref{eq:SVV_DM}), (\ref{eq:UDM_def}), (\ref{eq:PDM_sig}).

This form for the filter is qualitatively expected. The filter gives greater weight to the frequency bins that intrinsically contain more of the dark-matter signal (higher $d\rho_{\rm DM}/d\nu$), and lower weight to the frequency bins where noise is large (higher $B(\nu)$). For the optimal filter of (\ref{eq:f_opt}), the SNR is
\begin{align}
SNR^{\rm opt} & \left[ \nu_{\rm DM}^{0}, g_{\rm DM}, \frac{d\rho_{\rm DM}}{d\nu}(\nu,\nu_{\rm DM}^{0}), \tau \right] \nonumber \\
& \equiv SNR  \left[ \nu_{\rm DM}^{0}, g_{\rm DM}, \frac{d\rho_{\rm DM}}{d\nu} (\nu,\nu_{\rm DM}^{0}), \tau, f_{\rm rec}^{\rm opt} \left( \nu,\nu_{\rm DM}^{0}, \frac{d\rho_{\rm DM}}{d\nu} (\nu,\nu_{\rm DM}^{0}) \right) \right] \nonumber \\
&= \left( \tau \int d\nu \frac{ P_{\rm DM} \left( \nu, g_{\rm DM}, \frac{d\rho_{\rm DM}}{d\nu} (\nu,\nu_{\rm DM}^{0}) \right)^{2}}{B(\nu)^{2}} \right)^{1/2} \label{eq:SNR_1_opt}.
\end{align}
This form suggests another method for understanding the optimal filter of equation (\ref{eq:f_opt}). Consider a narrow frequency interval $[\nu,\nu+\delta\nu]$. The signal power in this bin is $\approx P_{\rm DM} \left( \nu, g_{\rm DM}, \frac{d\rho_{\rm DM}}{d\nu} (\nu,\nu_{\rm DM}^{0}) \right) \delta \nu$ and the noise power $\approx B(\nu) \delta \nu$. Thus, using eq. (\ref{eq:Dicke}), the SNR from only this bin is
\begin{equation} \label{eq:SNR_small_bin}
SNR_{bin}\left[ \nu_{\rm DM}^{0}, g_{\rm DM}, \frac{d\rho_{\rm DM}}{d\nu}(\nu,\nu_{\rm DM}^{0}), \tau \right]= \frac{P_{\rm DM}\left( \nu, g_{\rm DM}, \frac{d\rho_{\rm DM}}{d\nu} (\nu,\nu_{\rm DM}^{0}) \right)}{B(\nu)} \sqrt{\delta\nu \cdot \tau}.
\end{equation}
Comparison of (\ref{eq:SNR_1_opt}) and (\ref{eq:SNR_small_bin}) reveals that the optimal filter adds in quadrature the SNRs from all frequency bins in which there is dark-matter signal power. 


\subsubsection{SNR for multiple receiver configurations: combining data}
\label{sssec:SNR_scan}

We index the various receiver configurations with an index $i$. A quantity with the subscript or superscript $i$ relates to the $i$th receiver (for example, the $i$th resonance frequency setting of a single-pole resonator). We wish to test, with data from this set of receivers, whether the dark-matter rest mass is $m_{\rm DM}=\frac{h\nu_{\rm DM}^{0}}{c^{2}}$. Assume that for the $i$th receiver, we integrate for time $\tau_{i}$ and that this time is much longer than $t_{\rm rec,i}^{*}(\nu_{\rm DM}^{0})$.

We assume that the receiver timestreams represent approximately uncorrelated samples of the dark-matter signal. In a scan of a single tunable receiver, this is accomplished by waiting much longer than $t_{\rm rec,i}^{*}(\nu_{\rm DM}^{0})$ at each step, so that each integration exceeds the dark matter coherence time and so that the phases of the dark-matter excitations in the different receiver configurations are, to good approximation, distinct. Then, the appropriate combination of data for the total SNR at a dark-matter search frequency is
\begin{equation}\label{eq:data_sum}
\sum_{i} w_{i} d_{i},
\end{equation}
where $d_{i}$ is defined as in equation (\ref{eq:d_def}) and represents the time-averaged output power from $i$th receiver configuration. $\{ w_{i} \}$ are weights to the data that will be optimized. Note, in contrast, that, for two receivers scanning the same search frequency at the same time, the receiver timestreams represent correlated samples of the dark-matter signal. In this case, one would perform coherent addition of the signal amplitudes, rather than incoherent addition of the signal powers (\ref{eq:data_sum}). See Section \ref{sssec:BF_evade} for more discussion of such simultaneous scans.

For each receiver configuration, from the averaged output of the convolution filter, we obtain a signal power $d_{\rm DM,i}$ given by equation (\ref{eq:sig_1}) and a noise $\sigma_{d_{i}}$ given by equation (\ref{eq:noise_1}). The total SNR is
\begin{align}
SNR_{\rm tot} & \left[ \nu_{\rm DM}^{0}, g_{\rm DM,i}, \frac{d\rho_{\rm DM}}{d\nu}(\nu,\nu_{\rm DM}^{0}), \{ \tau_{i} \}, \left\{ f_{\rm rec,i} \left( \nu,\nu_{\rm DM}^{0}, \frac{d\rho_{\rm DM}}{d\nu}(\nu,\nu_{\rm DM}^{0}) \right) \right \}, \{ w_{i} \} \right] \nonumber \\ 
&= \frac{\sum_{i} w_{i} d_{\rm DM,i}  \left( \nu_{\rm DM}^{0}, g_{\rm DM,i}, \frac{d\rho_{\rm DM}}{d\nu} (\nu,\nu_{\rm DM}^{0}), f_{\rm rec,i} \left( \nu,\nu_{\rm DM}^{0}, \frac{d\rho_{\rm DM}}{d\nu} (\nu,\nu_{\rm DM}^{0}) \right) \right) }{ \left( \sum_{i} w_{i}^{2} \sigma_{d_{i}}\left(\tau_{i}, f_{\rm rec,i} \left( \nu,\nu_{\rm DM}^{0}, \frac{d\rho_{\rm DM}}{d\nu} (\nu,\nu_{\rm DM}^{0}) \right) \right)^{2} \right)^{1/2}}. \label{eq:weights_SNR}
\end{align}
We have allowed the dark-matter coupling $g_{\rm DM, i}$ to vary among receivers. This could be the result of possessing different geometrical factors $k(\nu_{\rm DM}^{0})$ in each configuration--see eq. (\ref{eq:gDM_def}). For any choice of weighting parameters $\{ w_{i} \}$,
\begin{align}
SNR_{\rm tot} & \left[ \nu_{\rm DM}^{0}, \{ g_{\rm DM,i} \}, \frac{d\rho_{\rm DM}}{d\nu}(\nu,\nu_{\rm DM}^{0}), \{ \tau_{i} \}, \left\{ f_{\rm rec,i} \left( \nu,\nu_{\rm DM}^{0}, \frac{d\rho_{\rm DM}}{d\nu}(\nu,\nu_{\rm DM}^{0}) \right) \right \}, \{ w_{i} \} \right] \nonumber \\
& \leq  \left(  \sum_{i} \frac{  d_{\rm DM,i}  \left( \nu_{\rm DM}^{0}, g_{\rm DM,i}, \frac{d\rho_{\rm DM}}{d\nu} (\nu,\nu_{\rm DM}^{0}), f_{\rm rec,i} \left( \nu,\nu_{\rm DM}^{0}, \frac{d\rho_{\rm DM}}{d\nu} (\nu,\nu_{\rm DM}^{0}) \right) \right)^{2} }{ \sigma_{d_{i}}\left( f_{\rm rec,i} \left( \nu,\nu_{\rm DM}^{0}, \frac{d\rho_{\rm DM}}{d\nu} (\nu,\nu_{\rm DM}^{0}) \right) \right)^{2} } \right)^{1/2}. \label{eq:weights_ineq}
\end{align}
The highest $SNR$ (equality in the above equation) is achieved if \begin{equation} \label{eq:weights_max}
w_{i} = \kappa \frac{d_{\rm DM,i}  \left( \nu_{\rm DM}^{0}, \{ g_{\rm DM,i} \}, \frac{d\rho_{\rm DM}}{d\nu} (\nu,\nu_{\rm DM}^{0}), f_{\rm rec,i} \left( \nu,\nu_{\rm DM}^{0}, \frac{d\rho_{\rm DM}}{d\nu} (\nu,\nu_{\rm DM}^{0}) \right) \right)}{\sigma_{d_{i}}\left( \tau_{i}, f_{\rm rec,i} \left( \nu,\nu_{\rm DM}^{0}, \frac{d\rho_{\rm DM}}{d\nu} (\nu,\nu_{\rm DM}^{0}) \right) \right)^{2}},
\end{equation}
where $\kappa$ is arbitrary. In particular, for the optimal filter (\ref{eq:f_opt}), eqs. (\ref{eq:noise_1}) and (\ref{eq:sig_1}) give the relation
\begin{align}
d_{\rm DM,i} & \left( \nu_{\rm DM}^{0}, g_{\rm DM,i}, \frac{d\rho_{\rm DM}}{d\nu} (\nu,\nu_{\rm DM}^{0}), f_{\rm rec,i}^{\rm opt} \left( \nu,\nu_{\rm DM}^{0}, \frac{d\rho_{\rm DM}}{d\nu} (\nu,\nu_{\rm DM}^{0}) \right) \right) \nonumber \\
&= g_{\rm DM,i}^{2} \tau_{i} \sigma_{d}\left( \tau_{i}, f_{\rm rec,i}^{\rm opt} \left( \nu,\nu_{\rm DM}^{0}, \frac{d\rho_{\rm DM}}{d\nu} (\nu,\nu_{\rm DM}^{0}) \right) \right)^{2} \label{eq:f_opt_weights}.
\end{align}
so that the weights 
\begin{equation}\label{eq:weights_opt}
w_{i}=g_{\rm DM,i}^{2}\tau_{i}
\end{equation}
correspond to $\kappa=1$ in (\ref{eq:weights_max}).

It follows from (\ref{eq:SNR_func_f}) that the SNR for the $i$th receiver configuration is the square root of the summand in (\ref{eq:weights_ineq}). Thus, when $SNR_{\rm tot}$ is maximized with respect to the weights, the total SNR is the quadrature sum of the SNRs from each receiver configuration. 
The maximized SNR over all configurations then results from the optimally filtered SNR for a single receiver configuration in (\ref{eq:SNR_1_opt}):
\begin{equation} \label{eq:SNR_opt_scan}
SNR_{\rm tot}^{\rm opt} \left[ \nu_{\rm DM}^{0}, \{ g_{\rm DM,i} \}, \frac{d\rho_{\rm DM}}{d\nu} (\nu,\nu_{\rm DM}^{0}), \{ \tau_{i} \} \right]= \left( \sum_{i} \tau_{i} \int d\nu \frac{P_{\rm DM,i} \left( \nu, g_{\rm DM,i}, \frac{d\rho_{\rm DM}}{d\nu} (\nu,\nu_{\rm DM}^{0}) \right)^{2}}{B_{i}(\nu)^{2}} \right)^{1/2}.
\end{equation}
We emphasize that this result applies not only to scan steps of a tunable resonator, but also to more general configuration changes involving non-resonant receivers, or to datastreams from entirely different instruments, subject to conditions discussed previously that the amplitude, frequency distribution, and (in the case of hidden photon) direction of the dark-matter field are fixed. The result (\ref{eq:SNR_opt_scan}) is central to our time allocation optimization in Section \ref{ssec:opt_time}.

\subsubsection{SNR of search with a quantum-limited amplifier} 
\label{sssec:SNR_QLA}



We assume that the imprecision and backaction noise modes of the amplifier are uncorrelated, so that
\begin{equation}\label{eq:c1c2n}
C_{12}(\nu)=C_{21}(\nu)^{*}=0.
\end{equation}
Such an assumption is part of the definition of the Standard Quantum Limit for phase-insensitive amplifiers operated in the scattering mode.\cite{clerk2010introduction} Additionally, uncorrelated backaction and imprecision is often achieved in practical setups by inserting a circulator, with one port terminated by a cold matched load (e.g. a 50 $\Omega$ resistor matched to a 50 $\Omega$ transmission line and held at temperature $kT_{\rm load} \ll h\nu$), between the matching network and amplifier input. For a two-port amplifier, the native backaction is absorbed by the matched load. In this case, the noise injected into the detector (circuit (1) in Fig. \ref{fig:ReactiveScattering}) from the matched load effectively plays the role of the backaction noise mode $c_{1}^{(2)}(\nu)$ in (\ref{eq:S_amp}). For a one-port amplifier, such as a resonant Josephson parametric amplifier \cite{castellanos2008amplification}, a circulator is used to embed the device in a two-port environment. Again, the backaction noise mode is effectively provided by the matched resistor, e.g. either from the fourth port on the embedding four-port circulator or from the terminated port on a separate, preceding three-port circulator. The circulator at the amplifier input also provides reverse isolation, protecting the resonator from any interference that propagates from the follow-on readout chain toward the detector. 

In Appendix \ref{sec:SM_QL}, we show that for a quantum-limited amplifier, the noise waves of equations (\ref{eq:S_amp}) and (\ref{eq:Cn_amp}) possess autocorrelations \cite{zmuidzinas2003thermal,clerk2010introduction}
\begin{equation}\label{eq:c1n}
C_{11}(\nu)=\frac{h\nu}{2} 
\end{equation}
\begin{equation}\label{eq:c2n}
C_{22}(\nu)=\frac{h\nu}{2}(|G(\nu)|-1).
\end{equation}
Note that in the particular case of a circulator inserted between the matching network and amplifier, it is critical that the termination on the circulator be cold, held at temperature $kT_{\rm load} \ll h\nu$, in order to obtain quantum-limited performance. The noise coming from the matched load (the termination), which effectively provides the backaction mode, possesses autocorrelation $C_{11}^{\rm load}(\nu)=h\nu(n_{\rm load}(\nu,T) +1/2)$, where $n_{\rm load}(\nu,T)=(\textrm{exp}(h\nu/kT_{\rm load})-1)^{-1}$ is the thermal occupation number of the load. Only when $kT_{\rm load} \ll h\nu$, so that $n_{\rm load}(\nu,T)<<1$, can we (approximately) achieve quantum-limited performance as governed by eqs. (\ref{eq:c1n}) and (\ref{eq:c2n}).

Combining equations (\ref{eq:a1n}) and (\ref{eq:c1c2n})-(\ref{eq:c2n}) then yields the following expression for $B(\nu)$:
\begin{align}
B(\nu) & =|G(\nu)||S_{21}^{(1)}(\nu)|^{2} h\nu \left( n(\nu)+ \frac{1}{2} \right) + |G(\nu)||S_{22}^{(1)}(\nu)|^{2} \frac{h\nu}{2} + \frac{h\nu}{2} (|G(\nu)|-1) \nonumber \\
& \approx h\nu |G(\nu)| \left( |S_{21}^{(1)}(\nu)|^{2} n(\nu) + 1 \right), \label{eq:B_QLamp}
\end{align}
where, since $S^{(1)}(\nu)$ represents a lossless, linear, passive receiver, we have used the power conservation relation resulting from unitarity (\ref{eq:S1_unitary})
\begin{align} \label{eq:ScattRelation}
|S_{21}^{(1)}(\nu)|^{2} + |S_{22}^{(1)}(\nu)|^{2} =1,
\end{align}
as well as the high-gain approximation $|G(\nu)| \gg 1$. $B(\nu)$ is readily interpreted as the mean noise power per unit bandwidth at the output of the amplifier. Here, $|S_{21}^{(1)}(\nu)|^{2} n(\nu)$ represents the thermal noise power transmitted from the signal source. The ``+1'' term in the approximate form of eqn. (\ref{eq:B_QLamp}) represents the quantum noise associated with the SQL. This noise incorporates 1/2 photon per unit bandwidth from the amplifier's added output noise (imprecision noise). The other 1/2 photon is split between the amplifier backaction and the zero-point energy in the signal source. The relative contributions are determined by the fraction of the zero-point energy transmitted through the matching network (with amplitude $|S_{21}^{(1)}|^2$) and the fraction of the backaction power reflected off of the impedance filter (with amplitude $|S_{22}^{(1)}|^2$). The two are related by eqn. (\ref{eq:ScattRelation}), summing to the 1/2 photon contribution in the SQL. Since the transmission is less than or equal to unity, we find that the signal-source-referred noise power per unit bandwidth is at least one photon plus the thermal noise. This is consistent with the definition of SQL given in Section \ref{ssec:amps}.

Furthermore, from (\ref{eq:B_QLamp}), we may also recognize that, for a quantum-limited amplifier, the amplifier noise impedance is equal to its input impedance $Z_{0}$, typically 50 $\Omega$ (see Fig. \ref{fig:ReactiveScattering}). When the source impedance is set equal to the input impedance, the transmission $|S_{21}|$ is unity, and the amplifier backaction does not contribute to the total noise. The total amplifier noise consists only of the imprecision noise, represented by the third term in the first line of equation (\ref{eq:B_QLamp}). This results in a quantum-limited noise temperature of $kT_{N}(\nu) = h\nu/2$. When the transmission is less than unity, the amplifier backaction, represented by the second term in (\ref{eq:B_QLamp}), contributes to the total noise power. This implies that we would need to increase $n(\nu)$ by more than 1/2 in order for the increase in noise from the source resistor to equal the amplifier noise. In this case, $kT_{N}(\nu)> h\nu/2$. The noise impedance may also be calculated directly from the techniques in ref. \cite{wedge1991computer}. \footnote{In fact, even for an amplifier that misses the SQL, the practical necessities of perfect input and output match and of a circulator at the amplifier input imply that the noise impedance equals the input impedance.}

We can define
\begin{equation}\label{eq:NoiseNumber_QLamp}
N_{\rm tot}(\nu, S_{21}^{(1)}(\nu), n(\nu)) \equiv \frac{|S_{21}^{(1)}(\nu)|^{2} n(\nu) + 1}{|S_{21}^{(1)}(\nu)|^{2}}
\end{equation}
as the total noise-equivalent number, which is a unitless measure of total system noise at frequency $\nu$, referred to the signal source. From (\ref{eq:PDM_sig}), (\ref{eq:SNR_1_opt}), and (\ref{eq:B_QLamp}), we may evaluate the SNR for the optimally filtered signal with a quantum-limited readout amplifier. For a measurement of duration $\tau$ from a single receiver configuration, the SNR is
\begin{align}
SNR^{\rm opt} & \left[ \nu_{\rm DM}^{0}, g_{\rm DM}, \frac{d\rho_{\rm DM}}{d\nu}(\nu,\nu_{\rm DM}^{0}), \tau \right] = 2\pi^{2} \label{eq:SNR_1_QLamp} \\
&\times \left( \tau  \int d\nu\ \left( \frac{\nu L_{\rm PU}}{R(\nu)} \frac{U_{\rm DM}(\nu,\nu_{\rm DM}^{0}, g_{\rm DM}, \frac{d\rho_{\rm DM}}{d\nu}(\nu,\nu_{\rm DM}^{0}))}{h} \frac{1}{N_{\rm tot}(\nu, S_{21}^{(1)}(\nu), n(\nu))} \right)^{2} \right)^{1/2}. \nonumber
\end{align}
It is evident from equation (\ref{eq:SNR_1_QLamp}) that $N_{\rm tot}$ contains all information and parameters about the impedance-matching network as they pertain to search sensitivity. The total noise-equivalent number must therefore play a principal role in determining the optimal matching network between signal source and amplifier, as is described in the first part of Section \ref{sec:scan_opt}. 

It is convenient, for use in the next section, to define a quantity $\psi$ which is the inverse of the total noise-equivalent number:
\begin{equation} \label{eq:InvN_Spectrum}
\psi(\nu, S_{21}^{(1)}(\nu), n(\nu)) \equiv N_{\rm tot}(\nu, S_{21}^{(1)}(\nu) , n(\nu))^{-1} = \frac{|S_{21}^{(1)}(\nu)|^{2}}{|S_{21}^{(1)}(\nu)|^{2} n(\nu) + 1},
\end{equation}
and parametrize the arguments of of the SNR in terms of $\psi$:
\begin{equation}\label{eq:SNR_psi_param}
SNR^{\rm opt} \left[ \nu_{\rm DM}^{0}, g_{\rm DM}, \frac{d\rho_{\rm DM}}{d\nu}(\nu,\nu_{\rm DM}^{0}), \tau \right] \rightarrow SNR^{\rm opt} \left[ \nu_{\rm DM}^{0}, g_{\rm DM}, \frac{d\rho_{\rm DM}}{d\nu}(\nu,\nu_{\rm DM}^{0}), \psi(\nu, S_{21}^{(1)}(\nu), n(\nu)), \tau \right].
\end{equation}
To facilitate further calculation, we also define a cutoff bandwidth $\Delta \nu_{\rm DM}^{c}(\nu_{\rm DM}^{0})$. We define this bandwidth such that, outside of the frequency band $\nu_{\rm DM}^{0} \leq \nu \leq \nu_{\rm DM}^{0} + \Delta \nu_{\rm DM}^{c}(\nu_{\rm DM}^{0})$, the dark-matter signal power can be neglected. In other words, the cutoff bandwidth sets the limits of integration in (\ref{eq:SNR_1_QLamp}). Because we are searching for non-relativistic dark matter, we take the cutoff bandwidth to satisfy
\begin{equation}\label{eq:delta_nuDMc}
    \Delta \nu_{\rm DM}(\nu_{\rm DM}^{0}) \ll \Delta \nu_{\rm DM}^{c}(\nu_{\rm DM}^{0}) \ll \nu_{\rm DM}^{0}.
\end{equation}
The exact value of the cutoff bandwidth is unimportant and may be chosen flexibly as long as this inequality is satisfied.

We now consider implications of the SNR analysis for a resonator search using quantum-limited readout. This discussion guides the matching network optimization in Section \ref{sec:scan_opt}. 

Consider placement of two resonators of the same Q: one resonator at frequency $\nu_{r1}=\nu_{\rm DM}^{0}$ and the other at a frequency $\nu_{r2}$ close to $\nu_{\rm DM}^{0}$ (i.e. $|\nu_{\rm DM}^{0}-\nu_{r2}| \ll \nu_{r2}$), but still several resonator bandwidths away. If, at each frequency $\nu$ of the integral in eq. (\ref{eq:SNR_1_QLamp}), the transmitted thermal noise power is much larger than the quantum noise so that $|S_{21}^{(1)}(\nu,\nu_{r})|^{2} n(\nu) \gg 1$, then the SNR is approximately the same for both circuits. In other words, the off-resonance receiver gives the same sensitivity to dark matter as the on-resonance receiver. \emph{Equivalently, one may observe that a single resonator is sensitive, without degradation from quantum noise, not only to the dark-matter search frequencies within the resonator bandwidth, but also to frequencies past the resonator rolloff.} 

\begin{figure}[htp] 
\includegraphics[width=\textwidth]{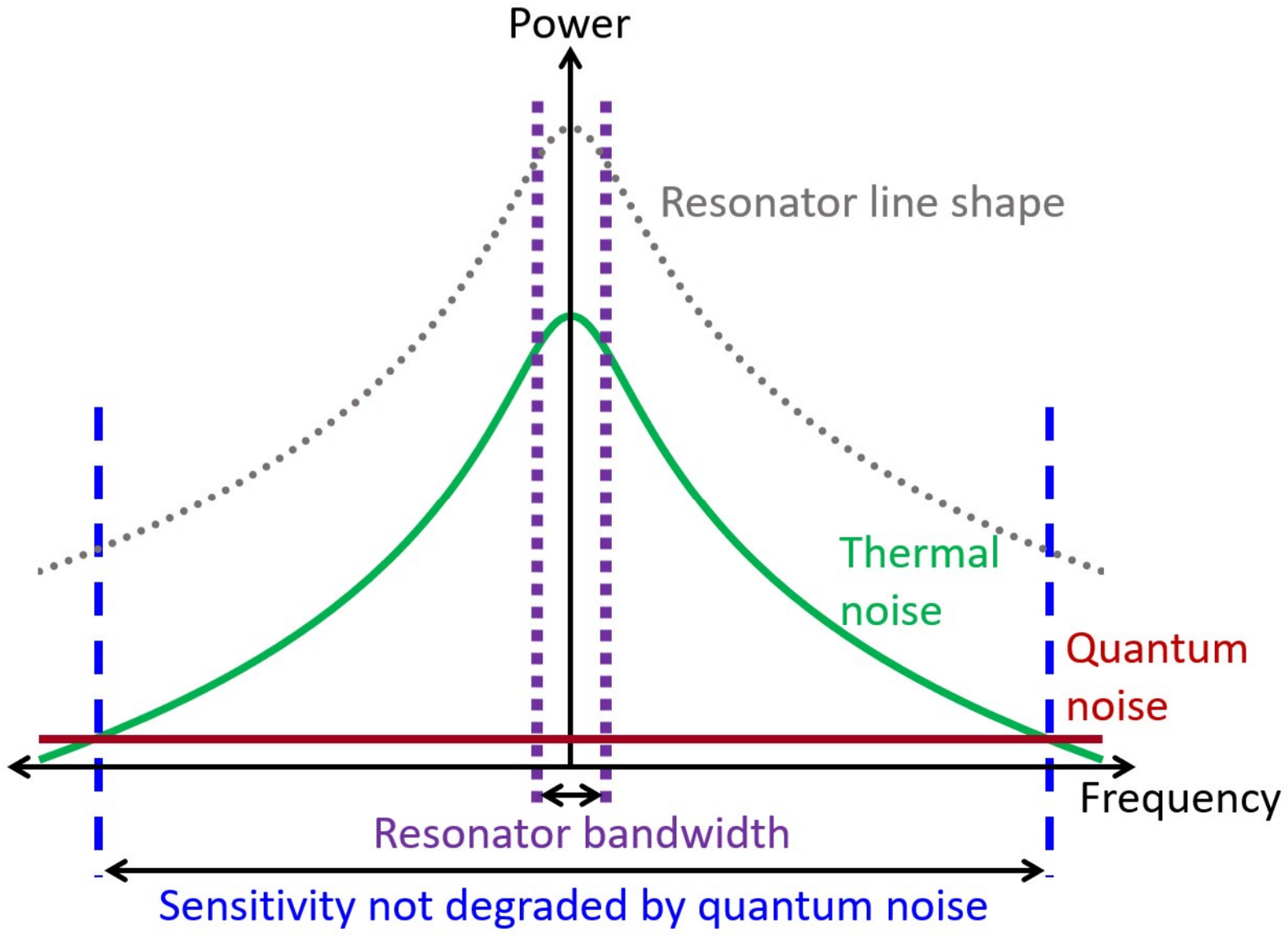}
\caption{Noise in a resonator. The x-axis is frequency (arbitrary units, linear scale), while the y-axis is power referred to the input of the amplifier (arbitrary units, logarithmic scale). The resonator line shape $|S_{21}^{(1)}(\nu,\nu_{r})|^{2}$, which does not represent a particular signal or noise source, is plotted as a gray dotted line. The resonance frequency coincides with the y-axis. Thermal noise is in green, and quantum noise (the ``+1" in eq. \ref{eq:B_QLamp}) is in dark red. The resonator bandwidth is bounded by the dashed purple lines. The resonator is sensitive to dark matter, without significant degradation from quantum noise, over the bandwidth bounded by the dashed blue lines. \label{fig:SensitivityBandwidth}}
\end{figure}

The physics of this process is displayed in Fig. \ref{fig:SensitivityBandwidth}, where we plot the noise as a function of frequency. For a dark-matter voltage drive of fixed strength (\ref{eq:SVV_DM}), the signal power at the amplifier input follows the resonator line shape (gray dotted line). The thermal noise (green line) possesses the same response shape. As such, the SNR remains approximately constant as a function of frequency (as dictated by the ratio of the gray and green lines), as long as the thermal noise is greater than the quantum noise (dark red line). The regime over which the thermal noise is larger is displayed between the dashed blue lines and can be considerably larger than the resonator bandwidth (dashed purple lines). We quantify the regime of undegraded sensitivity in Section \ref{sssec:res_match_opt} with the metric of sensitivity bandwidth.

The observation can also be explained from Fig. \ref{fig:DetectorBlockDiagram} in the introduction. If the amplifier noise is sub-dominant, then the filter characteristics of the impedance-matching network between the source and the readout, and in particular, the placement of the resonance frequency, do not affect SNR. The SNR is completely determined from the two parts of the signal source, the signal from the inductor and the noise from the resistor. The off-resonance contribution to the SNR was not considered in previous analyses, such as \cite{Chaudhuri:2014dla} and \cite{kahn2016broadband}. We show over the course of Sections \ref{sec:scan_opt} and \ref{ssec:QLA_sensitivity} that consideration of this effect adds substantially to the sensitivity of a search, especially at lower frequency.

Therefore, for a search over a wide band, the matching network optimization must consider SNR information available over the entire band. \emph{In other words, to determine what matching network is best (resonant or otherwise), we must develop a measure of frequency-integrated sensitivity.}


\section{Priors-Driven Scan Optimization}
\label{sec:scan_opt}

While it is clear that we must consider frequency-integrated sensitivity in optimizing the matching network, it is not readily clear how one rigorously quantifies such a concept. For a single-pole resonant matching network, one might consider using as the metric the bandwidth over which the thermal noise dominates quantum noise. One might also consider the ``scan rate''\cite{asztalos2010squid,brubaker2017first}, which quantifies the rate at which the resonator frequency is stepped in order to search a wide range of masses at a particular coupling strength (e.g. the DFSZ axion band). However, these concepts do not extend beyond the single-pole resonator, and as such, are inadequate for optimizing over the space of all possible linear, passive matching networks. For an arbitrary matching network, there may be arbitrarily-many, disconnected regions in frequency space in which the thermal noise is larger than the quantum noise. Moreover, the choice of matching network is independent of the choice to change receiver parameters in the course of a search (e.g. tune resonant frequency in a cavity). From our work in Section \ref{ssec:reactive_circuit}, we know that an active matching network can search the parameter space far faster than a single-pole resonant matching network, potentially eliminating the need for any receiver tuning. We have not yet established that a linear, passive matching network cannot possess the same advantage over a single-pole resonator. 

The key to producing an appropriate metric for matching network optimization is to recognize that, by saying we are interested in a ``search over a wide band'' or ``frequency-integrated sensitivity'', we are establishing prior probabilities on the dark-matter signal. There are three parameters on which we may establish priors, corresponding to three dark-matter parameters which are unknown:

\begin{itemize}
    \item The dark-matter rest-mass frequency $\nu_{\rm DM}^{0}$. We define a probability distribution $P(\nu_{\rm DM}^{0})$ for finding dark matter of mass $m_{\rm DM}=h\nu_{\rm DM}^{0}/c^{2}.$
    \item The dark-matter coupling to the receiver, quantified by $g_{\rm DM}$ in eq. (\ref{eq:gDM_def}). This coupling reflects the hidden-photon kinetic mixing angle $\varepsilon$ and the axion-photon coupling $\kappa_{a}=g_{a\gamma\gamma}\sqrt{\hbar c \epsilon_{0}}$. We define a conditional probability distribution $P(g_{\rm DM}|\nu_{\rm DM}^{0})$, which describes the probability that the coupling is $g_{\rm DM}$ given that the rest-mass frequency is $\nu_{\rm DM}^{0}$.
    \item The dark-matter frequency distribution $d\rho_{\rm DM}/d\nu$. We assume that, for each possible frequency $\nu_{\rm DM}^{0}$, there is a set of possible dark-matter distribution models:
    \begin{equation}
    \Bigg \{ \left( \frac{d\rho_{\rm DM}}{d\nu} \right)_{1}(\nu,\nu_{\rm DM}^{0}), \left( \frac{d\rho_{\rm DM}}{d\nu} \right)_{2}(\nu,\nu_{\rm DM}^{0}), \left( \frac{d\rho_{\rm DM}}{d\nu} \right)_{3}(\nu,\nu_{\rm DM}^{0}), ... \Bigg \}.
    \end{equation}
    The conditional probability associated with each model (i.e. the probability that model $ \left( \frac{d\rho_{\rm DM}}{d\nu} \right)_{j}$ is representative of the true dark-matter distribution) is denoted as $P \left( \left( \frac{d\rho_{\rm DM}}{d\nu} \right)_{j} (\nu,\nu_{\rm DM}^{0}) | \nu_{\rm DM}^{0} \right)$.\footnote{We have assumed implicitly that the probability distribution over dark-matter models is not conditional on the coupling $g_{\rm DM}$. In the event that one wishes to devise a search with such a condition, the modification in our probability distribution would be to make such a dependence explicit, e.g. $P \left( \left( \frac{d\rho_{\rm DM}}{d\nu} \right)_{j} (\nu,\nu_{\rm DM}^{0}) \Bigg | \nu_{\rm DM}^{0} \right) \rightarrow  P \left( \left( \frac{d\rho_{\rm DM}}{d\nu} \right)_{j} (\nu,\nu_{\rm DM}^{0}, g_{\rm DM}) \Bigg | \nu_{\rm DM}^{0}, g_{\rm DM} \right)$. \label{ValueAssumptionsNote}}
\end{itemize}

Priors are also important for accurately comparing single-pole resonant and reactive broadband searches. Unlike a broadband search, a resonant search is tuned to probe a wide mass range, and the sensitivity at a given rest-mass frequency is dependent on the scan time allocation. Thus, to compare the two search strategies and determine optimal scan time allocation in the resonant search, we require a language describing how the two searches possess the same, fixed objectives. This language is provided by priors. See Appendix \ref{sec:ResvBroad_SQUID} for details.

Consideration of priors is a central theme in this section, which analyzes scan optimization based on the SNR analysis of the previous section. We fix the lower and upper frequency limits of the search range, denoting them by $\nu_{l}$ and $\nu_{h}$, respectively. As summarized in the introduction, the optimization consists of two parts:

\begin{enumerate}
\item First, we optimize the matching network (the second element in the receiver diagram of Fig. \ref{fig:DetectorBlockDiagram}). \emph{This optimization applies across all single-moded, linear, passive receivers, resonant or otherwise, coupled to dark matter through a reactive element.} 

We introduce a value function for evaluating a given matching network. The function incorporates both the idea of sensitivity integrated across the search band, motivated in Fig. \ref{fig:SensitivityBandwidth}, as well as prior information regarding the dark-matter signal. 

We define in detail a ``log-uniform'' search. The log-uniform search assumes that dark matter is uniformly likely, on a logarithmic scale, to be anywhere in the search band. Under the log-uniform priors, and limiting our attention to the results of Section \ref{sssec:SNR_QLA} for a quantum-limited amplifier, we maximize the value function for a receiver inductively coupled to the dark-matter signal. In performing the optimization, we hold fixed the characteristics of the signal source (e.g. coupled energy, equivalent resistance and inductance, temperature of the loss). The maximization is subject to the Bode-Fano criterion \cite{bode1945network,fano1950theoretical}, which constrains the match between the complex-valued impedance of the signal source and the real-valued noise impedance of the quantum-limited amplifier. 

We optimize single-pole resonant matching networks for a log-uniform search and compare the result to the limit dictated by the Bode-Fano criterion. We find that the matching network is completely parametrized by the resonance frequency $\nu_{r}$ as well as the coupling coefficient $\xi=Q_{\rm int}/Q_{\rm cpl}$\cite{pozar2012microwave}. Holding the resonance frequency fixed, we maximize the value function with respect to $\xi$. Our results are readily interpreted in the context of amplifier noise-matching and measurement backaction. We find that the optimized resonant matching network is close to the Bode-Fano limit. An analogous result is found for detectors coupled capacitively to the dark-matter signal and for background AC magnetic fields. We thus establish the single-pole resonator as the near-ideal, single-moded technique for dark-matter detection.  (As explored in Section \ref{sssec:BF_evade}, the claim comes with some caveats associated with the limitations of the Bode-Fano criterion and the use of multi-moded receivers.) Because a resonator is a near-ideal matching network for a log-uniform search, as well as most other conceivable priors, we focus hereafter on scanning resonant searches.


We also briefly consider the optimization for validating a candidate signal that has been found in a previous search. This result, though different from the log-uniform search, is also closely related to the concept of noise matching.
 
\item An optimized resonant search requires tuning the resonance frequency across the search band. We assume that the total experiment time is fixed, i.e.
\begin{equation}
\sum_{i} \tau_{i} = T_{\rm tot}
\end{equation}
for some time $T_{\rm tot}$, where the sum is over all scan steps. We determine and optimize a value function for the distribution of time across scan steps. For a log-uniform search, the optimization corresponds to maximizing the area of the exclusion region in a log-log plot of mass $\nu_{\rm DM}^{0}$ vs. coupling $g_{\rm DM}$. To perform this maximization, we again consider a quantum-limited readout and assume a sufficiently dense scan, where each dark-matter frequency is probed by multiple resonance-frequency steps. We build upon the result of the first optimization step by assuming every resonator can be tuned to the optimal value of the coupling coefficient $\xi$. We also discuss other possible time-allocation value functions, such as those appropriate for QCD-axion dark-matter searches, as well as practical aspects of scan strategy.

\end{enumerate}

The analogous optimization for flux-to-voltage amplifiers is carried out in Appendices \ref{ssec:BodeFano_FV} and \ref{ssec:FtoV_opt_QL}. For a quantum-limited flux-to-voltage amplifier, we find that the optimal scan--in terms of both matching network and time allocation--is the same as that found here for scattering-mode amplifiers.

\subsection{Optimization of Matching Network}
\label{ssec:opt_impedance}

Under the above considerations of integrated sensitivity and priors, the generic value function for optimization of the matching network is the expected value of the square of the SNR. Explicitly, since the sensitivity, as it pertains to the matching network, is determined by the S-matrix--and specifically, the transmission $S_{21}^{(1)}(\nu)$-- the value functional is
\begin{align}
F\left[ S_{21}^{(1)}(\nu) \right] & \equiv \mathbf{E} \left[ SNR^{\rm opt} \left[ \nu_{\rm DM}^{0}, g_{\rm DM}, \frac{d\rho_{\rm DM}}{d\nu}(\nu,\nu_{\rm DM}^{0}), \psi(\nu, S_{21}^{(1)}(\nu), n(\nu)), \tau \right]^{2} \right] \label{eq:F_gen_def} \\
& = \int d\nu_{\rm DM}^{0} \int dg_{\rm DM}\ \sum_{j} SNR^{\rm opt} \left[ \nu_{\rm DM}^{0}, g_{\rm DM}, \frac{d\rho_{\rm DM}}{d\nu}(\nu,\nu_{\rm DM}^{0}), \psi(\nu, S_{21}^{(1)}(\nu), n(\nu)), \tau \right]^{2} \nonumber \\
& \hspace{4.75 cm} \times P \left( \left( \frac{d\rho_{\rm DM}}{d\nu} \right)_{j} (\nu,\nu_{\rm DM}^{0}) \Bigg | \nu_{\rm DM}^{0} \right)P(g_{\rm DM} | \nu_{\rm DM}^{0}) P(\nu_{\rm DM}^{0}).  \nonumber 
\end{align}
In the top line of this equation (see also (\ref{eq:SNR_1_QLamp})), we have implicitly assumed, by parametrizing SNR with $\psi$, that the matching network is lossless and read out by a quantum-limited amplifier. The value functional is readily extended to situations in which the network possesses loss and in which a quantum-limited amplifier is not used--see, for example, Section \ref{sssec:BF_evade} and Appendix \ref{ssec:FtoV_opt_Gen}. (One simply replaces the SNR formula (\ref{eq:SNR_1_QLamp}) with the appropriate expression.) Nevertheless, since the SQL represents a fundamental noise floor with a phase-insensitive amplifier, we fully work out that case, while laying the foundation of a more complex optimization with different noise and loss parameters. Strictly, the value functional depends on the thermal-occupation-number function $n(\nu)$. However, as this is determined by signal-source temperature, rather than the matching network properties, we omit it from the explicit representation as an argument of $F$. For similar reasons, we also omit the integration time $\tau$. Note that we have chosen the square of SNR, instead of simply SNR, in the definition of our value functional, since the SNR contributions from all receiver configurations add in quadrature.


The optimal matching network maximizes $F [S_{21}^{(1)}(\nu)]$.  $F [S_{21}^{(1)}(\nu)]$ measures the receiver sensitivity to dark matter, integrated over a wide search range and weighted by probability densities relating to its mass, its coupling to electromagnetism, and its frequency/velocity distribution. Clearly, $F[ S_{21}^{(1)}(\nu) ]$ is user defined, and sometimes qualitative. In that case, we may only make some qualitative statements about these probability distributions. 

The joint probability distribution of mass and coupling, given by $P(g_{\rm DM}|\nu_{\rm DM}^{0})P(\nu_{\rm DM}^{0})$, may have negligible weight at parts of the phase space that have been excluded, either by indirect astrophysical probes or by direct detection experiments. Additionally, for the axion, we may weight the joint distribution higher at mass-coupling ranges corresponding to the QCD axion. For the hidden photon, we may weight the distribution higher at mass-coupling ranges corresponding to an inflationary production mechanism.\cite{Graham:2015rva} We may choose to weight the distribution over rest-mass frequency very high at the frequencies of candidate signals found in previous searches. 


We know that the distribution $d\rho_{\rm DM}/d\nu$ has approximate width $\Delta \nu_{\rm DM}/\nu_{\rm DM}^{0} \sim 10^{-6}$ (as determined by virialization), but we do not know the precise value of the signal bandwidth or the particular fine structure.  The standard halo model, presented in Appendix \ref{sec:SH_Model}, is only one possible distribution. As shown in ref. \cite{sikivie1995velocity}, the narrowband distribution may contain even narrower peaks, representing dark matter that has fallen into the galactic gravitational potential well relatively recently, and therefore, has not yet virialized. These features may constitute a significant percentage of the dark-matter energy density.

\subsubsection{Value functional for matching network optimization in a log-uniform search}
\label{sssec:neutral_search_opt}

We compute the value functional under the following assumptions, holding the properties of the signal source fixed. These assumptions are for a ``log-uniform'' search--an uninformative prior for the dark-matter properties. In large regions of unexplored axion/hidden photon phase space, this is the most appropriate assumption.
\begin{enumerate}
\item The probability distribution over rest-mass frequency is log-uniform: 
\begin{equation}
P(\nu_{\rm DM}^{0}) \propto \frac{1}{\nu_{\rm DM}^{0}}.
\end{equation}
A log-uniform distribution is appropriate for a search in which we are ignorant about the mass scale of dark matter. It qualitatively matches exclusion plots, which are usually logarithmic, and it is also more natural than a linear-uniform distribution because the bandwidth of the dark-matter signal is proportional to the rest-mass frequency $\nu_{\rm DM}^{0}$.

\item Within the range of search frequencies, the probability distribution over $g_{\rm DM}$ is approximately independent of dark-matter mass. 

\item It is a fairly common assumption that the distribution over dark-matter velocity--and therefore, the distribution over speed-- has no explicit dependence on mass or coupling. \cite{baudis2012direct,freese2013colloquium,rosenberg2004searching} The standard halo model, as demonstrated in Appendix \ref{sec:SH_Model}, is one such model. Then, the distribution over frequency $\frac{d\rho_{\rm DM}}{d\nu} (\nu,\nu_{\rm DM}^{0})$ depends not separately on the two arguments, but on the parameter $u=(\nu-\nu_{\rm DM}^{0})/\nu_{\rm DM}^{0}= v^{2}/2c^2$, where $v$ is the speed corresponding to frequency $\nu$. (See equation (\ref{eq:disprel_nonrelativistic}).) We may then re-parameterize the energy-density distribution over frequency as

\begin{equation}
\left( \frac{d\rho_{\rm DM}}{d\nu} \right) (\nu,\nu_{\rm DM}^{0}) = \frac{1}{\nu_{\rm DM}^{0}} \frac{d\rho_{\rm DM}}{du}(u).
\end{equation}

We assume a single such distribution. In other words, the probability is unity for that distribution and zero for all others, and for each mass, $d\rho_{\rm DM}/du$ is the same. It is evident from equation (\ref{eq:SNR_1_QLamp}) that the limit on sensitivity, set by the coupling $g_{\rm DM,min}$ for which the total SNR is unity, depends on the particular dark-matter distribution. However, as we show, the optimal matching does not depend on the distribution. We define a constant based on the cutoff bandwidth (\ref{eq:delta_nuDMc})
\begin{equation} \label{eq:Deltamax_def}
\beta \equiv \rm{max}_{\nu_{l} \leq \nu_{\rm DM}^{0} \leq \nu_{h}} \frac{\Delta \nu_{\rm DM}^{c}(\nu_{\rm DM}^{0})}{\nu_{\rm DM}^{0}}.
\end{equation}
By construction, $\beta \ll 1$. 

\item We assume that the dwell time $\tau$ is much longer than $t_{\rm rec}^{*}(\nu_{\rm DM}^{0})$ for all rest-mass frequencies within the search band. Then, for any matching network, steady-state is reached after excitation from a dark-matter voltage signal and eq. (\ref{eq:SNR_1_QLamp}) applies. 

We note that, in practice, this assumption may not hold. For instance, if probing with a narrowband resonator with resonance frequency near the high end of the search range, the dark-matter coherence time at the low end of the search range may be much longer than the resonator ring-up time. This is especially true when performing a search over orders of magnitude in mass. It is inefficient to wait such long times because, as we show in Section \ref{sssec:res_match_opt}, the sensitivity bandwidth is narrowband and the resonator does not provide high-SNR information at frequencies far detuned from resonance. Such long integrations also leave the experiment more vulnerable to spurious electromagnetic interference. Practical limitations restrict the range of dark-matter frequencies at which we retrieve data from a particular resonance frequency. The follow-on room-temperature electronics include analog filters, passing data only in a narrow band around the resonance. This is necessary to reject out-of-band noise that can saturate amplifiers. We stress that the purpose of our assumption about the dwell time is simply to enable an apples-to-apples comparison of all possible matching networks, resonant or otherwise, rather than a reflection of these practical circumstances.

\end{enumerate}

Under the assumptions for a log-uniform search, we find, from (\ref{eq:UDM_def}) and (\ref{eq:SNR_1_QLamp}), that the value functional reduces to
\begin{align}
F_{\rm log}[S_{21}^{(1)}(\nu)] & = \gamma_{0} \int_{\nu_{l}}^{\nu_{h}} d\nu_{\rm DM}^{0}\ 
\int_{0}^{\beta} du\ \Bigg( \frac{(1+u)L_{\rm PU}}{R(\nu_{\rm DM}^{0}(1+u))} \frac{1}{h} \frac{d\rho_{\rm DM}}{du}(u) \label{eq:F_neu_simple} \\
& \hspace{1 cm} \psi(\nu_{\rm DM}^{0} (1+u), S_{21}^{(1)}(\nu_{\rm DM}^{0}(1+u)), n(\nu_{\rm DM}^{0}(1+u))) \Bigg)^{2} \nonumber \\
& = \gamma_{0} \int_{0}^{\beta} du\ \left( \frac{\sqrt{1+u}}{h} \frac{d\rho_{\rm DM}}{du}(u) \right)^{2} \int_{\nu_{l}(1+u)}^{\nu_{h}(1+u)} d\bar{\nu}\ \left( \frac{L_{\rm PU}}{R(\bar{\nu})} \psi(\bar{\nu},S_{21}^{(1)}(\bar{\nu}),n(\bar{\nu})) \right)^{2}, \nonumber 
\end{align}
where $\gamma_{0}$ is a constant, independent of the matching network. In the second line, we have exchanged the order of integration and have made the change of variable $\bar{\nu}=\nu_{\rm DM}^{0}(1+u)$. In any practical search, during the data-taking period, there is not a sharp cutoff to the search range; for example, if a resonator is placed close to $\nu_{h}$, it collects data not only on the hypothesis that $\nu_{\rm DM}^{0}=\nu_{h}$ but also on the hypothesis that $\nu_{\rm DM}^{0}=\nu_{h}(1+\delta)$, where $\delta \ll 1$. We may then omit the $(1+u)$ factors in the limits of the integral over $\bar{\nu}$. In fact, as we see in the maximization with the Bode-Fano constraint, doing so introduces a negligible fractional error of order at most $\beta \ll 1$. This decouples the integrals over $u$ and $\bar{\nu}$ and yields
\begin{equation}
F_{\rm log}[S_{21}^{(1)}(\nu)]= \gamma_{1} \int_{\nu_{l}}^{\nu_{h}} d\bar{\nu}\ \left( \frac{L_{\rm PU}}{R(\bar{\nu})} \psi(\bar{\nu},S_{21}^{(1)}(\bar{\nu}),n(\bar{\nu})) \right)^{2} \label{eq:F_gen_neutral},
\end{equation}
where $\gamma_{1}$ is a constant containing the integral over $u$.

\subsubsection{Matching network optimization for log-uniform search: the Bode-Fano constraint}
\label{sssec:Bode_Fano_opt}

Let us assume that the loss is frequency-independent, $R=R(\bar{\nu})$. Scaling out the constants of equation (\ref{eq:F_gen_neutral}), namely $\gamma_{1} (L_{\rm PU}/R)^{2}$, we finally arrive at 
\begin{align} \label{eq:F_gen_scaled}
\bar{F}_{\rm log}[S_{21}^{(1)}(\nu)] &= \int_{\nu_{l}}^{\nu_{h}} d\bar{\nu}\ \left(\psi(\bar{\nu},S_{21}^{(1)}(\bar{\nu}),n(\bar{\nu})) \right)^{2} \\
&= \int_{\nu_{l}}^{\nu_{h}} d\bar{\nu}\ N_{\rm tot}(\bar{\nu}, S_{21}^{(1)}(\bar{\nu}),n(\nu))^{-2} = \int_{\nu_{l}}^{\nu_{h}} d\bar{\nu}\ \left( \frac{|S_{21}^{(1)}(\bar{\nu})|^{2}}{|S_{21}^{(1)}(\bar{\nu})|^{2} n(\bar{\nu}) +1} \right)^{2}, \nonumber
\end{align}
which we refer to as the scaled, log-uniform-search value functional. By maximizing this measure of frequency-integrated sensitivity, we determine the best matching network. It is similar to the Wiener-filtered energy resolution in a calorimeter \cite{stahl2005cryogenic}, which is the integral of the inverse-squared noise-equivalent power (instead of the integral of inverse-squared noise-equivalent number). 

The signal source contains a complex impedance, with resistance $R$ and pickup inductance $L_{\rm PU}$, and the amplifier input/noise impedance is real. Let us assume that the matching network is linear, passive, lossless, and reciprocal. The Bode-Fano criterion \cite{fano1950theoretical,bode1945network,pozar2012microwave} then applies:
\begin{equation}\label{eq:Bode_Fano}
\int_{\nu_{l}}^{\nu_{h}} d\bar{\nu}\ ln \left( \frac{1}{|S_{22}^{(1)}(\bar{\nu})|} \right) \leq \frac{R}{2 L_{\rm PU}}.
\end{equation}
We discuss relaxation of the assumptions on matching network in Section \ref{sssec:BF_evade}. Using (\ref{eq:ScattRelation}), the criterion may be rewritten as
\begin{equation}
\int_{\nu_{l}}^{\nu_{h}} d\bar{\nu}\ ln \left( \frac{1}{1-|S_{21}^{(1)}(\bar{\nu})|^{2}} \right) \leq \frac{R}{ L_{\rm PU}}.
\end{equation}

We now derive the constraint that the Bode-Fano criterion imposes on the scaled, log-uniform-search value functional (\ref{eq:F_gen_scaled}). 
The derivation of the constraint requires the definition and manipulation of an auxiliary function $h$.

We define the auxiliary function of two variables
\begin{equation}\label{eq:BF_magic}
h(y,s):= ln \left( \frac{1}{1-y} \right) - s \left( \frac{y}{y n(\nu_{h}) +1} \right)^{2}
\end{equation}
on the domain $(y,s) \in [0,1] \times [0,\infty)$. This function has the following properties:
\begin{itemize}
\item $h(0,s)=0$ for all $s$.
\item $\lim_{y \to 1^{-}} h(y,s)= \infty$ for all $s$.
\item There exists a unique value of $s$, termed $s^{*}$, such that $h(y,s^{*})$ has precisely two zeros and is nonnegative for all $y$, $0\leq y \leq 1$. One of these zeros is located at $y_{0} >0$. The zero is also a minimum of $h(y,s^{*})$, so from $h$ and its first derivative we find
\begin{equation}
ln \left(\frac{1}{1-y_{0}} \right) -s^{*} \left( \frac{y_{0}}{y_{0}n(\nu_{h}) +1} \right)^{2}=0,
\end{equation}
and
\begin{equation}
\frac{1}{1-y_{0}} -2s^{*} \frac{y_{0}}{(y_{0}n(\nu_{h})+1)^{3}} =0.
\end{equation}
We may solve for $s^{*}$ in terms of $y_{0}$, which gives
\begin{equation}\label{eq:BF_magic_s}
s^{*}= \left( \frac{y_{0}n(\nu_{h})+1}{y_{0}} \right)^{2} ln \left( \frac{1}{1-y_{0}} \right).
\end{equation}
We also find an equation for $y_{0}$:
\begin{equation}\label{eq:BF_magic_y0}
ln \left( \frac{1}{1-y_{0}} \right) = \frac{y_{0}}{1-y_{0}} \frac{y_{0}n(\nu_{h})+1}{2}.
\end{equation}
\end{itemize}

Since $0\leq |S_{21}^{(1)}(\nu)|^{2} \leq 1$, we may substitute $|S_{21}^{(1)}(\nu)|^{2}$ for $y$ in $h(y,s^{*})$. Then, we find for any frequency $\bar{\nu}$, $\nu_{l} \leq \bar{\nu} \leq \nu_{h}$, 
\begin{equation}
ln \left( \frac{1}{1-|S_{21}^{(1)}( \bar{\nu})|^{2}} \right) \geq s^{*} \left( \frac{|S_{21}^{(1)}(\bar{\nu})|^{2}}{|S_{21}^{(1)}(\bar{\nu})|^{2} n(\nu_{h}) +1} \right)^{2}.
\end{equation}
Integrating both sides and using $n(\nu) \geq n(\nu_{h})$, we arrive at
\begin{align}
\bar{F}_{\rm log}[S_{21}^{(1)}(\nu)] &= \int_{\nu_{l}}^{\nu_{h}} d\bar{\nu}\ \left( \frac{|S_{21}^{(1)}(\bar{\nu})|^{2}}{|S_{21}^{(1)}(\bar{\nu})|^{2} n(\bar{\nu}) +1} \right)^{2} \nonumber \\
& \leq \int_{\nu_{l}}^{\nu_{h}} d\bar{\nu}\ \left( \frac{|S_{21}^{(1)}(\bar{\nu})|^{2}}{|S_{21}^{(1)}(\bar{\nu})|^{2} n(\nu_{h}) +1} \right)^{2} \nonumber \\
& \leq \frac{1}{s^{*}} \int_{\nu_{l}}^{\nu_{h}} d\bar{\nu}\ ln \left( \frac{1}{1-|S_{21}^{(1)}(\bar{\nu})|^{2}} \right) \nonumber \\
& \leq \left( \frac{y_{0}}{y_{0}n(\nu_{h})+1} \right)^{2} \left( ln \left( \frac{1}{1-y_{0}} \right) \right)^{-1} \frac{R}{L_{\rm PU}}, \label{eq:BF_magic_ineq}
\end{align}
where in the last inequality, we have used (\ref{eq:BF_magic_s}). The value $y_{0}$ satisfies (\ref{eq:BF_magic_y0}), so it depends on $n(\nu_{h})$, i.e. $y_{0}=y_{0}(n(\nu_{h}))$.

Is it possible to have equality, or approximate equality, in (\ref{eq:BF_magic_ineq})? To satisfy with equality, we must have the following:
\begin{itemize}
\item From the third inequality of (\ref{eq:BF_magic_ineq}), the matching network must satisfy the Bode-Fano constraint (\ref{eq:Bode_Fano}) with equality. From \cite{fano1950theoretical}, we find that equality is satsfied by networks with top-hat transmission profiles.
\item From the second inequality, $h(|S_{21}^{(1)}(\bar{\nu})|^{2},s^{*})=0$ for all frequencies $\bar{\nu}$. This means the transmission profile only takes on two values: $|S_{21}^{(1)}(\bar{\nu})|^{2}=0$ or $|S_{21}^{(1)}(\bar{\nu})|^{2}=y_{0}(n(\nu_{h}))$.
\item From the first inequality, we must have $\bar{\nu} \approx \nu_{h}$ for all frequencies where $|S_{21}^{(1)}(\bar{\nu})|^{2} \ne 0$.
\end{itemize}
Together, these imply that equality is achieved by a narrowband top-hat transmission profile centered near $\nu_{h}$. The height of the top-hat transmission is
\begin{equation}\label{eq:hat_height_BF}
|S_{21}^{(1)}|_{\rm BF,hat}^{2}=y_{0}(n(\nu_{h})),
\end{equation}
and the width is, from (\ref{eq:Bode_Fano}),
\begin{equation} \label{eq:Delta_nu_BF}
\Delta \nu_{\rm BF,hat}=\frac{R}{L_{\rm PU}} \left( ln \left( \frac{1}{1-y_{0}(n(\nu_{h}))} \right) \right)^{-1}.
\end{equation}
The top-hat can be considered narrowband if $\Delta \nu_{\rm BF,hat} \ll \nu_{h}$, which is the case for a sufficiently low-loss signal source. 

We analyze two limits of expressions (\ref{eq:BF_magic_ineq}), (\ref{eq:hat_height_BF}), and (\ref{eq:Delta_nu_BF}): low thermal occupation number $n(\nu_{h}) \ll 1$ and high thermal occupation number $n(\nu_{h}) \gg 1$. In the $n(\nu_{h}) \ll 1$ limit, we find
\begin{equation}
|S_{21}^{(1)}|_{\rm BF,hat}^{2} \approx 0.7,\  \Delta \nu_{\rm BF,hat} \approx 0.8 \frac{R}{L_{\rm PU}},
\end{equation}
and
\begin{equation}
\bar{F}_{\rm log}[S_{21}^{(1)}(\nu)] \lessapprox 0.4 \frac{R}{L_{\rm PU}}.
\end{equation}
In the $n(\nu_{h}) \gg 1$ limit, we find
\begin{equation}
|S_{21}^{(1)}|_{\rm BF,hat}^{2} \approx \frac{1}{n(\nu_{h})},\  \Delta \nu_{\rm BF,hat} \approx n(\nu_{h}) \frac{R}{L_{\rm PU}},
\end{equation}
and
\begin{equation}
\bar{F}_{\rm log}[S_{21}^{(1)}(\nu)] \lessapprox \frac{1}{4n(\nu_{h})} \frac{R}{L_{\rm PU}}.
\end{equation}

Of particular interest in these limits is the Bode-Fano bound, summarized here:
\begin{equation} \label{eq:BF_magic_ineq_limits}
\bar{F}_{\rm log}[S_{21}^{(1)}(\nu)] \lessapprox \begin{cases}
0.4 \frac{R}{L_{\rm PU}}, \hspace{1 cm} n(\nu_{h}) \ll 1 \\
\frac{1}{4n(\nu_{h})} \frac{R}{L_{\rm PU}}, \hspace{0.7 cm} n(\nu_{h}) \gg 1
\end{cases}
\end{equation}

We note that a constraint similar to (\ref{eq:Bode_Fano}) exists for RC (resistor-capacitor) signal sources.\cite{pozar2012microwave} Therefore, if one chose to couple to the observable electric field produced by the dark matter, inequalities analogous to (\ref{eq:BF_magic_ineq}) and (\ref{eq:BF_magic_ineq_limits}) would result. A construction of the Bode-Fano optimal top-hat may be performed using Chebyshev filters.\cite{fano1950theoretical} As one increases the number of LC poles in the filter, the transmission profile approaches a top-hat. Because these networks are multi-pole and would need to be tunable to search for dark matter over a wide band, they would be difficult to implement in an experiment. It is therefore important to ask how well a single-pole resonator compares to the Bode-Fano optimal top-hat and whether this simply tunable implementation is close enough to optimal to be used in practical searches.

\subsubsection{Matching network optimization for log-uniform search with single-pole resonator}
\label{sssec:res_match_opt}

Here, we optimize the matching network under the assumption that the transmission profile represents a single-pole resonator. The log-uniform-search value functional that we start with is (\ref{eq:F_gen_neutral}). Thus, unlike the Bode-Fano bound, our optimization of the single-pole resonator does not require the assumption of frequency-independent resistance. The properties of the signal source and the amplifier input impedance are held fixed in this analysis. The optimization occurs in three steps. First, we identify the specific resonator parameters (more specific than the entire transmission profile $S_{21}^{(1)}(\nu)$, which is already substantially constrained by restriction to a single-pole resonator) that govern matching between the signal source and the amplifier input. For the purpose of further analysis in Section \ref{sssec:matching_opt_discussion}, we relate these parameters back to Fig. \ref{fig:SensitivityBandwidth} and the concept of sensitivity beyond the resonator bandwidth, which led to our value functional. Second, fixing the resonance frequency $\nu_{r}$, we maximize the value functional with respect to the matching parameters. Third, setting the internal loss to be frequency-independent, varying the resonance frequency, and building on the results of the second step, we compare the optimized single-pole resonator to the Bode-Fano bound of (\ref{eq:BF_magic_ineq}).

\paragraph{Resonator matching parameters} \mbox{}

Let us hold the signal source properties $L_{\rm PU}$ and $R(\nu)$, as well as the amplifier input impedance $Z_{0}$ (matched to its input transmission line), fixed. The matching network (the capacitance $C$ and the coupling capacitance $C_{c}$ in Fig. \ref{fig:ResonatorScattering}) and, therefore, the scattering parameters, are then determined by the resonance frequency $\nu_{r}$ and the coupling coefficient\cite{pozar2012microwave}. The coupling coefficient is defined by
\begin{equation}\label{eq:xi_def}
\xi \equiv \frac{Q_{\rm int}}{Q_{\rm cpl}}.
\end{equation}
$\xi$ describes the coupling of the resonator to the output transmission line. For $\xi >1$, the resonator is overcoupled and losses through the transmission line dominate. For $\xi <1$, the resonator is undercoupled and losses intrinsic to the resonator dominate. For $\xi=1$, the resonator is critically coupled, and the loss is the same in both channels. The overall quality factor is 
\begin{equation}\label{eq:Q_xi}
Q= \frac{Q_{\rm int}}{1+ \xi}.
\end{equation}
From equations (\ref{eq:Qint_def}) and (\ref{eq:Qc_def}), $\xi$ can be rewritten as
\begin{equation}
\xi = \frac{(2\pi\nu_{r}L_{\rm PU})^{2} (2\pi\nu_{r}C_{c})^{2}Z_{0}}{R(\nu_{r})}.
\end{equation}
Recognizing $R(\nu_{r})/((2\pi\nu_{r} L_{\rm PU})^{2} (2\pi\nu_{r}C_{c})^{2})$ as the resonator impedance on resonance, as seen from the transmission line, we observe that $\xi$ is also the ratio of the amplifier noise impedance (equal to input impedance) to the resonator impedance. We therefore call $\xi$ the resonator matching parameter; it is of primary importance in the matching optimization. Additionally, note that the matching parameter can be adjusted (for purposes of optimization) in our equivalent circuit, Fig. \ref{fig:ResonatorScattering}, by varying the capacitance coupling the resonator to the output transmission line.

We rewrite the transmission profile and the inverse of the noise-equivalent number, $\psi$, in terms of the resonator matching parameter $\xi$. Eq. (\ref{eq:S21}) may be written as
\begin{equation} \label{eq:S21_xi}
|S_{21}^{(1)}(\nu,\nu_{r}, Q_{\rm int}, \xi)|^{2}= \frac{4\xi}{(1+\xi)^{2} + 4 (Q_{\rm int})^{2} \left( \frac{\nu}{\nu_{r}} -1 \right)^{2}},
\end{equation}
where we have made the approximation $|\nu-\nu_{r}| \ll \nu_{r}$. Eq. (\ref{eq:InvN_Spectrum}) then becomes
\begin{equation} \label{eq:InvN_Spectrum_xi}
\psi(\nu,\nu_{r},n(\nu), Q_{\rm int}, \xi) = \frac{4\xi}{4\xi n(\nu) + (1+\xi)^{2} + 4 (Q_{\rm int})^{2} \left( \frac{\nu}{\nu_{r}} -1 \right)^{2}}.
\end{equation}
Note that we have changed the parametrization of $\psi$ from (\ref{eq:InvN_Spectrum}). Instead of using $S_{21}^{(1)}(\nu)$ as the argument, we have inserted the relevant resonator variables $\nu_{r}$, $Q_{\rm int}$, and $\xi$.

Recall from Section \ref{sec:resonator_SNR} that we assume the total and internal Qs are both much greater 1, so $\xi \ll Q_{\rm int}$. As a consistency check, we show that the optimum value of the resonator matching parameter satisfies this inequality. We also assume, that for all resonant scan steps in our search band, $n(\nu_{r}) \ll Q_{\rm int}$--that is, the on-resonance thermal occupation number is much less than the internal quality factor. At dilution refrigerator temperatures of 10 mK and for internal quality factors on the order of one million, this is an accurate approximation down to a resonance frequency of $\sim$1 kHz. 

Under these assumptions,  (\ref{eq:InvN_Spectrum_xi}) shows that $\psi(\nu,\nu_{r},n(\nu),Q_{\rm int},\xi)$ is strongly peaked at $\nu=\nu_{r}$. In particular, for frequencies $\nu$, $|\nu-\nu_{r}| \ll \nu_{r}$,
\begin{align}
\psi(\nu,\nu_{r},n(\nu), Q_{\rm int}, \xi) 
& \approx \psi(\nu,\nu_{r},n(\nu_{r}), Q_{\rm int}, \xi) \nonumber \\ 
&= \frac{4\xi}{4\xi n(\nu_{r}) + (1+\xi)^{2}} \left( 1 + \frac{4(Q_{\rm int})^{2}}{4\xi n(\nu_{r}) + (1+\xi)^{2}} \left( \frac{\nu}{\nu_{r}} -1 \right)^{2} \right)^{-1}. \label{eq:InvN_Spectrum_Lorentzian}
\end{align}
$\psi$ thus behaves as a Lorentzian centered at $\nu_{r}$ with a maximum value of
\begin{equation}\label{eq:InvN_Spectrum_max}
\psi_{\max}(\nu_{r},n(\nu_{r}), \xi)= \frac{4\xi}{4\xi n(\nu_{r}) + (1+\xi)^{2}},
\end{equation}
and a quality factor of
\begin{equation}\label{eq:InvN_Spectrum_Q}
Q_{s}(\nu_{r},n(\nu_{r}), Q_{\rm int}, \xi)= \frac{Q_{\rm int}}{(4\xi n(\nu_{r}) + (1+\xi)^{2})^{1/2}} \gg 1.
\end{equation}
We refer to $Q_{s}(\nu_{r},n(\nu_{r}), Q_{\rm int}, \xi)$ as the sensitivity quality factor and $\nu_{r}/Q_{s}(\nu_{r},n(\nu_{r}), Q_{\rm int}, \xi)$ as the sensitivity bandwidth. The sensitivity bandwidth describes the set of frequencies over which the total noise-equivalent number (\ref{eq:NoiseNumber_QLamp}) is less than twice its on-resonance (minimum) value. It is over this range of frequencies that the resonator is sensitive to dark matter without significant degradation from quantum noise. Note that, if the transmitted thermal noise is larger than the quantum noise, i.e.
\begin{equation}
|S_{21}^{(1)}(\nu,\nu_{r},Q_{\rm int}, \xi) |^{2} n(\nu) \geq 1,
\end{equation}
then
\begin{equation}
|\nu-\nu_{r}| \leq \frac{\nu_{r}}{2Q_{s}(\nu_{r},n(\nu_{r}), Q_{\rm int}, \xi)}.
\end{equation}
Our definition of sensitivity bandwidth is thus consistent with the qualitative description of sensitivity beyond the resonator bandwidth developed in Fig. \ref{fig:SensitivityBandwidth}. Note that a smaller value of $Q_{s}$ implies a larger sensitivity bandwidth. 

It is also useful to make the change of variable $x=2Q_{\rm int}\left( \frac{\nu-\nu_{r}}{\nu_{r}} \right)$. $x$ represents the detuning from the resonance frequency as a fraction of the linewidth of the uncoupled resonator. Eq. (\ref{eq:InvN_Spectrum_Lorentzian}) becomes
\begin{align}
\psi(x,\nu_{r}, n(\nu_{r}), \xi) &= \frac{4\xi}{4\xi n(\nu_{r}) + (1+\xi)^{2}} \left( 1 + \frac{1}{4\xi n(\nu_{r}) + (1+\xi)^{2}} x^{2} \right)^{-1} \nonumber \\
& = \psi_{\max}(\nu_{r},n(\nu_{r}),\xi) \left(1 + \frac{Q_{s}(\nu_{r},n(\nu_{r}),\xi)^{2}}{Q_{\rm int}^{2}} x^{2} \right)^{-1}. \label{eq:InvN_Spectrum_detuning}
\end{align}

\paragraph{Optimization at fixed resonator frequency} \mbox{}


Fixing the resonance frequency $\nu_{r}$ somewhere in the search band, $\nu_{l} \leq \nu_{r} \leq \nu_{h}$, we optimize the value functional (\ref{eq:F_gen_neutral}) with respect to $\xi$. With resonance frequency and signal-source properties fixed (which, in turn, fixes internal quality factor), we observe from (\ref{eq:InvN_Spectrum_xi}) that this is equivalent to maximizing with respect to the transmission profile $S_{21}^{(1)}(\nu)$.

Since $\psi$ is a sharply-peaked Lorentzian of quality factor much greater than unity and the loss is assumed to be relatively slowly varying with frequency, we may approximate the resistance to take on its on-resonance value everywhere in the integration range: $R(\bar{\nu}) \approx R(\nu_{r})$. Additionally, changing variables $x=2Q_{\rm int}\frac{\bar{\nu}-\nu_{r}}{\nu_{r}}$, we find
\begin{equation}\label{eq:F_neu_res}
F_{\rm log}[\nu_{r}, \xi] = \gamma_{1} \left( \frac{L_{\rm PU}}{R(\nu_{r})} \right)^{2} \frac{\nu_{r}}{2Q_{\rm int}} \int_{-\infty}^{\infty} dx\ \psi(x,\nu_{r},n(\nu_{r}),\xi)^{2}.
\end{equation}
We have extended the limits of integration to $-\infty$ to $\infty$ because the search band contains multiple sensitivity bandwidths on each side of $\nu_{r}$. (As discussed below eq. (\ref{eq:F_neu_simple}), the search band cutoffs are ``soft'' and can be adjusted slightly for resonance frequencies on either edge.) Performing the integral using (\ref{eq:InvN_Spectrum_detuning}) yields
\begin{equation}\label{eq:F_neu_res_eval}
F_{\rm log}[\nu_{r}, \xi] = \gamma_{1} \left( \frac{L_{\rm PU}}{R(\nu_{r})} \right)^{2} \frac{4\pi \nu_{r}}{Q_{\rm int}} \frac{\xi^{2}}{(4\xi n(\nu_{r}) + (1+\xi)^{2})^{3/2}}.
\end{equation}
Maximizing with respect to $\xi$,
\begin{equation} \label{eq:xi_opt}
\xi^{\rm opt}(\nu_{r},n(\nu_{r}))= \frac{1}{2} \left( 2n(\nu_{r})+1 + \sqrt{(2n(\nu_{r})+1)^{2}+8} \right).
\end{equation}
In the high $n(\nu_{r}) \ll 1$ and low $n(\nu_{r}) \gg 1$ occupation limits, this expression reduces to
\begin{equation} \label{eq:xi_opt_limits}
\xi^{\rm opt}(\nu_{r},n(\nu_{r})) \approx \begin{cases} 2  & \mbox{if } n(\nu_{r}) \ll 1 \\ 
2n(\nu_{r}) & \mbox{if } n(\nu_{r}) \gg 1 \end{cases}
\end{equation}
As a consistency check, note that, because $n(\nu_{r}) \ll Q_{\rm int}$, our assumption of $\xi \ll Q_{\rm int}$ is satisfied by the optimum. Where convenient, we omit the arguments of $\xi^{\rm opt}$.

Additionally, we observe that our result for the optimum matching parameter can be derived with assumptions 1 and 2 above eq. (\ref{eq:F_neu_simple}) relaxed; we do not require a log-uniform probability distribution over mass, and we do not require the assumption that the probability distribution over $g_{\rm DM}$ be approximately independent of dark-matter mass. Under most conceivable priors in which the prior probability over mass is not sharply peaked (contrast with the candidate signal optimization in Section \ref{sssec:candidate_search_opt}), the resonator transmission profile is sufficiently sharply peaked that we may approximate $\nu_{\rm DM}^{0} P(\nu_{\rm DM}^{0}) P(g_{\rm DM} | \nu_{\rm DM}^{0}) \approx \nu_{r} P(\nu_{r}) P(g_{\rm DM} | \nu_{r})$. Then, we may absorb the integral over $g_{\rm DM}$, along with the constant $\nu_{r} P(\nu_{r})$, into the constant $\gamma_{0}$ in (\ref{eq:F_neu_simple}). Following the same steps as above, we again obtain equation (\ref{eq:xi_opt}). Equation (\ref{eq:xi_opt}) therefore represents the resonator optimization not only for a log-uniform search, but also the optimization when one generically desires to search over a wide range of frequencies. This observation proves useful when discussing the optimization of time allocation using priors in Section \ref{ssec:opt_time}. It is particularly relevant for the axion, where consideration of the well-motivated QCD band can result in probability distributions over coupling that have substantial dependence on dark-matter mass. 

\paragraph{Comparison of Bode-Fano optimal matching circuit and single-pole resonator}\mbox{}

We now compare the single-pole resonator to the Bode-Fano optimal matching network (the top-hat described in Section \ref{sssec:Bode_Fano_opt}). We find the resonance frequency and the matching parameter $\xi$ that maximize the scaled, log-uniform-search value functional (\ref{eq:F_gen_scaled}) for integrated sensitivity across $\nu_{l}\leq \nu \leq \nu_{h}$. In the previous section, we asked, given a resonance frequency, how do we optimize the matching parameter $\xi$ for integrated sensitivity. Here, we ask which resonance frequency we choose to best approach the Bode-Fano bound. 

We relax the assumption of fixed resonance frequency and observe that whatever the resonance frequency is, the optimal matching parameter is given by (\ref{eq:xi_opt}). We take the loss $R(\nu)=R$ to be frequency-independent. Using equation (\ref{eq:F_neu_res_eval}), the scaled, log-uniform-search value functional (\ref{eq:F_gen_scaled}) reads
\begin{equation}
\bar{F}_{\rm log}[\nu_{r}, \xi^{\rm opt}(\nu_{r},n(\nu_{r}))] \approx \frac{2(\xi^{\rm opt}(\nu_{r},n(\nu_{r})))^{2}}{(4\xi^{\rm opt}(\nu_{r},n(\nu_{r})) n(\nu_{r}) + (1+\xi^{\rm opt}(\nu_{r},n(\nu_{r})))^{2})^{3/2}} \frac{R}{L_{\rm PU}},
\end{equation}
where we have--similar to (\ref{eq:InvN_Spectrum_xi})--replaced the transmission profile with the relevant network parameters in the argument of $\bar{F}_{\rm log}$. The first fraction is an increasing function of $\nu_{r}$, so the value functional is optimized by taking $\nu_{r}=\nu_{h}$:
\begin{align}\label{eq:BF_res_opt}
\bar{F}_{\rm log}[\nu_{h}, \xi^{\rm opt}(\nu_{h},n(\nu_{h}))] &\approx \frac{2(\xi^{\rm opt}(\nu_{h},n(\nu_{h})))^{2}}{(4\xi^{\rm opt}(\nu_{h},n(\nu_{h})) n(\nu_{h}) + (1+\xi^{\rm opt}(\nu_{h},n(\nu_{h})))^{2})^{3/2}} \frac{R}{L_{\rm PU}} \nonumber \\
& \approx \begin{cases} \frac{8}{27} \frac{R}{L_{\rm PU}}  & \mbox{if } n(\nu_{h}) \ll 1 \\ 
\frac{1}{3\sqrt{3}} \frac{1}{n(\nu_{h})} \frac{R}{L_{\rm PU}} & \mbox{if } n(\nu_{h}) \gg 1 \end{cases}
\end{align}

We now discuss important features of the results for resonators--in particular, the optimum matching parameter (\ref{eq:xi_opt}) and the result (\ref{eq:BF_res_opt}) comparing the best resonator to the Bode-Fano bound.

\subsubsection{Discussion of matching optimization for single-pole resonator}
\label{sssec:matching_opt_discussion}

Interestingly, for the log-uniform search, noise matching to the amplifier on resonance, corresponding to $\xi=1$ and $|S_{21}^{(1)}(\nu_{r},\nu_{r},Q_{\rm int},\xi)|=1$, is not the optimum for detector sensitivity. This may seem counterintuitive. We are not transmitting maximum power from the resonator at the optimum and the optimal match is a noise mismatch!

To understand this result, 
we focus on the limit $n(\nu) \gg 1$. The argument for the limit $n(\nu) \ll 1$ is similar, but the difference in value function (between optimal matching and noise matching) is small, approximately 20\%. From (\ref{eq:F_neu_res}), the quantity of interest is the integral of the square of the inverse noise-equivalent number, $\psi(x,\nu_{r},n(\nu_{r}),\xi)^{2}$, over detuning. As $\psi$ is Lorentzian, the integral can be approximated parametrically as the maximum value of the integrand multiplied by the linewidth. From (\ref{eq:InvN_Spectrum_detuning}),
\begin{equation} \label{eq:psi_int_approx}
\int_{-\infty}^{\infty} dx\ \psi(x, \nu_{r}, n(\nu_{r}), \xi)^{2} \sim \psi_{\rm max}(\nu_{r},n(\nu_{r}), \xi)^{2} \frac{Q_{\rm int}}{Q_{s}(\nu_{r},n(\nu_{r}),Q_{\rm int},\xi)}.
\end{equation}

Using eqs. (\ref{eq:InvN_Spectrum_max}) and (\ref{eq:InvN_Spectrum_Q}), for the noise-matched (critically coupled) case $\xi=1$, the maximum value of $\psi$ is
\begin{equation} \label{eq:psimax_match}
\psi_{\max}(\nu_{r},n(\nu_{r}),\xi=1) \approx \frac{1}{n(\nu_{r})},
\end{equation}
and the sensitivity quality factor is
\begin{equation} \label{eq:Qs_match}
Q_{s}(\nu_{r},n(\nu_{r}),Q_{\rm int},\xi=1) \approx \frac{Q_{\rm int}}{2\sqrt{n(\nu_{r})}}.
\end{equation}
In contrast, for the optimally matched, overcoupled case $\xi=\xi^{\rm opt}\approx 2n(\nu_{r})$, the maximum value of $\psi$ is
\begin{equation} \label{eq:psimax_opt}
\psi_{\max}(\nu_{r},n(\nu_{r}), \xi = \xi^{\rm opt}) \approx \frac{2}{3n(\nu_{r})}.
\end{equation}
This corresponds to an SNR that is a factor of 2/3 worse than the noise-matched case for a signal on resonance. The sensitivity quality factor is
\begin{equation} \label{eq:Qs_opt}
Q_{s}(\nu_{r},n(\nu_{r}), Q_{\rm int}, \xi=\xi^{\rm opt}) \approx \frac{Q_{\rm int}}{2\sqrt{3}n(\nu_{r})},
\end{equation}
so the sensitivity bandwidth is larger in the optimally matched case by a factor of $\sqrt{3n(\nu_{r})}$. 

Now we see what has happened. In return for a small, order-unity sacrifice in on-resonance sensitivity, we have gained in \emph{integrated} sensitivity, increasing parametrically the frequency range over which the receiver is not degraded by quantum noise. The bandwidth is $\sim \sqrt{n(\nu_{r})}$ larger in the optimally matched case than the noise-matched case, so our value function should be $\sim \sqrt{n(\nu_{r})}$ larger. Indeed, equation (\ref{eq:F_neu_res_eval}) gives, in the thermal limit,
\begin{equation}
\frac{F_{\rm log}[\nu_{r},\xi^{\rm opt}]}{F_{\rm log}[\nu_{r},\xi=1]} \approx \left( \frac{16}{27} n(\nu_{r}) \right)^{1/2}. 
\end{equation}

\begin{figure}[htp] 
\includegraphics[width=\textwidth]{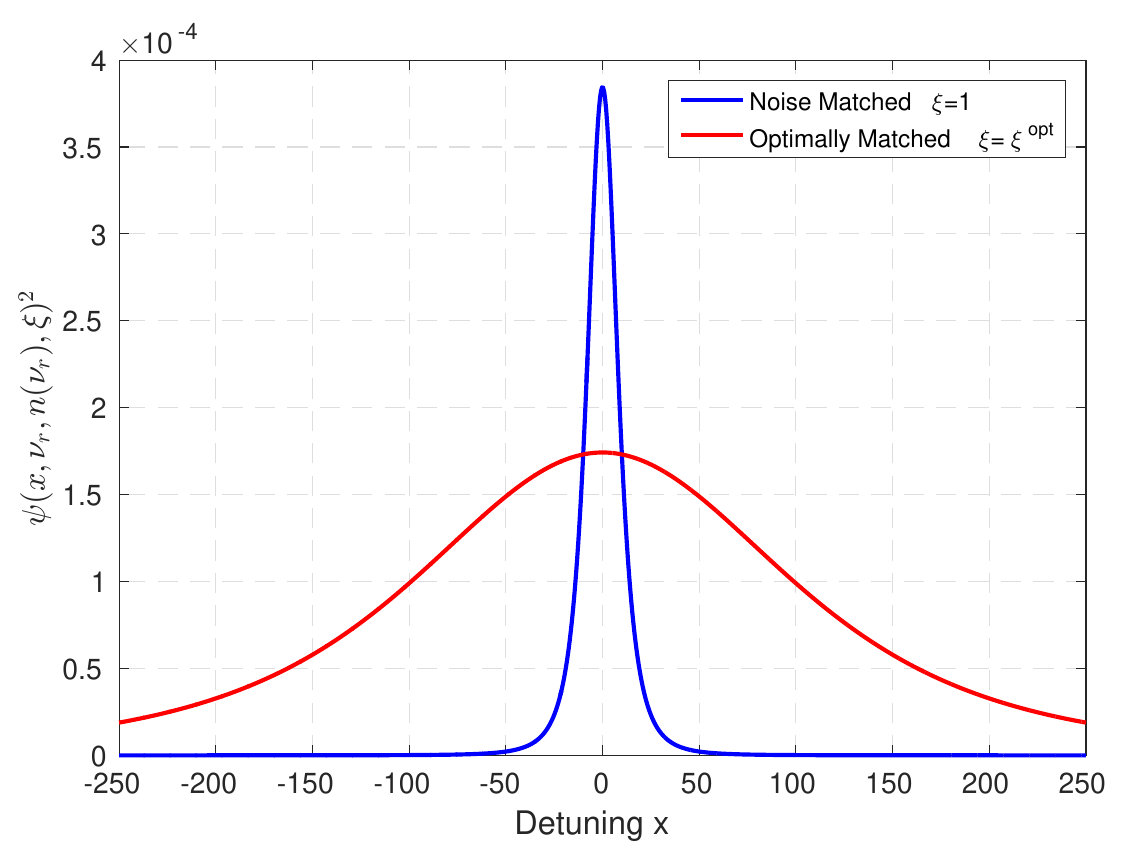}
\caption{ $\psi(x,\nu_{r},n(\nu_{r}),\xi)^{2}$ vs. the detuning from resonance $x=2Q_{\rm int} \left( \frac{\nu-\nu_{r}}{\nu_{r}} \right)$ for noise matching (blue) and optimal (red) matching, as determined by equation (\ref{eq:xi_opt}). We use $n(\nu_{r})=50$ for the on-resonance thermal occupation number. \label{fig:psivalue}}
\end{figure}

In Fig. \ref{fig:psivalue}, we pictorially demonstrate the advantage obtained with optimal matching. We plot the value of $\psi(x,\nu_{r},n(\nu_{r}),\xi)^{2}$ vs the detuning from resonance $x=2Q_{\rm int} \left( \frac{\nu-\nu_{r}}{\nu_{r}} \right)$ for the noise-matched and optimally matched resonators, using an on-resonance thermal occupation number of $n(\nu_{r})=50$. The maximum value of $\psi^{2}$ is smaller in the optimally matched case by a factor of $\approx$ 4/9. In the noise-matched case, $\psi^{2}$ falls to half of its maximum value at $|x|\approx 9$. In contrast, in the optimally matched case, $\psi^{2}$ has much larger width, falling to half of its maximum at $|x|\approx 112$. Therefore, the area under the curve, i.e. the integral of $\psi^{2}$, which measures integrated detector sensitivity and which directly relates to the value of $F_{\rm log}[\nu_{r}, \xi]$, is larger in the optimally matched resonator than in the noise-matched resonator.

It is useful to connect the result (\ref{eq:xi_opt}) back to the discussion of amplifier noise in the introduction. After all, it was that discussion, along with the analysis of the noise in Section \ref{sssec:SNR_QLA} and Fig. \ref{fig:SensitivityBandwidth}, that ultimately led to these results. We consider specifically how varying the matching parameter $\xi$ affects the signal-source-referred amplifier noise. Our analysis gives another method for understanding why the optimal resonator is noise-mismatched.

We begin with the total noise number of eq. (\ref{eq:NoiseNumber_QLamp}). From equations (\ref{eq:S21}), (\ref{eq:S22}), (\ref{eq:B_QLamp}), 
\begin{equation}
N_{\rm tot}(\nu,\nu_{r},n(\nu_{r}), Q_{\rm int}, \xi) =  n(\nu) + \frac{1}{2} + \frac{1}{|S_{21}^{(1)}(\nu,\nu_{r}, Q_{\rm int}, \xi)|^{2}} \frac{1}{2} + \frac{|S_{22}^{(1)}(\nu,\nu_{r}, Q_{\rm int}, \xi)|^{2}}{|S_{21}^{(1)}(\nu,\nu_{r}, Q_{\rm int}, \xi)|^{2}} \frac{1}{2},  \label{eq:B_QLamp_tradeoff}
\end{equation}
Rewriting this equation in terms of detuning $x$, for $|\nu-\nu_{r}| \ll \nu_{r}$,
\begin{equation}
N_{\rm tot}(x,\nu_{r},n(\nu_{r}),\xi) \approx n(\nu_{r}) + \frac{1}{2} + \left( \frac{(1+\xi)^{2}}{4\xi} + \frac{x^{2}}{4 \xi} \right) \frac{1}{2} + \left( \frac{(1-\xi)^{2}}{4\xi} + \frac{x^{2}}{4 \xi} \right) \frac{1}{2}. \label{eq:NoiseNumbert_QLamp_detuning}
\end{equation}
On the right-hand side of (\ref{eq:NoiseNumbert_QLamp_detuning}), the four terms, from left to right, represent thermal noise, zero-point fluctuation noise, amplifier quantum-imprecision noise, and amplifier quantum-backaction noise. By virtue of the fact that we are referring noise to the signal source, the dependence of noise number on matching parameter $\xi$ is given solely in the amplifier noise terms. We may readily read off the added noise number of the amplifier at detuning $x$:
\begin{equation}\label{eq:NAmp_detuning}
N_{A}(x,\xi) =  \left( \frac{(1+\xi)^{2}}{4\xi} + \frac{x^{2}}{4 \xi} \right) \frac{1}{2} + \left( \frac{(1-\xi)^{2}}{4\xi} + \frac{x^{2}}{4 \xi} \right) \frac{1}{2} =  \frac{1 + \xi^{2}}{4\xi} + \frac{x^{2}}{4 \xi}. 
\end{equation}
The above equation shows more directly, in the context of the definition of noise temperature provided in footnote \ref{TNnote} of Sec. \ref{sec:intro}, that the noise temperature only reaches the quantum limit when on-resonance and when critically coupled ($\xi=1$). 

The first term on the right-hand side of eq. (\ref{eq:NAmp_detuning}) is a frequency-independent contribution to the amplifier noise number, depending only on $\xi$. It is also the added noise number of the amplifier on resonance. The second term depends on both frequency detuning and resonator matching parameter. It represents the effect of resonator rolloff on signal-source-referred noise. At a given detuning/search frequency, as long as the frequency-dependent noise term is less than the sum of the thermal/zero-point noise and the frequency-independent part of amplifier noise (which, in sum, is equal to the total on-resonance noise), then the sensitivity is not degraded by rolloff. The sensitivity bandwidth (\ref{eq:InvN_Spectrum_Q}) gives precisely the range of detunings for which this is the case. 

From equation (\ref{eq:psi_int_approx}), to determine the optimal value of the matching parameter $\xi$ for integrated sensitivity, we must consider the product of the on-resonance sensitivity and the sensitivity bandwidth. As functions of $\xi$, the former is governed by the on-resonance total noise number $N_{\rm tot}(x=0,\nu_{r},n(\nu_{r}),\xi)$, while the latter is governed by the frequency-dependent part of the amplifier noise in (\ref{eq:NoiseNumbert_QLamp_detuning}) and (\ref{eq:NAmp_detuning}).
The on-resonance noise number is minimized when noise-matched, $\xi=1$. As $\xi$ is changed from unity, the frequency-dependent term in (\ref{eq:NoiseNumbert_QLamp_detuning}) and (\ref{eq:NAmp_detuning}), representing an increase in noise number away from resonance, is effectively rescaled $x/2 \rightarrow x/2\sqrt{\xi}$. 

For $\xi<1$, the rescaling squeezes the profile of $N_{\rm tot}(x,\nu_{r},n(\nu_{r}),\xi)$ vs $x$, relative to the profile that we would find if $x^{2}/4\xi$ in (\ref{eq:NoiseNumbert_QLamp_detuning}) was replaced by its $\xi=1$ value, $x^{2}/4$. As a result, there is a smaller sensitivity bandwidth. Combined with the increased on-resonance noise number, this implies that the integrated sensitivity must be strictly worse for $\xi<1$ than for $\xi=1$. Any value of $\xi$ less than unity therefore cannot be the optimum of $F_{\rm log}[\nu_{r},\xi]$. 

In contrast, for $\xi>1$, the rescaling stretches the profile. In this regime, there is then a tradeoff between the on-resonance amplifier noise and the sensitivity bandwidth. This observation is consistent with Fig. \ref{fig:psivalue}. We do not expect significant degradation in on-resonance SNR as long as the on-resonance amplifier noise is less than or comparable to the thermal noise:
\begin{equation} \label{eq:NoiseNumber_AmpNoise}
\left( \frac{(1+\xi)^{2}}{4\xi} + \frac{(1-\xi)^{2}}{4\xi} \right) \frac{1}{2} \lesssim n(\nu_{r}),
\end{equation}
which is equivalent to
\begin{equation} \label{eq:NoiseNumber_AmpNoise_small}
1+ (\xi)^{2} \lesssim 4\xi n(\nu_{r}).
\end{equation}
In combination with the stretching of the profile as $\xi$ increases, (\ref{eq:NoiseNumber_AmpNoise_small}) suggests that the optimal value of $\xi$ varies parametrically as
\begin{equation}
\xi^{\rm opt} \sim n(\nu_{r}),
\end{equation}
which is consistent with equation (\ref{eq:xi_opt}).

\begin{figure}[htp] 
\includegraphics[width=\textwidth]{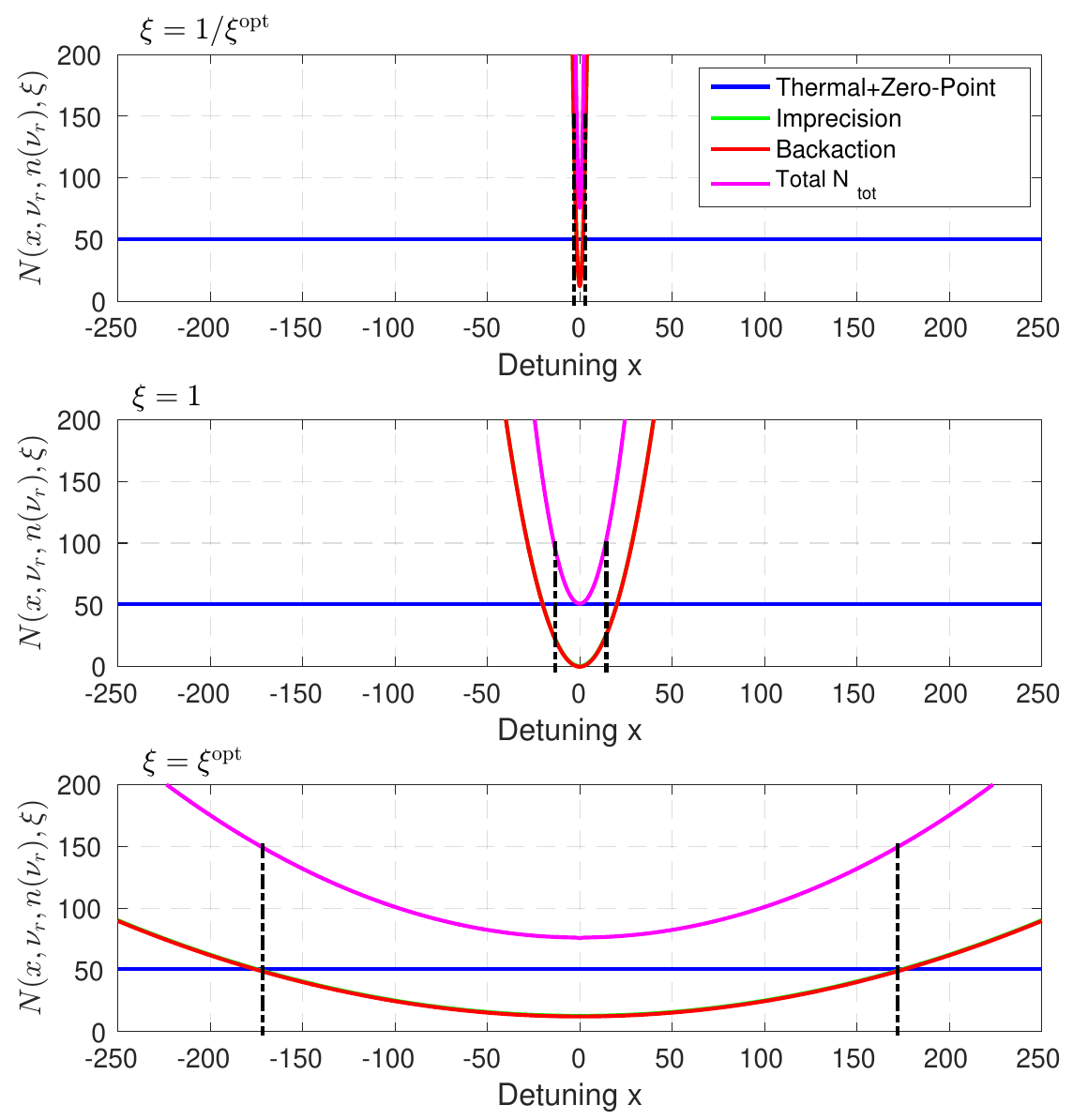}
\caption{ The noise terms of equation (\ref{eq:NoiseNumbert_QLamp_detuning}) plotted for three difference values of $\xi$: the inverse of optimal matching $\xi=1/\xi^{\rm opt}$ (top panel), noise matching $\xi=1$ (middle panel), and optimal matching $\xi=\xi^{\rm opt}$ (bottom panel). Thermal and zero-point noise is in blue, imprecision noise is in green, and backaction noise is in red. The sum of these three curves, which represents the total noise number $N_{\rm tot}(x,\nu_{r},n(\nu_{r}),\xi)$, is plotted in magenta.  As in Fig. \ref{fig:psivalue}, we use the thermal occupation number $n(\nu_{r})=50$. The imprecision curve is not visible because it overlaps with the backaction curve. The region within the dash-dotted black lines represents the sensitivity bandwidth.} \label{fig:NoiseNumber_comparison}
\end{figure}

To further demonstrate the effect of matching parameter $\xi$ on the noise, we plot each of the terms contributing to $N_{\rm tot}(x,\nu_{r},n(\nu_{r}),\xi)$ in Fig. \ref{fig:NoiseNumber_comparison}, using the same parameters as Fig. \ref{fig:psivalue}. We display plots for three different values of $\xi$: $\xi=1/\xi^{\rm opt}$, $\xi=1$, $\xi=\xi^{\rm opt}$. We group together the thermal and zero-point fluctuation noise (blue curve). The imprecision curve (green) is not visible because of the overlap with the backaction (red). The sum of these three curves gives the total noise number (magenta). As indicated in (\ref{eq:NoiseNumbert_QLamp_detuning}), the on-resonance noise number is invariant under the transformation $\xi \rightarrow  1/\xi$, so it should not change between the cases $\xi=1/\xi^{\rm opt}$ (top panel) and $\xi=\xi^{\rm opt}$ (bottom panel). This is observed in Fig. \ref{fig:NoiseNumber_comparison}, where the on-resonance noise number is approximately $75$, which is 50\% worse than the noise matched case $\xi=1$ (middle panel). 
As $\xi$ increases, we observe the stretching of the imprecision and backaction noise profiles, which in turn, stretches the profile of total noise number. 
Due to the stretching, whereas the sensitivity bandwidth (region between dash-dotted black lines in Fig. \ref{fig:NoiseNumber_comparison}) corresponds to detunings $|x| \lessapprox \sqrt{3}$ for $\xi=1/\xi^{\rm opt}$ and $|x| \lessapprox 14$ for $\xi=1$, the sensitivity bandwidth corresponds to the much wider $|x| \lessapprox 173$ for $\xi=\xi^{\rm opt}$. We see that the optimally matched resonator compensates for a small, order-unity penalty in on-resonance noise number with a parametrically larger sensitivity bandwidth, yielding the maximum integrated sensitivity. 

Additionally, the added noise number of the amplifier on resonance for optimal match is far above the quantum-limited value of 1/2:
\begin{equation}\label{eq:NAmp_opt}
N_{A}(x=0,\xi^{\rm opt}) \approx \frac{1}{2} n(\nu_{r}),
\end{equation}
where we have used the high-temperature value of (\ref{eq:xi_opt_limits}). The backaction and imprecision contributions to the amplifier noise number are approximately equal and both are comparable in size to the thermal noise. 

\emph{In summary, the optimal match for a wideband search with a single-pole resonator read out by a quantum-limited amplifier is a noise mismatch on resonance, which is accompanied by large backaction on the resonator and amplifier noise number comparable to the thermal occupation number.} A similar result, regarding measurement backaction and sensitivity outside of the resonator bandwidth, applies to flux-to-voltage amplifiers. This is discussed in Appendices \ref{ssec:FtoV_opt_QL} and \ref{ssec:FtoV_opt_Gen}.

One should note the practical advantages from exploiting the optimized sensitivity bandwidth. A large sensitivity bandwidth means that the resonator may be tuned more coarsely, significantly reducing requirements on the resolution of the tuning system. Suppose, for example, that the resonator is tuned so that the thermal occupation number is 50 (e.g. 4.2 MHz resonator at 10 mK), as in Figs. \ref{fig:psivalue} and \ref{fig:NoiseNumber_comparison}. Assume that the internal quality factor is one million. Instead of tuning the resonator at one part in $\sim Q_{\rm int}= 10^{6}$ for full coverage near the resonance frequency (without degradation from quantum noise), we may tune at one part in $\sim Q_{s} \approx 5800$. The benefits of large sensitivity bandwidths are revisited in Section \ref{ssec:QLA_sensitivity} after calculating the fundamental limits on scan sensitivity. There, we discuss the linear relationship between sensitivity bandwidth and scan rate.

Finally, we place the optimized single-pole resonator in the context of the Bode-Fano bound on single-moded inductive detectors (\ref{eq:BF_magic_ineq}). Comparing (\ref{eq:BF_res_opt}) and this bound, we find that the optimized resonator is approximately $75\%$ of the fundamental limit, regardless of the value of the thermal occupation number $n(\nu_{h})$. \emph{For any search range, the optimized single-pole resonator is close to the bound on integrated sensitivity given by the Bode-Fano criterion.} In combination with its superiority over radiatively-coupled searches, demonstrated in Sections \ref{sec:power_flow}-\ref{sec:DM_circuits}, this observation establishes the resonator as a near-ideal method for searching for dark matter over a wide bandwidth. Utilizing optimal matching is important as it provides a parametric $\sim \sqrt{n(\nu_{r})}$ boost to the value of the value function. An analogous Bode-Fano constraint exists for capacitive signal sources. Consequently, when capacitively coupling to electric fields induced by the dark matter (which one may do when the Compton wavelength of the dark matter is comparable to the detector size), the single-pole resonator is again $\approx$75\% of the Bode-Fano optimum. A similar result may be derived for background, monochromatic AC fields (as opposed to a DC magnetic field), showing that single-pole resonators are near-optimal in frequency-integrated sensitivity.

Because the single-pole resonator is practical to implement, and is close to the Bode-Fano limit, we focus on optimization and fundamental limits of a search with a tunable resonator. But first we digress to place the Bode-Fano constraint in broader context.

\subsubsection{Evading the Bode-Fano constraint}
\label{sssec:BF_evade}

The Bode-Fano constraint is derived subject to particular (stated) assumptions. Relaxation of these assumptions can in principle make it possible to outperform the constraint. 

In deriving the bound in eq. (\ref{eq:BF_magic_ineq}), we assume that the matching network is linear, passive, reciprocal, and lossless. These are key assumptions in the proof of the Bode-Fano criterion, as derived in \cite{fano1950theoretical}. Nonlinear detection schemes may be used to outperform our derived constraint. However, it should be remembered that nonlinearities can provide considerable practical challenges in detector operation, specifically in calibration and data interpretation. Active and nonreciprocal matching networks can also be used to evade the Bode-Fano limit. Broadband matching limitations for such circuits are documented in refs. \cite{youla1964new,nie2015broadband}. Exploration of these schemes, including active feedback, is left to future work.

We establish that the Bode-Fano constraint (\ref{eq:BF_magic_ineq}) holds if the assumption of a lossless network is relaxed. It is not surprising that a lossy network does not outperform a lossless network. Broadband matching criteria allowing for network loss have been derived in a manner similar to the Bode-Fano criterion. Ref. \cite{nie2015broadband} gives
\begin{equation}
\int_{\nu_{l}}^{\nu_{h}} d\bar{\nu}\ ln \left( \frac{1}{1-|S_{21}^{(1)}(\bar{\nu})|^{2}} \right) \leq \frac{R}{ L_{\rm PU}}.
\end{equation}
For a lossless network, this expression reduces to the classic criterion of eq. (\ref{eq:Bode_Fano}). (See eq. (\ref{eq:ScattRelation}).) We assume that any network loss is at the same temperature as the loss $R$ in the signal source. The total noise-equivalent number is not that given by (\ref{eq:NoiseNumber_QLamp}), but instead, the more general expression
\begin{equation} \label{eq:NoiseNumber_Lossy}
N_{\rm tot}^{\rm lossy}(\nu, S^{(1)}(\nu), n(\nu))= \frac{(1-|S_{22}^{(1)}(\nu)|^{2})n(\nu)+1}{|S_{21}^{(1)}(\nu)|^{2}},
\end{equation}
which may be derived using the techniques in Ref. \cite{zmuidzinas2003thermal}. $|S_{21}^{(1)}(\nu)|^{2}+ |S_{22}^{(1)}(\nu)|^{2} < 1$ for a lossy network, so the noise-equivalent number is naturally greater when the matching network possesses loss. The analogue of the scaled, log-uniform-search value functional is the integral of the inverse-squared of equation (\ref{eq:NoiseNumber_Lossy}):
\begin{equation}
\bar{F}_{\rm log}^{\rm lossy}[S^{(1)}(\nu)]= \int_{\nu_{l}}^{\nu_{h}} d\bar{\nu}\ N_{\rm tot}^{\rm lossy}(\bar{\nu}, S^{(1)}(\bar{\nu}), n(\bar{\nu}))^{-2} = \int_{\nu_{l}}^{\nu_{h}} d\bar{\nu} \left( \frac{|S_{21}^{(1)}(\bar{\nu})|^{2}}{(1-|S_{22}^{(1)}(\bar{\nu})|^{2})n(\bar{\nu})+1} \right)^{2}.
\end{equation}
Note that
\begin{equation}
\bar{F}_{\rm log}^{\rm lossy}[S^{(1)}(\nu)] \leq \int_{\nu_{l}}^{\nu_{h}} d\bar{\nu} \left( \frac{|S_{21}^{(1)}(\bar{\nu})|^{2}}{|S_{21}^{(1)}(\bar{\nu})|^{2} n(\bar{\nu})+1} \right)^{2}.
\end{equation}
We may then proceed as in Section \ref{sssec:Bode_Fano_opt} to obtain a bound on integrated sensitivity $\bar{F}_{\rm log}^{\rm lossy}[S^{(1)}(\nu)]$, which is identical to (\ref{eq:BF_magic_ineq}). The bound can only be achieved with equality when the conditions laid out below eq. (\ref{eq:BF_magic_ineq}) are satisfied and when the matching network is lossless. In other words, optimized integrated sensitivity requires no network loss. This leaves us with just the three assumptions of linear, passive, and reciprocal networks.

It should be remembered that a search can be accelerated by using multiple receivers simultaneously and combining their information with the techniques described in Section \ref{sec:resonator_SNR}. In the language of Fig. 1 and equivalent circuits, such multiple-receiver approaches can be represented as possessing multiple signal sources, each of which couples to the dark matter. A multiple-receiver approach may naturally have multiple readout ports, i.e. more than one amplifier reading out the signals. Suppose that the experiment contains $N>1$ signal sources and $M \geq 1$ readout ports. Adapting the approaches of Section \ref{ssec:reactive_scattering}, we may describe the detector using a scattering matrix of size $(N+M)\times (N+M)$. We can determine the detector SNR by adding the output timestreams of the $M$ readouts coherently or incoherently, the latter having been performed in Section \ref{sssec:SNR_scan}.

A multi-wavelength structure, a frequency-division multiplexer, and a multi-moded cavity resonator may be described using such a formalism. In a multi-wavelength structure such as coupled cavities comprised of stacks of dielectrics \cite{caldwell2017dielectric,baryakhtar2018axion}, each half-wavelength section effectively acts as a signal source.

In a frequency-division multiplexer, the fixed experimental volume is divided into smaller sub-volumes that are scanned at different frequencies with approximately non-overlapping frequency response. However, one can show that, if the size of the setup is less than a coherence length in all dimensions, at all frequencies being probed, then this multiplexing approach suffers a disadvantage. We compare a tunable single-pole cavity resonator of volume $V_{c}$ to an $N$-element multiplexer, with each resonator element possessing volume $V_{c}/N$. (See Appendix \ref{sec:cav_circuit}.) With the multiplexer, one is able to integrate at each search frequency $\sim N$ times longer than the single, larger resonator, but the energy coupled is $\sim N$ times smaller. We assume that the noise power is identical for the large resonator and each resonator in the multiplexer. Then, the SNR at every frequency in the search range is reduced (relative to the large resonator) by $\sim N/\sqrt{N} = \sqrt{N}$. The spatial coherence of the dark matter allows the SNR to increase linearly in volume, which leads to the advantage for the larger, single-pole resonator. Alternatively, instead of frequency division multiplexing sub-volumes operated at non-overlapping frequencies, one may consider building multiple identical cavities in sub-volumes tuned simultaneously to the same frequency.\cite{Sho14} Adding the output timestreams coherently then gives the same SNR as one would find with the single, larger cavity. So there is no advantage over the single, larger cavity.

In a multi-moded cavity resonator, each mode can be modeled as an equivalent RLC circuit coupling reactively to the dark-matter-induced electromagnetic fields. Unlike the multiplexer, each mode may possess the full physical volume $V_{c}.$ Multi-mode circuits thus naturally reduce the scan time in (i.e. increase the sensitivity of) a light-field dark-matter search, and consequently, the development of such circuits is an active area of inquiry in the axion detection community. \cite{graham2015experimental}

Nevertheless, like their single-signal-source counterparts, detection schemes with multiple signal-source elements are constrained by fundamental impedance matching statements. There are constraints on sensitivity from generalized broadband matching criteria (i.e. generalizations of the Bode-Fano criterion), such as those in the aforementioned reference \cite{nie2015broadband}. One must consider fundamental limitations on signal combining. For instance, it is a well-known theorem that no three-port power combiner can be matched, reciprocal, and lossless.\cite{pozar2012microwave} There are also practical limitations in multiple-receiver approaches; the signal sources interact (e.g. mode interactions in a multi-moded cavity), which can provide practical challenges in calibration.

In summary, the Bode-Fano limit (\ref{eq:BF_magic_ineq}) can in principle be evaded by relaxing the assumption of a linear, passive, reciprocal system, using for example, active impedance-matching. Furthermore, the optimization framework developed in this paper can be extended to a multiple-receiver approach (such as multi-wavelength structures and multi-moded cavities) by extending to scattering matrices with larger dimensions and making optimal use of coherent and incoherent information. However, the optimized multiple-receiver approach is still limited by generalized broadband matching criteria.

\subsubsection{Matching network optimization for candidate signal}
\label{sssec:candidate_search_opt}

We return to the value function of equation (\ref{eq:F_gen_def}) and consider the situation in which there is a candidate signal to validate. Is it still optimal to noise mismatch, in accordance with equation (\ref{eq:xi_opt})? Suppose that the frequency of the signal is $\nu_{\rm DM}^{0*}$. Assume the resonator is centered at frequency $\nu_{\rm DM}^{0*}$ with $Q_{s}(\nu_{r}, n(\nu_{r}), Q_{\rm int}, \xi) \simlt 10^{6}$ so that the signal is maximally resonantly enhanced and so that the dark-matter signal, of width $\Delta\nu_{\rm DM}/\nu_{\rm DM}^{0*}\sim 10^{-6}$, is within the sensitivity bandwidth. This condition may be enforced by setting the internal quality factor to $Q_{\rm int} \simlt 10^{6}$. In such a situation, rather than assuming a log-uniform frequency probability distribution, we adopt a prior that weights the probability high at frequency $\nu_{\rm DM}^{0*}$. Keeping the other three assumptions introduced for the log-uniform search, this weighting reduces the value function (\ref{eq:F_gen_def}) to
\begin{equation}
F_{\rm cand}[\xi]\approx \int dg_{\rm DM}\  SNR^{\rm opt}\left[ \nu_{\rm DM}^{0*}, g_{\rm DM}, \left( \frac{d\rho_{\rm DM}}{d\nu} \right) (\nu,\nu_{\rm DM}^{0*}), \psi(\nu,\nu_{r},n(\nu), Q_{\rm int}, \xi) , \tau \right]^{2} P(g_{\rm DM}|\nu_{\rm DM}^{0*}).
\end{equation}
For example, we could set the distribution to a delta function
\begin{equation}
P(\nu_{\rm DM}^{0})=\delta(\nu_{\rm DM}^{0} - \nu_{\rm DM}^{0*}).
\end{equation}
Using (\ref{eq:SNR_1_QLamp}) and (\ref{eq:InvN_Spectrum_Lorentzian}),
\begin{equation}\label{eq:F_cand}
F_{\rm cand}[\xi] \approx \gamma_{2} \psi(\nu_{\rm DM}^{0*}=\nu_{r},\nu_{r},n(\nu_{r}), Q_{\rm int}, \xi)^{2} = \gamma_{2} \left( \frac{4\xi}{4\xi n(\nu_{r}) + (1+\xi)^{2}} \right)^{2},
\end{equation}
where $\gamma_{2}$ is a constant. $F_{\rm cand}[\xi]$ is maximized when $\xi=1$. Thus, when probing a fixed frequency, the resonator should be noise matched on-resonance to the amplifier, such that the signal power is fully transmitted. That we should noise match is expected because for a fixed frequency probe, obtaining maximum sensitivity at that frequency is important; obtaining maximum \textit{integrated} sensitivity, as we did for the log-uniform search, is not a factor. \emph{In summary, the optimization of the impedance-matching network differs for a wideband, scanning search and a search with the prior of a candidate signal. It is optimal to be noise-mismatched on resonance, as per (\ref{eq:xi_opt}), in the former case, and noise matched in the latter case.}

\subsubsection{Comparison to other calculations}
\label{sssec:other_calcs_comparison}

As a final remark for Section \ref{ssec:opt_impedance}, we compare our calculations to similar results.

It was first observed by Krauss, Moody, and Wilczek in \cite{krauss1985calculations} that the resonator coupling coefficient (matching parameter) that maximizes integrated sensitivity is different from that which maximizes on-resonance SNR. However, the authors did not consider the spectral shape of the thermal noise, which is filtered through the matching network and which is important to forming accurate conclusions regarding the optimization. If one does not take the shape into consideration, then $|S_{21}^{(1)}(\nu)|^{2}n(\nu)$ simply becomes $n(\nu)$ in the denominator of (\ref{eq:InvN_Spectrum}). In this case, the thermal noise factor can be pulled out of the value functional (\ref{eq:F_gen_neutral}) for a resonant detector, which yields the same optimization as that found for $n(\nu) \ll 1$: $\xi^{\rm opt} \approx 2$. This is precisely what was found in Ref. \cite{krauss1985calculations}.

A similar conclusion regarding integrated sensitivity was found in \cite{lehnert2016quantumaxion} for quantum squeezing. In this approach, the quantum limit is evaded by injecting a squeezed state in the resonator. The optimized coupling coefficient is $\xi \sim G_{S}$, where $G_{S}$ is the squeezer gain. The authors explain the effect in terms of sensitivity outside of the bare (uncoupled) resonator linewidth, as we have done here. However, squeezer gains are often of order $\sim$10, while at low frequencies, the thermal occupation number in a 10 mK dilution refrigerator is much larger than that. For experiments using near-quantum-limited SQUID amplifiers, this implies a much larger sensitivity bandwidth, relative to the uncoupled resonator bandwidth. Finally, we note that our observations regarding sensitivity outside of the resonator bandwidth may apply to searches for which the dark matter does not couple electromagnetically, especially if an electromagnetic readout element is used. For example, our insights may apply to CASPEr, which searches for the axion coupling to gluons and spin using nuclear magnetic resonance techniques \cite{Budker:2013hfa}. If the spin projection noise, which follows the resonant line shape, is larger than the SQUID readout noise, as well as any thermal noise from loss in tuned/untuned input circuits, then the bandwidth over which the NMR sample provides maximal sensitivity can be substantially larger than the NMR linewidth, set by the transverse relaxation time $T_{2}$.\cite{sleator1985nuclear} 


An analysis of noise matching at low resonant frequencies $h\nu_{r} \ll kT$ has previously been conducted for Weber bars, fixed mechanical oscillators used for gravitational wave detection.\cite{thorne1980gravitational} This analysis has some features in common with the analysis in this paper. However, the Weber bar analysis a priori assumes a resonant impedance matching. In contrast, we conduct a general analysis in the context of all possible reactively coupled receivers (resonant or otherwise). Furthermore, the fixed-frequency Weber bar analysis cannot be applied directly to scanned resonant electromagnetic dark-matter detectors, where noise matching must be optimized for integrated sensitivity over a frequency scan, with a comprehensive consideration of priors on the dark matter signal.

We have shown that the optimized single-pole resonator is close to the fundamental limit on matching networks, imposed by the Bode-Fano criterion. Note, importantly, that while the Bode-Fano criterion is typically used to constrain frequency-integrated signal transfer, we have used it to constrain frequency-integrated signal-to-noise, which is the appropriate figure of merit for an axion or hidden-photon dark-matter search. To our knowledge, this result is new and demonstrates the efficiency of resonant searches.

\subsection{Optimization of Time Distribution}
\label{ssec:opt_time}

We have seen from the Bode-Fano analysis that a narrowband matching network is more sensitive than a broadband search. (See also Section \ref{ssec:ResvBroad} and Appendix \ref{sec:ResvBroad_SQUID}.) It is necessary to tune the narrowband matching network to different central frequencies to search over a large mass range. In this part of the paper, we determine the optimal time distribution of a scanned system. We consider a single-pole resonator, which is almost as efficient as the Bode-Fano-optimal tophat filter, but is much more practical in a tuned experiment. We first introduce the value function for the scan optimization, motivated by our previous discussions on SNR and priors. Then, in Section \ref{sssec:log_uniform_scan}, we calculate the optimal scan strategy (optimal distribution of resonator frequencies and corresponding integration times) given prior probabilities that are log-uniform in both mass and coupling. In Section \ref{sssec:scan_opt_other}, we consider the optimal scan strategy in other situations, such as those that would be relevant to a search targeted at QCD axion models. We conclude in Section \ref{sssec:scan_practical}, where we interpret our results in terms of a practical scan strategy. 

For optimizing allocation of scan time, the potential for a significant off-resonance contribution to the SNR, described in Sections \ref{sssec:SNR_QLA} and \ref{ssec:opt_impedance}, underscores the importance of combining SNR information from multiple resonant scan steps. Let $\{ \nu_{r}^{i} \}$ represent the set of resonator scan frequencies; in the language of Section \ref{sssec:SNR_scan}, $\nu_{r}^{i}$ is the resonance frequency of the $i$th configuration in the scan. For dark matter at rest-mass frequency $\nu_{\rm DM}^{0}$, the SNR from the $i$th scan step is, from Section \ref{sssec:SNR_QLA},
\begin{align}
& SNR^{\rm opt} \left[ \nu_{\rm DM}^{0}, g_{\rm DM}, \left( \frac{d\rho_{\rm DM}}{d\nu} \right) (\nu,\nu_{\rm DM}^{0}), \psi(\nu,\nu_{r}^{i},n(\nu), Q_{\rm int}^{i}, \xi^{i}) , \tau_{i} \right]^{2} \label{eq:SNR_res_1} \\
& = (2\pi^{2})^{2} \tau_{i} \int_{\nu_{\rm DM}^{0}}^{\nu_{\rm DM}^{0} + \Delta\nu_{\rm DM}^{c}} d\nu\ \left( \frac{\nu L_{\rm PU}}{R(\nu)}  \frac{U_{\rm DM}(\nu,\nu_{\rm DM}^{0}, g_{\rm DM}, \frac{d\rho_{\rm DM}}{d\nu}(\nu,\nu_{\rm DM}^{0}))}{h} \psi(\nu,\nu_{r}^{i},n(\nu), Q_{\rm int}^{i}, \xi^{i}) \right)^{2} \nonumber \\
& \approx (4\pi Q_{\rm int}^{i})^{2} \tau_{i} \int_{\nu_{\rm DM}^{0}}^{\nu_{\rm DM}^{0} + \Delta\nu_{\rm DM}^{c}} d\nu\ \left( \frac{\nu}{\nu_{r}^{i}} \frac{g_{\rm DM}^{2}V_{\rm PU} \frac{d\rho_{\rm DM}}{d\nu}(\nu,\nu_{\rm DM}^{0})}{h} \frac{\xi^{i}}{4\xi^{i}n(\nu) + (1+\xi^{i})^{2} + 4 (Q_{\rm int}^{i})^{2} \left( \frac{\nu}{\nu_{r}^{i}} -1 \right)^{2}} \right)^{2}, \nonumber
\end{align}
where $\psi$ is as defined in (\ref{eq:InvN_Spectrum_Lorentzian}). We have fixed the pickup volume $V_{\rm PU}$, the geometric factor $k(\nu_{\rm DM}^{0})$, and for axions, the background magnetic field, which, from eq. (\ref{eq:gDM_def}), implies that $g_{\rm DM}$, the effective dimensionless coupling to dark matter, is independent of the scan step (e.g. subscript $i$ is not necessary).
We discuss deviations from these assumptions in Section \ref{sssec:scan_opt_other}; a case of interest that violates the constant volume assumption is a GHz-scale cavity search that scans over multiple octaves, with the volume of the cavity decreasing as frequency increases. In the third line we assume that $|\nu_{\rm DM}^{0} - \nu_{r}^{i} | \ll \nu_{r}^{i}$; otherwise, the dark-matter search frequency is far from the sensitivity bandwidth (which is a narrow bandwidth around resonance), and the resonator does not provide significant sensitivity. We can then approximate $R(\nu) \approx R(\nu_{r}^{i})$. Additionally, implicit in our SNR expression is the assumption that the dwell time $\tau_{i}$ is much longer than $t_{\textrm{rec},i}^{*}(\nu_{\rm DM}^{0})$, defined in (\ref{eq:tstar_def}). Because we restrict to resonators, we define the timescale specifically as
\begin{equation}\label{eq:t_star_res}
t_{\textrm{rec},i}^{*}(\nu_{\rm DM}^{0}) \rightarrow t^{*}(\nu_{\rm DM}^{0}, \nu_{r}^{i}) \equiv \textrm{max} (10^{6}/\nu_{\rm DM}^{0}, Q_{\textrm{int}}^{i} / (\nu_{r}^{i} (1+\xi^{i})) ).
\end{equation}
From Section \ref{sssec:SNR_scan}, the total SNR from all scan steps is the quadrature sum of the SNR from individual scan steps (\ref{eq:SNR_res_1}):
\begin{align}
SNR^{\rm opt}_{\rm tot} & \left[ \nu_{\rm DM}^{0}, g_{\rm DM}, \left( \frac{d\rho_{\rm DM}}{d\nu} \right) (\nu,\nu_{\rm DM}^{0}), \{\psi(\nu,\nu_{r}^{i},n(\nu), Q_{\rm int}^{i}, \xi^{i}) \} , \{\tau_{i}\} \right]^{2}  \label{eq:SNR_res_scan} \\
&= \sum_{i} SNR^{\rm opt} \left[ \nu_{\rm DM}^{0}, g_{\rm DM}, \left( \frac{d\rho_{\rm DM}}{d\nu} \right) (\nu,\nu_{\rm DM}^{0}), \psi(\nu,\nu_{r}^{i},n(\nu), Q_{\rm int}^{i}, \xi^{i}) , \tau_{i} \right]^{2}. \nonumber 
\end{align}

We know from the previous section how to use the SNR formulae to optimize the matching network given a resonance frequency. However, how do we use the formulae to choose the time allocation at each resonance frequency? More specifically, given fixed total integration time $T_{\rm tot}$, and a search band $\nu_{l} \leq \nu_{\rm DM}^{0} \leq \nu_{h}$, what is the optimal scan strategy? What is the optimal distribution of resonator frequencies and dwell times?

Of course, the answer to this question depends on how one defines ``optimal.'' In a scan, one generically aims to exclude as much of the mass-coupling parameter space as possible. However, like the matching-network optimization, we must take into account priors. For example, there may be portions of the search range $\nu_{l} \leq \nu_{\rm DM}^{0} \leq \nu_{h}$ where strong constraints have been set, e.g. by previous direct detection searches, and where allocating much scan time is unfavorable. Conversely, there may be regions of parameter space that are well-motivated theoretically and that we wish to probe deeply. Our value function for the scan strategy is then the weighted area in the mass-coupling parameter space excluded by the search:
\begin{equation} \label{eq:A_gen_def}
A[\{ \nu_{r}^{i} \}, \{ \tau_{i} \}] \equiv \int_{\nu_{l}}^{\nu_{h}} d\nu_{\rm DM}^{0} \int_{g_{\rm DM,min}(\nu_{\rm DM}^{0}, \{ \nu_{r}^{i} \} , \{ \tau_{i} \})}^{g_{\rm DM,max}(\nu_{\rm DM}^{0})} dg_{\rm DM}\ W(\nu_{\rm DM}^{0}, g_{\rm DM})
\end{equation}
where $W(\nu_{\rm DM}^{0},g_{\rm DM})$ is the weighting function for the area. The weighting function can be based solely on the probability distributions over mass and coupling, for instance
\begin{equation} \label{eq:area_weight_prob}
W(\nu_{\rm DM}^{0}, g_{\rm DM}) \propto P(g_{\rm DM}| \nu_{\rm DM}^{0}) P(\nu_{\rm DM}^{0}),
\end{equation}
where the constant of proportionality is arbitrary. The weighting function may also be based on the desire to probe a well-motivated region of parameter space, for instance, the QCD band for axions or the mass range corresponding to inflationary production for hidden photons. The upper limit on the integral over coupling $g_{\rm DM,max}(\nu_{\rm DM}^{0})$ is flexibly defined as a value for which dark matter at frequency $\nu_{\rm DM}^{0}$ has already been strongly excluded. The lower limit $g_{\rm DM, \rm min}(\nu_{\rm DM}^{0}, \{ \nu_{r}^{i} \} , \{ \tau_{i} \})$ is the minimum coupling to which our experiment is sensitive, defined by the coupling value for which 
\begin{equation}
SNR^{\rm opt}_{\rm tot} \left[ \nu_{\rm DM}^{0}, g_{\rm DM}, \left( \frac{d\rho_{\rm DM}}{d\nu} \right) (\nu,\nu_{\rm DM}^{0}), \{\psi(\nu,\nu_{r}^{i},n(\nu), Q_{\rm int}^{i}, \xi^{i}) \} , \{\tau_{i}\} \right]^{2} 
\end{equation}
of eq. (\ref{eq:SNR_res_scan}) is unity. $g_{\rm DM,min}$ depends on the set of scan steps $\{ \nu_{r}^{i} \}$ and dwell times $\{ \tau_{i} \}$ and is assumed to be less than $g_{\rm DM,max}(\nu_{\rm DM}^{0})$; it is here where the scan strategy enters into the value function. The objective of the optimization is to determine the scan steps and dwell times such that $A[\{ \nu_{r}^{i} \}, \{ \tau_{i} \}]$ is maximized.

\subsubsection{Optimizing time allocation for a log-uniform search}
\label{sssec:log_uniform_scan}

Here, we explore in more detail the log-uniform search which is not only log-uniform in mass (as in the matching network optimization of Section \ref{ssec:opt_impedance}), but also log-uniform in coupling. As per equation (\ref{eq:area_weight_prob}), we choose the weighting function
\begin{equation}
W(\nu_{\rm DM}^{0}, g_{\rm DM}) =  \frac{1}{\nu_{\rm DM}^{0}} \frac{1}{g_{\rm DM}}
\end{equation}
so that the value function becomes
\begin{equation} \label{eq:A_def}
A_{\rm log}[ \{ \nu_{r}^{i} \} , \{ \tau_{i} \} ] \equiv -\int_{\nu_{l}}^{\nu_{h}} \frac{d\nu_{\rm DM}^{0}}{\nu_{\rm DM}^{0}} \ln g_{\rm DM,min}(\nu_{\rm DM}^{0}, \{ \nu_{r}^{i} \} , \{ \tau_{i} \})  
\end{equation}
We have omitted a constant term, which is the frequency integral of $\ln g_{\rm DM,max}(\nu_{\rm DM}^{0})$. Since $g_{\rm DM}$ is proportional to the $\varepsilon$ for hidden photons and $g_{a\gamma \gamma}$ for axions (see eq. (\ref{eq:gDM_def})), our optimization function maximizes the exclusion area in a log-log plot of mass vs $\varepsilon$ / $g_{a\gamma \gamma}$. We discuss other possible value functions in Section \ref{sssec:scan_opt_other}. We now proceed to maximize $A_{\rm log}$ and determine the optimal time allocation for a log-uniform scan, subject to fixed total search time.

Define
\begin{align} 
\varphi & \left( \nu_{\rm DM}^{0},\nu_{r}^{i}, \frac{d\rho_{\rm DM}}{d\nu} (\nu,\nu_{\rm DM}^{0}) , Q_{\rm int}^{i} ,  \xi^{i} \right) \nonumber \\
& \equiv g_{\rm DM}^{-4} \tau_{i}^{-1} SNR^{\rm opt} \left[ \nu_{\rm DM}^{0}, g_{\rm DM}, \left( \frac{d\rho_{\rm DM}}{d\nu} \right) (\nu,\nu_{\rm DM}^{0}), \psi(\nu,\nu_{r}^{i},n(\nu), Q_{\rm int}^{i}, \xi^{i}) , \tau_{i} \right]^{2}, \label{eq:u_def}
\end{align}
which is independent of $g_{\rm DM}$ (see equation (\ref{eq:SNR_res_1})) . $\varphi \left( \nu_{\rm DM}^{0},\nu_{r}^{i}, \frac{d\rho_{\rm DM}}{d\nu} (\nu,\nu_{\rm DM}^{0}) , Q_{\rm int}^{i} ,  \xi^{i} \right)$ represents the square of the SNR contribution from the resonator at $\nu_{r}^{i}$ per unit time per unit $g_{\rm DM}^{4}$ when $\tau_{i} \gg t^{*}(\nu_{\rm DM}^{0},\nu_{r}^{i})$. It is therefore an important quantity for time allocation optimization. We have made explicit the dependence of $\varphi$ on the resonator parameters $Q_{\rm int}^{i}$ and $\xi^{i}$. These two parameters are discussed below in the assumptions for the optimization.  For each search frequency $\nu_{\rm DM}^{0}$, we only consider scan steps for which the dwell time is much longer than $t^{*}(\nu_{\rm DM}^{0},\nu_{r}^{i})$; otherwise, our SNR treatment from Section \ref{sec:resonator_SNR} is not valid and equation (\ref{eq:u_def}) cannot be interpreted as stated. We then obtain, from (\ref{eq:SNR_res_scan}),
\begin{equation} \label{eq:gDM_min_time}
\ln g_{\rm DM,min} (\nu_{\rm DM}^{0}, \{ \nu_{r}^{i} \},\{ \tau_{i} \}) = -\frac{1}{4} \ln \left( \sum_{i_{k}(\nu_{\rm DM}^{0})} \tau_{i_{k}(\nu_{\rm DM}^{0})} \varphi \left( \nu_{\rm DM}^{0}, \nu_{r}^{i_{k}(\nu_{\rm DM}^{0})}, \frac{d\rho_{\rm DM}}{d\nu} (\nu,\nu_{\rm DM}^{0}), Q_{\rm int}^{i_{k}(\nu_{\rm DM}^{0})}, \xi^{i_{k}(\nu_{\rm DM}^{0})} \right) \right),
\end{equation}
and
\begin{equation} \label{eq:A_opt_u}
A_{\rm log}[ \{ \nu_{r}^{i} \} , \{ \tau_{i} \} ] = \frac{1}{4} \int_{\nu_{l}}^{\nu_{h}} \frac{d\nu_{\rm DM}^{0}}{\nu_{\rm DM}^{0}} \ln \left( \sum_{i_{k}(\nu_{\rm DM}^{0})} \tau_{i_{k}(\nu_{\rm DM}^{0})} \varphi \left( \nu_{\rm DM}^{0}, \nu_{r}^{i_{k}(\nu_{\rm DM}^{0})}, \frac{d\rho_{\rm DM}}{d\nu} (\nu,\nu_{\rm DM}^{0}), Q_{\rm int}^{i_{k}(\nu_{\rm DM}^{0})}, \xi^{i_{k}(\nu_{\rm DM}^{0})} \right) \right) ,
\end{equation}
where $\{ i_{k}(\nu_{\rm DM}^{0}) \}$ is the index set for the subsequence of scan steps satisfying 
\begin{equation}
\tau_{i_{k}(\nu_{\rm DM}^{0})} \gg t^{*}(\nu_{\rm DM}^{0}, \nu_{r}^{i_{k}(\nu_{\rm DM}^{0})}).
\end{equation}
Equation (\ref{eq:A_opt_u}) would, in general, have to be evaluated and optimized numerically. However, in practice, the scan takes steps much smaller than a sensitivity bandwidth over the entire range. Thus, multiple scan frequencies are sensitive to the signal at any particular dark-matter search frequency without degradation. Such a scan pattern ensures that there are no gaps in coverage of the search band. We refer to this class of scans as ``dense'' scans. For dense scans, we may discuss the scan strategy not in terms of discrete steps, but rather in terms of an approximately continuous process. As we demonstrate, the transformation from a discrete to a continuous scan enables us to solve for the optimal scan strategy.

We adopt the assumptions from the optimization of the impedance-matching network for a log-uniform search, as carried out in Section \ref{ssec:opt_impedance}. We make five additional assumptions regarding the scan:
\begin{enumerate}
\item The internal Q is fixed during the scan. We revisit this assumption in Sections \ref{sssec:scan_opt_other} and \ref{ssec:QLA_sensitivity}. We also assume that the temperature is fixed.

\item For each scan step, the matching parameter $\xi^{i}$ is set to its optimal value $\xi^{i}=\xi^{\rm opt}(\nu_{r}^{i},n(\nu_{r}^{i}))$. We omit the dependence on thermal occupation number for brevity in this section; it is implied.

\item For each receiver configuration in the scan, $\tau_{i}$ is much longer than $t^{*}(\nu_{r}^{i},\nu_{r}^{i})$. This implies that, within any narrow band around resonance, the dwell time is much longer than $t^{*}(\nu_{\rm DM}^{0},\nu_{r}^{i})$, and (\ref{eq:u_def}) applies.

\item Suppose that the scan step sequence $\{ \nu_{r}^{i} \}$ is in ascending order, $\nu_{r}^{1} \leq \nu_{r}^{2} \leq  ...$. We assume that the set of scan steps that provide information on the hypothesis of dark matter at $\nu_{\rm DM}^{0}$ form an integer interval of indices. In other words, if resonators at $\nu_{r}^{i_{1}}$ and $\nu_{r}^{i_{2}}$ collect information on dark matter at $\nu_{\rm DM}^{0}$ and $i_{1} < i_{2}$, then, for any $i_{m}$, $i_{1} \leq i_{m} \leq i_{2}$, $\nu_{r}^{i_{m}}$ also collects information on $\nu_{\rm DM}^{0}$. This is a sensible assumption, given that we expect all resonant frequencies ``close'' to the search frequency to give high SNR information. 

We denote by $\nu_{r,l}(\nu_{\rm DM}^{0})$ the lowest resonance frequency which collects SNR data on $\nu_{\rm DM}^{0}$ and by $\nu_{r,h}(\nu_{\rm DM}^{0})$ the highest resonance frequency. In any practical search, we are able to tune the resonance over a wide bandwidth. In particular, we are able to cover a range around $\nu_{\rm DM}^{0}$ that contains several sensitivity bandwidths $\nu_{\rm DM}^{0}/Q_{s}(\nu_{\rm DM}^{0},n(\nu_{\rm DM}^{0}),Q_{\rm int}, \xi^{\rm opt}(\nu_{\rm DM}^{0}))$, as well as several dark-matter cutoff bandwidths $\beta\nu_{\rm DM}^{0}$. Therefore, we may assume that
$\nu_{r,l}(\nu_{\rm DM}^{0})$ and $\nu_{r,h}(\nu_{\rm DM}^{0})$ satisfy
\begin{equation} \label{eq:Delta_nur_int}
\frac{\nu_{\rm DM}^{0}}{Q_{s}(\nu_{\rm DM}^{0}, n(\nu_{\rm DM}^{0}), Q_{\rm int}, \xi_{\rm opt}(\nu_{\rm DM}^{0}))}, \beta \nu_{\rm DM}^{0} \ll |\nu_{r,l}(\nu_{\rm DM}^{0})-\nu_{\rm DM}^{0}|,\ |\nu_{r,h}(\nu_{\rm DM}^{0})-\nu_{\rm DM}^{0}| \ll \nu_{\rm DM}^{0}.
\end{equation}
We thus rewrite eq. (\ref{eq:A_opt_u}) as
\begin{align} 
A_{\rm log}[ \{ \nu_{r}^{i} \} , \{ \tau_{i} \} ] &= \frac{1}{4} \int_{\nu_{l}}^{\nu_{h}} \frac{d\nu_{\rm DM}^{0}}{\nu_{\rm DM}^{0}} \label{eq:A_opt_varphi_restricted} \\ 
& \ln \left( \sum_{i:\ \nu_{r,l}(\nu_{\rm DM}^{0}) \leq \nu_{r}^{i} \leq \nu_{r,h}(\nu_{\rm DM}^{0}) }  \tau_{i} \varphi \left( \nu_{\rm DM}^{0}, \nu_{r}^{i}, \frac{d\rho_{\rm DM}}{d\nu} (\nu,\nu_{\rm DM}^{0}), Q_{\rm int}, (\xi^{i})^{\rm opt} \right) \right) \nonumber
\end{align} 
We note that, though we make the restriction (\ref{eq:Delta_nur_int}), the values of $\nu_{r,l}(\nu_{\rm DM}^{0})$ and $\nu_{r,h}(\nu_{\rm DM}^{0})$ can be chosen flexibly.

\item The total experiment time $T_{\rm tot}$ is long enough that we may take small steps:
\begin{equation} \label{eq:small_step}
\frac{\Delta \nu_{r}^{i}}{\nu_{r}^{i}} = \frac{\nu_{r}^{i+1}-\nu_{r}^{i}}{\nu_{r}^{i}} \ll \frac{1}{Q_{s}(\nu_{r}^{i}, n(\nu_{r}^{i}), Q_{\rm int}, \xi^{\rm opt}(\nu_{r}^{i}) )} = \frac{\left( (1 + \xi^{\rm opt}(\nu_{r}^{i}))^{2} +4\xi^{\rm opt}(\nu_{r}^{i}) n(\nu_{r}^{i}) \right)^{1/2}}{Q_{\rm int}}
\end{equation}
We refer to this assumption as the ``dense scan'' assumption. It can be met under the condition that
\begin{equation} \label{eq:T_lower_constraint}
T_{\rm tot} \gg \int_{\nu_{l}}^{\nu_{h}} \frac{d\nu_{r}}{\nu_{r}} t^{*}(\nu_{r},\nu_{r}) \frac{Q_{\rm int}}{ \left( (1 + \xi^{\rm opt}(\nu_{r}))^{2} +4\xi^{\rm opt}(\nu_{r})n(\nu_{r})  \right)^{1/2} },
\end{equation}
where, from (\ref{eq:t_star_res}), we have $t^{*}(\nu_{r},\nu_{r})= \rm{max}(10^{6}/\nu_{r}, Q_{\rm int}/(\nu_{r}(1+\xi^{\rm opt}(\nu_{r}))))$. For reference, for internal quality factors on the order of one million, $\nu_{l}=1$ kHz, and $\nu_{h}=100$ MHz, the integral in (\ref{eq:T_lower_constraint}) evaluates to several hours.

\end{enumerate}


As discussed above, the dense scan assumption implies that, for each dark-matter frequency, there are multiple resonant frequencies that are sensitive to the signal without degradation from quantum noise. We may then turn the sum in equation (\ref{eq:A_opt_varphi_restricted}) into an integral. A useful proxy for dwell time $\tau_{i}$ is a positive, differentiable scan density function $\tau(\nu_{r})$. Here, $\tau(\nu_{r})$ is defined such that the time to scan a frequency band of width $d\nu_{r}$ centered at frequency $\nu_{r}$ is $\tau(\nu_{r}) \frac{d\nu_{r}}{\nu_{r}}$. By choosing a scan density function $\tau(\nu_{r})$ and steps $\{\nu_{r}^{i} \}$, we choose dwell times:
\begin{equation}
\tau_{i}= \tau(\nu_{r}^{i}) \frac{\nu_{r}^{i+1}-\nu_{r}^{i}}{\nu_{r}^{i}}.
\end{equation}
One may note that turning a sum into a integral usually means that we are taking the limit of infinitesimally small frequency steps: $\Delta \nu_{r}^{i}=\nu_{r}^{i+1}-\nu_{r}^{i} \rightarrow 0$. This may seem odd given that, physically, for a fixed scan time $T_{\rm tot}$, the dwell time $\tau_{i}$ at each step would also go to zero; this would imply that we are not dwelling longer than the resonator ring-up/dark-matter-coherence times. When not dwelling for a coherence time, the SNR formulae presented above are no longer valid, and therefore, it would seem that our argument is not self-consistent. However, as long as the physical frequency scan steps (each with dwell time longer than the ring-up/coherence times) are sufficiently small, then mathematically, the sum may be approximated as an integral. We thus distinguish between the physical requirement that the dwell time be longer than the ring-up/coherence times for each resonant scan step (which still holds) and the mathematical approximation of turning the sum of (\ref{eq:A_opt_varphi_restricted}) into an integral for a dense scan.\footnote{For resonator-based searches, one often quantifies the stepping of central frequency in terms of scan rate $d\nu_{r}/dt$, which is related to scan density function by $d\nu_{r}/dt=\nu_{r}/\tau(\nu_{r})$.\cite{brubaker2018first} We do not utilize the metric of scan rate for a number of reasons. In comparison to scan rate, scan density function interfaces more directly with the SNR formulae and with the constraint of fixed total search time. Unlike scan density function, scan rate, as used in the literature, often implicitly or explicitly encodes a specific prior.\cite{asztalos2010squid}} 

We then obtain
\begin{equation} \label{eq:A_tau_cont}
A_{\rm log}[\tau(\nu_{r})]= \frac{1}{4} \int_{\nu_{l}}^{\nu_{h}} \frac{d\nu_{\rm DM}^{0}}{\nu_{\rm DM}^{0}}\ \ln \left( \int_{\nu_{r,l}(\nu_{\rm DM}^{0}) }^{ \nu_{r,h}(\nu_{\rm DM}^{0}) } \frac{d\nu_{r}}{\nu_{r}}\ \tau(\nu_{r}) \varphi \left(\nu_{\rm DM}^{0}, \nu_{r}, \frac{d\rho_{\rm DM}}{d\nu}(\nu,\nu_{\rm DM}^{0}), Q_{\rm int}, \xi^{\rm opt}(\nu_{r}) \right) \right),
\end{equation}
Consider a fixed search frequency $\nu_{\rm DM}^{0}$. We can take the scan-density function $\tau(\nu_{r})$ to vary slowly on the scale of the dark-matter cutoff bandwidth $\beta \nu_{\rm DM}^{0}$ because any pair of resonators separated by less than this bandwidth probe the same dark-matter signal. Additionally, we can take the scan density function $\tau(\nu_{r})$ to vary slowly on the scale of the sensitivity bandwidth $\nu_{\rm DM}^{0}/Q_{s}(\nu_{\rm DM}^{0},n(\nu_{\rm DM}^{0}),Q_{\rm int}, \xi^{\rm opt}(\nu_{\rm DM}^{0}))$. Consider two resonators in a dense scan and assume that their resonance frequencies are located within the sensitivity bandwidth, which is centered at $\nu_{\rm DM}^{0}$. Suppose we allocate equal time (equal $\tau(\nu_{r})$) at the two steps. If we allocated more time at one resonance frequency than at the other (represented by a higher value of $\tau(\nu_{r})$ at the former frequency), then the contribution at $\nu_{\rm DM}^{0}$ to the value functional $A[\tau(\nu_{r})]$ would not change appreciably from the equal-time situation. The dark-matter signal rings up both resonators without degradation from quantum noise and hence, both resonators contribute approximately equally to the sensitivity on coupling.\cite{brubaker2018first} We thus assume 
\begin{equation} \label{eq:tau_slow}
\left| \frac{d\tau}{d \nu_{r}}(\nu_{\rm DM}^{0}) \right| \ll \frac{\tau(\nu_{\rm DM}^{0})}{\nu_{r,h}(\nu_{\rm DM}^{0})-\nu_{r,l}(\nu_{\rm DM}^{0}) }.
\end{equation}
From eq. (\ref{eq:SNR_res_1}), we find that equation (\ref{eq:A_tau_cont}) can then be approximated as 
\begin{equation} \label{eq:A_tau_cont_slow}
A_{\rm log}[\tau(\nu_{r})]\approx \frac{1}{4} \int_{\nu_{l}}^{\nu_{h}} \frac{d\nu_{\rm DM}^{0}}{\nu_{\rm DM}^{0}}\ \ln \left( \tau(\nu_{\rm DM}^{0}) \int_{\nu_{r,l}(\nu_{\rm DM}^{0}) }^{ \nu_{r,h}(\nu_{\rm DM}^{0}) } \frac{d\nu_{r}}{\nu_{r}}\ \varphi \left(\nu_{\rm DM}^{0}, \nu_{r}, \frac{d\rho_{\rm DM}}{d\nu}(\nu,\nu_{\rm DM}^{0}), Q_{\rm int}, \xi_{\rm opt}(\nu_{r}) \right) \right).
\end{equation}

Maximizing the functional $A_{\rm log}[ \tau(\nu_{r}) ] $ with respect to the constraint of fixed search time
\begin{equation}\label{eq:Ttot_constraint}
T_{\rm tot}= \int_{\nu_{l}}^{\nu_{h}} \frac{d\nu_{r}}{\nu_{r}} \tau(\nu_{r})
\end{equation}
is now straightforward. The solution is that $\tau(\nu)$ should be constant:
\begin{equation} \label{eq:t_opt}
\tau_{\rm opt}(\nu)=\frac{T_{\rm tot}}{\ln(\nu_{h}/\nu_{l})}.
\end{equation}
As such, the optimum scan distributes time logarithmically among bins. The time spent at a resonance frequency should be proportional to the size of the frequency step $\Delta \nu_{r}$. It is not better, from the standpoint of maximizing the area of the log-log exclusion curve, to spend more time in one decade of frequency than another. 

A constant scan-density function is expected. In the log-space of coupling $g_{\rm DM}$, one does not gain faster as a function of time at some set of search frequencies over another set. The single-step SNR increases with time as $t^{1/2}$ regardless of frequency, as long as the dwell time is much longer than the dark-matter coherence/resonator ring-up times.

Note that though we solved for the optimal scan-density function, we said nothing about the frequencies  $\{ \nu_{r}^{i} \}$ at which we should step. For a sufficiently dense scan, the particular frequencies are inconsequential.\cite{brubaker2018first} It is easy to see that if we double the number of steps, and cut the dwell time in half, the value of the functional $A_{\rm log}[ \{\nu_{r}^{i} \} , \{ \tau_{i} \} ]$ changes negligibly (assuming that we still wait longer than the dark-matter coherence time and resonator-ring-up time).

\subsubsection{Optimizing time allocation for other priors}
\label{sssec:scan_opt_other}

The time allocation of equation (\ref{eq:t_opt}) is appropriate when we use log-uniform priors for the dark-matter mass and coupling. However, we can also consider scan strategies/time distributions that take into account strongly favored regimes. For example, for axion dark matter, the QCD band is a coupling regime that is well-motivated theoretically by the strong CP problem.\cite{peccei1977cp}  In order to search for axion dark matter that also solves the strong CP problem, it is desirable to integrate down to the QCD band at each frequency, rather than following the scan strategy function (\ref{eq:t_opt}). How far one integrates, e.g. whether to the KSVZ line or to the DFSZ line \cite{kim1979weak,shifman1980can,dine1981simple,zhitnitskij1980possible,graham2015experimental}, depends specifically on one's priors and whether one applies additional weighting in the function $W$ that makes integrating to the DFSZ hypothesis more favorable (e.g. a ``bonus'' for reaching a particular part of parameter space).

Nevertheless, if we still assume a dense scan that is slowly varying and satisfies eq. (\ref{eq:tau_slow}), we may use much of the framework developed in Section \ref{sssec:log_uniform_scan} to compute the optimal time allocation in wideband searches. Using equations (\ref{eq:u_def}) and (\ref{eq:tau_slow}), we find
\begin{align}
g_{\rm DM,min}[ \nu_{\rm DM}^{0}, \tau(\nu_{r}) ]^{-4} &= \int_{\nu_{r,l}(\nu_{\rm DM}^{0})}^{\nu_{r,h}(\nu_{\rm DM}^{0})} \frac{d\nu_{r}}{\nu_{r}} \tau(\nu_{r}) \varphi \left( \nu_{\rm DM}^{0}, \nu_{r}, \frac{d\rho_{\rm DM}}{d\nu}(\nu,\nu_{\rm DM}^{0}), Q_{\rm int}, \xi^{\rm opt}(\nu_{r}) \right) \nonumber \\
& \approx \tau(\nu_{\rm DM}^{0}) \int_{\nu_{r,l}(\nu_{\rm DM}^{0})}^{\nu_{r,h}(\nu_{\rm DM}^{0})} \frac{d\nu_{r}}{\nu_{r}} \varphi \left( \nu_{\rm DM}^{0}, \nu_{r}, \frac{d\rho_{\rm DM}}{d\nu}(\nu,\nu_{\rm DM}^{0}), Q_{\rm int}, \xi^{\rm opt}(\nu_{r}) \right) \label{eq:gDMmin_gen}
\end{align}
where we have replaced the resonant frequencies and dwell times in the argument of $g_{\rm DM,min}$ with the scan density function and have adopted the five assumptions that are listed in Section \ref{sssec:log_uniform_scan}. Note, in particular, that we have adopted assumption 2, which sets the matching parameter to the optimal value of eq. (\ref{eq:xi_opt}). Such an assumption is appropriate even when not assuming log-uniform priors; see the discussion following eq. (\ref{eq:xi_opt_limits}).

It is possible that, during a search, the volume $V_{\rm PU}$ changes. In other words, the pickup volume can be a function of the resonance frequency. This would occur in a many-octave scan using a cavity mode. In such a wideband scan, the cavity is typically exchanged (i.e. between runs of the detector) at periodic intervals; as the mass/frequency increases, the size of the cavity becomes smaller because the size of the cavity cannot be much larger than $(\lambda_{\rm DM}^{0})^{3}$. One can compensate for the reduced pickup volume from a single cavity by using an array of cavities all tuned to the same resonant frequency\cite{Sho14}. However, if one does not do this (e.g. due to practical constraints), then the pickup volume decreases. Similarly, the internal quality factor and the geometric factor $k(\nu_{\rm DM}^{0})$ could change during the search. In fact, as shown in eq. (\ref{eq:cav_factor}), it is natural for the geometric factor, related to the cavity overlap factor, to change if the cavity is exchanged. In such a situation, $g_{\rm DM}$ can be a function of the receiver configuration; it can depend on the configuration index ``$i$.'' Then, we more conveniently define the value functions with respect to the candidate-specific couplings $g_{a\gamma \gamma}$ and $\varepsilon$, which are related to $g_{\rm DM}$ in eq. (\ref{eq:gDM_def}). Specifically, for the axion we can write the value function
\begin{equation}\label{eq:A_axion}
A_{a}[ \{ \nu_{r}^{i} \} , \{ \tau_{i} \} ] \equiv \int_{\nu_{l}}^{\nu_{h}} d\nu_{\rm DM}^{0} \int_{g_{a \gamma \gamma,\rm min}(\nu_{\rm DM}^{0}, \{ \nu_{r}^{i} \} , \{ \tau_{i} \})}^{g_{a\gamma \gamma,\rm max}(\nu_{\rm DM}^{0})} dg_{a\gamma \gamma}\ W_{a}(\nu_{\rm DM}^{0}, g_{a\gamma\gamma})
\end{equation}
where the weighting function $W_{a}(\nu_{\rm DM}^{0}, g_{a\gamma \gamma})$ is defined analogous to $W(\nu_{\rm DM}^{0}, g_{\rm DM})$. $g_{a\gamma \gamma, \rm max}$ is defined analogous to $g_{\rm DM,max}$, and $g_{a\gamma \gamma, \rm min}(\nu_{\rm DM}^{0},\{ \nu_{r}^{i} \}, \{ \tau_{i} \})$, dependent on the scan strategy, is defined analogous to $g_{\rm DM,min}(\nu_{\rm DM}^{0},\{ \nu_{r}^{i} \}, \{ \tau_{i} \})$. For the hidden photon, we may similarly write
\begin{equation}\label{eq:A_HP}
A_{\gamma'}[ \{ \nu_{r}^{i} \} , \{ \tau_{i} \} ] \equiv \int_{\nu_{l}}^{\nu_{h}} d\nu_{\rm DM}^{0} \int_{\varepsilon_{\rm min}(\nu_{\rm DM}^{0}, \{ \nu_{r}^{i} \} , \{ \tau_{i} \})}^{\varepsilon_{\rm max}(\nu_{\rm DM}^{0})} d\varepsilon \ W_{\gamma'}(\nu_{\rm DM}^{0}, \varepsilon)
\end{equation}
where $W_{\gamma '}(\nu_{\rm DM}^{0},\varepsilon)$ is the weighting function for the parameter space of mass and mixing angle. Similar to eq. (\ref{eq:u_def}), we may define
\begin{align} 
\varphi_{a} & \left( \nu_{\rm DM}^{0},\nu_{r}^{i}, \frac{d\rho_{\rm DM}}{d\nu} (\nu,\nu_{\rm DM}^{0}) , Q_{\rm int}^{i} , \xi^{i}, k^{i}(\nu_{\rm DM}^{0}), V_{\rm PU}^{i} \right) \nonumber \\
& \equiv g_{a\gamma \gamma }^{-4} \tau_{i}^{-1} SNR^{\rm opt} \left[ \nu_{\rm DM}^{0}, g_{a\gamma \gamma}, \left( \frac{d\rho_{\rm DM}}{d\nu} \right) (\nu,\nu_{\rm DM}^{0}), \psi(\nu,\nu_{r}^{i},n(\nu), Q_{\rm int}^{i}, \xi^{i}), k^{i}(\nu_{\rm DM}^{0}), V_{\rm PU}^{i}, \tau_{i} \right]^{2}, \label{eq:u_a_def}
\end{align}
\begin{align} 
\varphi_{\gamma'} & \left( \nu_{\rm DM}^{0},\nu_{r}^{i}, \frac{d\rho_{\rm DM}}{d\nu} (\nu,\nu_{\rm DM}^{0}) , Q_{\rm int}^{i} , \xi^{i}, k^{i}(\nu_{\rm DM}^{0}), V_{\rm PU}^{i} \right) \nonumber \\
& \equiv \varepsilon^{-4} \tau_{i}^{-1} SNR^{\rm opt} \left[ \nu_{\rm DM}^{0}, \varepsilon, \left( \frac{d\rho_{\rm DM}}{d\nu} \right) (\nu,\nu_{\rm DM}^{0}), \psi(\nu,\nu_{r}^{i},n(\nu), Q_{\rm int}^{i}, \xi^{i}), k^{i}(\nu_{\rm DM}^{0}), V_{\rm PU}^{i}, \tau_{i} \right]^{2}, \label{eq:u_gamma_def}
\end{align}
and derive, for a continuous scan density function $\tau(\nu_{r})$,
\begin{equation} \label{eq:g_a_min}
g_{a\gamma \gamma, \rm min}(\nu_{\rm DM}, \tau(\nu_{r}))^{-4} = \int_{\nu_{r,l}(\nu_{\rm DM}^{0})}^{\nu_{r,h}(\nu_{\rm DM}^{0})} \frac{d\nu_{r}}{\nu_{r}} \tau(\nu_{r}) \varphi_{a} \left( \nu_{\rm DM}^{0}, \nu_{r}, \frac{d\rho_{\rm DM}}{d\nu}, Q_{\rm int}(\nu_{r}), k(\nu_{\rm DM}^{0},\nu_{r}), V_{\rm PU}(\nu_{r}), \xi^{\rm opt}(\nu_{r}) \right) 
\end{equation}
and similarly for the hidden photon. We have explicitly allowed the internal quality factor $Q_{\rm int}$, the geometric factor $k(\nu_{\rm DM}^{0})$, and the pickup volume $V_{\rm PU}$ to vary with resonant frequency in eqs. (\ref{eq:u_a_def}), (\ref{eq:u_gamma_def}), and (\ref{eq:g_a_min}).

We revisit the computational aspects of scan optimization in Section \ref{ssec:QLA_sensitivity}, where we calculate the quantum limits on optimized resonant scan sensitivity for log-uniform priors and discuss calculation of minimum coupling (for example, calculation of the integrals in (\ref{eq:gDMmin_gen}) and (\ref{eq:g_a_min})) for generic scan density functions (e.g. for other priors).

\subsubsection{Aspects of a practical scan strategy}
\label{sssec:scan_practical}

Multiple aspects of a practical scan must be observed in interpreting our result for time allocation optimization.

For instance, in assumption 4 above, we assume that all resonant frequencies ``close'' to the search frequency give high SNR information. However, it is possible that a resonance frequency close to the search frequency gives low SNR. The existence of parasitic mechanical resonances and electromagnetic interference is inevitable at some level in any practical system. The coupling between electrical and mechanical modes can produce a sideband at the dark-matter search frequency, or an electromagnetic pickup line can interfere at the search frequency. Such data would need to be considered on a case-by-case basis and may result in discarding the data from the total dataset. For our optimization, for simplicity, we ignore these practical problems, and assume that they are managed with more sophisticated signal processing and experimental protocols.

Furthermore, no scan is entirely continuous, as assumed above. At each step, one must wait for the system to settle. If one is tuning the resonator with a dielectric structure (as is the case in DM Radio, ADMX, and HAYSTAC), moving the dielectric inevitability causes vibrations. One must wait for these vibrations to damp to a negligible amplitude before acquiring low-noise data. The settling time may be dependent on resonance frequency, and a more detailed time optimization would be required to take this effect into account. For a flux-to-voltage amplifier (see Appendix \ref{ssec:FtoV_opt_QL}), after some number of scan steps, when the frequency has changed enough that the thermal occupation number is appreciably different, the noise impedance of the amplifier (i.e. the coupling to the amplifier) needs to be re-optimized. The re-optimization can be conducted, e.g., by tuning a variable transformer based on Josephson junctions, or by tuning the bias parameters of some amplifiers. There is inevitably a settling time associated with this tuning. 

Additionally, searches over many octaves require significant hardware changes over the scan because of the limited frequency range of any tuning system. Such a hardware change could entail, for example, changing the pickup inductor in a lumped-element system (e.g. adding turns to the coil to lower the range of resonance frequencies achieved with a capacitively tunable resonator\cite{Chaudhuri:2014dla}) or switching cavities. These changes might occur, for example, every octave in frequency. In addition, the (flux-to-voltage) amplifier would need to be swapped with an amplifier with different coupling strength to enable a larger tuning range in noise impedance. These changes of course cannot be made while acquiring search data.   

One must also consider how searches are affected by yearly scan schedules and ongoing technological improvements. Consider, for example, a two-year scan. Suppose that in the first year, a search covers a particular mass range, and in the second year, an identical scan is conducted (same dwell time at the same resonance frequencies as the first year). Assuming no change in resonator sensitivity, i.e. no change in amplifier noise or quality factor, the doubling of data results in a $\sqrt{2}$ improvement in SNR, or equivalently, a $2^{1/4}$ improvement in sensitivity to photon-dark matter coupling. A single two-year scan, in which one doubles the single-year dwell time at each resonant frequency, would, in theory, give the same SNR improvement. However, the single scan is, in practice, more efficient, because it halves the number of scan steps and thus, reduces total settling time. At the same time, one must consider that searches are constantly improved as the underlying technologies are developed. It is possible that, in the second year of the scan, a higher-Q resonator or a lower-noise amplifier becomes available. One may find that the benefits of improved resonator sensitivity outweigh the loss in integration time from additional scan steps. 

It is straightforward to include the lost integration time due to settling, cryogenic cycling, and apparatus modification in the optimization analysis above, if these costs are quantitatively known. However, these time costs are often sufficiently ad hoc that they should probably be implemented on a case-by-case basis. We note that the idealized optimization here still provides a baseline for planning scans that can be modified due to practical constraints.
\section{Scan Sensitivity of A Resonant Search}
\label{sec:resonator_sensitivity}

Through our analysis in Sections \ref{sec:power_flow}-\ref{sec:scan_opt}, we have now conducted a broad optimization of single-moded, linear, passive receivers subject to the Standard Quantum Limit on phase-insensitive amplification, covering all three boxes of the receiver block diagram Fig. \ref{fig:DetectorBlockDiagram}. We have shown a generic disadvantage for radiative couplings relative to reactive couplings in the signal source, owing to a mismatch betwen the dark-matter source impedance and the free-space impedance. Using the Bode-Fano criterion, we have demonstrated that a single-pole resonator constitutes the near-optimal linear, passive matching network subject to the SQL. We have optimized noise matching and time allocation in a scan. In this section, we combine these insights to yields limits on search sensitivity. In Section \ref{ssec:QLA_sensitivity}, we calculate the fundamental limit on resonator sensitivity read out by a quantum-limited amplifier. In Section \ref{ssec:FOM}, we interpret our fundamental limit in the context of practical tradeoffs that may be necessary in an experimental campaign. 

\subsection{Resonator Read Out by a Quantum-Limited Amplifier}
\label{ssec:QLA_sensitivity}

As in Sections \ref{ssec:opt_impedance} and \ref{ssec:opt_time}, we first calculate sensitivity for a log-uniform search. We discuss two aspects of the calculation: the parametric dependence on quality factor and the implications for scan rate at low frequency. We then consider sensitivity for other priors and the relation of the sensitivity calculation to the time allocation optimization in Section \ref{ssec:opt_time}.

\subsubsection{Sensitivity for a log-uniform search}

Assuming a sufficiently dense scan lasting at least several hours, and assuming fixed internal $Q$, pickup volume $V_{\rm PU}$, and geometric factor $k(\nu_{\rm DM}^{0})$, equations (\ref{eq:SNR_res_1}) and (\ref{eq:SNR_res_scan}) give, with optimized log-uniform matching network (\ref{eq:xi_opt}) and time allocation (\ref{eq:t_opt}),
\begin{align}
&SNR^{\rm opt}\left[ \nu_{\rm DM}^{0}, g_{\rm DM},  \frac{d\rho_{\rm DM}}{d\nu} (\nu,\nu_{\rm DM}^{0})  \right]^{2} = ( 4\pi Q_{\rm int})^{2}  \int \frac{d\nu_{r}}{\nu_{r}} \tau_{\rm opt}(\nu_{r}) \label{eq:SNR_opt_total} \\ 
& \hspace{0.25 cm} \times \int_{\nu_{\rm DM}^{0}}^{\nu_{\rm DM}^{0} + \Delta\nu_{\rm DM}^{c}} d\nu\ \left( \frac{\nu}{\nu_{r}} \frac{g_{\rm DM}^{2}V_{\rm PU} \frac{d\rho_{\rm DM}}{d\nu}(\nu,\nu_{\rm DM}^{0})}{h} \frac{\xi^{\rm opt}(\nu_{r})}{4\xi^{\rm opt}(\nu_{r})n(\nu) + (1+\xi^{\rm opt}(\nu_{r}))^{2} + 4(Q_{\rm int})^{2} \left(\frac{\nu}{\nu_{r}}-1 \right)^{2} } \right)^{2} \nonumber \\
& \hspace{6cm} \approx 4\pi^{3} Q_{\rm int} \frac{T_{\rm tot}}{\ln(\nu_{h}/\nu_{l})} \frac{\xi^{\rm opt}(\nu_{\rm DM}^{0})^{2}}{ (4\xi^{\rm opt}(\nu_{\rm DM}^{0})n(\nu_{\rm DM}^{0}) + (1+\xi^{\rm opt}(\nu_{\rm DM}^{0}))^{2})^{3/2} } \nonumber \\
& \hspace{6.5cm} \times \int_{\nu_{\rm DM}^{0}}^{\nu_{\rm DM}^{0} + \Delta\nu_{\rm DM}^{c}} d\nu \left( \frac{g_{\rm DM}^{2} V_{\rm PU} \frac{d\rho_{\rm DM}}{d\nu} (\nu,\nu_{\rm DM}^{0}) }{h} \right)^{2}, \nonumber
\end{align}
where the first integral is over the range of data-taking frequencies flexibly-defined in the assumptions of Section \ref{ssec:opt_time}. It may be evaluated in a manner similar to the Lorentzian integral over $\bar{\nu}$ in equations (\ref{eq:F_neu_res}) and (\ref{eq:F_neu_res_eval}). Additionally, note that $\xi^{\rm opt}(\nu_{\rm DM}^{0})$ is implicitly dependent on the thermal occupation number $n(\nu_{\rm DM}^{0})$. The limit on sensitivity is determined by the coupling $g_{\rm DM}$ for which the SNR is unity. Therefore, the detector is sensitive to couplings
\begin{align}
g_{\rm DM} & \geq \left( Q_{\rm int} \frac{\nu_{\rm DM}^{0} T_{\rm tot}}{\ln(\nu_{h}/\nu_{l})} \frac{ 4\pi^{3}  \xi^{\rm opt}(\nu_{\rm DM}^{0})^{2}}{ (4\xi^{\rm opt}(\nu_{\rm DM}^{0})n(\nu_{\rm DM}^{0}) + (1+\xi^{\rm opt}(\nu_{\rm DM}^{0}))^{2})^{3/2} } \left( \frac{\rho_{\rm DM}V_{\rm PU}}{h\nu_{\rm DM}^{0}}\right)^{2} \right)^{-1/4} \nonumber\\
& \left( \int_{\nu_{\rm DM}^{0}}^{\nu_{\rm DM}^{0} + \Delta\nu_{\rm DM}^{c}} \frac{d\nu}{\nu_{\rm DM}^{0}} \left( \frac{\nu_{\rm DM}^{0}}{\rho_{\rm DM}} \frac{d\rho_{\rm DM}}{d\nu} (\nu,\nu_{\rm DM}^{0})  \right)^{2} \right)^{-1/4}. \label{eq:g_QLamp}
\end{align}
For a top-hat dark-matter spectrum of width $\Delta \nu_{\rm DM} =\nu_{\rm DM}^{0}/Q_{\rm DM}$, where $Q_{\rm DM} = 10^{6}$ is the characteristic quality factor for the dark-matter bandwidth, this can be further simplified to
\begin{equation}
g_{\rm DM} \geq \left( Q_{\rm DM} Q_{\rm int} \frac{ \nu_{\rm DM}^{0} T_{\rm tot}}{\ln(\nu_{h}/\nu_{l})} \frac{ 4\pi^{3}  \xi_{\rm opt}(\nu_{\rm DM}^{0})^{2}}{ (4\xi_{\rm opt}(\nu_{\rm DM}^{0})n(\nu_{\rm DM}^{0}) + (1+\xi_{\rm opt}(\nu_{\rm DM}^{0}))^{2})^{3/2} } \right)^{-1/4} \left( \frac{ \rho_{\rm DM} V_{\rm PU} }{h \nu_{\rm DM}^{0} } \right)^{-1/2}. \label{eq:g_QLamp_flat}
\end{equation}
We may also evaluate the limit for the standard halo model using the results in Appendix \ref{sec:SH_Model}. The result is 
\begin{equation}
g_{\rm DM} \geq \left( (4.4 \times 10^{5}) Q_{\rm int} \frac{ \nu_{\rm DM}^{0} T_{\rm tot}}{\ln(\nu_{h}/\nu_{l})} \frac{ 4\pi^{3}  \xi_{\rm opt}(\nu_{\rm DM}^{0})^{2}}{ (4\xi_{\rm opt}(\nu_{\rm DM}^{0})n(\nu_{\rm DM}^{0}) + (1+\xi_{\rm opt}(\nu_{\rm DM}^{0}))^{2})^{3/2} } \right)^{-1/4} \left( \frac{ \rho_{\rm DM} V_{\rm PU} }{h \nu_{\rm DM}^{0} } \right)^{-1/2}. \label{eq:g_QLamp_SHM}
\end{equation}
Using (\ref{eq:kappa_def}) and (\ref{eq:gDM_def}), we may evaluate the sensitivity to hidden photon mixing angle $\varepsilon$ and axion-photon coupling $g_{a\gamma\gamma}$. As alluded to in Section \ref{ssec:amps} and discussed in detail in Appendix \ref{ssec:FtoV_opt_QL}, these sensitivity formulae are also valid for quantum-limited flux-to-voltage amplifiers. 

Note that the calculated sensitivity for the hidden-photon search is based on a resonator that happens to be perfectly aligned with each Fourier component of the effective current density. (See Section \ref{ssec:reactive_circuit}.) To minimize the SNR degradation due to detector misalignment, we require three identical resonators pointed in mutually orthogonal directions. We calculate in Appendix \ref{ssec:direction_variation} that such a configuration limits the degradation in SNR to $\sqrt{3}$ and the degradation in sensitivity to mixing angle to $\sqrt[4]{3}$.

Additionally, one may note that we could use Bode-Fano optimal matching networks, as described in Section \ref{sssec:Bode_Fano_opt}, instead of single-pole resonators for the best search. However, these are narrowband and would also need to be scanned. The optimal scan-density function (\ref{eq:t_opt}) would still represent the best time allocation for a log-uniform search, except instead of resonant frequencies, the argument would represent the center frequency of the top-hat. The total squared SNR at the dark-matter search frequency would be $\sim 1/3$ better than that for a resonator. The limit on dark-matter coupling would only be several percent better. \emph{Given a fixed receiver volume, loss, magnetic field (for axions), and temperature, the quantum-limited scan with a single-pole resonator is nearly optimal relative to all single-moded linear, passive receivers using a phase-insensitive amplifier subject to the SQL.}

\subsubsection{Dependence of sensitivity on resonator quality factor}

One important feature of the sensitivity expressions (\ref{eq:g_QLamp}) and (\ref{eq:g_QLamp_flat}) is the dependence on internal quality factor. The sensitivity to dark matter grows as $Q_{\rm int}^{-1/4}$, independent of whether the overall $Q$ satisfies $Q<Q_{\rm DM}$ or $Q>Q_{\rm DM}$. This parametric dependence differs from previous sensitivity estimates given in \cite{Sik85} and \cite{Chaudhuri:2014dla}.\footnote{ Due to the optimization of matching, which fixes $\xi$ to its optimal value, $Q_{\rm int}$, $Q$, and $Q_{s}$ are all proportional to each other. Thus, a scaling of sensitivity as $Q_{\rm int}^{-1/4}$ implies a scaling as $Q^{-1/4}$ and $Q_{s}^{-1/4}$. See equations (\ref{eq:Q_xi}) and (\ref{eq:InvN_Spectrum_Q}) for scattering-mode amplifiers, as well as the treatments of flux-to-voltage amplifiers in Appendices \ref{sec:FtoV_Amps} and \ref{ssec:FtoV_opt_QL}. \label{Qnote} }  We now explain this difference. 

It is useful to reference the SNR equation (\ref{eq:Dicke}), rexpressed here in terms of the system noise temperature $T_{S}$ \cite{asztalos2010squid}:
\begin{equation}\label{eq:Dicke_TS}
SNR \approx \frac{P_{\rm sig}}{kT_{S}\Delta \nu} \sqrt{\Delta\nu \cdot t}.
\end{equation}
As there are two system bandwidths here--namely, the resonator bandwidth and the dark-matter-signal bandwidth--one must carefully choose the value of $\Delta \nu$. The proper choice is provided by our insights in Section \ref{sssec:SNR_one}. We should use the bandwidth of the optimally filtered signal, which is obtained from the integrand of (\ref{eq:sig_1}) with $f=f^{\rm opt}$. For dark matter at rest-mass frequency $\nu_{\rm DM}^{0}$ and the resonator centered nearby at $\nu_{r}^{i} \approx \nu_{\rm DM}^{0}$, the bandwidth is approximately the minimum of the dark-matter bandwidth and sensitivity bandwidth
\begin{equation}\label{eq:delta_nuDMopt}
\Delta \nu_{\rm DM}^{\rm opt}(\nu_{\rm DM}^{0},\nu_{r}^{i}) \approx \textrm{min}(\nu_{\rm DM}^{0}/Q_{\rm DM},\nu_{r}^{i}/Q_{s}(\nu_{r}^{i},n(\nu_{r}^{i}),Q_{\rm int},\xi^{\rm opt}(\nu_{r}^{i}) ).
\end{equation}
We have used the variable $Q_{\rm DM}$, rather than the specific value of $10^{6}$ above eq. (\ref{eq:g_QLamp_flat}), in order to show that the effects discussed below are actually independent of what value we set for the dark-matter bandwidth.

In the case of hidden-photon dark matter, in evaluating the limits for $Q<Q_{\rm DM}$, the authors of reference \cite{Chaudhuri:2014dla} find that the sensitivity to $g_{\rm DM}$ varies as $Q^{-1/2}$. Here we find that it varies more weakly with quality factor--as $Q^{-1/4}$.  The authors use a bandwidth for the noise power of $\Delta \nu \sim \nu_{\rm DM}^{0}/Q$, rather than a smaller bandwidth of $\sim \nu_{\rm DM}^{0}/Q_{\rm DM}$. Using the larger bandwidth is not appropriate. Since $Q_{s} \leq Q$, the bandwidth of the optimal filter (\ref{eq:delta_nuDMopt}) is approximately the bandwidth of the dark-matter signal. Using more filter bandwidth than the optimal results in greater noise power, while keeping the signal power the same. This implies that reference \cite{Chaudhuri:2014dla} overestimates the optimal uncertainty in noise power, $\sigma_{P}=kT_{S}\Delta\nu/\sqrt{\Delta\nu \cdot t}$, by a factor of $\sqrt{Q_{\rm DM}/Q}$. In turn, the authors underestimate the SNR by a factor of $\sqrt{Q_{\rm DM}/Q}$. As the SNR is proportional to $g_{\rm DM}^{2}$, this resolves the discrepancy in Q-dependence between the previous estimate for hidden photon sensitivity and the one contained here. Observe that the use of the incorrect bandwidth does not affect the sensitivity limits claimed in \cite{Chaudhuri:2014dla}, as the authors take $Q=Q_{\rm DM}$.

More significantly, the seminal work of ref. \cite{Sik85} presents a sensitivity for resonant axion searches which scales as $Q^{-1/4}$ for $Q \leq Q_{\rm DM}$ and is independent of Q for $Q \geq Q_{\rm DM}$; there is no gain in sensitivity when the resonator bandwidth is smaller than the dark-matter-signal bandwidth. In contrast, in our calculation, sensitivity scales as $Q^{-1/4}$ in both regimes. \emph{We benefit from using a resonator whose bandwidth is smaller than that of the dark-matter signal.}

To understand this sensitivity result, we first observe that the most appropriate delineation of regimes is not one which compares the resonator bandwidth and the dark-matter-signal bandwidth. Rather, it is the one which compares the \emph{sensitivity} bandwidth and the dark-matter-signal bandwidth. For the scattering-mode amplifiers discussed in sections \ref{sec:resonator_SNR} and \ref{sec:scan_opt}, this distinction may seem superfluous because $Q$ and $Q_{s}$ differ by a factor of order unity. (See eqs. (\ref{eq:Q_xi}) and (\ref{eq:InvN_Spectrum_Q}) with $\xi=\xi^{\rm opt}(\nu_{r})$, determined in (\ref{eq:xi_opt}).) However, for the quantum-limited flux-to-voltage amplifiers discussed in Appendices \ref{sec:FtoV_Amps} and \ref{sec:FtoV_opt}, the low damping from the amplifier input results in the overall $Q$ and internal $Q$ being approximately the same; as a result, at low frequencies where $n(\nu_{r}) \gg 1$, $Q$ is much larger than $Q_{s}$. In such a situation, it is clear that, with a $Q \gg 10^{6}$ resonator centered at $\nu_{r}=\nu_{\rm DM}^{0}$, we have maximal sensitivity, without degradation from quantum noise, to the entire dark-matter signal as long as $Q_{s} \lesssim Q_{\rm DM}$. We thus consider two cases: (i) the resonator is sensitive to the whole dark-matter signal, $Q_{s} \lesssim Q_{\rm DM}$ and (ii) the resonator is only sensitive to part of the dark-matter signal, $Q_{s} \gtrsim Q_{\rm DM}$. We determine the scaling with $Q_{s}$ in both cases using equation (\ref{eq:Dicke_TS}). It will be important to recall that $Q_{s}$ is proportional to the overall quality factor $Q$, with the constant of proportionality determined by the matching parameter and thermal occupation number. 

We fix the system noise temperature. (Note that our analysis below does not require the amplifier to be quantum-limited.) We assume that the scan step size is 
\begin{equation}\label{eq:step_size}
\Delta \nu_{r}/\nu_{r} = 1/(NQ_{s}),
\end{equation}
where $N \gg 1$ is a constant, and use the scan density function (\ref{eq:t_opt}). The concrete scan pattern assists in the explanation, although other scan patterns may arrive at the same conclusions.

\begin{enumerate}
\item Case (i): $Q_{s}\lesssim Q_{\rm DM}$ 

Consider a tunable resonator of sensitivity quality factor less than the dark-matter quality factor. Suppose we use the resonator to scan over the dark-matter band $\nu_{\rm DM}^{0} \leq \nu \leq \nu_{\rm DM}^{0} + \nu_{\rm DM}^{0}/Q_{\rm DM}$. We consider the signal power at a single scan step. The bandwidth of dark-matter frequency components that can drive the circuit at maximum sensitivity (without degradation from amplifier noise) is independent of $Q_{s}$ because the entire dark-matter signal fits within the sensitivity bandwidth. The ratio of the resonator energy $E_{r}$ to the drive energy varies as $Q^{2}$, and the power into the readout varies as $E_{r}/Q$, so the signal power varies as $Q_{s}$.

We next consider the noise power in the receiver, denoted by $P_{n}$. The bandwidth of the optimally filtered signal is the dark-matter bandwidth and the noise temperature is fixed, so the total noise power is independent of $Q_{s}$.

Consider the number of independent samples of the noise power obtained from the data at the single scan step. This is indicated by the factor $\Delta\nu \cdot t$ in (\ref{eq:Dicke_TS}) and is representative of averaging fluctuations in the noise power to increase measurement SNR. $\Delta\nu$, as we have discussed, is independent of $Q_{s}$. The amount of integration time at this step is proportional to $1/Q_{s}$, so that the number of samples is proportional to $1/Q_{s}$. Then, $\sqrt{\Delta\nu \cdot t}$ scales as $Q_{s}^{-1/2}$.

If we step at one part in $Q_{s}$, then we are sensitive to the dark-matter signal over $\sim$1 step. If we step at one part in $NQ_{s}$, then we are sensitive to the dark-matter signal over $\sim N$ steps, independent of $Q_{s}$.

Putting these contributions all together, with separate scan step SNRs adding in quadrature, we find that the total SNR is proportional to $Q_{s} \times Q_{s}^{-1/2}=Q_{s}^{1/2}$. Therefore, the minimum coupling to which one is sensitive is proportional to $Q_{s}^{-1/4} \propto Q_{\rm int}^{-1/4} \propto Q^{-1/4}$.

Our comparison is similar to that of resistive absorbers and half-wave cavities in Section \ref{ssec:ResvBroad}. The power received by the readout linearly with $Q$, while we take a $Q^{-1/2}$ ``penalty'' for resonant scanning, resulting in an SNR that is proportional to $Q^{1/2}$.

\item Case (ii): $Q_{s}\gtrsim Q_{\rm DM}$

Now, consider a tunable resonator of sensitivity quality factor larger than the dark-matter quality factor. Suppose that it scans over the dark-matter signal band.

The bandwidth of dark-matter frequency components driving the circuit at maximal sensitivity now decreases as $1/Q_{s}$ because only part of the signal sits within the sensitivity bandwidth. Proceeding then as in Case (i), the signal power from a measurement at a single scan step is independent of $Q_{s}$. This is consistent with estimates in \cite{Sik85} and \cite{asztalos2010squid}. 

The appropriate optimal-filter bandwidth for evaluating the noise power of the single-scan-step measurement is the sensitivity bandwidth. Therefore, the noise power decreases as $1/Q_{s}$.

The number of independent samples of noise power at the scan step decreases as $1/Q_{s}^{2}$ because the bandwidth decreases as $1/Q_{s}$ and the integration time decreases as $1/Q_{s}$. Then, $\sqrt{\Delta\nu \cdot t}$ scales as $1/Q_{s}$. Combining the three contributions (signal power, noise power, and number of independent samples), we find that the SNR at a single scan step is proportional to $Q_{s} \times 1/Q_{s} \propto 1$. That is, it is independent of $Q_{s}$. So far, we are still consistent with previous works.

However, there is one contribution that we are yet to consider, which was not considered in previous works. The contribution is that from \emph{the scan.} If we were to step at 1 part in $Q_{s}$, we would be sensitive to dark matter over more than one scan step! We would be sensitive over $\sim Q_{s}/Q_{\rm DM}$ scan steps. At a step size of 1 part in $NQ_{s}$, we are sensitive over $\sim NQ_{s}/Q_{\rm DM}$ steps. Adding SNR contributions in quadrature then tells us that the total SNR at the dark-matter signal band varies as $Q_{s}^{1/2}$, giving a sensitivity scaling of $Q_{s}^{-1/4} \propto Q_{\rm int}^{-1/4} \propto Q^{-1/4}$.

\end{enumerate}

We have now explain how the scan sensitivity to dark matter increases with quality factor even when the sensitivity quality factor exceeds the dark-matter quality factor. \emph{In particular, an experiment benefits fundamentally from a quality factor above one million.} 

One caveat is that there must be a limit to this scaling due to power conservation. This is related directly to the impedance-matching concept discussed in Sections \ref{sec:power_flow} and \ref{sec:DM_circuits}. There, we found that at extremely high cavity quality factors, the apparatus backacts on the dark-matter source. These quality factors are impractical, so we may ignore them for any realizable experiment.

Of course, there are practical aspects that must be taken into consideration. Obtaining an internal Q larger than $10^{6}$ is a difficult task, especially for axion searches in which the large magnetic field makes the use of superconducting materials challenging. Depending on the frequency and the Q, integrating for a resonator ring-up time (which is critical to obtaining the enhanced sensitivity with increased Q) may be impractical. Also, because of the possibility of dark-matter clumping, the dark-matter signal may disappear or may be substantially reduced in the middle of the longer integration (see Appendix \ref{sec:variation}). However, this could also happen at quality factors lower than one million. We simply illustrate here that there is no known, fundamental property of the dark-matter field that prevents one from obtaining higher sensitivity at quality factors above $Q_{\rm DM}$ for a resonant scanning search. The coherence time of the dark matter and the intrinsic bandwidth of the dark-matter signal do not fundamentally set an ``optimal Q'' for detection.

In summary, for a scanning search to probe a wide range of dark-matter masses, it is beneficial to obtain as high a quality factor as possible, as long as the dwell time at each resonance frequency is longer than the dark-matter coherence time at the resonance frequency, as well as the resonator ring-up time. The minimum coupling to which the resonator is sensitive scales with quality factor as $Q^{-1/4}$. \\

\subsubsection{Implications for scan rate in a low-frequency tunable resonator search}

Some previous work, such as ref. \cite{Chaudhuri:2014dla}, computed the sensitivity of a tunable resonator search only considering the information within the resonator bandwidth. As we have shown, at low frequencies $h\nu_{\rm DM}^{0} \ll kT$, such a consideration is not appropriate and does not reflect an optimal scan strategy. Sensitivity, undegraded by amplifier noise, is available far outside of the resonator bandwidth. Here, we consider the enhancement in scan rate available at low frequencies from exploiting the entire sensitivity bandwidth.

In terms of sensitivity bandwidth, for a resonator at frequency $\nu_{r}$, we are able to scan $\sim n(\nu_{r})$ (uncoupled) resonator bandwidths simultaneously. As we move across the range, we scan every dark-matter search frequency with multiple resonators, each giving an independent measurement of the dark-matter signal. The total SNR at each frequency then increases, relative to the resonator-bandwidth-only search, by $\sim \sqrt{n(\nu_{\rm DM}^{0})}$. The limit on dark-matter-to-photon couplings, quantified by the minimum $\varepsilon$ or $g_{a\gamma \gamma}$ to which an experiment is sensitive, is deeper by a factor of $\sim n(\nu_{\rm DM}^{0})^{1/4}$. For a $\sim$1 kHz resonator at $\sim$10 mK read out with a quantum-limited amplifier, this corresponds to a limit that is better by approximately 1.25 orders of magnitude.

Alternatively, we may consider the scan time required to reach a particular limit on coupling. The scan rate is linearly proportional to the sensitivity bandwidth, so the integration is faster by a factor of $\sim n(\nu_{\rm DM}^{0})$. For the $\sim$1 kHz resonator above, the scan rate is enhanced by five orders of magnitude. Additionally, the scan rate should be further enhanced in practice because of the fewer scan steps required (see Section \ref{sssec:matching_opt_discussion}) when exploiting the optimized sensitivity bandwidth. Fewer tuning steps means a reduction in total settling time and an increased duty cycle. 

\emph{The optimal scan strategy allows a reduction in integration times by a few orders of magnitude at frequencies $h\nu_{\rm DM}^{0} \ll kT$.} The insights into the optimal scan strategy provided in this paper should thus have a substantial impact on the sensitivity of low-frequency axion and hidden-photon dark matter searches, such as DM Radio. \\


\subsubsection{Sensitivity calculation and optimal time allocation for other priors}

We digress briefly to consider the computational aspects of sensitivity, as they relate to priors which are not log-uniform and as they relate to the time optimization of Section \ref{ssec:opt_time}.

The expression on the right-hand side of inequalities (\ref{eq:g_QLamp})-(\ref{eq:g_QLamp_SHM}) represents $g_{\rm DM,min}$, as defined previously. Our sensitivity has been derived for the log-uniform search of eq. (\ref{eq:t_opt}). It is useful to ask how the sensitivity formulae are modified for priors that are not log-uniform and how the modified formulae may be used to determine optimal time allocation under such priors. For an arbitrary slowly varying scan-density function satisfying (\ref{eq:tau_slow}), we would replace $\tau_{\rm opt}(\nu_{r})=T_{\rm tot}/\ln(\nu_{h}/\nu_{l})$ with $\tau(\nu_{\rm DM}^{0})$ in eq. (\ref{eq:g_QLamp})-(\ref{eq:g_QLamp_SHM}). The resulting minimum coupling can be used for a more generic optimization in (\ref{eq:A_gen_def}). 

As we have discussed, it is possible that the volume and geometric factor $k(\nu_{\rm DM}^{0})$ change during the scan. In this case (see Section \ref{sssec:scan_opt_other}), we describe sensitivity directly in terms of axion-photon coupling constant and hidden photon mixing angle. Additionally, if the internal quality factor, volume, and geometric form factor vary slowly as a function of resonant frequency during the scan, we may effectively make them a function of dark matter frequency on the right-hand side of (\ref{eq:g_QLamp}):  $Q_{\rm int}\rightarrow Q_{\rm int}(\nu_{\rm DM}^{0})$, $V_{\rm PU} \rightarrow V_{\rm PU}(\nu_{\rm DM}^{0})$, and $k(\nu_{\rm DM}^{0}) \rightarrow k(\nu_{\rm DM}^{0},\nu_{\rm DM}^{0})$. For each search frequency $\nu_{\rm DM}^{0}$, these functions represent the corresponding resonator parameters for all resonant frequencies that are within several sensitivity bandwidths/dark matter cutoff bandwidths, i.e. all resonant frequencies that are sufficiently close. In this manner, the formula in eq. (\ref{eq:g_QLamp}), in combination with (\ref{eq:gDM_def}), may be used for generic time allocation optimizations that relax aforementioned assumptions. For example, for axions, we obtain from (\ref{eq:kappa_def}) and (\ref{eq:g_QLamp}),
\begin{align}
g_{a\gamma \gamma} & \geq \left( Q_{\rm int}(\nu_{\rm DM}^{0}) \nu_{\rm DM}^{0}\tau(\nu_{\rm DM}^{0}) \frac{ 4\pi^{3}  \xi^{\rm opt}(\nu_{\rm DM}^{0})^{2}}{ (4\xi^{\rm opt}(\nu_{\rm DM}^{0})n(\nu_{\rm DM}^{0}) + (1+\xi^{\rm opt}(\nu_{\rm DM}^{0}))^{2})^{3/2} } \left( \frac{\rho_{\rm DM}V_{\rm PU}(\nu_{\rm DM}^{0})}{h\nu_{\rm DM}^{0}}\right)^{2} \right)^{-1/4} \nonumber\\
& \hspace{0.5cm} \times \frac{2\pi \nu_{\rm DM}^{0}}{\sqrt{\hbar c \epsilon_{0}} k(\nu_{\rm DM}^{0},\nu_{\rm DM}^{0}) c^{2} B_{0}} \left( \int_{\nu_{\rm DM}^{0}}^{\nu_{\rm DM}^{0} + \Delta\nu_{\rm DM}^{c}} \frac{d\nu}{\nu_{\rm DM}^{0}} \left( \frac{\nu_{\rm DM}^{0}}{\rho_{\rm DM}} \frac{d\rho_{\rm DM}}{d\nu} (\nu,\nu_{\rm DM}^{0})  \right)^{2} \right)^{-1/4}, \label{eq:g_a_QLamp}
\end{align}
with the right-hand side giving $g_{a\gamma \gamma, \rm min}[\nu_{\rm DM}^{0}, \tau(\nu_{\rm DM}^{0})]$. We then plug the expression for minimum coupling into the value function (\ref{eq:A_axion}) and optimize with respect to $\tau(\nu_{\rm DM}^{0})$.

\subsection{Performance Figure of Merit: Practical Tradeoffs in a Search for Axion and Hidden-Photon Dark Matter}
\label{ssec:FOM}

In the previous section, we calculated a quantum limit on the sensitivity of a resonant search, based on fundamental considerations. It is clear that the sensitivity increases as the volume, quality factor, integration time, and (for axions) background magnetic field strength increase and as the physical temperature decreases. However, we must consider the sensitivity in the context of the practical limitations of a search. One cannot produce arbitrarily large volumes and quality factors or arbitrarily low temperatures. Additionally, achieving in-situ quantum-limited noise can be challenging, so it is important to consider sensitivity with excess amplifier noise, as we do in Appendix \ref{ssec:FtoV_opt_Gen}. There are also often practical tradeoffs between parameters, e.g. improving some design parameters often comes at the cost of degrading other design parameters.

In this section, we quantify these practical aspects of an experimental campaign with a figure of merit and discuss how this figure of merit may be used to inform a design. For simplicity, we focus on axions and assume a log-uniform search. 

We use eqs. (\ref{eq:gDM_def}), (\ref{eq:g_QLamp}), and (\ref{eq:SNR_PhiV_Gen_Simple}), the last of which extends our results to amplifiers operating in excess of the quantum limit. The minimum axion-photon coupling to which an experiment is sensitive (set by an SNR of unity) can be decomposed into a product of a receiver parameter function $\mathcal{R}(m_{\rm DM})$ and the dark-matter model parameters:
\begin{equation}
    g_{a \gamma \gamma, \rm min}(m_{\rm DM})^{-1} = \mathcal{R}(m_{\rm DM}) \sqrt{\hbar c^{3} \epsilon_{0}} \left( \int_{\nu_{\rm DM}^{0}}^{\nu_{\rm DM}^{0} + \Delta\nu_{\rm DM}^{c}} d\nu \left( \frac{1}{h} \frac{d\rho_{\rm DM}}{d\nu} (\nu,\nu_{\rm DM}^{0})  \right)^{2} \right)^{1/4}
\end{equation}
where 
\begin{equation}\label{eq:PFOM}
    \mathcal{R}(m_{\rm DM})=\frac{k(m_{\rm DM})}{m_{\rm DM}c/\hbar} \left( \int d^{3} \vec{x}\ |\vec{B}_{b}(\vec{x})|^{2} \right)^{1/2} \left( 4\pi^{3}Q_{\rm int} \frac{T_{\rm tot}}{\ln(\nu_{h}/\nu_{l})} \bar{\mathcal{G}}(\alpha_{1}^{\rm opt} (\nu_{\rm DM}^{0}),\eta(\nu_{\rm DM}^{0})) \right)^{1/4}.
\end{equation}
We have replaced the rest-mass frequency $\nu_{\rm DM}^{0}$ with the mass $m_{\rm DM}$ in the argument of the geometric factor $k$ and the minimum axion-photon coupling for convenience. The parameters $\bar{\mathcal{G}}(\alpha_{1}^{\rm opt} (\nu_{\rm DM}^{0}),\eta(\nu_{\rm DM}^{0})$, $\alpha_{1}^{\rm opt} (\nu_{\rm DM}^{0})$, and $\eta(\nu_{\rm DM}^{0})$ are defined in (\ref{eq:eta_def}), (\ref{eq:G_PhiV_red}), and (\ref{eq:alpha_opt}). $\eta(\nu_{\rm DM}^{0})$ is the ratio of amplifier noise temperature to the half-photon quantum-limited noise temperature at the search frequency. It is equal to unity at the quantum limit, in which case $\alpha_{1}^{\rm opt}(\nu_{\rm DM}^{0})$, which accounts for optimal noise matching and integrated resonator sensitivity, reduces to the inverse of $\xi^{\rm opt}(\nu_{\rm DM}^{0})$ in (\ref{eq:g_QLamp}). The larger the receiver parameter function, the smaller the coupling that may be probed, so the objective of an experimental design for a resonator search is to maximize this parameter function. We consider $\mathcal{R}(m_{\rm DM})$ under two different limits.\\

\paragraph{Thermal-Noise-Dominated Limit}\mbox{}\\

We evaluate the $\bar{\mathcal{G}}$ parameter in the limit that the physical temperature $T_{\rm ph}$ is larger than the minimum amplifier noise temperature, $n(\nu_{\rm DM}^{0}) \gg \eta(\nu_{\rm DM}^{0}) \geq 1$:
\begin{equation}
    \bar{\mathcal{G}}(\alpha_{1}^{\rm opt} (\nu_{\rm DM}^{0}),\eta(\nu_{\rm DM}^{0}) \approx \frac{1}{6\sqrt{3} n(\nu_{\rm DM}^{0}) \eta(\nu_{\rm DM}^{0})}
\end{equation}
in which case the parameter function varies as
\begin{equation}
    \mathcal{P}_{FOM} \equiv \frac{k(m_{\rm DM}) (\int d^{3} \vec{x}\ |\vec{B}_{b}(\vec{x})|^{2})^{1/2} Q_{\rm int}^{1/4} T_{\rm tot}^{1/4}}{(m_{\rm DM}c/\hbar) T_{ph}^{1/4} \eta^{1/4}} \propto \mathcal{R}(m_{\rm DM})\label{eq:PFOM_1}
\end{equation}
where we refer to $\mathcal{P}_{FOM}$ as the performance figure of merit. Maximizing $\mathcal{P}_{FOM}$ will maximize the receiver parameter function in the thermal-noise-dominated limit.\\

\paragraph{Amplifier-Noise-Dominated Limit}\mbox{}\\

We can also evaluate the limit in which the thermal occupation number is much smaller than $\eta$, assumed to be much greater than unity:
\begin{equation}
    \bar{\mathcal{G}}(\alpha_{1}^{\rm opt} (\nu_{\rm DM}^{0}),\eta(\nu_{\rm DM}^{0}) \approx \frac{2}{3\sqrt{3} \eta^{2}}
\end{equation}
in which case
\begin{equation}
    \mathcal{P}_{FOM} = \frac{k(m_{\rm DM}) (\int d^{3} \vec{x}\ |\vec{B}_{b}(\vec{x})|^{2})^{1/2} Q_{\rm int}^{1/4} T_{\rm tot}^{1/4}}{(m_{\rm DM}c/\hbar) \eta^{1/2}} \propto \mathcal{R}(m_{\rm DM}).\\ \label{eq:PFOM_2}
\end{equation}

There are a large number of practical tradeoffs specific to a particular design, and discussing particular designs is outside the scope of this paper. Here, we list a few tradeoffs that are common to many experimental searches:
\begin{itemize}
    \item From (\ref{eq:PFOM}), it is clear that we should maximize the receiver volume. However, larger volumes are more difficult to cool and may naturally result in a higher physical temperature.
    \item One generally desires to maximize the DC magnetic field strength. However, creating high fields over large volumes can be prohibitive from a cost perspective.
    \item It is optimal to reduce the physical temperature as much as possible. Commercial dilution refrigerators can achieve a temperature of $\sim$10 mK. However, while certain sources of loss are reduced as temperature decreases, e.g. quasiparticle loss in superconducting materials, other sources of loss become larger at lower temperature, e.g. two level systems.\cite{tinkham2004introduction,gao2008physics,romanenko2020three}
    \item We should maximize the geometric factor $k(m_{\rm DM})$. For a resonant cavity search (see eq. (\ref{eq:cav_factor})), this entails both maximizing the cavity overlap factor and maximizing the fill factor of the cavity within the magnet bore. A single cavity has dimensions on the scale of the Compton wavelength, so at high masses, achieving a large fill factor may become challenging. One possible avenue for achieving a larger volume and larger fill of the bore is to use cavity arrays, in which all cavities are tuned to the same frequency.\cite{Sho14} Additionally, while the geometric factor may be order unity for a cavity search, it is suppressed for a quasi-static search such as DMRadio or ABRACADABRA\cite{Chaudhuri:2014dla,kahn2016broadband,Sik14}. We may write
    \begin{equation}\label{eq:cPU_param}
        k(m_{\rm DM})=c_{\rm PU} \frac{m_{DM}c}{\hbar} V_{\rm PU}^{1/3} \ll 1 
    \end{equation}
    where $c_{\rm PU}$ is another geometric factor. Using the DC magnetic-field energy parametrization in eq. (\ref{eq:DC_energy_param}), the performance figure of merit in eqs. (\ref{eq:PFOM_1}) and (\ref{eq:PFOM_2}) then varies as
    \begin{equation}
        \mathcal{P}_{FOM} \propto \begin{cases} V_{\rm PU}^{5/6},\hspace{1 cm} \textrm{quasi-static limit} \\
        V_{\rm PU}^{1/2}, \hspace{1 cm} \textrm{otherwise}
        \end{cases}
    \end{equation}
    In particular, in the quasi-static, thermal-noise-dominated case, we have
    \begin{equation}
    \mathcal{P}_{FOM} \equiv \frac{c_{\rm PU} V_{\rm PU}^{1/3} (\int d^{3} \vec{x}\ |\vec{B}_{b}(\vec{x})|^{2})^{1/2} Q_{\rm int}^{1/4} T_{\rm tot}^{1/4}}{ T_{ph}^{1/4} \eta^{1/4}} \propto \mathcal{R}(m_{\rm DM}).\label{eq:PFOM_QSL}
    \end{equation}
    One may avoid the suppression using an upconversion scheme in a background AC field in a superconducting radio-frequency (SRF) cavity, instead of a background DC field.\cite{sikivie2010superconducting,Berlin:2019ahk} As described throughout this paper, an analogous optimization---showing the disadvantage of radiative couplings relative to reactive couplings and demonstrating the single-pole resonator as the near-optimal single-moded reactive coupling (using the Bode-Fano criterion)---can be conducted for background AC fields. An analogous receiver parameter function to that given in (\ref{eq:PFOM}) is found in this case. In terms of practical tradeoffs, relative to the background DC field case, an upconversion scheme with a background AC field may achieve higher quality factors at the expense of lower background-field strength and higher operating temperature. An upconversion scheme with an SRF cavity also possesses greater cryogenic requirements ($\sim$1 kW of cooling power at 2K vs several watts at 4K with a commercial dilution refrigerator) and is sensitive to frequency noise, such as that from mechanical vibrations and piezoelectric (tuning) fluctuations and and two-level systems in surface dielectrics.
    \item The derivation of the figure of merit assumes optimal noise matching at each scan step. However, because the optimal coupling to the amplifier at fixed temperature is a function of resonance frequency, as shown in eq. (\ref{eq:xi_opt}), one must tune the coupling at each scan step to maximize sensitivity. This reduces the duty cycle of the search and can result in loss of integration time.
\end{itemize}

In summary, we have calculated a Standard Quantum Limit on the sensitivity of a search for light-field dark matter. However, it is important to consider this fundamental sensitivity limit in the context of practical tradeoffs in producing an experimental design.
\section{Conclusions}
\label{sec:conclude}

In this work, we considered the electromagnetic detection of axion and hidden-photon dark matter from a fundamental perspective, conducting a broad, first-principles receiver optimization and setting a Standard Quantum Limit on the sensitivity of a measurement with a single-moded, linear, passive receiver. We began with the definition of interfaces in a receiver. The three components of the receiver were described as the signal source, the matching network, and the readout. We introduced the various considerations required for a receiver optimization: the impedance match to dark matter, the set of possible frequency response functions, periodically varied receiver parameters, irreducible noise sources, and priors.

Using complex-power flow equations to describe the impedance match to dark matter, we identified three  generic categories of receiver-coupling to the dark matter excitation: radiative coupling, inductive coupling, and capacitive coupling. The second and third categories were grouped together and identified collectively as reactive coupling. Generalizing upon a comparison of toy receivers and mapping the complex-power flow equations to equivalent-circuit representations, we demonstrated that radiatively coupled receivers, read out by amplifiers subject to the SQL, are typically disadvantaged relative to reactively coupled receivers, owing to the mismatch between the dark-matter source impedance and the free-space impedance. We thereafter focused on single-moded, reactively coupled receivers. We then compared inductive and capacitive coupling. We demonstrated that, in the quasi-static, subwavelength limit, inductive coupling is superior. In the limit that the detector size is comparable to or larger than a Compton wavelength, the two coupling methods are comparable, and detectors using a cavity mode can be modeled as an equivalent LC circuit, with coupling to the dark matter handled as an effective inductive coupling. (See Appendix \ref{sec:cav_circuit}.) Thereafter, we thus considered inductively coupled receivers without loss of generality.

Having optimized the coupling element in the signal source, we then simultaneously optimized the matching network and the readout. We found that when thermal noise dominates amplifier noise, high-SNR information is available away from the resonator bandwidth. This motivated the notion of integrated sensitivity for a dark-matter search (resonant or otherwise) over wide bandwidth. Building from the concept of integrated sensitivity, we introduced the search optimization. We assumed the use of quantum-limited amplifiers to maximize sensitivity. We created a comprehensive framework for consideration of prior probabilities on the dark-matter signal. This framework takes into account previous astrophysical and direct detection constraints, as well as theoretically well-motivated parameter space. Combining the notion of integrated sensitivity with priors was central to defining the value functions for our priors-based optimization.

In the first part of the priors-based optimization, we optimized the matching network. We introduced a search with log-uniform priors. We showed that in this case, the value function for a matching network reduces to integrated, inverse-squared noise-equivalent number. In both the scattering-mode and op-amp-mode (flux-to-voltage) measurement schemes, the Bode-Fano criterion constrains the match between the equivalent-circuit LR source, with complex-valued impedance, and the amplifier, with real-valued noise impedance. We used the Bode-Fano criterion to set a bound on the log-uniform-search value function. We then optimized a single-pole resonator using the same value function, maximizing with respect to coupling coefficient/matching parameter. We found that, while a single-pole resonator does not satisfy the Bode-Fano bound with equality (equality being satisfied by a multi-pole LC Chebyshev filter), its integrated sensitivity is approximately $75\%$ of the bound. An analogous result holds for RC signal sources. In combination with the insights regarding the optimizing coupling element in Sections \ref{sec:power_flow} and \ref{sec:DM_circuits}, we thus established that the resonator is the near-ideal single-moded detector for dark matter. We found that the optimized resonator at any given frequency is not noise-matched to the amplifier and explained the result in terms of measurement backaction and sensitivity beyond the resonator bandwidth.

In the second part of the priors-based optimization, we focused on single-pole resonators, since they are both practical and close to the Bode-Fano bound. We optimized the time allocation in the scan. To do so, we realized that any practical scan is sufficiently dense as to be ``continuous.'' We introduced the concept of a scan density function and showed that the optimal scan for a log-uniform search spends equal time in each decade of frequency. We also discussed scan strategy optimization under different sets of priors, including those optimized for probing QCD axion models. We analyzed aspects of a practical scan strategy.

We used our results to derive a limit on the performance of tunable resonant dark-matter detectors. We showed that the sensitivity continues to increase even for quality factors above one million, the characteristic quality factor associated with the virialized dark-matter-signal bandwidth. The result was interpreted in terms of sensitivity to dark matter at a single rest-mass frequency over multiple scan steps. Finally, we showed that, at low frequencies, use of the optimized scan can reduce required scan times by a few orders of magnitude, for a fixed limit on dark-matter-to-photon coupling. We interpreted our limit in the context of practical tradeoffs for an experimental campaign. While our optimization analysis broadly informs electromagnetic searches for axion and hidden-photon dark matter, it is the basis for DMRadio, a new DOE-funded program in axion and hidden-photon dark matter detection. 

This paper determines the Standard Quantum Limit (assuming fixed receiver volume, temperature, loss, background magnetic field, and integration time, which are all practically constrained) for single-moded electromagnetic scans for dark-matter axions and hidden photons with linear, passive impedance matching. As illustrated in \cite{chaudhuri2019optimal}, given practical constraints on receiver parameters, no dark-matter receiver subject to the limits established in this paper may probe the QCD axion band below $\lesssim$1 MHz, corresponding to masses below a few neV. This highlights the importance of developing techniques to measure better than the limits. 

One option is to evade the Bode-Fano constraint (Sections \ref{sssec:Bode_Fano_opt}, \ref{sssec:BF_evade}). We may build nonlinear or active impedance-matching networks, as opposed to the linear, passive networks analyzed in this work. Such devices naturally come with additional calibration challenges. We may build receivers with multiple equivalent-circuit elements coupling to the dark matter, such as multi-moded cavities. Doing so requires fine control of parasitics and fabrication processing. The exploration of such schemes requires one to pay particular attention to the application of broadband matching criteria (similar to the Bode-Fano criterion) that constrain the performance of active and multiple-signal-source receivers.

Another option is to evade the SQL for phase-insensitive amplification by non-classical techniques. These techniques include squeezing and entanglement approaches, variational readout, backaction-evading measurements, and photon counting. Over the last decade, owing to rapid progress in the quantum metrology community, such techniques have been realized and may be used in practical high-precision measurements.  Experimental work in this direction has already begun \cite{lehnert2016quantumaxion,brubaker2018first,backes2020quantum,dixit2020searching,chaudhuri2019dark}. The results of this paper provide strong motivation for the broad use of quantum measurement techniques in the search for axion and hidden-photon dark matter. Analysis and implementation of quantum measurement techniques, in the context of this generalized optimization framework, is a promising and exciting direction for future work.

\begin{acknowledgments}
This research is funded in part by the Gordon and Betty Moore Foundation. Additional support was provided by the Heising-Simons Foundation, and some material is based upon work supported by the U.S. Department of Energy, Office of Science, National Quantum Information Science Research Centers. We thank Asimina Arvanitaki, Hsiao-Mei Cho, Carl Dawson, Michel Devoret, Stephen Kuenstner, Dale Li, Harvey Moseley, Lyman Page, Arran Phipps, Jamie Titus, Tony Tyson, Karl van Bibber, Betty Young, and Jonas Zmuidzinas for useful discussions.
\end{acknowledgments}

\appendix

\section{Coherence Properties of the Axion and Hidden-Photon Dark-Matter Field}
\label{sec:variation}

In this section, we discuss some properties of the axion and hidden-photon dark-matter fields that are important in understanding detection. We first discuss the concepts of conherence length and coherence time. We then describe how one may account for variations in amplitude, direction (for hidden photons), and velocity distribution of the dark-matter fields in optimizing search strategy.

\subsection{Coherence Length and Time}
\label{ssec:DM_coh}

As with any other classical field, the dark-matter field possesses spatial and temporal coherence, which is imprinted on the detector signal. These coherence properties are intimately related to the nonzero dark-matter velocity. To understand this, we Fourier transform the free axion and hidden-photon fields:
\begin{equation}\label{eq:axion_fft}
a(\vec{x},t)= Re \left( \int d^{3}\vec{k}\ a(\vec{k}, \nu_{\rm DM}^{0}) \textrm{exp}(2\pi i \nu(k) t - i \vec{k} \cdot \vec{x})  \right)
\end{equation}
\begin{equation}\label{eq:HP_fft}
A'_{\mu}(\vec{x},t)= Re \left( \int d^{3}\vec{k}\ A'_{\mu}(\vec{k}, \nu_{\rm DM}^{0})\textrm{exp}(2\pi i \nu(k) t -i \vec{k} \cdot \vec{x}) \right),
\end{equation}
In this paper, we consider the hidden-photon potential $A'_{\mu}$ that results from rotating into the interaction basis, as opposed to the potential that results from rotating into the mass basis.\cite{Chaudhuri:2014dla} $\nu(k)$ is given by the dispersion relation
\begin{equation}\label{eq:DM_disprel}
(2\pi\nu(k))^{2}= (2\pi\nu_{\rm DM}^{0})^{2} + (\vec{k}c)^{2},
\end{equation}
where $\vec{k}$ is the wavevector of the dark-matter component and may be related to its velocity $\vec{v}(\vec{k})$ by $\vec{v}(\vec{k})= \hbar \vec{k}/m_{\rm DM}$. $k=|\vec{k}|$ is the wavenumber. In the nonrelativistic limit, the dispersion relation can be rewritten as
\begin{equation}\label{eq:disprel_nonrelativistic}
h\nu(k)= h\nu_{\rm DM}^{0} + \frac{1}{2}m_{\rm DM} \vec{v}^{2}.
\end{equation}
We have thus represented the dark-matter fields as a continuous distribution of independent oscillators in momentum/frequency space. 

Expansions (\ref{eq:axion_fft}) and (\ref{eq:HP_fft}) do not account for slow variations in the Fourier amplitudes that may occur due to interactions, e.g. electromagnetic or gravitational. We discuss such variations in the context of detection below. The lower bound on the expansions is zero velocity (and wavenumber), corresponding to frequency $\nu_{\rm DM}^{0}$. The upper bound is not explicitly stated here, but is determined by the highest dark-matter velocities. Virialization endows dark matter with a $\sim 10^{-3}c$ speed in the galactic rest frame, and the speed of the Earth (and therefore, the detector) in this rest frame is also $\sim 10^{-3}c$. The resultant characteristic velocity determines the width of the integration in $k$-space to be $\Delta k \sim 10^{-3} m_{\rm DM}c/\hbar$. Then, from (\ref{eq:disprel_nonrelativistic}), the width in frequency space is $\Delta\nu_{\rm DM} \sim 10^{-6} \nu_{\rm DM}^{0}$. We now rewrite the expansions as
\begin{equation}\label{eq:axion_fft_slow}
a(\vec{x},t)=  Re \left( \textrm{exp}(2\pi i \nu_{\rm DM}^{0}t) \int d^{3}\vec{k}\ a(\vec{k}, \nu_{\rm DM}^{0}) \textrm{exp}(2\pi i (\nu(k) - \nu_{\rm DM}^{0}) t - i \vec{k} \cdot \vec{x})  \right)
\end{equation}
\begin{equation}\label{eq:HP_fft_slow}
A'_{\mu}(\vec{x},t)=  Re \left( \textrm{exp}(2\pi i \nu_{\rm DM}^{0}t) \int d^{3}\vec{k}\ A'_{\mu}(\vec{k}, \nu_{\rm DM}^{0})\textrm{exp}(2\pi i (\nu(k) - \nu_{\rm DM}^{0}) t -i \vec{k} \cdot \vec{x}) \right)  
\end{equation}
The integral expression now contains both slow spatial components and slow temporal components. The characteristic length scale for spatial variation, or \emph{coherence length}, is
\begin{equation} \label{eq:l_coh}
\lambda_{\rm coh} \sim \frac{2\pi}{\Delta k} \sim 10^{3} \frac{h}{m_{\rm DM}c} \approx 1240 \ \textrm{km} \frac{1\ \textrm{neV}}{m_{\rm DM}c^{2}},
\end{equation}
Over the coherence length, the axion and hidden-photon fields can be treated as spatially uniform. That is, the field has approximately the same amplitude and phase everywhere within the coherence length at all times. For the hidden photon, the field also has the same direction.

Suppose we are sitting at a fixed position $\vec{x}_{0}$ and observing the field over a time interval $[t_{0}, t_{0}+\delta t]$. Eqs. (\ref{eq:axion_fft_slow}) and (\ref{eq:HP_fft_slow}) indicate that the dark-matter field is effectively an oscillator at frequency $\nu_{\rm DM}^{0}$ with time-dependent complex phasor represented by the integral expression. On timescales $(\nu_{\rm DM}^{0})^{-1} \ll \delta t \ll (\Delta \nu_{\rm DM})^{-1}$, the integral can be treated as a constant, so that the dark-matter field behaves as a phase-coherent, monochromatic wave at frequency $\nu_{\rm DM}^{0}$. On timescales $\delta t \gtrsim (\Delta \nu_{\rm DM})^{-1}$, the complex phasor has changed amplitude and phase, relative to its value at $t_{0}$. The coherence time of the field is therefore
\begin{equation} \label{eq:t_coh}
t_{coh} \sim \frac{1}{\Delta \nu_{\rm DM}} \sim \frac{10^{6}}{m_{\rm DM}c^{2}/h} \approx 4 \ \textrm{sec} \frac{1\ \textrm{neV}}{m_{\rm DM}c^{2}}.
\end{equation}

The temporal and spatial coherence properties of light-field dark-matter provide key discrimination mechanisms for determining whether a detection of dark matter has been made. One of the major technical challenges in any coherent detection experiment is mitigation of electromagnetic interference. It is inevitable, regardless of the quality of the shielding and grounding, that spurious pickup from the environment occurs, given sufficiently long integration times. If a signal is detected, we may autocorrelate the timestream to determine whether the signal possesses the appropriate coherence time needed to be dark matter. Furthermore, if a signal is detected, a second experiment may be constructed to validate the signal as dark matter or reject it as a false signal. If that second experiment is located well within the coherence length (\ref{eq:l_coh}) corresponding to the signal frequency, then the output timestreams should be highly correlated, with the signals carrying the same amplitude and phase in the two experiments (and in the case of the hidden photon, implying the same direction).

\subsection{Amplitude Variation} \label{ssec:amplitude_variation}

The local dark matter is likely to have some substructure.  This substructure can affect the optimal scan strategy for a resonant experiment, as we discuss in this appendix.

Substructure can be created just through the gravitational interactions of dark matter, as is seen in N-body simulations.  This applies to any dark-matter particle so long as its de Broglie wavelength, $h(m v)^{-1}$, is small (much less than the length scale on which we are interested in substructure).  Over much of the frequency range of interest to axion and hidden-photon detectors, the de Broglie wavelength--the coherence length (\ref{eq:l_coh})--is indeed small and such substructure can be relevant to direct detection experiments.

Additionally, more dramatic substructure may have been created by the original dark-matter production mechanism. 
For example, the most natural production mechanism for light vector dark matter is quantum fluctuations of the vector field during inflation \cite{Graham:2015rva}.  This mechanism naturally produces a power spectrum with a peak at intermediate length scales (which depends on the Hubble scale of inflation and the mass of the vector).  Such a vector has substructure on scales which turn out to be relevant for direct detection experiments. Clumps of the vector dark-matter can be many orders of magnitude more dense than the average local dark-matter density.  However, since the Hubble scale of inflation is currently unknown, even given this production mechanism we cannot know the length scale for these clumps.  Thus in this section we consider the optimal scan strategy if the substructure length scale is unknown.  Of course if we did know the clump size, we would use the optimal strategy for that size.

To see that dark-matter substructure can affect the optimal scan strategy, consider the case that an order one fraction of dark matter is in clumps of significant overdensity compared to the average.  Suppose that the Earth (or the direct detection experiment) passes through a clump every $10^5$ s and spends $10$ s inside each clump.  Then the optimal strategy would be to scan the entire frequency range every $10$ s and continually repeat this scan to use up the entire experimental integration time.  If we are inside a clump, we desire to scan all frequencies while we are inside that clump in order to maximize the chance of detecting dark matter.  A caveat to this scan strategy is that it is (likely) still optimal to spend time at each frequency equal to the dark-matter coherence time (or the resonator ring-up time, if that is longer), and the maximum practical scan speed is likely to be limited by constraints on the experimental apparatus. Depending on the actual time spent inside a clump and the frequency range being considered, it may not be possible to cover the entire range during that time and still integrate to the coherence time at each frequency. In this case it would be optimal to cover only part of the range.

Since we do not know the substructure length scale, the actual optimal scan strategy is simply to scan as fast as possible over all frequencies while still spending a coherence time at each frequency, and then continually repeat that scan.
This optimal scan strategy assumes that we do not lose any time while changing frequencies.  In a practical implementation, there is a settling time while changing from one frequency to another. The length of this settling time compared to the other time scales affects the optimal scan strategy. Furthermore, it may be necessary to change coil sets every $\sim$ decade in frequency, or have multiple copies of the experiment at different frequency ranges.

\subsection{Direction Variation} \label{ssec:direction_variation}
For axion searches, the direction of the effective current density--and therefore, the direction of resultant oscillating electromagnetic fields-- is set by the direction of the applied DC magnetic field. We may always align our resonant detector to couple optimally to the axion-induced electromagnetic fields; for example, in the quasi-static limit, if the DC magnetic field direction is set so that any axion-induced magnetic field lies in the $\hat{z}$ direction, then we may build a physically-lumped LC circuit whose inductor couples to magnetic fields in the $\hat{z}$ direction. 

In contrast, for the hidden photon, which is a vector particle, the direction of the effective current density cannot be defined by the experimentalist. It is unknown. During the course of a scan, the direction of the hidden-photon field may change. It is natural that a single detector only has sensitivity to dark-matter current density fields in one particular direction (i.e. it can only couple energy from currents in one direction) and no sensitivity along either of the two orthogonal directions  (see e.g. \cite{Chaudhuri:2014dla}). It thus follows that an optimal search for the hidden photon will involve a minimum of three identical detectors, pointing in mutually orthogonal directions. (Of course, a robust limit can be set with a single detector scanned at multiple times, with proper consideration of the statistics.)

We calculate the loss in search SNR due to detector misalignment in such a three-receiver configuration. For concreteness, we consider the case in which each receiver takes the form of a free-space cylindrical cavity and receives through the $TM_{010}$ mode\cite{pozar2012microwave}. Similar arguments may be made for other receiver structures, such as physically-lumped-element pickup inductors as well as other cavity geometries. We assume that the three receivers are located within a coherence length of each other, so that for a given Fourier component of the hidden-photon field, the direction at each cavity is the same. Suppose, without loss of generality, that the central axes of the three cavities point along the $\hat{x}$, $\hat{y}$, and $\hat{z}$ directions, respectively. Suppose that the Fourier component $\vec{J}_{HP}^{\rm eff}(\nu,\nu_{\rm DM}^{0})$ of the hidden-photon effective electromagnetic current density points in the direction 
\begin{equation}
\hat{u}(\nu,\nu_{\rm DM}^{0})=(\mathrm{cos}(\theta_{\rm DM}^{(1)}(\nu,\nu_{\rm DM}^{0})),\ \mathrm{cos}(\theta_{\rm DM}^{(2)}(\nu,\nu_{\rm DM}^{0})),\ \mathrm{cos}(\theta_{\rm DM}^{(3)}(\nu,\nu_{\rm DM}^{0})))
\end{equation}
where $0\leq \theta_{\rm DM}^{(1,2,3)}(\nu,\nu_{\rm DM}^{0}) \leq \pi$. For each of the three receivers, the cavity overlap factor (\ref{eq:cav_overlap}) is
\begin{equation}
    \mathcal{C}^{(k)} = \bar{\mathcal{C}}  \mathrm{cos}^{2}(\theta_{\rm DM}^{(i)}(\nu,\nu_{\rm DM}^{0}))
\end{equation}
where $\bar{\mathcal{C}}$ is the cavity overlap factor in the case that the vector direction of the hidden-photon Fourier component points along the axis of cavity (the maximal situation). We assume that the volume of each of the three receivers is the same as the volume of a single receiver that is maximally aligned with the hidden-photon direction. (The total volume of the three-receiver configuration is three times larger than the single, maximally-aligned cavity.) Then, the multiplicative reduction in cavity overlap factor results in an identical reduction in coupled energy in eq. (\ref{eq:UDM_def}).

Since we do not know the relationship of the phases of the dark-matter equivalent-circuit voltage drives in the three receivers, the timestreams of the three receivers must be added incoherently. The SNRs from the three receivers then add in quadrature. The multiplicative reduction in coupled energy results in a factor of $\mathrm{cos}^{4}(\theta^{(i)}(\nu,\nu_{\rm DM}^{0}))$ in the integrand of (\ref{eq:SNR_1_QLamp}). The maximum loss in search SNR is then given by minimizing the sum
\begin{equation}
    \sum_{i} \mathrm{cos}^{4}(\theta^{(i)}(\nu,\nu_{\rm DM}^{0}))
\end{equation}
subject to the constraint
\begin{equation}
    \sum_{i} \mathrm{cos}^{2}(\theta^{(i)}(\nu,\nu_{\rm DM}^{0}))=1
\end{equation}
The minimum is $1/3$, so the maximum degradation in search SNR, from the three combined receivers, is a factor of $\sqrt{3}$, relative to a receiver which happens to be perfectly aligned with each Fourier component of the hidden-photon field ($\mathrm{cos}^{2}(\theta_{\rm DM}(\nu,\nu_{\rm DM}^{0}))=1$ for all frequencies $\nu$). 

For practical purposes, one may benefit from having two sets of three mutually orthogonal detectors, with the sets spaced by less than a coherence length and misaligned in space. For example, if one set of three is aligned to sense hidden-photon currents in the $\hat{x}$, $\hat{y}$, and $\hat{z}$ directions, then the other set of three could be aligned in the $\hat{x}+\hat{y}+\hat{z}$, $\hat{x}-2\hat{y}+\hat{z}$, and $\hat{x}-\hat{z}$ directions. If a signal is detected in both sets of three detectors at the same frequency, then one may determine the hidden-photon direction that would be required to explain the signal in each set. Because the sets are spaced by less than a coherence length, the direction inferred from each set must be the same. This provides a strong discrimination technique for any potential signal.

\subsection{Distribution Variation} \label{ssec:distribution_variation}

The distribution of dark matter with frequency/velocity can vary during the search. In fact, owing to the Earth's orbit around the Sun and the Sun's motion in the galactic rest frame, we expect some form of annual modulation in the dark-matter signal. In this section, we describe this annual modulation and how it may be used as another discrimination mechanism for candidate signals.

Annual modulation was first described in the context of WIMP direct detection \cite{drukier1986detecting, freese2013colloquium}. In the galactic rest frame, in which the bulk (average) velocity of the dark matter is zero, the Sun is moving at a speed of $v_{S} \approx$ 220 km/s. In the rest frame of the Sun, this produces a ``wind'' of dark matter. When the Earth is orbiting against the wind, the flux of dark matter through the detector is highest, leading to an enhanced event rate. When the Earth is orbiting in the same direction as the wind, the flux of dark matter is lowest, leading to a reduced event rate. Thus, there is an annual modulation in the detection event rate.

Because ultralight dark matter is better described as a field, rather than a particle, it is not necessarily appropriate to discuss event rate. Rather, there is annual modulation in the bandwidth of the signal. We may write the orbital speed of the Earth--and the detector-- in the galactic rest frame as \cite{freese2013colloquium,kelso2012toward}
\begin{equation}\label{eq:det_velocity}
v_{d}(t)=v_{S} + v_{O} \cos \gamma \cos(\omega_{O}(t-t_{0})),
\end{equation}
where $\omega_{O}=2\pi/$ year is the orbital frequency, $v_{O} \approx$ 29.8 km/s is the orbital speed of the Earth around the Sun, and $\cos \gamma \sim$ 0.51 is a numerical factor arising from the relative orientations of the orbital plane and the velocity of the Sun. $t_{0}$ represents the time when the velocity $v_{d}(t)$ is maximal (Earth's orbital velocity aligns maximally with the Sun's motion) and is the time at which the flux of dark matter would be largest. 

In the galactic rest frame, the maximum speed of the dark-matter is set approximately by galactic escape velocity. In the Milky Way, this escape velocity is $v_{\rm esc}=544$ km/sec. Then, in the rest frame of the detector, the maximum speed is $\sim v_{\rm esc}+v_{d}(t)$. From equation (\ref{eq:disprel_nonrelativistic}), this corresponds to a dark-matter-signal bandwidth of
\begin{equation}\label{eq:bandwidth_annual_mod}
\frac{\Delta\nu_{\rm DM}}{\nu_{\rm DM}^{0}} \sim \frac{1}{2} \frac{(v_{\rm esc}+v_{d}(t))^{2}}{c^{2}}.
\end{equation}
We thus expect a $\sim 7.6\%$ peak-to-peak variation in signal bandwidth. At time $t=t_{0}$, the bandwidth is largest, owing to the largest maximum speed $\sim v_{\rm esc}+v_{d}(t_{0})$ for dark-matter particles in the detector frame. The width is smallest 1/2 year later, owing to the smallest maximum speed.


If we detect a signal in a receiver, we may monitor it over the course of one year. If the signal does indeed represent dark matter, we should observe an approximately sinusoidal variation in bandwidth.
\section{The Standard Halo Model}
\label{sec:SH_Model}

As discussed in \cite{rosenberg2004searching,krauss1985calculations}, the standard halo model (SHM) describes the velocity distribution of virialized dark matter. Under this model, it is assumed that the dark matter follows a isothermal, isotropic phase-space distribution, yielding, in the galactic rest frame, a Maxwellian velocity distribution cut off at the galactic escape velocity $v_{\rm esc}$:
\begin{equation}
Q(\vec{v})=\frac{K}{\sigma_{v}^{3}} \exp \left( -\frac{\vec{v}^{2}}{2\sigma_{v}^{2}} \right)
\end{equation}
for $|\vec{v}| \leq v_{\rm esc}$ and $Q(\vec{v})=0$ for $|\vec{v}| > v_{\rm esc}$. For the Milky Way, the velocity dispersion is $\sigma_{v}=\sqrt{3/2}v_{c}$, $v_{c}=220$ km/s being the circular speed. $K$ is a normalization constant defined such that
\begin{equation}
\int Q(\vec{v}) d^{3}\vec{v} =1.
\end{equation}

Let $\vec{v}_{d}$ be the velocity of the galaxy in the detector rest frame (the opposite of the detector velocity in the galaxy rest frame, whose magnitude is given in equation (\ref{eq:det_velocity})). We calculate the mass density distribution $\frac{d\rho_{\rm DM}}{d\nu}$, which determines the signal power density. (See equations (\ref{eq:UDM_def}) and (\ref{eq:PDM_sig}).) The dark-matter distribution over velocity relative to the detector is
\begin{equation}
\bar{Q}(\vec{v}_{r})=\frac{K}{\sigma_{v}^{3}} \exp \left( -\frac{(\vec{v_{r}} + \vec{v}_{d})^{2}}{2\sigma_{v}^{2}} \right)
\end{equation}
for $|\vec{v}_{r} + \vec{v}_{d}| \leq v_{\rm esc}$ and $\bar{Q}(\vec{v}_{r})=0$ for $|\vec{v}_{r} + \vec{v}_{d}| > v_{\rm esc}$. Letting $\theta$ be the polar angle between the relative velocity and the detector velocity, and integrating over azimuthal and polar angles yields the distribution $q(v_{r})$ over speed $v_{r}=|\vec{v}_{r}|$:
\begin{align}
1= \int \bar{Q}(\vec{v}_{r}) d^{3}\vec{v}_{r} =& 2\pi \frac{K}{v_{d}\sigma_{v}} \Bigg( \int_{0}^{v_{\rm esc}-v_{d}} dv_{r}\ v_{r} \left( \exp \left( -\frac{(v_{r}- v_{d})^{2}}{2 \sigma_{v}^{2}} \right) -  \exp \left( -\frac{(v_{r}+ v_{d})^{2}}{2 \sigma_{v}^{2}} \right) \right) \nonumber \\
& \hspace{1 cm} +\int_{v_{\rm esc} - v_{d}}^{v_{\rm esc}+v_{d}} dv_{r}\ v_{r} \left( \exp \left( -\frac{(v_{r}- v_{d})^{2}}{2 \sigma_{v}^{2}} \right) -  \exp \left( -\frac{v_{\rm esc}^{2}}{2 \sigma_{v}^{2}} \right) \right) \Bigg)
\end{align}
so 
\begin{equation} \label{eq:SHM_speed}
q(v_{r})= \frac{2\pi K}{v_{d}\sigma_{v}} \times \begin{cases}
v_{r} \left( \exp \left( -\frac{(v_{r}- v_{d})^{2}}{2 \sigma_{v}^{2}} \right) -  \exp \left( -\frac{(v_{r}+ v_{d})^{2}}{2 \sigma_{v}^{2}} \right) \right) & \qquad v_{r}<v_{-}(v_{d}) \\
v_{r} \left( \exp \left( -\frac{(v_{r}- v_{d})^{2}}{2 \sigma_{v}^{2}} \right) -  \exp \left( -\frac{v_{\rm esc}^{2}}{2 \sigma_{v}^{2}} \right) \right)  
 & \qquad v_{-}(v_{d}) \leq v_{r} \leq v_{+}(v_{d}) \\
0 & \qquad v_{r} > v_{+}(v_{d}) 
\end{cases}
\end{equation}
where $v_{\pm}(v_{d})=v_{\rm esc} \pm v_{d}$. From (\ref{eq:disprel_nonrelativistic}),
\begin{equation}
v_{r} = c \sqrt{\frac{2(\nu-\nu_{\rm DM}^{0})}{\nu_{\rm DM}^{0}}}.
\end{equation}
The mass distribution over frequency is therefore given by
\begin{equation}\label{eq:SHM_freq}
\frac{d\rho_{\rm DM}}{d\nu} (\nu,\nu_{\rm DM}^{0})= \rho_{\rm DM} q(v_{r}) \frac{dv_{r}}{d\nu} = \rho_{\rm DM} q \left( c \sqrt{\frac{2(\nu-\nu_{\rm DM}^{0})}{\nu_{\rm DM}^{0}}} \right) \frac{c}{\sqrt{2\nu_{\rm DM}^{0}(\nu-\nu_{\rm DM}^{0})}},
\end{equation}
where $\rho_{\rm DM}$ is the local dark-matter density, measured to be $\sim 0.45\ \textrm{GeV/cm}^{3}$.

Finally, we evaluate the integral
\begin{equation}
\int_{\nu_{\rm DM}^{0}}^{\nu_{\rm DM}^{0} + \Delta\nu_{\rm DM}^{c}} \frac{d\nu}{\nu_{\rm DM}^{0}} \left( \frac{\nu_{\rm DM}^{0}}{\rho_{\rm DM}} \frac{d\rho_{\rm DM}}{d\nu} (\nu,\nu_{\rm DM}^{0})  \right)^{2}
\end{equation} 
relevant to the detector sensitivity of equation (\ref{eq:g_QLamp}). Taking the detector speed (\ref{eq:det_velocity}) to be its maximum value $v_{d}(t)=v_{d}(t_{0})$, this integral evaluates to \footnote{If we assume a dark-matter density that stays constant throughout the year, then the value of this integral will vary annually. We will see an annual modulation in the value of the integral, ranging from $4.4 \times 10^{5}$ to $4.8 \times 10^{5}$. The modulation corresponds to a $\sim 3.6$\% variation in SNR. However, it is possible that such a small effect will not be observable in an experiment due to variations in system noise or variations in density. \label{SNR_AM note} }
\begin{equation}\label{eq:SHM_sensitivity}
\int_{\nu_{\rm DM}^{0}}^{\nu_{\rm DM}^{0} + \Delta\nu_{\rm DM}^{c}} \frac{d\nu}{\nu_{\rm DM}^{0}} \left( \frac{\nu_{\rm DM}^{0}}{\rho_{\rm DM}} \frac{d\rho_{\rm DM}}{d\nu} (\nu,\nu_{\rm DM}^{0})  \right)^{2} \approx 4.4 \times 10^{5}.
\end{equation}
\section{Equivalent-Circuits of High-Q Resonant Cavities}
\label{sec:cav_circuit}

In this appendix, we develop an equivalent series-RLC circuit for an abitrary high-Q resonant cavity mode, valid for frequencies near resonance. We thus generalize the equivalent-circuit construction of the toy cavity analyzed in Section \ref{ssec:toy_circuits}. We begin by introducing the governing equations for electromagnetic cavities. We then write a Maxwellian complex-power flow statement for a cavity mode, and similar to Section \ref{ssec:toy_circuits}, determine the equivalent-circuit parameters via a mapping between Kirchhoff's Laws for the RLC circuit and the power flow statment. See ref. \cite{montgomery1987principles}, which presents a related derivation of the cavity-mode equivalent-circuit using the electromagnetic Lagrangian. We relate the equivalent-circuit voltage excitation to the cavity overlap factor, a mathematical construct used to calculate the dark-matter drive in previous literature\cite{Sik85}. We also relate the voltage to the mathematical formalism of eq. (\ref{eq:VDM_mono}). As a particularly instructive example of the equivalent-circuit formalism and the concepts introduced here, we revisit the toy cavity of Section \ref{ssec:toy_circuits} at the end of the appendix.

For simplicity, we assume that the completely-closed cavity, of volume $V_{c}$, is filled uniformly with a medium of permittivity possessing frequency-independent real part $\epsilon$ and permeability possessing frequency-independent real part $\mu$. We assume that the associated dielectric and magnetic loss tangents are much less than unity (as required for a high-Q cavity). For a cavity filled with vacuum, $\epsilon=\epsilon_{0}$ and $\mu=\mu_{0}$. For the cavities of interest in electromagnetic axion searches, the effect of kinetic inductance is small, so we neglect the reactance contribution from free electrons. This means that the free-electron term in (\ref{eq:Pelec}) only possesses real part, representing dissipation. We index the modes of the cavity by the subscript $n$. Each mode is characterized by a resonance frequency $\omega_{n}$, a wavenumber $k_{n}=\omega_{n} \sqrt{\mu\epsilon}$, and dimensionless, complex-valued electric-field and magnetic-field mode functions, $\vec{\mathcal{E}}_{n}(\vec{x})$ and $\vec{\mathcal{B}}_{n}(\vec{x})$. These mode functions satisfy the differential equations
\begin{equation}\label{eq:E_mode_PDE}
    \nabla^{2} \vec{\mathcal{E}}_{n} =-k_{n}^{2} \vec{\mathcal{E}}_{n},\ \vec{\nabla} \cdot \vec{\mathcal{E}}_{n}=0
\end{equation}
\begin{equation}\label{eq:B_mode_PDE}
    \nabla^{2} \vec{\mathcal{B}}_{n} =-k_{n}^{2} \vec{\mathcal{B}}_{n},\ \vec{\nabla} \cdot \vec{\mathcal{B}}_{n}=0
\end{equation}
\begin{equation}\label{eq:EB_mode_relation}
    \vec{\nabla} \times \vec{\mathcal{E}}_{n} = -k_{n} \vec{\mathcal{B}}_{n},\ \vec{\nabla} \times \vec{\mathcal{B}}_{n} = k_{n} \vec{\mathcal{E}}_{n}
\end{equation}
within the cavity volume, as well as the boundary conditions
\begin{equation}
    \hat{n} \times \vec{\mathcal{E}}_{n}=0
\end{equation}
\begin{equation}
    \hat{n} \cdot \vec{\mathcal{B}}_{n}=0
\end{equation}
at the surfaces of the cavity walls. Here, $\hat{n}$ is the surface normal vector. The mode functions form an orthogonal basis set for complex-valued vector fields within the cavity, with the normalization
\begin{equation}\label{eq:E_mode_norm}
    \int_{V_{\rm c}} d^{3}\vec{x}\ \vec{\mathcal{E}}^{*}_{m}(\vec{x}) \cdot \vec{\mathcal{E}}_{n}(\vec{x}) = V_{\rm c} \delta_{mn}
\end{equation}
\begin{equation}\label{eq:B_mode_norm}
    \int_{V_{\rm c}} d^{3}\vec{x}\ \vec{\mathcal{B}}^{*}_{m}(\vec{x}) \cdot \vec{\mathcal{B}}_{n}(\vec{x}) = V_{\rm c} \delta_{mn}.
\end{equation}
The electric field $\vec{E}_{c}(\vec{x},t)$ and magnetic field $\vec{B}_{c}(\vec{x},t)$ within the cavity can be expanded in terms of the mode functions:\cite{montgomery1987principles,hill2009electromagnetic}
\begin{equation}\label{eq:E_mode_expand}
    \vec{E}_{c}(\vec{x},t)= \frac{1}{\epsilon}Re \left( \sum_{n} k_{n}^{2}q_{n}(t)\vec{\mathcal{E}}_{n}(\vec{x}) \right)
\end{equation}
and
\begin{equation}\label{eq:B_mode_expand}
    \vec{B}_{c}(\vec{x},t)= \mu Re \left( \sum_{n} k_{n} \frac{dq_{n}}{dt}(t)\vec{\mathcal{B}}_{n}(\vec{x}) \right).
\end{equation}
Observe that $q_{n}(t)$ has the dimension of charge. As we will see, $q_{n}(t)$ describes the charge on the capacitor for the equivalent circuit model of the $n$th mode, while $I_{n}(t)=dq_{n}/dt$ describes the current through the equivalent inductor.

Suppose that the cavity is driven by a monochromatic dark-matter signal at frequency $\omega_{\rm DM}^{0}$ and that the cavity is in steady-state.  Suppose additionally, that the modes are well-separated so that we may ignore mode-interactions and that the drive frequency is close to the $n$th mode frequency, $|\omega_{\rm DM}^{0} -\omega_{n}| \ll \omega_{n}$. Then, we may ignore the excitation of all other modes and write
\begin{equation}\label{eq:E_mode_res}
    \vec{E}_{c}(\vec{x},t)= \frac{1}{\epsilon}Re \left( k_{n}^{2}q_{n}(\omega_{\rm DM}^{0})\vec{\mathcal{E}}_{n}(\vec{x})  \textrm{exp}(+i\omega_{\rm DM}^{0}t) \right)
\end{equation}
\begin{equation}\label{eq:B_mode_res}
    \vec{B}_{c}(\vec{x},t)= \mu Re \left( k_{n} I_{n}(\omega_{\rm DM}^{0})\vec{\mathcal{B}}_{n}(\vec{x}) \textrm{exp}(+i\omega_{\rm DM}^{0}t) \right),
\end{equation}
where we have substituted
\begin{equation}\label{eq:qi_mono}
    q_{n}(t)=q_{n}(\omega_{\rm DM}^{0})\textrm{exp}(+i\omega_{\rm DM}^{0}t),\ I_{n}(t)=I_{n}(\omega_{\rm DM}^{0})\textrm{exp}(+i\omega_{\rm DM}^{0}t)
\end{equation}
with $I_{n}(\omega_{\rm DM}^{0})=i\omega_{\rm DM}^{0}q_{n}(\omega_{\rm DM}^{0})$.

With regards to complex power-flow from the dark-matter field, at frequencies near a high-Q resonance, the receiver fields are much larger than the dark-matter-induced drive fields, and the electric fields at the cavity surface are negligible. This means that we can ignore the surface terms in the complex-power flow (\ref{eq:DM_Poynting}). Moreover, Lorentz reciprocity\cite{pozar2012microwave} yields, for the input complex-power flow (\ref{eq:workin}),
\begin{equation}
    \tilde{P}_{in}(\omega_{\rm DM}^{0})= \frac{1}{2} \int_{V_{c}} \vec{J}^{*}_{\rm rec}(\vec{x},\omega_{\rm DM}^{0}) \cdot \vec{E}_{\rm DM}(\vec{x},\omega_{\rm DM}^{0}) = -\frac{1}{2} \int_{V_{c}}  \vec{J}_{\rm DM}(\vec{x},\omega_{\rm DM}^{0}) \cdot \vec{E}^{*}_{\rm rec}(\vec{x},\omega_{\rm DM}^{0})
\end{equation}
where we identify the receiver fields (with subscript ``rec'') with the cavity fields in eqs. (\ref{eq:E_mode_res}) and (\ref{eq:B_mode_res}), and the effective dark-matter current density is given in (\ref{eq:J_axion}) for axions and (\ref{eq:J_HP}) for hidden photons. Eq. (\ref{eq:DM_Poynting}) then gives
\begin{align}
    -\frac{1}{2} \int_{V_{c}}  \vec{J}_{\rm DM}(\vec{x},\omega_{\rm DM}^{0}) \cdot \vec{E}^{*}_{\rm rec}(\vec{x},\omega_{\rm DM}^{0}) &= P_{\rm diss}(\omega_{\rm DM}^{0}) \label{eq:power_flow_cavity} \\
    &+ \frac{i\omega_{\rm DM}^{0}}{2\mu} \int_{V_{c}} |\vec{B}_{\rm rec}(\vec{x},\omega_{\rm DM}^{0})|^{2} - \frac{i \omega_{\rm DM}^{0} \epsilon}{2} \int_{V_{c}} |\vec{E}_{\rm rec}(\vec{x},\omega_{\rm DM}^{0})|^{2} \nonumber
\end{align}
Note that we have grouped together the reactive electric polarization term in the complex-power flow with the volume integral over receiver electric-field energy. We have done similarly for the magnetization term and the volume integral over magnetic-field energy. The dissipation $P_{\rm diss}(\omega_{\rm DM}^{0})$ represents loss in the free electrons, as well as loss in the dielectric and magnetic medium filling the cavity. 
Plugging eqs. (\ref{eq:E_mode_res})-(\ref{eq:qi_mono}) into (\ref{eq:power_flow_cavity}) yields
\begin{align}
    \frac{1}{2} \frac{k_{n}^{2}}{\epsilon} \frac{I_{n}(\omega_{\rm DM}^{0})^{*}}{ i\omega_{\rm DM}^{0}} \int_{V_{c}}  \vec{J}_{\rm DM}(\vec{x},\omega_{\rm DM}^{0}) \cdot \mathcal{\vec{E}}^{*}_{n}(\vec{x}) &= P_{\rm diss}(\omega_{\rm DM}^{0}) \label{eq:power_flow_mapping} \\
    &+ \frac{i\omega_{\rm DM}^{0}}{2} \mu k_{n}^{2} V_{c} |I_{n}(\omega_{\rm DM}^{0})|^{2} + \frac{1}{i\omega_{\rm DM}^{0}} \frac{1}{\epsilon} k_{n}^{4} V_{c} |I_{n}(\omega_{\rm DM}^{0})|^{2} \nonumber
\end{align}

We may model the $n$th cavity mode as an equivalent series-RLC circuit with frequency-dependent resistance denoted $R_{n}(\omega_{\rm DM}^{0})$, inductance $L_{n}$, capacitance $C_{n}$, and voltage phasor $V_{\rm DM,n}(\omega_{\rm DM}^{0})$ driving current $I_{n}(\omega_{\rm DM}^{0})$. The inductance and capacitance represent magnetic and electric field-energy stored within the cavity (including electric polarization and magnetization), while the resistance represents cavity loss. The power-flow statement derived from Kirchhoff's Laws for this circuit is
\begin{equation}\label{eq:KL_mode_mapping}
    \frac{1}{2}I_{n}^{*}(\omega_{\rm DM}^{0}) V_{\rm DM,n}(\omega_{\rm DM}^{0}) = \frac{1}{2}|I_{n}(\omega_{\rm DM}^{0})|^{2} R_{n}(\omega_{\rm DM}^{0}) + \frac{1}{2}i\omega_{\rm DM}^{0}L_{n}|I_{n}(\omega_{\rm DM}^{0})|^{2} + \frac{1}{2} \frac{1}{i\omega_{\rm DM}^{0}C_{n}} |I_{n}(\omega_{\rm DM}^{0})|^{2}.
\end{equation}
To determine the values of the equivalent-circuit parameters, we may map (\ref{eq:KL_mode_mapping}) to (\ref{eq:power_flow_mapping}), which yields
\begin{equation}\label{eq:mode_voltage}
    V_{\rm DM,n}(\omega_{\rm DM}^{0}) \rightarrow \frac{k_{n}^{2}}{\epsilon} \frac{1}{ i\omega_{\rm DM}^{0}} \int_{V_{c}}  \vec{J}_{\rm DM}(\vec{x},\omega_{\rm DM}^{0}) \cdot \mathcal{\vec{E}}^{*}_{n}(\vec{x})
\end{equation}
\begin{equation}\label{eq:mode_circuit}
    \frac{1}{2}|I_{n}(\omega_{\rm DM}^{0})|^{2} R_{n}(\omega_{\rm DM}^{0}) \rightarrow P_{diss}(\omega_{\rm DM}^{0}),\ L_{n} \rightarrow \mu k_{n}^{2} V_{c},\ C_{n} \rightarrow \frac{\epsilon}{k_{n}^{4}V_{c}}
\end{equation}
The quality factor of the $n$th mode is
\begin{equation}
    Q_{n}=\frac{\omega_{n}L_{n}}{R_{n}(\omega_{n})}
\end{equation}

As described in the main text, the cavity may physically possess both inductive and capacitive couplings. However, mathematically, we may treat the voltage excitation as an inductive coupling or capacitive coupling, i.e. put the voltage source in series with the inductor or in series with the capacitor. For the former case, we find
\begin{align}
    |V_{\rm DM,n}(\omega_{\rm DM}^{0})|^{2} &= \frac{k_{n}^{4}}{\epsilon^{2}} \frac{1}{(\omega_{\rm DM}^{0})^{2}} \left| \int_{V_{c}}  \vec{J}_{\rm DM}(\vec{x},\omega_{\rm DM}^{0}) \cdot \mathcal{\vec{E}}^{*}_{n}(\vec{x}) \right|^{2} \nonumber \\ 
    &= \frac{L_{n}}{\mu \epsilon^{2}} \frac{k_{n}^{2}}{(\omega_{\rm DM}^{0})^{2}} \frac{1}{V_{c}} \left| \int_{V_{c}}  \vec{J}_{\rm DM}(\vec{x},\omega_{\rm DM}^{0}) \cdot \mathcal{\vec{E}}^{*}_{n}(\vec{x}) \right|^{2} \nonumber \\
    & \approx L_{n} \mathcal{C} \frac{1}{\epsilon_{0}} \int_{V_{c}} | \vec{J}_{\rm DM}(\vec{x},\omega_{\rm DM}^{0})|^{2}
\end{align}
where we have used the approximation that the dark-matter frequency is close to the resonance frequency and defined the cavity overlap factor
\begin{equation} \label{eq:cav_overlap}
    \mathcal{C} \equiv \frac{\epsilon_{0}}{\epsilon} \frac{ \left| \int_{V_{c}}  \vec{J}_{\rm DM}(\vec{x},\omega_{\rm DM}^{0}) \cdot \mathcal{\vec{E}}^{*}_{n}(\vec{x}) \right|^{2}}{V_{c}\int_{V_{c}} | \vec{J}_{\rm DM}(\vec{x},\omega_{\rm DM}^{0})|^{2}}.
\end{equation}
This overlap factor, taking value between 0 and 1, is independent of dark-matter frequency because the dependencies in the numerator and denominator of (\ref{eq:cav_overlap}) cancel. The geometry factor $k(\omega_{\rm DM}^{0})$, introduced in eq. (\ref{eq:VDM_mono}), may be related to the cavity overlap factor. For axions, we have
\begin{equation}\label{eq:cav_factor}
    k(\omega_{\rm DM}^{0})^{2} = \frac{1}{2}\mathcal{C} \frac{\int_{V_{c}} |\vec{B}_{b}(\vec{x})|^{2}}{\int_{V} |\vec{B}_{b}(\vec{x})|^{2}}
\end{equation}
where $V$ is a volume surrounding the DC magnetic field of finite spatial extent. For hidden photons, if we take the shielding volume to equal the cavity volume (as cavities are self-shielding), we find
\begin{equation}
    k(\omega_{\rm DM}^{0})^{2}= \frac{1}{2} \mathcal{C}.
\end{equation}

As a particularly instructive example of the framework introduced in this appendix, we revisit the toy cavity of Section \ref{ssec:ResvBroad} and \ref{ssec:toy_circuits} and show that the construction here reproduces the power dissipated in the cavity for a monochromatic dark-matter signal near resonance, given in eq. (\ref{eq:Pdiss_cavity_Lorentzian}). The volume of the cavity is $V_{c}=Al$, where $A$ is the area of each sheet and $l$ is the separation between the sheets. The fundamental mode has resonance frequency $\omega_{1}=\pi c/l$, corresponding to wavenumber $k_{1}=\pi /l$. The electric-field mode function is given by
\begin{equation}
    \vec{\mathcal{E}}_{1}(\vec{x})= \sqrt{2} \mathrm{sin}(k_{1}y) \hat{z}
\end{equation}
where the normalization is determined by equation (\ref{eq:E_mode_norm}). From (\ref{eq:mode_circuit}), the equivalent inductance and capacitance of the fundamental mode are
\begin{equation}
    L_{1}=\mu_{0} \frac{\pi^{2}A}{l},\ C_{1}=\epsilon_{0} \frac{l^{3}}{\pi^{4} A}
\end{equation}
From (\ref{eq:Pdiss_cavity_Lorentzian}), the quality factor of the cavity mode is $Q_{1}=\frac{\pi Z_{fs}}{2 Z_{c}}$, so the equivalent resistance is 
\begin{equation}
    R_{1}= \frac{\omega_{1} L_{1}}{Q_{1}}= \frac{2\pi^{2}A}{l^{2}}Z_{c}
\end{equation}
Using eqs. (\ref{eq:J_axion}), (\ref{eq:J_HP}), (\ref{eq:Efield_DM}), and (\ref{eq:mode_voltage}) , we find that the frequency-domain, dark-matter-induced voltage drive is, for frequencies $\nu_{\rm DM}^{0}$ close to resonance $\nu_{1}$, $|\nu_{\rm DM}^{0} - \nu_{1}| \ll \nu_{1}$,
\begin{equation}
    \tilde{V}_{\rm DM}(\omega_{\rm DM}^{0})=-2\sqrt{2} A \frac{m_{\rm DM}c}{\hbar}\tilde{E}_{\rm DM},
\end{equation}
where $\tilde{E}_{\rm DM}$ is given in (\ref{eq:Efield_hidden}) for hidden photons and (\ref{eq:Efield_axion}) for axions. Plugging this voltage drive into (\ref{eq:KL_mode_mapping}) yields the current produced in the equivalent RLC circuit.
\begin{equation}
    \tilde{I}_{\rm DM}(\omega_{\rm DM}^{0})=\frac{\tilde{V}_{\rm DM}}{R_{1}^{\rm eq}} \frac{1}{1+2iQ_{1} (\nu_{\rm DM}^{0}-\nu_{1})/\nu_{1})}.
\end{equation}
We may then derive the cavity power dissipation as
\begin{align}
    P_{c}(\omega_{\rm DM}^{0}) &= \frac{1}{2} |\tilde{I}_{\rm DM}|^{2}R_{1}^{\rm eq}= A \frac{2|\tilde{E}_{\rm DM}|^{2}}{Z_{c}} \frac{1}{|1+2iQ_{1} (\omega_{\rm DM}^{0}-\omega_{1})/\omega_{1})|^{2}} \nonumber \\
    &= A \frac{4}{\pi} \frac{|\tilde{E}_{\rm DM}|^{2}}{Z_{fs}} \frac{Q}{|1+2iQ_{1} (\omega_{\rm DM}^{0}-\omega_{1})/\omega_{1})|^{2}} \label{eq:Pdiss_cavity_circuit}
\end{align}
which matches (\ref{eq:Pdiss_cavity_Lorentzian}) with the substitutions $Q_{1} \rightarrow Q$ and $\omega_{1} \rightarrow \omega_{r}$. Note that the values of the equivalent inductance, capacitance, and resistance found here do not match those found in Section \ref{ssec:toy_circuits}. This is simply a result of our choice of mode normalization in eqs. (\ref{eq:E_mode_norm}) and (\ref{eq:B_mode_norm}). If we choose $l^{3}/2\pi^{2}$ as the normalization factor instead of $V_{c}$, then we recover the equivalent-circuit values of Section \ref{ssec:toy_circuits}. In general, there are infinitely many equivalent circuits for a cavity mode, each corresponding to a different normalization, that reproduce the same physical power dissipation and that are physically equivalent.
\section{Noise Correlations for Quantum-Limited, Phase-Insensitive Amplifiers in the Scattering Mode}
\label{sec:SM_QL}

Here, we derive the noise correlations (\ref{eq:c1c2n})-(\ref{eq:c2n}) for a quantum-limited, phase-insensitive amplifier in the scattering mode representation. For more information on these correlations, as they pertain to our work, see \cite{zmuidzinas2003thermal} and \cite{caves1982quantum}. We assume, as we did in the main text, that the amplifier possesses high gain $|G(\nu)| \gg 1$, perfect input and output match, and reverse isolation. The two-port amplifier scattering matrix is then
\begin{equation}\label{eq:S_amp_nonunitary}
S_{\rm amp} = \left[ \begin{array}{cc} 0 & 0 \\ \sqrt{G(\nu)} & 0 \end{array} \right]
\end{equation}
Because this matrix is non-unitary, the amplifier circuit must add noise. We denote the noise modes classically as $c_{1}^{(2)}(\nu)$ and $c_{2}^{(2)}(\nu)$ in the main text.

To derive the noise correlations for a quantum-limited amplifier, we promote the amplitudes in equation (\ref{eq:S_amp}) to operators, e.g. $a_{1}^{(2)}(\nu) \rightarrow \sqrt{h\nu} a_{1}^{\dagger}(\nu)$. The prefactor $\sqrt{h\nu}$ relates to the energy of a single photon at frequency $\nu$ and keeps the units consistent. Thus, we may write the scattering relation:
\begin{equation} \label{eq:S_amp_Quantum}
\left[ \begin{array}{c} b_{1}^{\dagger}(\nu) \\
b_{2}^{\dagger}(\nu) \end{array} \right] = \left[ \begin{array}{cc} 0 & 0 \\ \sqrt{G(\nu)} & 0 \end{array} \right]
\left[ \begin{array}{c} a_{1}^{\dagger}(\nu) \\
a_{2}^{\dagger}(\nu) \end{array} \right] 
+ \left[ \begin{array}{c} c_{1}^{\dagger}(\nu) \\
c_{2}^{\dagger}(\nu) \end{array} \right].
\end{equation}
and similarly for the operators $a_{1,2}(\nu)$, $b_{1,2}(\nu)$, $c_{1,2}(\nu)$, with $G(\nu)$ replaced by $G(\nu)^{*}$. Because the input ($a_{1,2}^{\dagger}(\nu)$) and output operators  ($b_{1,2}^{\dagger}(\nu)$) of the amplifier represent signals consisting of photons, they must obey the canonical bosonic commutation relations:
\begin{equation}
[a_{1}(\nu),a_{1}^{\dagger}(\nu')]=\delta(\nu-\nu'),
\end{equation}
and similarly for $a_{2}^{\dagger}(\nu)$, $b_{1,2}^{\dagger}(\nu)$. Plugging the commutation relations into equation (\ref{eq:S_amp_Quantum}), we obtain
\begin{equation}\label{eq:c1_commute}
[c_{1}(\nu),c_{1}^{\dagger}(\nu')]=\delta(\nu-\nu')
\end{equation}
and
\begin{equation} \label{eq:c2_commute}
[c_{2}(\nu),c_{2}^{\dagger}(\nu')]=(1-|G(\nu)|) \delta(\nu-\nu').
\end{equation}
As promised, the commutator for the imprecision noise mode $c_{2} (\nu)$ is nonzero as long as $|G(\nu)|>1$; if the amplifier gains the input signal, the noise mode must be nontrivial.

The symmetrized quantum noise correlator obeys
\begin{equation}\label{eq:commute_inequality}
< \{ c_{1}^{\dagger}(\nu),c_{1}(\nu') \}>\ \ \geq \ |<[c_{1}(\nu),c_{1}^{\dagger}(\nu')]>|,
\end{equation}
and similarly for $c_{2}^{\dagger}(\nu)$. Here, $<>$ represents an expectation value, and $\{, \}$ represents the anti-commutator.  The inequality becomes an equality, thus minimizing the correlator, when $c_{1}^{\dagger}(\nu)$ and $c_{2}^{\dagger}(\nu)$ each represent a single degree of freedom in its vacuum state:
\begin{equation}
c_{1}(\nu)=g(\nu)
\end{equation}
\begin{equation}
c_{2}(\nu)= \sqrt{|G(\nu)|-1} h^{\dagger}(\nu),
\end{equation}
where $g(\nu)$ and $h^{\dagger}(\nu)$ obey the canonical bosonic commutation relations. These conditions represent the SQL for a phase-insensitive amplifier operated in scattering mode. \cite{caves1982quantum},\cite{clerk2010introduction}

The classical noise correlators may be related to the symmetrized quantum noise correlators via \cite{zmuidzinas2003thermal}
\begin{equation} \label{eq:Q_to_C}
<c_{1,2}^{(2)}(\nu) (c_{1,2}^{(2)}(\nu'))^{*}>= \frac{h\nu}{2} <\{ c_{1,2}^{\dagger}(\nu),c_{1,2}(\nu') \}>.
\end{equation}
Assume that the imprecision and backaction noise modes are uncorrelated. As discussed in the main text, this is typical in many experimental realizations of quantum-limited scattering-mode amplifiers \cite{castellanos2008amplification} and is part of the definition of SQL for the scattering mode (see the treatment of the minimal two-port scattering amplifier in \cite{clerk2010introduction}). Combining equations (\ref{eq:c1_commute})-(\ref{eq:Q_to_C}) then yields equations (\ref{eq:c1c2n})-(\ref{eq:c2n}).
\section{Flux-to-Voltage Amplifiers}
\label{sec:FtoV_Amps}

In the main text, we discussed the optimization of noise matching/matching network and time allocation in a single-moded receiver read out with an amplifier operated in the scattering mode. In a scattering-mode receiver system, the input impedance of the amplifier is matched to an input transmission line. The noise impedance and input impedance of the amplifier are identical and real-valued. Such an amplifier is appropriate for a free-space cavity detector, as used in ADMX and HAYSTAC. (See Sections \ref{ssec:amps}, \ref{ssec:resonator_scattering}, and \ref{sssec:SNR_QLA}.) 

In this appendix, we discuss op-amp mode readouts and the representative example of flux-to-voltage amplifiers. Such amplifiers are used in the DM Radio and ABRACADABRA searches. The most common example of a flux-to-voltage amplifier is a dc SQUID. The current in the input coil of a dc SQUID couples flux into the device, resulting in an amplified voltage on its output. The dc SQUID typically possesses a noise impedance far greater than its input impedance.\cite{falferi1998back, hilbert1985measurements} In a sense, it is the low-impedance dual of a standard voltage op-amp: a low-impedance current amplifier reading out a relatively high impedance source. Moreover, owing to correlations between the imprecision noise mode and the backaction noise mode, the noise impedance of a dc SQUID tends to be complex-valued.

Numerous sophisticated models for noise in dc SQUIDs have been developed\cite{clarke2006squid}, but these models typically only apply to specific architectures in specific regimes. Moreover, the last few decades has seen the advent of numerous Josephson-junction-based flux-to-voltage amplifiers which are not described by these preexisting models. One example, discussed in refs. \cite{mates2008demonstration, Chaudhuri:2014dla}, is a dissipationless rf SQUID coupled to a lithographed microwave resonator. The microwave resonator is coupled to a feedline and interrogated with a $\sim$GHz probe tone. The SQUID acts as a flux-variable inductor. When a flux is applied to the SQUID loop, the inductance of the SQUID changes, causing the resonance frequency of the coupled microwave circuit to change. The change in resonance frequency is read out as a phase shift in the probe tone. This phase shift may be modeled as an equivalent microwave-frequency voltage source in series with the SQUID loop, which is converted to a low-frequency, near-DC voltage source at a follow-on homodyne mixer. Owing to the use of unshunted junctions, these dissipationless rf SQUIDs enable a higher quality factor in the dark-matter detector than is possible with a dc SQUID. 

As such, rather than adopting a pre-existing model for a particular architecture, in this appendix, we develop a broad framework for understanding imprecision and backaction noise in idealized flux-to-voltage amplifiers. We first discuss these noise modes classically and define noise temperature (Appendix \ref{ssec:FtoV_classcial}). We then use a linear response approach, adapted from ref. \cite{clerk2010introduction}, to place a Standard Quantum Limit on the performance of these amplifiers (Appendix \ref{ssec:FtoV_QL}). The next appendix, Appendix \ref{sec:FtoV_opt}, is dedicated to optimization of receivers read out by flux-to-voltage amplifiers. 

\subsection{Classical Description of Noise in Flux-to-Voltage Amplifiers}
\label{ssec:FtoV_classcial}

\begin{figure}[htp] 
\includegraphics[width=\textwidth]{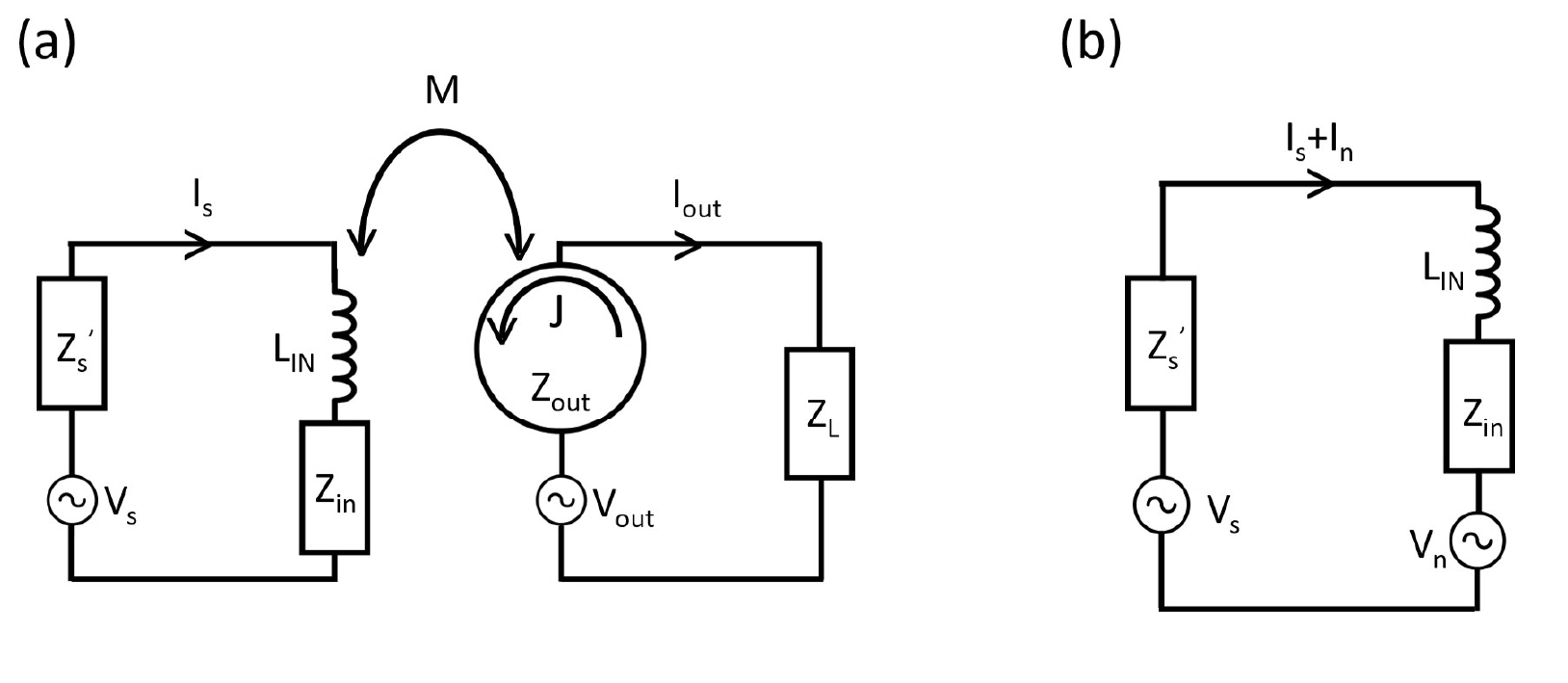}
\caption{Flux-to-voltage amplifier equivalent circuits. (a) Equivalent circuit including input circuit, amplifier, and output circuit. (b) Equivalent circuit showing the effect of noise sources in the amplifier on the input circuit. All noises and impedances are permitted to be frequency-dependent, but we have suppressed the dependence for brevity. \label{fig:FtoV}}
\end{figure}

An equivalent circuit for the flux-to-voltage amplifier is shown in Fig. \ref{fig:FtoV}a. The source of voltage $V_{s}(\nu)$ and impedance $Z_{s}^{'}(\nu)$ drives current $I_{s}(\nu)$ through the amplifier input coil of inductance $L_{\rm IN}$. In response to the flux $\Phi_{a}(\nu)=MI_{s}(\nu)$ applied to the amplifier loop, the amplifier, presenting input impedance $Z_{\rm in}(\nu)$ to the source, produces circulating screening current $J(\nu)$. The circulating current, in turn, produces a voltage $V_{\rm out}(\nu)$ in the output circuit. \footnote{In a dissipationless rf SQUID coupled to a microwave resonator, the voltage is the open-circuit voltage produced at the output of a homodyne mixer, which reads out the phase change of the probe tone.\label{rfSQUIDnote}} This voltage drives currents across the series combination of the amplifier output impedance $Z_{\rm out}(\nu)$ and the load impedance $Z_{L}(\nu)$. The load impedance is typically the input of a second-stage amplifier. We assume that the gain of the flux-to-voltage amplifier is large enough that the noise from the later amplification stages is negligible.

There are two noise sources in the amplifier. First, there are intrinsic fluctuations in the output voltage. Through the flux-to-voltage transfer function, the output voltage fluctuation can be referred as a fluctuation in the current through the input circuit. This noise current is denoted as $I_{n}(\nu)$ in Fig. \ref{fig:FtoV}b and represents the imprecision noise of the amplifier. Second, there are fluctuations in the circulating current. This circulating current, via Faraday's Law, produces a noise voltage in the input current. This noise voltage is denoted as $V_{n}(\nu)$ in Fig. \ref{fig:FtoV}b and represents the backaction noise of the amplifier. It drives additional noise currents through the input coil, resulting in additional noise at the output of the amplifier. The total noise current in the input circuit is thus
\begin{equation}\label{eq:In_tot}
I_{n, \rm tot}(\nu)= I_{n}(\nu) + \frac{V_{n}(\nu)}{Z_{s}'(\nu) + 2\pi i\nu L_{\rm IN} + Z_{in}(\nu)}.
\end{equation}

We now define the noise temperature of the flux-to-voltage amplifier. We define the following single-sided, classical noise spectral densities:
\begin{equation} \label{eq:S_VV_SQ}
<V_{n}(\nu) V_{n}^{*}(\nu')> \equiv \mathcal{S}_{VV}^{FV}(\nu) \delta(\nu-\nu')
\end{equation}
\begin{equation} \label{eq:S_II_SQ}
<I_{n}(\nu) I_{n}^{*}(\nu')> \equiv \mathcal{S}_{II}^{FV}(\nu) \delta(\nu-\nu')
\end{equation}
\begin{equation} \label{eq:S_IV_SQ}
<I_{n}(\nu) V_{n}^{*}(\nu')> \equiv \mathcal{S}_{IV}^{FV}(\nu) \delta(\nu-\nu')
\end{equation}


We assume that these noise spectral densities are intrinsic to the amplifier and have no dependence on the input circuit. Such is typical for an amplifier with adequate linearity to be used in a practical measurement. We define the total source impedance as
\begin{equation} \label{eq:Zs_def}
Z_{s}(\nu) \equiv Z_{s}'(\nu) + 2\pi i\nu L_{\rm IN}.
\end{equation}
This is effectively the source impedance as seen by the flux-to-voltage amplifier. From this point, we refer to this quantity as ``the source impedance.'' The total impedance of the input circuit, as seen by any voltage source in the input circuit, is
\begin{equation} \label{eq:ZT_def}
Z_{T}(\nu) \equiv Z_{s}(\nu) + Z_{in}(\nu).
\end{equation}

The one-sided, classical noise spectral density for the total noise current is
\begin{equation}\label{eq:SQ_n_total}
\mathcal{S}_{II,\rm tot}^{FV}(\nu) = \mathcal{S}_{II}^{FV}(\nu) + \frac{\mathcal{S}_{VV}^{FV}(\nu)}{|Z_{T}(\nu)|^{2}} + 2 Re \left( \frac{\mathcal{S}_{IV}^{FV}(\nu)}{Z_{T}^{*}(\nu)} \right).
\end{equation}
We define the noise temperature $T_{N}(\nu)$ by
\begin{equation}
\frac{4kT_{N}(\nu)Re(Z_{T}(\nu))}{|Z_{T}(\nu)|^{2}} \equiv \mathcal{S}_{II,\rm tot}^{FV}(\nu).
\end{equation}
Qualitatively, suppose the real part of the total impedance is represented as a physical, equilibrium resistor at temperature $kT_{0} \gg h\nu$. The noise temperature is the amount that we would have to increase the physical temperature of the resistor to increase the Johnson current noise spectral density by an amount $\mathcal{S}_{II,\rm tot}^{FV}(\nu)$. Writing $Z_{T}(\nu)=|Z_{T}(\nu)|e^{i\phi(\nu)}$,
\begin{equation}
kT_{N}(\nu)= \frac{1}{4 \cos \phi(\nu)} \left( \frac{\mathcal{S}_{VV}^{FV}(\nu)}{|Z_{T}(\nu)|} + \mathcal{S}_{II}^{FV}(\nu)|Z_{T}(\nu)| + 2 Re \left( \mathcal{S}_{IV}^{FV}(\nu) e^{i\phi(\nu)} \right) \right).
\end{equation}
Minimizing with respect to $Z_{T}(\nu)$, we obtain a bound on the noise temperature
\begin{equation}\label{eq:TN_def}
kT_{N}(\nu) \geq \frac{1}{2} \left( \sqrt{\mathcal{S}_{VV}^{FV}(\nu) \mathcal{S}_{II}^{FV}(\nu) - (Im\ \mathcal{S}_{IV}^{FV}(\nu))^{2}} + Re\ \mathcal{S}_{IV}^{FV}(\nu) \right) ,
\end{equation}
where the minimum is achieved for the noise impedance $Z_{N}(\nu)=|Z_{N}(\nu)| e^{i\phi_{N}(\nu)}$, given by
\begin{equation} \label{eq:Zn_mag}
|Z_{N}(\nu)|=\sqrt{\mathcal{S}_{VV}^{FV}(\nu)/\mathcal{S}_{II}^{FV}(\nu)}
\end{equation}
\begin{equation} \label{eq:Zn_phase}
\sin \phi_{N}(\nu) = Im\ \mathcal{S}_{IV}^{FV}(\nu)/\sqrt{\mathcal{S}_{VV}^{FV}(\nu) \mathcal{S}_{II}^{FV}(\nu)}
\end{equation}
Note that if the noise spectral densities appearing on the right-hand side of equations (\ref{eq:Zn_mag}) and (\ref{eq:Zn_phase}) are dependent on the input circuit, it may not be possible to reach the minimum noise temperature (\ref{eq:TN_def}). We avoid this scenario, as it is usually not representative of a linear amplifier.

\subsection{Standard Quantum Limit on Noise in a Flux-to-Voltage Amplifier}
\label{ssec:FtoV_QL}

As discussed in \cite{clerk2010introduction}, the scattering-mode description of quantum limits, which appears in ref. \cite{caves1982quantum} and which is the subject of Appendix \ref{sec:SM_QL}, does not directly apply to flux-to-voltage amplifiers. To describe the Standard Quantum Limit on noise in a flux-to-voltage amplifier, we utilize the linear response approach of refs. \cite{clerk2004quantum, clerk2010introduction}, where a related analysis is carried out for high-impedance voltage op-amps. The linear response approach enables us to express the quantum limits in terms of the noise spectral densities (\ref{eq:S_VV_SQ})-(\ref{eq:S_IV_SQ}). 

In the linear response approach, a Hamiltonian formalism and first-order perturbation theory are used to develop a description of the amplifier and its interaction with input and output circuits. We derive response coefficients that describe how system currents and voltages are affected by the interaction. In particular, we characterize the amplifier's input and output impedance and power gain. The response coefficients are then used in expression of the quantum limit.

Defining interaction Hamiltonians for the input and output requires us to promote the voltages and currents in Fig. \ref{fig:FtoV} to Hermitian operators. The input Hamiltonian describes a flux from the input circuit which couples to the circulating currents of the amplifier \cite{yurke1984quantum}:
\begin{equation}\label{eq:Hint_in}
H_{\rm int}= M\hat{I}_{s}\hat{J},
\end{equation}
where $\hat{I}_{s}$ is the interaction-picture operator for the current in the input circuit, and $\hat{J}$ is the operator for the circulating current. The output Hamiltonian describes a current in the output circuit due to the voltage signal from the amplifier:
\begin{equation}\label{eq:Hint_out}
H_{\rm int}^{'}=\hat{Q}_{\rm out} \hat{V}_{\rm out},
\end{equation}
where $\hat{I}_{\rm out}= -\frac{d\hat{Q}_{\rm out}}{dt}$ is the current flowing in the output circuit, and $\hat{V}_{\rm out}$ is the operator for voltage at the amplifier output. We assume that the coupling between the input circuit and the amplifier is sufficiently weak such that the voltage response is linear in the input current. 

We characterize the amplifier input and output impedances and power gain in terms of the amplifier operators $\hat{J}$ and $\hat{V}_{\rm out}$. For what follows, we assume that when the amplifier is uncoupled from the input and output circuits,  the expectation values of the circulating current in the flux-to-voltage amplifier loop and the output voltage vanish. If the expectation values are nonzero, we subtract them from the operators (i.e. subtract the product of the expectation and the identity operator) and proceed as below. Assuming that the interaction Hamiltonians are turned on adiabatically at time $t=-\infty$, the expectation value of the circulating current in the flux-to-voltage amplifier due to the input signal $\hat{I}_{s}(t)$ is given by the Kubo formula
\begin{equation}
<\hat{J}(t)>= M \int_{-\infty}^{\infty} dt'\ \chi_{JJ}(t-t') <\hat{I}_{s}(t')>,
\end{equation}
where the $J-J$ susceptibility is defined by
\begin{equation} \label{eq:JJ_susc}
\chi_{JJ}(t) \equiv -\frac{i}{\hbar} \theta(t) <[\hat{J}(t),\hat{J}(0)]>_{0},
\end{equation}
and $\theta(t)$ is the Heaviside function. The subscript zero indicates that the expectation value of the commutator is taken with respect to the density matrix of the uncoupled amplifier. Fourier transforming the currents and the $J-J$ susceptibility
\begin{equation} \label{eq:JJ_susc_FFT}
\chi_{JJ}(\nu)=-\frac{i}{\hbar} \int_{0}^{\infty} dt\ <[\hat{J}(t),\hat{J}(0)]>_{0} \textrm{exp}(+2\pi i \nu t)
\end{equation}
yields
\begin{equation}
<\hat{J}(\nu)>= M \chi_{JJ}(\nu) <\hat{I}_{s}(\nu)>.
\end{equation}
The circulating current induces a voltage $\hat{V}(\nu)=2\pi i \nu M \hat{J}(\nu)$ in the input circuit, so we obtain an input impedance of
\begin{equation}\label{eq:Z_in}
Z_{\rm in}(\nu)=2\pi i \nu M^{2} \chi_{JJ}(\nu)
\end{equation}
The real part of the input impedance determines the damping of the input circuit due to the coupling to the amplifier. Similarly, we may relate the output voltage to the output current via the $V_{\rm out}-V_{\rm out}$ susceptibility.
\begin{equation}
<\hat{V}_{\rm out}(t)>=\int_{-\infty}^{\infty} dt'\ \chi_{V_{\rm out}V_{\rm out}}(t-t') <Q_{\rm out}(t')>
\end{equation}
\begin{equation}
\chi_{V_{\rm out} V_{\rm out}}(t) \equiv -\frac{i}{\hbar} \theta(t) <[\hat{V}_{\rm out}(t'),\hat{V}_{\rm out}(0)]>_{0}
\end{equation}
Fourier transforming, we find
\begin{equation}
Z_{\rm out}=\frac{i\chi_{V_{\rm out}V_{\rm out}}(\nu)}{2\pi \nu}.
\end{equation}

We define a power-gain $G_{P}(\nu)$ in the frequency domain, which  is the ratio of the power delivered to the load impedance $Z_{L}$ to the power drawn by the amplifier input, maximized over the load impedance. To compute the power gain, we first note that the output voltage response to an input current signal is
\begin{equation}
<\hat{V}_{\rm out}(t)>=M \int_{-\infty}^{\infty} dt'\ \chi_{V_{\rm out}J}(t-t') <\hat{I}_{s}(t')>,
\end{equation}
where the $V_{\rm out}-J$ susceptibility is defined by
\begin{equation}
\chi_{V_{\rm out}J}(t) \equiv -\frac{i}{\hbar}\theta(t) <[\hat{V}_{\rm out}(t),\hat{J}(0)]>_{0}.
\end{equation}
Identifying $\hat{\Phi}_{a}=M\hat{I}_{s}$ as the flux applied to the flux-to-voltage amplifier input, we may identify the Fourier component $\chi_{V_{\rm out}J}(\nu)$ as the more familiar flux-to-voltage transfer function:
\begin{equation}
\chi_{V_{\rm out}J}(\nu) \leftrightarrow V_{\Phi}(\nu)=\frac{V_{\rm out}(\nu)}{\Phi_{a}(\nu)}.
\end{equation}

The maximum power delivered to the load is achieved for $Z_{L}=Z_{\rm out}^{*}$, so we obtain a power gain
\begin{equation}\label{eq:G_P}
G_{P}(\nu)=\frac{M^{2} |\chi_{V_{\rm out}J}(\nu)|^{2}}{4 Re(Z_{\rm in}(\nu)) Re(Z_{\rm out}(\nu))} = \frac{|\chi_{V_{\rm out}J}(\nu)|^{2}}{4 Im(\chi_{JJ}(\nu)) Im(\chi_{VV}(\nu)) }.
\end{equation}
When the power gain is greater than unity, both quadratures of the signal are amplified (the amplifier discussed here being phase-insensitive) and Heisenberg's uncertainty principle dictates that noise must be added by the measurement.

We are now ready to set the quantum limit on the added noise of the idealized flux-to-voltage amplifier. We assume that the amplifier possesses no reverse gain: a signal coupled into the amplifier through the output voltage does not produce input flux. Furthermore, we assume that the in-phase correlation of the current imprecision noise and voltage backaction noise vanishes, $Re\ \mathcal{S}_{IV}^{FV}(\nu)=0$. (Such an assumption is compatible with some dc SQUID models. See, for example, \cite{clarke1979optimization}.) Heisenberg's uncertainty principle places a constraint on the symmetrized quantum noise spectral densities of the flux-to-voltage amplifier circulating current and the output voltage. Identifying the symmetrized quantum spectral densities with the one-sided classical noise spectral densities (i.e. $\hat{F}=\hat{J}$, $\hat{I}=\hat{V}_{\rm out}$, and $\bar{S}_{JJ} \rightarrow \mathcal{S}_{JJ}/2$ in equation (4.11) of \cite{clerk2010introduction}), the constraint can be written as
\begin{equation}\label{eq:QNC_canonical}
\mathcal{S}_{V_{\rm out}V_{\rm out}}(\nu) \mathcal{S}_{JJ}(\nu) - |\mathcal{S}_{V_{\rm out}J}(\nu)|^{2} \geq |\hbar \chi_{V_{\rm out}J}(\nu) |^{2} \left( 1 + \Xi \left( \frac{\mathcal{S}_{V_{\rm out} J}(\nu)}{\hbar \chi_{V_{\rm out}J}(\nu)} \right) \right),
\end{equation}
where $\Xi(z)$ is a complex function defined as
\begin{equation}
\Xi(z) = \frac{|1+z^{2}|-(1+ |z|^{2})}{2}.
\end{equation}
We may relate the output voltage and flux-to-voltage amplifier circulating current to the current and induced voltage in the input coil:
\begin{equation}
I(\nu)=\frac{1}{M \chi_{V_{\rm out}J}(\nu)} V_{\rm out}(\nu)
\end{equation}
\begin{equation}
V(\nu)=2\pi i \nu M J(\nu)
\end{equation}
so 
\begin{equation}\label{eq:QNC_input}
\mathcal{S}_{VV}^{FV}(\nu) \mathcal{S}_{II}^{FV}(\nu) - |\mathcal{S}_{IV}^{FV}(\nu)|^{2} \geq |h \nu |^{2} \left( 1 + \Xi \left( \frac{i\mathcal{S}_{IV}^{FV}(\nu)}{h\nu} \right) \right).
\end{equation}
Since we assume that $Re\ \mathcal{S}_{IV}^{FV}=0$, $\Xi$ vanishes. From (\ref{eq:TN_def}) and (\ref{eq:QNC_input}), we now obtain the quantum limit on the noise temperature of a flux-to-voltage amplifier:
\begin{equation}\label{eq:TN_QL}
kT_{N}(\nu) \geq \frac{1}{2} \sqrt{\mathcal{S}_{VV}^{FV}(\nu) \mathcal{S}_{II}^{FV}(\nu) - (Im \ \mathcal{S}_{IV}^{FV}(\nu))^{2}} \geq \frac{h\nu}{2}.
\end{equation}
\emph{The quantum limit on the added noise of the amplifier is equal to the zero-point fluctuation noise of the equivalent source resistance $Re\ Z_{T}(\nu)$.} The quantum limit on noise temperature is achieved when:
\begin{enumerate}
\item The quantum noise constraint (\ref{eq:QNC_input}) for noise spectral densities $\mathcal{S}_{VV}^{FV}$, $\mathcal{S}_{II}^{FV}$, and $\mathcal{S}_{IV}^{FV}$ is satisfied (i.e. equality is obtained).
\item The impedance $Z_{T}(\nu)$, which is the sum of source and input impedances, is the noise impedance $Z_{N}(\nu)$ determined by the spectral densities. That is, the input circuit is noise-matched to the amplifier.
\end{enumerate}

Before moving on to the optimization of searches using flux-to-voltage amplifiers, we discuss a few of their properties. First, when the quantum noise constraint (\ref{eq:QNC_canonical}) is satisfied and the power gain is much larger than unity, one may show that $\mathcal{S}_{V_{\rm out}J}(\nu)$ is in phase with $\chi_{VJ}(\nu)$. (See Appendix I.3 of \cite{clerk2010introduction}.) This implies that $Re\ \mathcal{S}_{IV}^{FV}(\nu)=0$, which is consistent with our original assumptions. Second, combining equations (\ref{eq:Zn_mag}) and (\ref{eq:Z_in}), we find
\begin{equation}\label{eq:IN_damp}
\left| \frac{Re\ Z_{in}(\nu)}{Z_{N}(\nu)} \right| = \left| \sqrt{\frac{\mathcal{S}_{VV}(\nu)}{\mathcal{S}_{JJ}(\nu)}} \frac{Im\ \chi_{JJ}(\nu)}{\chi_{VJ}(\nu)} \right| = \frac{1}{2\sqrt{G_{P}(\nu)}} \sqrt{\frac{\mathcal{S}_{VV}(\nu)}{\mathcal{S}_{JJ}(\nu)} \frac{Im\ \chi_{JJ}(\nu)}{Im\ \chi_{VV}(\nu)}} = \frac{1}{2\sqrt{G_{P}(\nu)}} \ll 1,
\end{equation}
where, in the last equality, we have used the proportionality condition (I13)-(I15) from \cite{clerk2010introduction}.  Assuming that the noise impedance is real-valued (no correlations between imprecision and backaction noise modes), eq. (\ref{eq:IN_damp}) demonstrates that, when the input circuit is noise-matched to the amplifier, the damping resulting from coupling to the amplifier is much less than the internal damping of the input circuit. Then, $Z_{T}(\nu) \approx Z_{s}(\nu) + Im\ Z_{in}(\nu)$. In Appendix \ref{ssec:FtoV_opt_QL}, we actually require the stronger condition that $G_{P}(\nu) \gg \xi^{\rm opt}(\nu,n(\nu))^{2}/4$, where $\xi$ is defined in (\ref{eq:xi_opt}), in order to ignore input damping. The purpose of eq. (\ref{eq:IN_damp}) is to show that sufficiently high power gain is consistent with negligible damping from the amplifier. Without such a property, additional complexities emerge in the scan optimization.

Equations (\ref{eq:QNC_input}) and (\ref{eq:TN_QL}) and conditions 1 and 2 above describe a generalized quantum noise constraint. The \emph{Standard} Quantum Limit for flux-to-voltage readouts, as will be used in this paper, possesses the additional constraint that the imprecision and backaction noise are uncorrelated, $\mathcal{S}_{IV}^{FV}=0$. We thus adopt the definition of SQL used in refs. \cite{caves1980quantum,caves1981quantum,jaekel1990quantum}. In Appendices \ref{sec:FtoV_opt} and \ref{sec:ResvBroad_SQUID}, we refer to flux-to-voltage amplifiers that operate at the SQL (when noise-matched) as quantum-limited flux-to-voltage amplifiers.
\section{Scan Optimization for Searches with Flux-to-Voltage Amplifiers}
\label{sec:FtoV_opt}

With the results on quantum limits in hand, we now discuss the optimization of searches with flux-to-voltage amplifiers. Our treatment mirrors that in Sections \ref{sec:resonator_SNR} and \ref{sec:scan_opt} of the main text.

In Appendix \ref{ssec:FV_SNR}, we present a brief SNR analysis for a readout with a flux-to-voltage amplifier, which is analogous to the scattering-mode SNR analysis in Section \ref{sec:resonator_SNR}. Then, in Appendix \ref{ssec:BodeFano_FV}, we establish a Bode-Fano constraint on integrated sensitivity in a log-uniform search read out by a quantum-limited amplifier. The constraint is analogous to that in Section \ref{sssec:Bode_Fano_opt}.

We then turn our attention to the optimization of single-pole resonant dark matter searches with flux-to-voltage amplifiers. We discuss two examples. In the first example (Appendix \ref{ssec:FtoV_opt_QL}), we consider a quantum-limited flux-to-voltage amplifier, which possesses real-valued noise impedance and uncorrelated imprecision and backaction noise. Such an amplifier is analogous to the scattering-mode amplifier described by equations (\ref{eq:c1c2n})-(\ref{eq:c2n}) of the main text. We show that, like the scattering-mode case, a single-pole resonator is close to the Bode-Fano limit. In the second example (Appendix \ref{ssec:FtoV_opt_Gen}), we relax the assumption of a minimum noise temperature equal to one-half photon and the assumption of uncorrelated imprecision and backaction noise.

\subsection{Signal-To-Noise Ratio of Search With Flux-To-Voltage Amplifier}
\label{ssec:FV_SNR}

\begin{figure}[htp] 
\includegraphics[width=12cm]{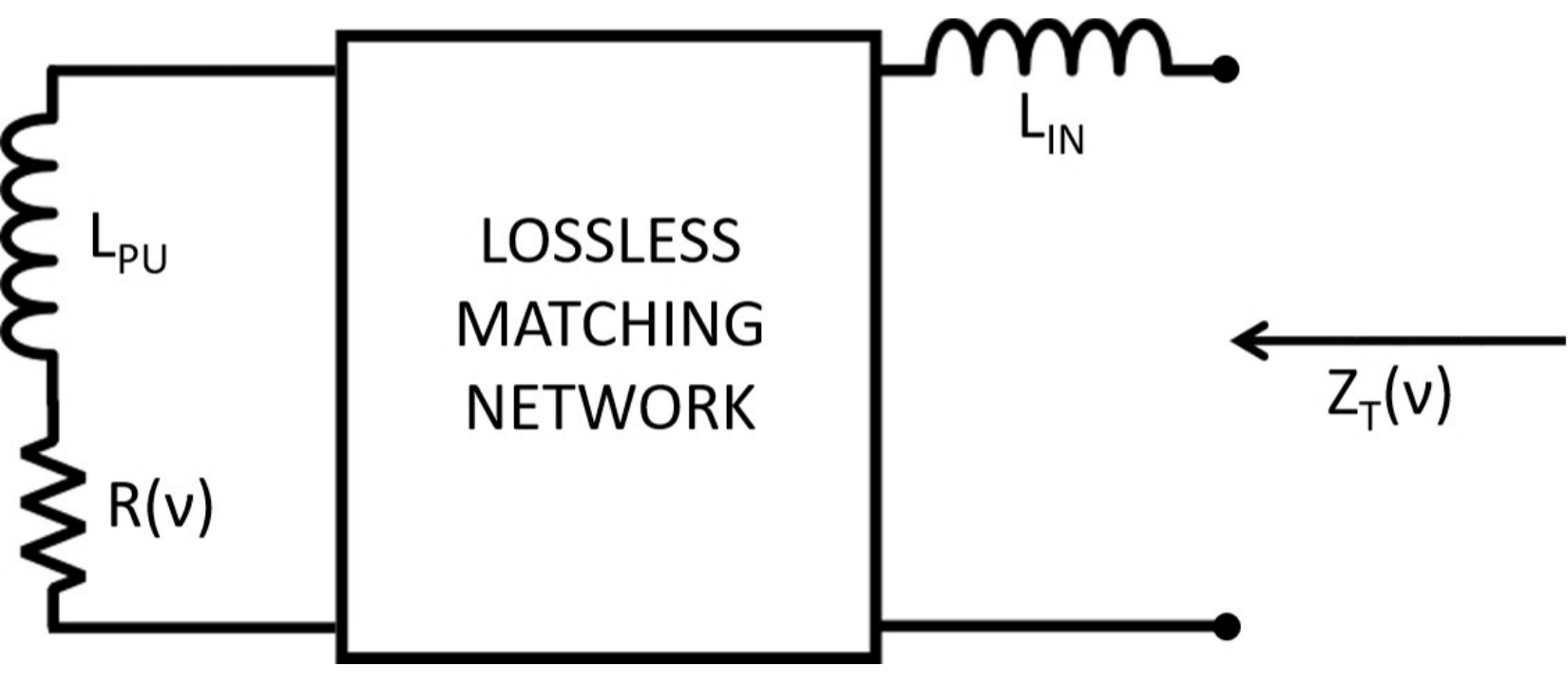}
\caption{Impedance seen by the flux-to-voltage amplifier. \label{fig:FtoV_Network}}
\end{figure}

Here, we evaluate the SNR of a search with a flux-to-voltage amplifier. We evaluate the sensitivity, not in terms of scattering parameters, but rather, in terms of the impedance seen by the amplifier. This quantity is denoted as $Z_{T}(\nu)$ in eq. (\ref{eq:ZT_def}), and is displayed schematically, with the LR signal source, in Fig. \ref{fig:FtoV_Network}. We assume that the damping resulting from coupling to the amplifier is negligible, which is consistent with eq. (\ref{eq:IN_damp}). We have lumped the imaginary part of the amplifier input impedance into the impedance $Z_{T}(\nu)$. We assume that the matching network between the signal source and amplifier is lossless and contains only linear, passive, and reciprocal elements. 

The SNR is determined by the voltage measured at the output of the amplifier, which may be referred to a current through the input inductor $L_{IN}$. This is analogous to the determination of the SNR by analyzing the timestream of the wave at the output of the scattering-mode amplifier.

Referencing Fig. \ref{fig:FtoV}, as well as our discussion in Section \ref{sec:resonator_SNR}, there are two voltage sources in the signal source that drive currents in the amplifier input coil. First, the dark-matter signal coupling to the equivalent inductor produces a drive voltage $V_{\rm DM} \left( \nu, g_{\rm DM}, \frac{d\rho_{\rm DM}}{d\nu}(\nu,\nu_{\rm DM}^{0}) \right)$, defined in equation (\ref{eq:SVV_DM}), across the inductor. Second, the loss in the receiver produces an equivalent noise voltage $V_{FD}(\nu)$ across the resistor $R(\nu)$, as dictated by the fluctuation-dissipation theorem. The noise spectral density is given by 
\begin{equation}
<V_{FD}(\nu) V_{FD}(\nu')>= \mathcal{S}_{V_{FD}V_{FD}}(\nu) \delta(\nu-\nu')
\end{equation}
where
\begin{equation}\label{eq:SVV_FD} 
\mathcal{S}_{V_{FD}V_{FD}}(\nu) =4h\nu R(\nu) (n(\nu) + 1/2)
\end{equation}
In order to determine the SNR, we must transform these series voltages in the signal source to equivalent voltages between the terminals at the right-hand side of Fig. \ref{fig:FtoV_Network}. For the thermal/zero-point noise, the transformed noise voltage $V_{FD}^{t}(\nu)$ possesses power spectral density
\begin{equation}
<V_{FD}^{t}(\nu) V_{FD}^{t}(\nu')>= \mathcal{S}_{V_{FD}^{t}V_{FD}^{t}}(\nu) \delta(\nu-\nu')
\end{equation}
where
\begin{equation}\label{eq:SVV_FDT} 
\mathcal{S}_{V_{FD}^{t}V_{FD}^{t}}(\nu) =4h\nu Re(Z_{T}(\nu)) (n(\nu) + 1/2)
\end{equation}
The dark-matter signal voltage spectral density must transform similarly because impedance transformations do not change signal-to-thermal noise. We denote note the transformed Fourier voltages as $V_{DM}^{t}\left( \nu, g_{\rm DM}, \frac{d\rho_{\rm DM}}{d\nu}(\nu,\nu_{\rm DM}^{0}) \right)$. In analogy with eq. (\ref{eq:SVV_DM}), these Fourier voltages possess spectral density
\begin{equation} \label{eq:SVV_DMT}
\mathcal{S}_{VV}^{\rm DM,t} \left(\nu,g_{\rm DM}, \frac{d\rho_{\rm DM}}{d\nu}(\nu,\nu_{\rm DM}^{0}) \right) = \frac{Re(Z_{T}(\nu))}{R(\nu)} \mathcal{S}_{VV}^{\rm DM} \left(\nu,g_{\rm DM}, \frac{d\rho_{\rm DM}}{d\nu}(\nu,\nu_{\rm DM}^{0}) \right).
\end{equation}

The flux-to-voltage amplifier produces an equivalent current imprecision noise and voltage backaction noise, denoted by $I_{n}(\nu)$ and $V_{n}(\nu)$ in equation (\ref{eq:In_tot}), with noise spectral densities given by equations (\ref{eq:S_VV_SQ})-(\ref{eq:S_IV_SQ}). The total equivalent current fed into the amplifier is then
\begin{equation}\label{eq:I_in_SQ}
I(\nu)=\frac{V_{\rm DM}^{t} \left( \nu, g_{\rm DM}, \frac{d\rho_{\rm DM}}{d\nu}(\nu,\nu_{\rm DM}^{0}) \right) + V_{FD}^{t}(\nu) + V_{n}(\nu)}{Z_{T}(\nu)} + I_{n}(\nu)
\end{equation}
Note that, if the thermal noise dominates the flux-to-voltage amplifier noise, the SNR is independent of the value of the impedance $Z_{T}(\nu)$. In the case of a resonant impedance, this gives rise to sensitivity at frequencies outside of the resonator bandwidth. 

Using the same optimal-filtering approach as in the main text, and replacing the amplifier output $b(\nu)$ with the input current $I(\nu)$ (which can be referred from the flux-to-voltage amplifier output voltage), we find that the SNR of a measurement of integration time $\tau$ much longer than the dark-matter coherence time and receiver pole time ($t_{\rm pole}$, introduced above (\ref{eq:tstar_def})) is
\begin{align}\label{eq:SNR_func_f_PhiV}
SNR & \left[ \nu_{\rm DM}^{0},g_{\rm DM}, \frac{d\rho_{\rm DM}}{d\nu}(\nu,\nu_{\rm DM}^{0}), \tau, f_{\rm rec} \left( \nu,\nu_{\rm DM}^{0}, \frac{d\rho_{\rm DM}}{d\nu}(\nu,\nu_{\rm DM}^{0}) \right) \right] \\
&= \frac{\int d\nu |f_{\rm rec} \left( \nu,\nu_{\rm DM}^{0}, \frac{d\rho_{\rm DM}}{d\nu}(\nu,\nu_{\rm DM}^{0}) \right) |^{2} \mathcal{S}_{VV}^{\rm DM,t} \left(\nu,g_{\rm DM}, \frac{d\rho_{\rm DM}}{d\nu}(\nu,\nu_{\rm DM}^{0}) \right)/|Z_{T}(\nu)|^{2}}{(\tau^{-1} \int d\nu |f_{\rm circ} \left( \nu,\nu_{\rm DM}^{0}, \frac{d\rho_{\rm DM}}{d\nu}(\nu,\nu_{\rm DM}^{0}) \right)|^{4} \mathcal{S}_{II}^{\rm tot} (\nu)^{2})^{1/2}} \nonumber 
\end{align}
where $\mathcal{S}_{II}^{\rm tot}(\nu)$ is the total noise spectral density
\begin{equation} \label{eq:SII_tot_def}
\mathcal{S}_{II}^{\rm tot}(\nu)= \frac{ \mathcal{S}_{V_{FD}^{t}V_{FD}^{t}}(\nu) + \mathcal{S}_{VV}^{FV}(\nu) }{|Z_{T}(\nu)|^{2}} + \mathcal{S}_{II}^{FV}(\nu) + 2Re \left( \frac{\mathcal{S}_{IV}^{FV}(\nu)}{Z_{T}^{*}(\nu)} \right)
\end{equation}
The filter which maximizes SNR is the Wiener filter
\begin{equation}
\Bigg| f_{\rm rec}^{\rm opt} \left( \nu,\nu_{\rm DM}^{0}, \frac{d\rho_{\rm DM}}{d\nu}(\nu,\nu_{\rm DM}^{0}) \right) \Bigg|^{2} = \frac{\mathcal{S}_{VV}^{\rm DM,t} \left(\nu,g_{\rm DM}, \frac{d\rho_{\rm DM}}{d\nu}(\nu,\nu_{\rm DM}^{0}) \right)}{g_{\rm DM}^2 |Z_{T}(\nu)|^{2} \mathcal{S}_{II}^{\rm tot}(\nu)^{2}} 
\end{equation}
for which the SNR is
\begin{align}\label{eq:SNR_f_opt_PhiV}
SNR^{\rm opt} & \left[ \nu_{\rm DM}^{0},g_{\rm DM}, \frac{d\rho_{\rm DM}}{d\nu}(\nu,\nu_{\rm DM}^{0}), \tau \right] \\
& \equiv SNR^{\rm opt} \left[ \nu_{\rm DM}^{0},g_{\rm DM}, \frac{d\rho_{\rm DM}}{d\nu}(\nu,\nu_{\rm DM}^{0}), \tau, f_{\rm rec}^{\rm opt} \left( \nu,\nu_{\rm DM}^{0}, \frac{d\rho_{\rm DM}}{d\nu}(\nu,\nu_{\rm DM}^{0}) \right) \right] \nonumber \\
&= \left( \tau \int d\nu \frac{\mathcal{S}_{VV}^{\rm DM,t} \left(\nu,g_{\rm DM}, \frac{d\rho_{\rm DM}}{d\nu}(\nu,\nu_{\rm DM}^{0}) \right)^{2}}{|Z_{T}(\nu)|^{4} \mathcal{S}_{II}^{\rm tot}(\nu)^{2}} \right)^{1/2} \nonumber
\end{align}
In a scan in which multiple receiver configurations are used, each with a different impedance $Z_{T}(\nu)$, one may calculate the total SNR by adding the single-configuration SNRs in quadrature.

\subsection{Bode-Fano Constraint on Integrated Sensitivity}
\label{ssec:BodeFano_FV}

Having an expression for the SNR in a search with a flux-to-voltage amplifier, we establish a Bode-Fano constraint analogous to the scattering-mode treatment of Section \ref{sssec:Bode_Fano_opt}.

We assume that the flux-to-voltage amplifier is quantum-limited and thus possesses a real-valued noise impedance $Z_{N}$ (eq. (\ref{eq:Zn_phase}) with $\mathcal{S}_{IV}^{FV}=0$). We assume that the noise impedance is frequency-independent; we revisit this assumption at the end of this section. The real-valued, frequency-independent noise impedance mirrors that in the scattering mode case, where the noise impedance is equal to the real-valued, frequency-independent input impedance. We express the voltage backaction and current imprecision noise spectral densities as
\begin{equation}
\mathcal{S}_{VV}^{FV}(\nu)=h\nu Z_{N}
\end{equation}
\begin{equation}
\mathcal{S}_{II}^{FV}(\nu)=\frac{h\nu}{Z_{N}}
\end{equation}
Combining the amplifier noise spectral densities with equations (\ref{eq:SVV_DM}), (\ref{eq:SVV_FDT}), (\ref{eq:SVV_DMT}) yields
\begin{align}
SNR^{\rm opt} & \left[ \nu_{\rm DM}^{0}, g_{\rm DM}, \frac{d\rho_{\rm DM}}{d\nu}(\nu,\nu_{\rm DM}^{0}), \tau \right] = 2\pi^{2} \label{eq:SNR_1_QLamp_FV} \\
&\times \left( \tau  \int_{\nu_{\rm DM}^{0}}^{\nu_{\rm DM}^{0} + \Delta\nu_{\rm DM}^{c}(\nu_{\rm DM}^{0})} d\nu\ \left( \frac{\nu L_{\rm PU}}{R(\nu)} \frac{U_{\rm DM}(\nu, g_{\rm DM}, \frac{d\rho_{\rm DM}}{d\nu}(\nu,\nu_{\rm DM}^{0}))}{h} \frac{1}{N_{\rm tot}^{FV}(\nu, Z_{T}(\nu), n(\nu))} \right)^{2} \right)^{1/2}. \nonumber
\end{align}
where
\begin{equation}\label{eq:NoiseNumber_FV}
N_{\rm tot}^{FV}(\nu, Z_{T}(\nu), n(\nu))= \frac{Re(Z_{T}(\nu))(n(\nu)+ \frac{1}{2}) + \frac{1}{4} \left( \frac{|Z_{T}(\nu)|^{2}}{Z_{N}} + Z_{N} \right)}{Re(Z_{T}(\nu))} 
\end{equation}
plays the role of noise-equivalent number, similar to eq. (\ref{eq:NoiseNumber_QLamp}). Define
\begin{equation}
\zeta(Z_{T}(\nu))=1 - \left| \frac{Z_{T}(\nu)-Z_{N}}{Z_{T}(\nu)+Z_{N}} \right|^{2}
\end{equation}
$\zeta(Z_{T}(\nu))$ describes the match, at frequency $\nu$, between the impedance seen by the flux-to-voltage amplifier and the noise impedance. $\zeta$ takes on values between zero and one; when it is equal to one, the input circuit is noise-matched to the amplifier. 

One may note the similarity between transmission in the scattering system of the main text and the definition of $\zeta$. $\zeta$ plays the role of $|S_{21}^{(1)}(\nu)|^{2}$ in the evaluation of SNR and the determination of the Bode-Fano limit. In terms of $\zeta$, the noise equivalent number of equation (\ref{eq:NoiseNumber_FV}) may be rewritten as
\begin{equation} \label{eq:NoiseNumber_zeta}
N_{\rm tot}^{FV}(\nu, Z_{T}(\nu), n(\nu))= \frac{\zeta(Z_{T}(\nu))n(\nu) +1}{\zeta(Z_{T}(\nu))}
\end{equation}
which is identical in form to (\ref{eq:NoiseNumber_QLamp}) with $|S_{21}^{(1)}(\nu)|^{2}$ replaced by $\zeta(Z_{T}(\nu))$. Similar to Section \ref{sssec:SNR_QLA}, the ``+1'' in the numerator represents the quantum noise associated with the SQL. Eq. (\ref{eq:NoiseNumber_zeta}) shows that only when the amplifier is noise-matched, $\zeta=1$, can we achieve one-photon of noise (referred to the signal source) in excess of the thermal noise of $n(\nu)$ photons.

Following Section \ref{sssec:neutral_search_opt}, we may define a scaled log-uniform-search value functional, which is the frequency-integrated inverse squared of noise-equivalent number. Assuming a frequency-independent signal source resistance, $R=R(\nu)$, the functional is
\begin{equation}\label{eq:F_neu_FV}
\bar{F}_{\rm log}^{FV}[\zeta(Z_{T}(\nu))] = \int_{\nu_{l}}^{\nu_{h}} d\nu \ \left( \frac{\zeta(Z_{T}(\nu))}{\zeta(Z_{T}(\nu))n(\nu) +1} \right)^{2}
\end{equation}
where $\nu_{l}\leq \nu \leq \nu_{h}$ is the search band. The Bode-Fano criterion constrains the match between the impedance $Z_{T}(\nu)$ and the real noise impedance $Z_{N}$:
\begin{equation}
\int_{\nu_{l}}^{\nu_{h}} d\nu\ ln \frac{1}{\left| \frac{Z_{T}(\nu)-Z_{N}}{Z_{T}(\nu) + Z_{N}} \right|} \leq \frac{R}{2L_{PU}}
\end{equation}
which implies
\begin{equation}
\int_{\nu_{l}}^{\nu_{h}} d\nu\ ln \frac{1}{1 - \zeta(Z_{T}(\nu))} \leq \frac{R}{L_{PU}}
\end{equation}
Substituting $\zeta$ for $|S_{21}^{(1)}(\nu)|^{2}$ in Sec. \ref{sssec:Bode_Fano_opt}, we may proceed identically and establish an identical limit on the value functional. The result is reproduced below:
\begin{equation} \label{eq:BF_FV_ineq}
\bar{F}_{\rm neu}^{FV}[\zeta(Z_{T}(\nu))] \lessapprox \begin{cases}
0.4 \frac{R}{L_{\rm PU}}, \hspace{1 cm} n(\nu_{h}) \ll 1 \\
\frac{1}{4n(\nu_{h})} \frac{R}{L_{\rm PU}}, \hspace{1 cm} n(\nu_{h}) \gg 1
\end{cases}
\end{equation}

Equality is achieved when $\zeta(Z_{T}(\nu))$ is a top-hat with respect to frequency. In other words, outside of a narrow band, the input circuit is maximally mismatched to the amplifier noise impedance. Such a top hat may be constructed using a multi-pole LC Chebyshev filter. Note that the Bode-Fano limit is independent of the noise impedance. We will find that the single-pole resonator, with optimized noise impedance, is close to the fundamental Bode-Fano limit.


Having set the Bode-Fano limit, we revisit the assumption made at the beginning of the section that the noise impedance is frequency-independent. A flux-to-voltage amplifier, such as a dc SQUID, tends to possess a noise impedance that increases linearly with frequency \cite{falferi1998back}. In other words, the noise impedance is not frequency-independent. However, we expect that the frequency-dependence has little effect on the limit on integrated sensitivity. If the Bode-Fano limit (\ref{eq:BF_FV_ineq}) pointed toward the optimal circuit possessing a broadband noise match, then the frequency-dependence of noise impedance would be important. However, the Bode-Fano limit (\ref{eq:BF_FV_ineq}) points to a narrowband match. In a narrow band, the noise impedance can be approximated as constant. We thus conclude that, even with a frequency-dependent real-valued noise impedance, the single-pole resonator is close to ideal for single-moded dark matter detection.

We now turn our attention to the optimization of single-pole resonant searches. Suppose that the input equivalent-circuit is resonant. The impedance seen by a voltage source is the sum of circuit and amplifier input impedances:
\begin{equation}\label{eq:ZT_res}
Z_{T}(\nu) = R(\nu) + 2\pi i \nu (L_{\rm IN} + L_{\rm PU}) + \frac{1}{2\pi i \nu C}.
\end{equation}
For a quantum-limited flux-to-voltage amplifier, the intrinsic damping of the input circuit is much greater than the damping due to the amplifier (equation (\ref{eq:IN_damp})) so that the resistance $R$ is solely the intrinsic resistance (dielectric losses,  etc). As such, from this point, we ignore the real part of the amplifier input impedance. We have lumped the imaginary part of the input impedance into the reactive part of the circuit impedance $L_{\rm IN}$.\footnote{We have implicitly assumed here that the magnitude of the imaginary part of the input impedance is small enough that it acts to reduce or increase the input coil inductance. A small reduction is commonplace; it represents the screening of the input inductance by the input loop in a practical flux-to-voltage amplifier.\label{InputImpedanceNote}} Denoting the total equivalent-circuit inductance as $L=L_{\rm IN} + L_{\rm PU}$, the resonance frequency of this circuit is $\nu_{r}=1/2\pi \sqrt{LC}$ and the quality factor is $Q(\nu_{r})=2\pi \nu_{r}L/R(\nu_{r})$. We have denoted the quality factor as a function of resonance frequency because in general, the loss may vary as we tune the resonator. For quantum-limited flux-to-voltage amplifiers, unlike quantum-limited scattering-mode amplifiers, the internal quality factor and overall quality factor are approximately the same, so we use the two interchangeably. This is due to the fundamentally different damping properties of the quantum-limited amplifiers, as dictated by the input impedance. For frequencies $|\nu -\nu_{r}| \ll \nu_{r}$, we may then write the impedance as
\begin{equation}\label{eq:Ztot_res}
Z_{T}(\nu,\nu_{r}) \approx R(\nu_{r}) \left( 1 + 2iQ(\nu_{r})\frac{\nu-\nu_{r}}{\nu_{r}} \right).
\end{equation}
A scan is comprised of measurements conducted at a set of resonance frequencies ${\nu_{r}^{i}}$, with corresponding quality factors $Q^{i} \equiv Q(\nu_{r}^{i})$. In accordance with equation (\ref{eq:t_star_res}), we assume that the dwell time $\tau_{i}$ at frequency $\nu_{r}^{i}$ is much longer than $t^{*}(\nu_{\rm DM}^{0},\nu_{r}^{i})=\textrm{max}(10^{6}/\nu_{\rm DM}^{0}, Q^{i}/\nu_{r}^{i})$ at all frequencies $\nu_{\rm DM}^{0}$ for which (\ref{eq:Ztot_res}) is a good approximation. Equation (\ref{eq:SNR_f_opt_PhiV}) gives the SNR for a measurement at a single resonance frequency; the total SNR, integrated over the scan, is obtained from adding in quadrature the SNRs from each resonance frequency.

\subsection{Resonator Scan Optimization for Flux-to-Voltage Amplifiers Operating at the Standard Quantum Limit}
\label{ssec:FtoV_opt_QL}

We assume that the noise impedance of the amplifier can be changed or tuned at each scan step, so that the noise spectral densities of the amplifier depend both on frequency and resonance frequency: $\mathcal{S}_{VV}^{FV}(\nu) \rightarrow \mathcal{S}_{VV}^{FV}(\nu, \nu_{r}^{i})$ and similarly for the current noise. From Appendix \ref{sec:FtoV_Amps}, for a quantum-limited amplifier,
\begin{equation}\label{eq:SVV_QL}
\mathcal{S}_{VV}^{FV}(\nu,\nu_{r}^{i})=h\nu Z_{N}(\nu,\nu_{r}^{i})
\end{equation}
and
\begin{equation}\label{eq:SII_QL}
\mathcal{S}_{II}^{FV}(\nu, \nu_{r}^{i})=\frac{h\nu}{Z_{N}(\nu,\nu_{r}^{i})},
\end{equation}
where $Z_{N}$ is real-valued. We define
\begin{equation} \label{eq:xiV_def}
\xi_{FV}^{i} \equiv \frac{Z_{N}(\nu=\nu_{r}^{i},\nu_{r}^{i})}{R(\nu_{r}^{i})}
\end{equation}
as the ratio of the on-resonance noise impedance to the signal-source-resistance. For $\xi_{FV}^{i} < 1$, the flux-to-voltage amplifier noise is dominated on-resonance by the imprecision noise, while for $\xi_{FV}^{i}>1$, the flux-to-voltage amplifier noise is dominated by the backaction. At $\xi_{V}^{i}=1$, the readout is noise-matched; the two noise sources contribute equally to the total flux-to-voltage amplifier noise. Equation (\ref{eq:xiV_def}) is the analogue of equation (\ref{eq:xi_def}). 

Combining spectral densities (\ref{eq:SVV_QL}) and (\ref{eq:SII_QL}) with equations (\ref{eq:SVV_DM}), (\ref{eq:UDM_def}), (\ref{eq:SVV_FD}), (\ref{eq:SII_tot_def}), (\ref{eq:SNR_f_opt_PhiV}), (\ref{eq:SNR_1_QLamp_FV}), (\ref{eq:Ztot_res}) yields the SNR for a single resonance frequency
\begin{align}
SNR^{\rm opt} &\left[ \nu_{\rm DM}^{0},g_{\rm DM}, \frac{d\rho_{\rm DM}}{d\nu}(\nu,\nu_{\rm DM}^{0}),\nu_{r}^{i}, \xi_{FV}^{i}, \tau_{i} \right]^{2} = (4\pi Q^{i})^{2} \tau_{i} \label{eq:SNR_1_QLamp_PhiV} \\
& \times \int_{\nu_{\rm DM}^{0}}^{\nu_{\rm DM}^{0} + \Delta \nu_{\rm DM}^{c}} d\nu \left( \frac{\nu}{\nu_{r}^{i}} \frac{L_{\rm PU}}{L} \frac{g_{\rm DM}^{2}V_{\rm PU}\frac{d\rho_{\rm DM}}{d\nu}(\nu,\nu_{\rm DM}^{0})}{h} \frac{\xi_{FV}^{i}}{4\xi_{FV}^{i}n(\nu) + (1+\xi_{FV}^{i})^{2} + 4(Q^{i})^{2} \left( \frac{\nu}{\nu_{r}^{i}} -1 \right)^{2} } \right)^{2} \nonumber
\end{align}

In equation (\ref{eq:SNR_1_QLamp_PhiV}), we have assumed that the dark-matter frequency is sufficiently close to the resonance frequency and the noise impedance is sufficiently slowly varying, so that it may be taken as a constant over the integration range: $Z_{N}(\nu,\nu_{r}^{i}) \approx Z_{N}(\nu_{r}^{i},\nu_{r}^{i})$. In particular, we require that this be a good approximation within a few sensitivity bandwidths of the resonance, as defined by the Q-factor (\ref{eq:InvN_Spectrum_Q}). 

Similar to Section \ref{sec:scan_opt}, our optimization procedure consists of two parts:
\begin{enumerate}
\item For each scan frequency $\nu_{r}^{i}$, we maximize the expectation value of the square of the SNR with respect to $\xi_{FV}^{i}$. In performing this optimization, we hold the internal resistance fixed, so we are effectively optimizing with respect to noise impedance. The value function for this optimization is $F[\nu_{r}^{i},\xi_{FV}^{i}]$, as defined in (\ref{eq:F_gen_def}); we have replaced the scattering transmission $S_{21}^{(1)}(\nu)$ with the resonator matching parameters $\nu_{r}^{i},\xi_{FV}^{i}$ of the flux-to-voltage detection scheme. This function measures the resonator sensitivity to dark matter, integrated over a wide search range, and weighted by probability densities associated with dark-matter properties. In this step of the optimization, we ask, for a particular resonator design, with a particular quality factor (i.e. resistance), what is the optimal noise impedance for the amplifier. Is it optimal to be noise-matched? 
\item Assume that the total experiment time is fixed, i.e. $\sum_{i} \tau_{i} = T_{\rm tot}$. We find the distribution of time over scan steps that maximizes the weighted area of the search's exclusion region in mass $\nu_{\rm DM}^{0}$-coupling $g_{\rm DM}$ parameter space. The value function for this optimization is the integral $A[ \{\nu_{r}^{i} \}, \{ \tau_{i} \} ]$ of (\ref{eq:A_gen_def}).
\end{enumerate}

We assume a log-uniform search, as defined in Section \ref{sec:scan_opt} for both the matching optimization and the time allocation optimization. It is evident from eq. (\ref{eq:SNR_1_QLamp_PhiV}) that, to maximize SNR, we should choose the pickup inductance to be much larger than the input inductance, $L_{\rm PU} \gg L_{\rm IN}$, so that $L_{\rm PU} \approx L$. Note the difference between this inductance optimization and more common inductance optimizations in which the backaction noise is negligible and the SQUID noise is characterized solely in terms of intrinsic energy resolution (based on spectral density $\mathcal{S}_{V_{\rm out}V_{\rm out}})$. In the latter case, one desires to maximize energy transfer to the SQUID for maximum sensitivity to the input signal, and as such, the pickup and input inductances are matched. However, integrated sensitivity, rather than energy transfer, is the figure of merit, and as we have seen in the main text, measurement backaction plays a substantial role in the optimization with a quantum-limited amplifier. Holding the coupled energy $U_{\rm DM}$, quality factor, and resonance frequency fixed, it is evident that we should maximize the participation of the pickup inductor to maximize the voltage signal and thus, SNR at each search frequency. This means taking $L_{\rm PU} \gg L_{\rm IN}$, so that $L_{\rm PU}$ is the dominant inductance. 

Then, comparing equation (\ref{eq:SNR_1_QLamp_PhiV}) with equation (\ref{eq:SNR_res_1}), we note that the two are identical if we replace $Q^{i}$ with $Q_{\rm int}^{i}$ and $\xi^{i}$ with $\xi_{FV}^{i}$. The first identification is simply a reflection of our treatment of the flux-to-voltage amplifier and in particular, the fact that the overall Q and internal Q are approximately the same. This is in contrast to the scattering mode picture, where the input impedance of the amplifier strongly damps the resonator, so that when the thermal occupation number of the resonator is much greater than unity, the optimum value of $\xi^{i}$ is such that overall Q is much less than the internal Q. The second identification reflects that both parameters are the ratio of noise impedance to internal resonator equivalent-resistancce. The similarity in the SNR expressions should not be surprising. In both cases, a resonator is being read out by a quantum-limited amplifier, which means that the minimum noise added by the amplifier is equal to the zero-point noise in the system. (See equation (\ref{eq:B_QLamp}).) Note that the sensitivity limits from the flux-to-voltage amplifier are identical to those shown in equations (\ref{eq:g_QLamp_flat}) and (\ref{eq:g_QLamp_SHM}).

Recognizing the similarity in SNR expressions, we may make the following statements about an optimized search with a flux-to-voltage amplifier, which are identical to their scattering-mode counterparts:
\begin{enumerate}
\item The optimum value of $\xi_{FV}^{i}$ in the log-uniform search is
\begin{equation}
\xi^{\rm opt}(\nu_{r}^{i},n(\nu_{r}^{i}))=\frac{1}{2} \left( 2n(\nu_{r}^{i}) + 1 + \sqrt{(2n(\nu_{r}^{i})+1)^{2}+8} \right).
\end{equation}
\item The optimum scan strategy in the log-uniform search is described by the density function
\begin{equation}
\tau_{\rm opt}(\nu)=\frac{T_{\rm tot}}{\ln(\nu_{h}/\nu_{l})}.
\end{equation}
The particular placement of resonant frequencies is not consequential, for the scan is sufficiently dense.
\end{enumerate}
The second of these observations follows the same reasoning as the scattering case. The first implies that, in a resonant scan across a large frequency regime, it is optimal not to be noise-matched, but rather, dominated by back-action! We explore this further, focusing on the thermal limit $n(\nu) \gg 1$. 

For a circuit with resonance frequency $\nu_{r}^{i}$ and quality factor $Q^{i}$, the total current noise from the resistor and the quantum-limited flux-to-voltage amplifier is described by the spectral density
\begin{equation} \label{eq:S_II_tot_xi}
\mathcal{S}_{II}^{\rm tot}(\nu,\nu_{r}^{i},\xi_{FV}^{i}) \approx \frac{h\nu_{r}^{i}}{R(\nu_{r}^{i})} \left( \frac{1}{\xi_{FV}^{i}} + \frac{\xi_{FV}^{i} + 2(2n(\nu_{r}^{i})+1)}{1+4(Q^{i})^{2} \left( (\nu-\nu_{r}^{i})/\nu_{r}^{i} \right)^{2}} \right)
\end{equation}
for frequencies sufficiently close to resonance, $|\nu-\nu_{r}^{i}| \ll \nu_{r}^{i}$. Here, we have used equations (\ref{eq:SVV_FD}), (\ref{eq:SII_tot_def}), (\ref{eq:SVV_QL}),(\ref{eq:SII_QL}), along with the approximation $R(\nu) \approx R(\nu_{r}^{i})$. The first term represents the current imprecision noise, the second represents the voltage backaction, and the third represents the noise from the resistor. The thermal/zero-point noise and the backaction noise--like the dark matter signal--are filtered by the resonant impedance, while the imprecision noise is independent of the impedance. Therefore, if the second and third terms dominate the first, then the SNR, assuming a dark-matter voltage signal of fixed strength, is approximately independent of the detuning from resonance. This gives rise to significant sensitivity outside of the resonator bandwidth. As we did in the main text, we may quantify the sensitivity outside of the resonator bandwidth using the sensitivity quality factor, defined in eq. (\ref{eq:InvN_Spectrum_Q}) of Section \ref{sssec:res_match_opt} and denoted here as $Q_{s}(\nu_{r},n(\nu_{r}),Q,\xi_{FV})$. Within the sensitivity bandwidth, of width $\nu_{r}/Q_{s}(\nu_{r},n(\nu_{r}),Q,\xi_{FV})$, the noise-equivalent number of equation (\ref{eq:NoiseNumber_zeta}) is less than twice its on-resonance (minimum) value. Note that, in the limit $\xi_{FV} \geq 1$, $n(\nu_{r}) \gg 1$, the sensitivity bandwidth corresponds approximately to the regime over which imprecision accounts for less than one-half of the total noise; this is consistent with the earlier qualitative description of resonator sensitivity. We assume that, for each scan step, $Q_{s}(\nu_{r}^{i},n(\nu_{r}^{i}),Q^{i},\xi_{FV}^{i}) \gg 1$. This assumption was validated in Section \ref{sssec:res_match_opt}.

Suppose the readout is noise-matched to the input circuit on resonance. The current noise power spectral density at resonance is
\begin{equation}
\mathcal{S}_{II}^{\rm tot}(\nu=\nu_{r}^{i}, \nu_{r}^{i}, \xi_{FV}^{i}=1)\approx \frac{4h\nu_{r}^{i}}{R}n(\nu_{r}^{i}).
\end{equation}
$Q_{s}(\nu_{r}^{i},n(\nu_{r}^{i}),Q^{i},\xi_{FV}^{i}=1)\approx Q^{i}/ (2\sqrt{n(\nu_{r}^{i})})$, so the sensitivity of the resonator to dark matter is approximately undegraded by imprecision noise for all frequencies
\begin{equation}
|\nu-\nu_{r}^{i}| \lesssim \frac{\nu_{r}^{i}}{Q^{i}} \sqrt{n(\nu_{r}^{i})}.
\end{equation}

Now suppose the readout is noise-mismatched so that $\xi_{V}^{i}=\xi^{\rm opt}(\nu_{r}^{i},n(\nu_{r}^{i}))$ and the voltage noise backaction dominates. $\xi_{FV}^{i} \approx 2n(\nu_{r}^{i}) \gg 1$, so the current noise power spectral density on resonance is
\begin{equation}
\mathcal{S}_{II}^{\rm tot}(\nu=\nu_{r}^{i}, \nu_{r}^{i}, \xi_{FV}^{i}=\xi^{\rm opt}(\nu_{r}^{i},n(\nu_{r}^{i}))) \approx \frac{6h\nu_{r}^{i}}{R} n(\nu_{r}^{i}).
\end{equation}
$Q_{s}(\nu_{r}^{i},n(\nu_{r}^{i}),Q^{i}, \xi_{FV}^{i}=\xi^{\rm opt}(\nu_{r}^{i},n(\nu_{r}^{i})))\approx Q^{i}/ ( 2\sqrt{3}n(\nu_{r}^{i}))$, so the resonance possesses undegraded sensitivity for all detunings
\begin{equation}
|\nu-\nu_{r}^{i}| \lesssim \frac{\nu_{r}^{i}}{Q^{i}} n(\nu_{r}^{i}) \sqrt{3}.
\end{equation}
In return for a modest penalty in on-resonance sensitivity, the optimal coupling to the amplifier achieves increased frequency-integrated sensitivity (which is the figure of merit in the log-uniform search), with a parametrically larger frequency range over which the imprecision noise is subdominant. The bandwidth is $\sim \sqrt{n(\nu_{r}^{i})}$ larger in the optimal, noise-mismatched case than the noise-matched case. 

\begin{figure}[htp] 
\centering
\includegraphics[width=11cm]{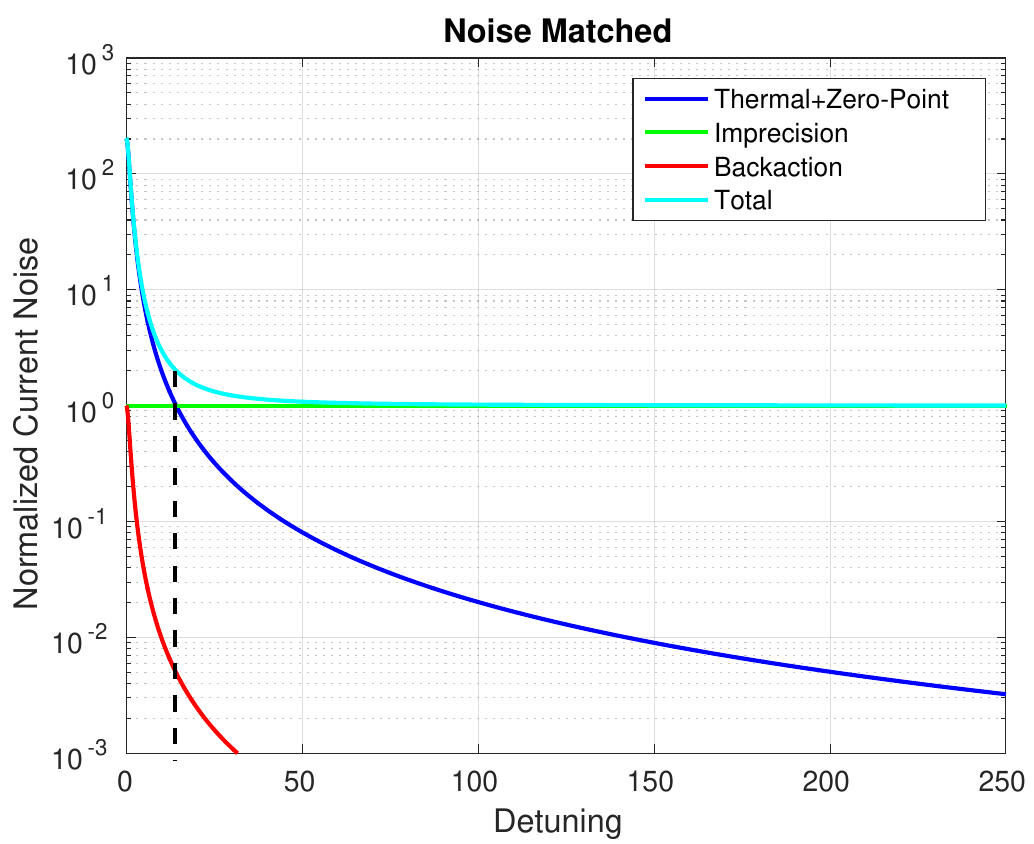}
\includegraphics[width=11cm]{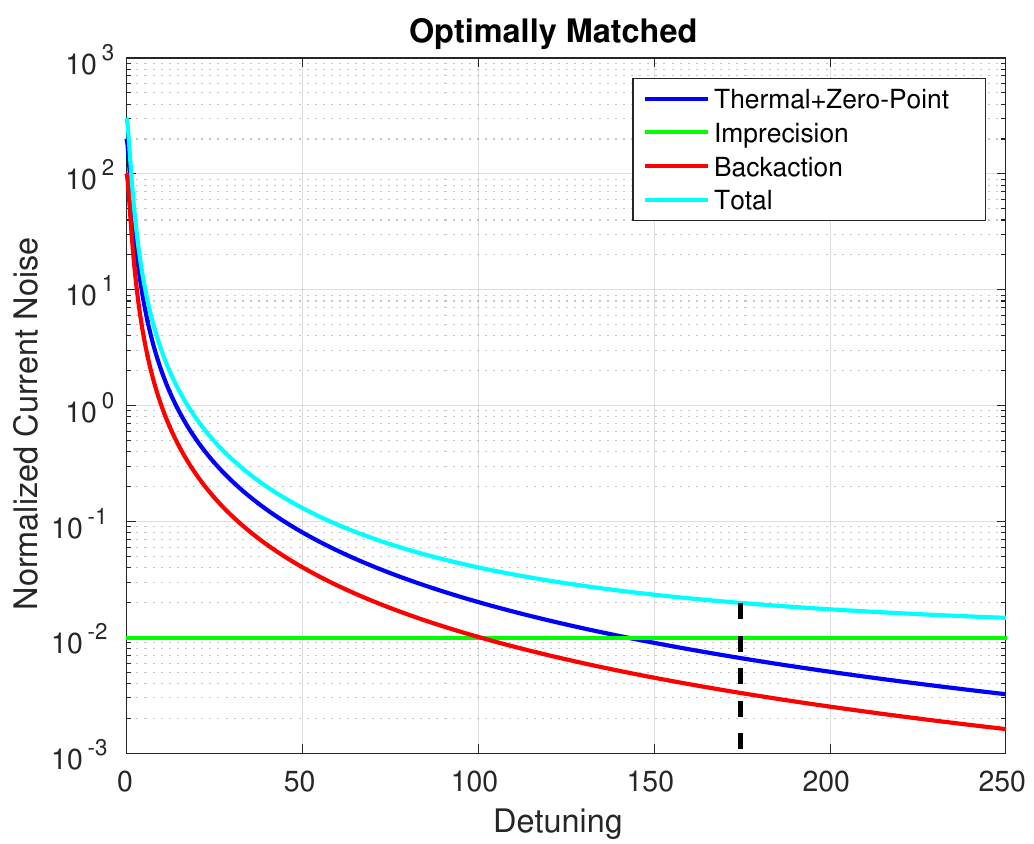}
\caption{ The noise terms of equation (\ref{eq:S_II_tot_xi}) plotted for the noise matched (top) and optimally matched (bottom) networks. The x-axis is the detuning from resonance $x=2Q^{i} \frac{\nu-\nu_{r}^{i}}{\nu_{r}^{i}}$, while the y-axis is a normalized noise spectral density $S_{II} /(h\nu_{r}^{i}/R(\nu_{r}^{i}))$. We restrict to positive detunings as the spectrum is symmetric about $x=0$. We assume a thermal occupation number of $n(\nu_{r}^{i})=50$. Thermal and zero-point noise are lumped together in blue, imprecision noise is in green, and backaction noise is in red. The total noise is in cyan. All detunings below the vertical dashed black line are within the sensitivity bandwidth. \label{fig:PhiVcrit_v_opt}}
\end{figure}

A pictorial representation of this result, displaying the relative contributions of current imprecision noise, voltage backaction noise, and thermal/zero-point noise of the resonator, is shown in Fig. \ref{fig:PhiVcrit_v_opt}. We plot each noise term on the right-hand side of equation (\ref{eq:S_II_tot_xi}), normalized to the current noise spectral density scale $h\nu_{r}^{i}/R(\nu_{r}^{i})$, as a function of detuning $x=2Q^{i} \frac{\nu-\nu_{r}^{i}}{\nu_{r}^{i}}$. We assume a thermal occupation number of $n(\nu_{r}^{i})=50$ and consider both the noise-matched and optimally matched scenarios.

In the noise-matched case (top panel of Fig. \ref{fig:PhiVcrit_v_opt}), the imprecision noise and backaction noise are sub-dominant to the thermal noise on resonance by a factor of $\approx 1/4n(\nu_{r}^{i})$. In the optimally noise-mismatched case (bottom panel of Fig. \ref{fig:PhiVcrit_v_opt}), we have a larger $\approx 2n(\nu_{r}^{i})$ voltage back-action contribution relative to the noise-matched amplifier. The backaction noise on resonance is $\approx$1/2 of the thermal noise, resulting in an amplifier added noise number of $\approx n(\nu_{r})/2$ (same as in the scattering mode case in eq. \ref{eq:NAmp_opt}) and a 50\% SNR penalty. 

However, for the quantum-limited flux-to-voltage amplifier, which possesses no current-voltage noise correlations, the current-noise imprecision and voltage-noise back-action are inversely proportional, as per equation (\ref{eq:TN_QL}). This means the imprecision noise is reduced by a factor of $\approx 1/2n(\nu_{r}^{i})$ compared to the noise-matched case, resulting in sensitivity further away from the resonance frequency. In the case shown in the figure, the sensitivity bandwidth (represented by the dashed black line) corresponds to detunings $|x| \lessapprox 14$ when the readout is noise-matched. In contrast, when the readout is optimally-mismatched, the sensitivity bandwidth corresponds to detunings $|x| \lessapprox 173$. We thus again find that the penalty in on-resonance sensitivity is order-unity, while the sensitivity bandwidth is parametrically $\sim \sqrt{n(\nu_{r}^{i})}$ much larger. 

In summary, in the thermal limit, it is beneficial to be noise-mismatched and backaction-dominated, with noise impedance a factor of $\approx 2n(\nu_{r}^{i})$ larger than the on-resonance resistance. This is analogous to the scattering mode picture in the main text, where the backaction is increased to an order-unity fraction of the thermal noise in order to provide sensitivity over a larger frequency range. See Fig. \ref{fig:NoiseNumber_comparison}.

We may compare the single-pole resonator with optimized noise impedance to the Bode-Fano limit of eq. (\ref{eq:BF_FV_ineq}). Evaluating the log-uniform search value functional (\ref{eq:F_neu_FV}) for a resonator at frequency $\nu_{r}$ with amplifier coupling $\xi^{\rm opt}(\nu_{r},n(\nu_{r}))$ (i.e. with frequency-independent, real-valued noise impedance given by the value of $R\xi^{\rm opt}(\nu_{r},n(\nu_{r}))$)  yields
\begin{equation}
\bar{F}_{\rm log}^{FV}[\nu_{r},\xi^{\rm opt}(\nu_{r},n(\nu_{r}))] = \frac{2(\xi^{\rm opt}(\nu_{r},n(\nu_{r})))^{2}}{(4\xi^{\rm opt}(\nu_{r},n(\nu_{r})) n(\nu_{r}) + (1+\xi^{\rm opt}(\nu_{r},n(\nu_{r})))^{2})^{3/2}} \frac{R}{L_{\rm PU} + L_{\rm IN}} 
\end{equation}
The first fraction is an increasing function of $\nu$, so the value functional is optimized by taking $\nu_{r}=\nu_{h}$. The resonant circuit that best approaches the Bode-Fano bound in (\ref{eq:BF_FV_ineq}) is one at the top of the search range, for which the value of the optimization function is
\begin{align}\label{eq:BF_FV_res_opt}
\bar{F}_{\rm log}^{FV}[\nu_{h}, \xi^{\rm opt}(\nu_{h},n(\nu_{h}))] &= \frac{2(\xi^{\rm opt}(\nu_{h},n(\nu_{h})))^{2}}{(4\xi^{\rm opt}(\nu_{h},n(\nu_{h})) n(\nu_{h}) + (1+\xi^{\rm opt}(\nu_{h},n(\nu_{h})))^{2})^{3/2}} \frac{R}{L_{\rm PU} + L_{\rm IN}} \nonumber \\
& \approx \begin{cases} \frac{8}{27} \frac{R}{L_{\rm PU}+L_{\rm IN}}  & \mbox{if } n(\nu_{h}) \ll 1 \\ 
\frac{1}{3\sqrt{3}} \frac{1}{n(\nu_{h})} \frac{R}{L_{\rm PU} + L_{\rm IN}} & \mbox{if } n(\nu_{h}) \gg 1 \end{cases}
\end{align}
Using $L_{\rm PU} \gg L_{\rm IN}$ (see discussion under the description of the two-part optimization), comparison of eqs. (\ref{eq:BF_FV_ineq}) and (\ref{eq:BF_FV_res_opt}) demonstrates that the optimized single-pole resonator is approximately 75\% of the Bode-Fano limit; the result is identical to the scattering-mode analysis of the main text. \emph{We find that in a log-uniform search with quantum-limited flux-to-voltage amplification, the single-pole resonator is a near-ideal single-moded dark matter detector.}

As a final remark, we ask: What if we decide to sit at a single resonance frequency, rather than scanning? This is what one would do if a signal was found, and a longer integration was needed to validate the signal. In that case, the resonator would sit at the frequency of the candidate signal and the value function reduces to the analogue of (\ref{eq:F_cand}), proportional to
\begin{equation}
\frac{\xi_{FV}^{i}}{4\xi_{FV}^{i}n(\nu_{r}^{i}) + (1+\xi_{FV}^{i})^{2}},
\end{equation}
which is maximized when noise-matched on-resonance, $\xi_{FV}^{i}=1$. Therefore, the optimal noise impedance differs between a resonant scan and an integration at a single frequency. The fundamental difference lies in the fact that, in the former, one is concerned about sensitivity integrated over a wide bandwidth, whereas in the latter, one is concerned about obtaining the largest possible sensitivity at a single frequency.

\subsection{Optimization for Imperfect Flux-to-Voltage Amplifiers}
\label{ssec:FtoV_opt_Gen}

Here, we relax two of the assumptions made in Appendix \ref{ssec:FtoV_opt_QL}. We permit the amplifier to possess a minimum noise temperature greater than the quantum limit of equation (\ref{eq:QNC_input}). We also permit correlations between the current noise and voltage noise, $\mathcal{S}_{IV}^{FV}(\nu,\nu_{r}) \ne 0$. However, to compare our results with the previous section, we later set the real part of the correlation to zero.  We also assume, in keeping with appendices \ref{ssec:FtoV_QL} and \ref{ssec:FtoV_opt_QL}, that we may ignore the damping from the input impedance of the amplifier and that the damping is dominated by intrinsic sources.

The SNR for a dark-matter signal at frequency $\nu_{\rm DM}^{0}$ from a resonance at $\nu_{r}^{i}$ is, for $|\nu_{\rm DM}^{0}-\nu_{r}^{i}| \ll \nu_{r}^{i}$,
\begin{align}
&SNR \left[ \nu_{\rm DM}^{0},g_{\rm DM}, \frac{d\rho_{\rm DM}}{d\nu}(\nu,\nu_{\rm DM}^{0}), \nu_{r}^{i}, \mathcal{S}_{VV}^{FV}(\nu,\nu_{r}^{i}), \mathcal{S}_{II}^{FV}(\nu,\nu_{r}^{i}), \mathcal{S}_{IV}^{FV}(\nu,\nu_{r}^{i}) \right]^{2} \label{eq:SNR_PhiV_1_Gen}  \\
&= (4\pi Q^{i})^{2} \left( \frac{L_{\rm PU}}{L} \right)^{2} \tau_{i} \int_{\nu_{\rm DM}^{0}}^{\nu_{\rm DM}^{0} + \Delta \nu_{\rm DM}} d\nu \left( \frac{\nu}{\nu_{r}^{i}} \frac{g_{\rm DM}^{2}V_{\rm PU} \frac{d\rho_{\rm DM}(\nu,\nu_{\rm DM}^{0})}{d\nu}}{h} \right)^{2} \times \nonumber \\
& \left( \frac{1}{4 (n(\nu) + 1/2) + \frac{\mathcal{S}_{VV}^{FV}(\nu,\nu_{r}^{i})}{h\nu R(\nu)} + \frac{\mathcal{S}_{II}^{FV}(\nu,\nu_{r}^{i}) R(\nu)}{h\nu}  \left( 1 + 4(Q^{i})^{2} (\frac{\nu-\nu_{r}^{i}}{\nu_{r}^{i}})^{2} \right) +2 Re(\frac{\mathcal{S}_{IV}^{FV}(\nu,\nu_{r}^{i})}{h\nu} (1 + 2iQ^{i}\frac{\nu-\nu_{r}^{i}}{\nu_{r}^{i}} )) } \right)^{2} \nonumber
\end{align}
For optimum SNR, we take the pickup equivalent-inductance to be much larger than the input inductance, so that $L_{\rm PU} \approx L$. The SNR of a search, consisting of scan frequencies $\{ \nu_{r}^{i} \}$, is given by adding the single-resonance-frequency SNRs in quadrature:
\begin{align}
&SNR_{\rm tot} \left[ \nu_{\rm DM}^{0}, g_{\rm DM}, \frac{d\rho_{\rm DM}}{d\nu}(\nu,\nu_{\rm DM}^{0}), \{ \nu_{r}^{i} \}, \{ \mathcal{S}_{VV}^{FV}(\nu,\nu_{r}^{i}) \}, \{ \mathcal{S}_{II}^{FV}(\nu,\nu_{r}^{i}) \}, \{ \mathcal{S}_{IV}^{FV}(\nu,\nu_{r}^{i}) \} \right]^{2} \nonumber \\
&=\sum_{i} SNR \left[ \nu_{\rm DM}^{0}, g_{\rm DM}, \frac{d\rho_{\rm DM}}{d\nu}(\nu,\nu_{\rm DM}^{0}), \nu_{r}^{i}, \mathcal{S}_{VV}^{FV}(\nu,\nu_{r}^{i}), \mathcal{S}_{II}^{FV}(\nu,\nu_{r}^{i}), \mathcal{S}_{IV}^{FV}(\nu,\nu_{r}^{i}) \right]^{2} \label{eq:SNR_PhiV_Gen}.
\end{align}

The four terms in the denominator of (\ref{eq:SNR_PhiV_1_Gen}) represent the various noise sources in the system. The first term is the thermal and zero-point noise of the resonator. The second and third terms are the backaction and imprecision noise of the amplifier, respectively. The last term is the noise due to correlations between the backaction and imprecision noise. The amplifier noise spectral densities are functions both of the frequency and of the resonance frequency, as we assume that the amplifier can be re-optimized at each scan frequency. If detuning-dependent (i.e. dependent on $\nu-\nu_{r}^{i}$) terms are smaller than the other terms in the denominator, then the sensitivity is not degraded by the resonator rolloff. At these detunings, the signal current rolls off, but the noise current does as well, keeping the SNR approximately the same. Similar to the treatment in the main text and Appendix \ref{ssec:FtoV_opt_QL}, the frequencies over which this occurs are the frequencies over which the resonator is maximally sensitive. An optimization of the integrated sensitivity entails maximizing this range of search frequencies without significantly degrading the sensitivity on/near resonance.

Assume that at each resonance frequency, all noise spectral densities can be taken to be constants over frequencies $|\nu-\nu_{r}^{i}| \ll \nu_{r}^{i}$, e.g. 
\begin{equation}\label{eq:S_Amp_const_L}
\mathcal{S}_{VV}^{FV}(\nu,\nu_{r}^{i}) \approx \mathcal{S}_{VV}^{FV}(\nu_{r}^{i},\nu_{r}^{i})
\end{equation}
\begin{equation}\label{eq:S_Amp_const_R}
\mathcal{S}_{VV}^{FV}(\nu_{r}^{i},\nu) \approx \mathcal{S}_{VV}^{FV}(\nu_{r}^{i},\nu_{r}^{i})
\end{equation}
and similarly for $\mathcal{S}_{II}^{FV}$ and $\mathcal{S}_{IV}^{FV}$. The precise range over which we need this approximation to hold for self-consistency is discussed below. Let
\begin{equation}\label{eq:alpha1}
\alpha_{1}(\nu_{r})= \frac{\mathcal{S}_{II}^{FV}(\nu_{r},\nu_{r})R(\nu_{r})}{h\nu_{r}}
\end{equation}
\begin{equation}\label{eq:alpha2}
\alpha_{2}(\nu_{r})= \frac{Im\ \mathcal{S}_{IV}^{FV}(\nu_{r},\nu_{r})}{h\nu_{r}}
\end{equation}
\begin{equation}\label{eq:alpha3}
\alpha_{3}(\nu_{r})= 4(n(\nu_{r})+1/2) + \frac{\mathcal{S}_{VV}^{FV}(\nu_{r},\nu_{r})}{h\nu_{r} R(\nu_{r})} + \frac{\mathcal{S}_{II}^{FV}(\nu_{r},\nu_{r}) R(\nu_{r})}{h\nu_{r}} + 2\frac{Re\ \mathcal{S}_{IV}^{FV}(\nu_{r},\nu_{r})}{h\nu_{r}}
\end{equation}
As in previous appendices, define $x=2Q \frac{\nu-\nu_{r}^{i}}{\nu_{r}^{i}}$, as the detuning from the resonance frequency as a fraction of the resonator bandwidth. The fraction in the last line of equation (\ref{eq:SNR_PhiV_1_Gen}), which represents the SNR as a function of detuning, then reads
\begin{align}
\bar{\mathcal{F}}(x,\alpha_{1}(\nu_{r}^{i}),\alpha_{2}(\nu_{r}^{i}),\alpha_{3}(\nu_{r}^{i})) & \approx \frac{1}{\alpha_{1}(\nu_{r}^{i})x^{2} - 2\alpha_{2}(\nu_{r}^{i})x + \alpha_{3}(\nu_{r}^{i})} \nonumber \\
&= \alpha_{1}(\nu_{r}^{i})^{-1} \left( \left( x - \frac{\alpha_{2}(\nu_{r}^{i})}{\alpha_{1}(\nu_{r}^{i})} \right)^{2} + \frac{\alpha_{1}(\nu_{r}^{i})\alpha_{3}(\nu_{r}^{i})-\alpha_{2}(\nu_{r}^{i})^{2}}{\alpha_{1}(\nu_{r}^{i})^{2}} \right)^{-1}. \label{eq:barF_def}
\end{align}
The SNR is not degraded by the resonator rolloff, and the resonator is maximally sensitive to the dark-matter signal at detuning $x$, as long as
\begin{equation}\label{eq:OB_sensitivity_CrossCorr}
\left| x - \frac{\alpha_{2}(\nu_{r}^{i})}{\alpha_{1}(\nu_{r}^{i})} \right| \lesssim \sqrt{\frac{\alpha_{1}(\nu_{r}^{i})\alpha_{3}(\nu_{r}^{i})-\alpha_{2}(\nu_{r}^{i})^{2}}{\alpha_{1}(\nu_{r}^{i})^{2}}}.
\end{equation}
Equivalently, in the language of the main text, the sensitivity Q, describing the bandwidth of maximal sensitivity, is
\begin{equation}\label{eq:Qeff_def_Gen}
Q_{s}(\nu_{r}^{i}, Q^{i}, \alpha_{1}(\nu_{r}^{i}),\alpha_{2}(\nu_{r}^{i}), \alpha_{3}(\nu_{r}^{i})) = Q^{i} \left( \frac{\alpha_{1}(\nu_{r}^{i})\alpha_{3}(\nu_{r}^{i})-\alpha_{2}(\nu_{r}^{i})^{2}}{\alpha_{1}(\nu_{r}^{i})^{2}} \right)^{-1/2}. 
\end{equation}
We require that the dark-matter frequencies that lie within a few sensitivity bandwidths of the resonance satisfy the approximation of a constant noise spectral density in (\ref{eq:S_Amp_const_L})-(\ref{eq:S_Amp_const_R}). We assume, as we did in the main text, that
\begin{equation}\label{eq:OB_bandassump_CrossCorr}
Q_{s}(\nu_{r}^{i}, Q^{i}, \alpha_{1}(\nu_{r}^{i}),\alpha_{2}(\nu_{r}^{i}), \alpha_{3}(\nu_{r}^{i})) \gg 1,
\end{equation}
and that the frequency range of equation (\ref{eq:OB_sensitivity_CrossCorr}) is not comparable in scale to the resonance frequency itself. Implicit in this calculation is the assumption that $\alpha_{1}(\nu_{r}^{i})\alpha_{3}(\nu_{r}^{i})-\alpha_{2}(\nu_{r}^{i})^{2} \geq 0$. If $Re\ \mathcal{S}_{IV}^{FV}(\nu_{r}^{i},\nu_{r}^{i}) \geq 0$, this is readily satisfied because $\mathcal{S}_{VV}^{FV}(\nu_{r}^{i},\nu_{r}^{i})\mathcal{S}_{II}^{FV}(\nu_{r}^{i},\nu_{r}^{i})-|\mathcal{S}_{IV}^{FV}(\nu_{r}^{i},\nu_{r}^{i})|^{2} \geq 0$ (see eq. (\ref{eq:QNC_input})).

We see from equation (\ref{eq:barF_def}) that the presence of out-of-phase current-voltage correlations, represented by $Im\ S_{IV}$, has introduced an asymmetry in the resonator sensitivity to dark matter. If $Im\ S_{IV}^{FV}(\nu_{r}^{i},\nu_{r}^{i})>0$, then for detunings $x>0$, the correlation reduces the noise, relative to the detuning $-x$; this results in a larger range of positive detunings where the SNR is not degraded (relative to the range of negative detunings). If $Im\ S_{IV}^{FV}(\nu_{r}^{i},\nu_{r}^{i})<0$, the opposite happens, and there is a larger range of negative detunings at which the SNR is not degraded. 

We optimize the resonator readout and noise matching at each resonance frequency for a log-uniform search. We maximize the expectation value of the square of the SNR. The value function evaluates to, in analogy with equation (\ref{eq:F_neu_simple}),
\begin{align}
F_{\rm log}(\alpha_{1,2,3}(\nu_{r}^{i}) & ) \approx \gamma_{0}^{FV} \int_{0}^{\beta} du \left( \frac{\sqrt{1+u}}{h} \frac{d\rho_{\rm DM}}{du}(u) \right)^{2} \label{eq:PhiV_Gen_simple} \\ 
&\times \int_{\nu_{l}(1+u)}^{\nu_{h}(1+u)} d\bar{\nu} \left( \frac{1}{4(Q^{i})^{2} \alpha_{1}(\nu_{r}^{i}) \left( \frac{\bar{\nu}}{\nu_{r}^{i}} -1 \right)^{2} -4Q^{i}\alpha_{2}(\nu_{r}^{i}) \left( \frac{\bar{\nu}}{\nu_{r}^{i}} -1 \right) + \alpha_{3}(\nu_{r}^{i})} \right)^{2} \nonumber,
\end{align}
where $\gamma_{0}^{FV}$ is a constant. Note that, instead of optimizing with respect to a single noise matching parameter $\xi_{FV}^{i}$, we are optimizing with respect to three parameters, $\alpha_{1,2,3}(\nu_{r}^{i})$, owing to the two additional degrees of freedom---noise temperature and correlations. Far away from resonance, the approximation (\ref{eq:Ztot_res}) breaks down; strictly speaking, the integrand of the $\bar{\nu}$-integral in (\ref{eq:PhiV_Gen_simple}) is then not an appropriate representation for the SNR. However, at these frequencies far detuned from resonance, there is negligible sensitivity, and the contribution to the integral is nearly zero. As such, it is an appropriate approximation, for the purpose of evaluating the integral, to use this integrand expression at all frequencies. The integrand is sharply peaked near $\bar{\nu} \approx \nu_{r}^{i}$, so using the change of variables $x=2Q^{i} (\bar{\nu}-\nu_{r}^{i})/\nu_{r}^{i}$ and extending the limits of integration to $\pm \infty$ (similar to (\ref{eq:F_neu_res_eval})), we find
\begin{equation}
F_{\rm log}(\alpha_{1,2,3}(\nu_{r}^{i}) ) \approx \gamma_{1}^{FV}\frac{\alpha_{1}(\nu_{r}^{i})}{(\alpha_{1}(\nu_{r}^{i}) \alpha_{3}(\nu_{r}^{i}) - \alpha_{2}(\nu_{r}^{i})^{2})^{3/2}}, 
\end{equation}
where $\gamma_{1}^{FV}$ is a constant, containing the integral over $u$. The expression 
\begin{equation}\label{eq:G_PhiV_def}
\bar{\mathcal{G}}(\alpha_{1,2,3}(\nu_{r}^{i}))=\frac{\alpha_{1}(\nu_{r}^{i})}{(\alpha_{1}(\nu_{r}^{i}) \alpha_{3}(\nu_{r}^{i}) - \alpha_{2}(\nu_{r}^{i})^{2})^{3/2}}
\end{equation}
represents the effect of thermal, zero-point, and amplifier noise integrated over the search band. In analogy with the optimization over $\xi$ performed for resonator readout with quantum-limited scattering-mode and flux-to-voltage amplifiers (see Section \ref{sssec:res_match_opt} and Appendix \ref{ssec:FtoV_opt_QL}), we optimize $\bar{\mathcal{G}}$ with respect to the three noise matching parameters $\alpha_{1,2,3}(\nu_{r}^{i})$, assuming fixed physical temperature and fixed minimum noise temperature. 

Assume that $Re\ \mathcal{S}_{IV}^{FV}=0$, so that we may compare our results to the previous section and utilize the quantum limits derived in Appendix \ref{ssec:FtoV_QL}. This assumption is also in keeping with SQUID models, such as that in \cite{clarke1979optimization}. Let $kT_{N}^{min}(\nu_{r}^{i})$ be the minimum noise temperature for the amplifier corresponding to the step at resonance frequency $\nu_{r}^{i}$:
\begin{equation}\label{eq:S_TN_min_PhiV}
kT_{N}^{min}(\nu_{r}^{i}) \equiv \frac{1}{2}\sqrt{\mathcal{S}_{VV}^{FV}(\nu_{r}^{i},\nu_{r}^{i})\mathcal{S}_{II}^{FV}(\nu_{r}^{i},\nu_{r}^{i}) - (Im\ \mathcal{S}_{IV}^{FV}(\nu_{r}^{i},\nu_{r}^{i}))^{2}}.
\end{equation}
Let
\begin{equation}\label{eq:eta_def}
\eta(\nu_{r}^{i}) \equiv \frac{kT_{N}^{min}(\nu_{r}^{i})}{h\nu_{r}^{i}/2} \geq 1.
\end{equation}
Then, (\ref{eq:G_PhiV_def}) can be rewritten as
\begin{equation}\label{eq:G_PhiV_red}
\bar{\mathcal{G}}(\alpha_{1}(\nu_{r}^{i}),\eta(\nu_{r}^{i}))= \frac{\alpha_{1}(\nu_{r}^{i})}{ (\alpha_{1}(\nu_{r}^{i})^{2} + 2(2n(\nu_{r}^{i})+1) \alpha_{1}(\nu_{r}^{i}) + \eta(\nu_{r}^{i})^{2})^{3/2}}.
\end{equation}
With noise temperature fixed, the maximum of $\bar{\mathcal{G}}$ over $\alpha_{1}(\nu_{r}^{i})$ occurs at
\begin{equation}\label{eq:alpha_opt}
\alpha_{1}^{\rm opt}(\nu_{r}^{i})= \frac{2\eta(\nu_{r}^{i})^{2}}{ 2n(\nu_{r}^{i})+1 + \sqrt{(2n(\nu_{r}^{i}) +1)^{2} + 8\eta(\nu_{r}^{i})^{2} }},
\end{equation}
which corresponds to a current noise spectral density of
\begin{equation}\label{eq:SII_opt_PhiV}
\mathcal{S}_{II}(\nu_{r}^{i},\nu_{r}^{i})= \frac{h\nu_{r}^{i} \alpha_{1}^{\rm opt}(\nu_{r}^{i})}{R(\nu_{r}^{i})}.
\end{equation}
Note that at $\eta=1$, corresponding to a quantum-limited readout, $\alpha_{1}^{\rm opt}(\nu_{r}^{i}) = \xi^{\rm opt}(\nu_{r}^{i},n(\nu_{r}^{i}))^{-1}$, and we recover the results of Appendix \ref{ssec:FtoV_opt_QL}. We find that the optimal amplifier, given fixed noise temperature (e.g. the quantum-limited noise temperature $kT_{N}(\nu)=h\nu/2$ given in eq. (\ref{eq:TN_QL})), is not unique. The conditions for the optimum only precisely constrain the current noise; there are infinitely many solutions for amplifier noise spectral densities $\mathcal{S}_{VV}^{FV}, \mathcal{S}_{IV}^{FV}$ that satisfy equation (\ref{eq:S_TN_min_PhiV}), given the constraint (\ref{eq:SII_opt_PhiV}) on $\mathcal{S}_{II}^{FV}$. In particular, for fixed noise temperature, a readout with out-of-phase correlations between current imprecision noise and voltage backaction noise can perform (in terms of maximizing the value function $F_{\rm log}(\alpha_{1,2,3}(\nu_{r}^{i}))$) as well as a readout with uncorrelated noise sources. This is of practical importance because these noise sources are often the result of the same physical fluctuations (e.g. Johnson noise from resistive shunts in a dc SQUID) and therefore, are usually correlated. 

We may have guessed that correlations do not prevent us from obtaining the optimum from analysis of equation (\ref{eq:barF_def}). This equation may be rewritten in terms of the noise temperature parameter $\eta(\nu_{r}^{i})$
\begin{equation}
\bar{\mathcal{F}}(x,\alpha_{1}(\nu_{r}^{i}),\alpha_{2}(\nu_{r}^{i}), \eta(\nu_{r}^{i}))= \alpha_{1}(\nu_{r}^{i})^{-1}  \left( \left( x - \frac{\alpha_{2}(\nu_{r}^{i})}{\alpha_{1}(\nu_{r}^{i})} \right)^{2} + \frac{\alpha_{1}(\nu_{r}^{i})^{2} + 2(2n(\nu_{r}^{i})+1) \alpha_{1}(\nu_{r}^{i}) + \eta(\nu_{r}^{i})^{2}}{\alpha_{1}(\nu_{r}^{i})^{2}} \right)^{-1}.
\end{equation}
The second fraction in the outer parentheses represents the number of bandwidths over which the resonator has maximal sensitivity to dark matter; it is the value of $(Q^{i}/Q_{s})^{2}$. This term depends only on the value of $\alpha_{1}$ and $\eta$. The maximum value of $\bar{\mathcal{F}}$ with respect to detuning, which is a metric of the maximum sensitivity of the resonator, also depends only on $\alpha_{1}$. As our value function $F_{\rm log}(\alpha_{1,2,3}(\nu_{r}^{i}))$ is roughly proportional to the maximum of $\bar{\mathcal{F}}^{2}$ multiplied by the frequency range of maximum sensitivity, it follows that our value function, and the value of $\bar{\mathcal{G}}$ only depends on these parameters. This is precisely what we see in (\ref{eq:G_PhiV_red}).  The only relevant effect of correlations between imprecision and backaction noise is to shift the frequency at which $\bar{\mathcal{F}}$ is maximized, i.e. introduce an asymmetry in resonator sensitivity at positive and negative detunings. 

One may also understand this result by looking at the total noise spectral density referred to a voltage excitation in the resonator:
\begin{equation}\label{eq:SVV_tot_Corr}
\frac{\mathcal{S}_{VV}^{\rm ref,tot}(x,\nu_{r}^{i})}{h\nu_{r}^{i}R(\nu_{r}^{i})} \approx 4(n(\nu_{r}^{i}) + 1/2) + \frac{ \alpha_{2}(\nu_{r}^{i})^{2} + \eta(\nu_{r}^{i})^{2}}{\alpha_{1}(\nu_{r}^{i})}  - 2x\alpha_{2}(\nu_{r}^{i}) + \alpha_{1}(\nu_{r}^{i}) (1+x^{2}).
\end{equation}
The first term represents the thermal/zero-point noise from the resonator, the second term represents the voltage-backaction-noise spectral density of the amplifier, the third term represents the correlation between voltage backaction noise and current imprecision noise, and the fourth term represents the current-imprecision-noise spectral density of the amplifier. 

\begin{figure}[htp]
\centering
\includegraphics[width=0.5\textwidth]{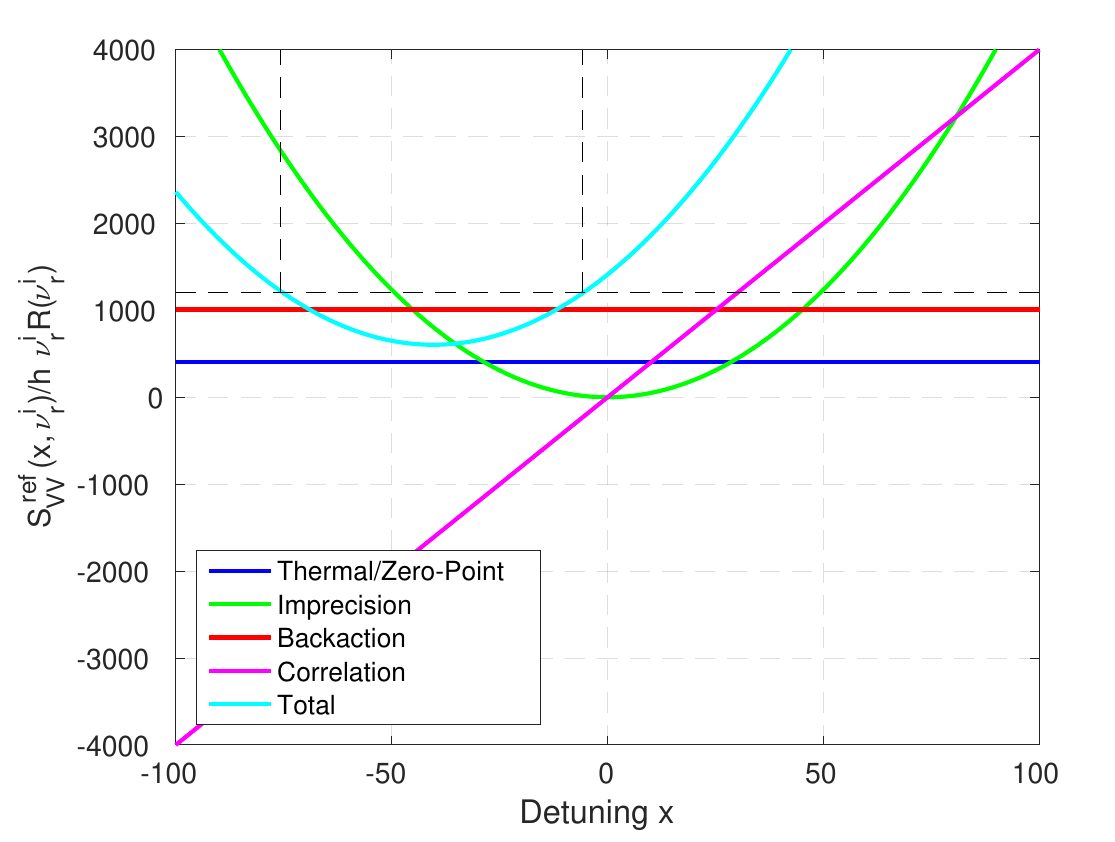}\\
\includegraphics[width=0.5\textwidth]{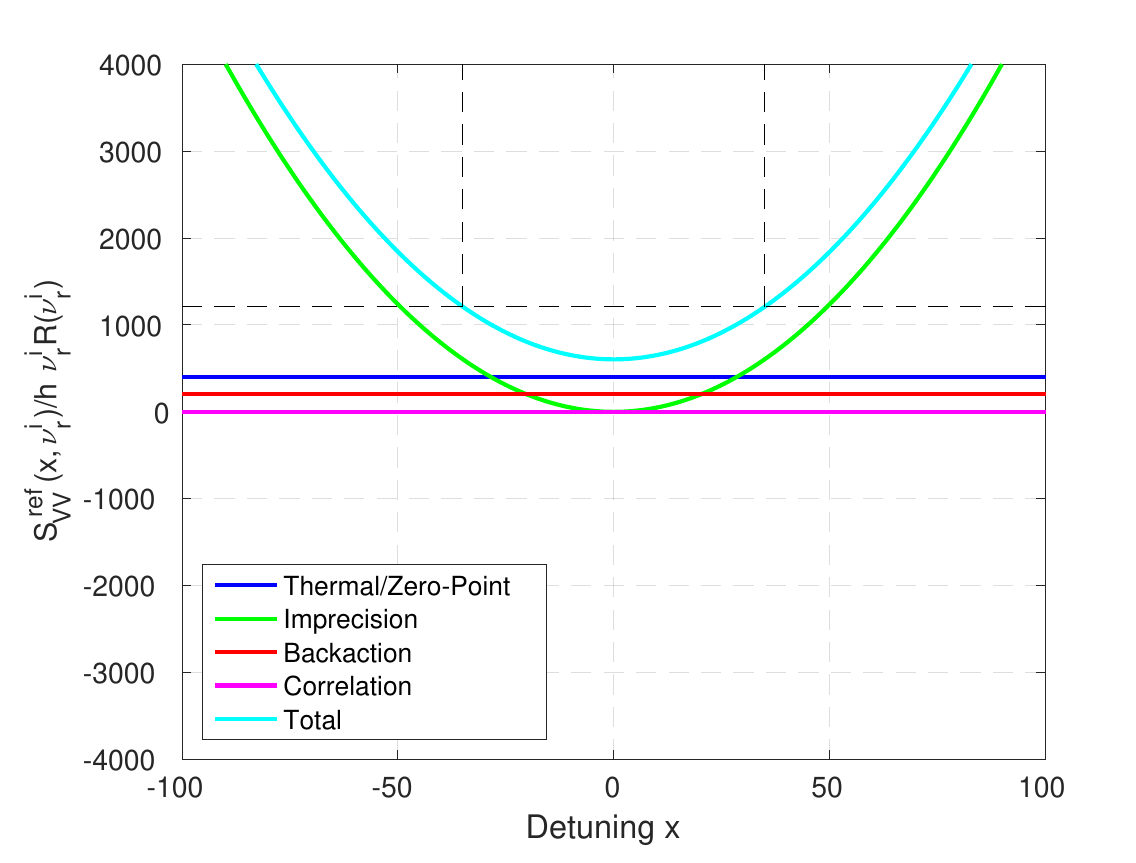}\\
\includegraphics[width=0.5\textwidth]{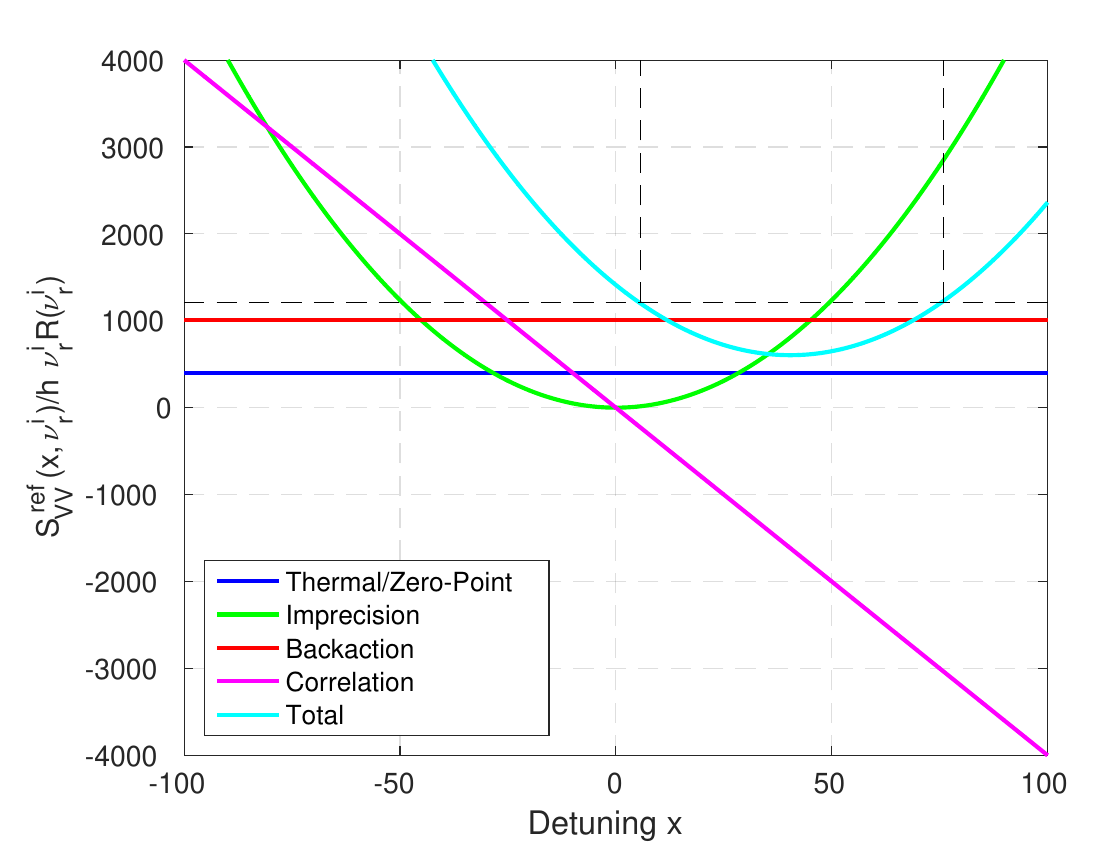}
\caption{The various contributions to the noise, referred to an excitation voltage in the input circuit, as a function of detuning $x=2Q^{i} \frac{\nu-\nu_{r}^{i}}{\nu_{r}^{i}}$ from resonance frequency. $\eta(\nu_{r}^{i})=10$, $n(\nu_{r}^{i})=100$, $\alpha_{1}(\nu_{r}^{i})=\alpha_{1}^{\rm opt}(\nu_{r}^{i})$. The top, middle, and bottom figures represent three different values of $\alpha_{2}(\nu_{r}^{i})$: $\alpha_{2}(\nu_{r}^{i})=-20,\ 0,\ +20$, respectively. The region of the total noise curve below the horizontal dashed line represents the detunings at which the total noise is less than a factor of 2 from its minimum value. This region, bounded by the two vertical dashed lines, is the sensitivity bandwidth.} \label{fig:SVV_diffalpha}
\end{figure}

\begin{figure}[htp]
\centering
\includegraphics[width=.9\textwidth]{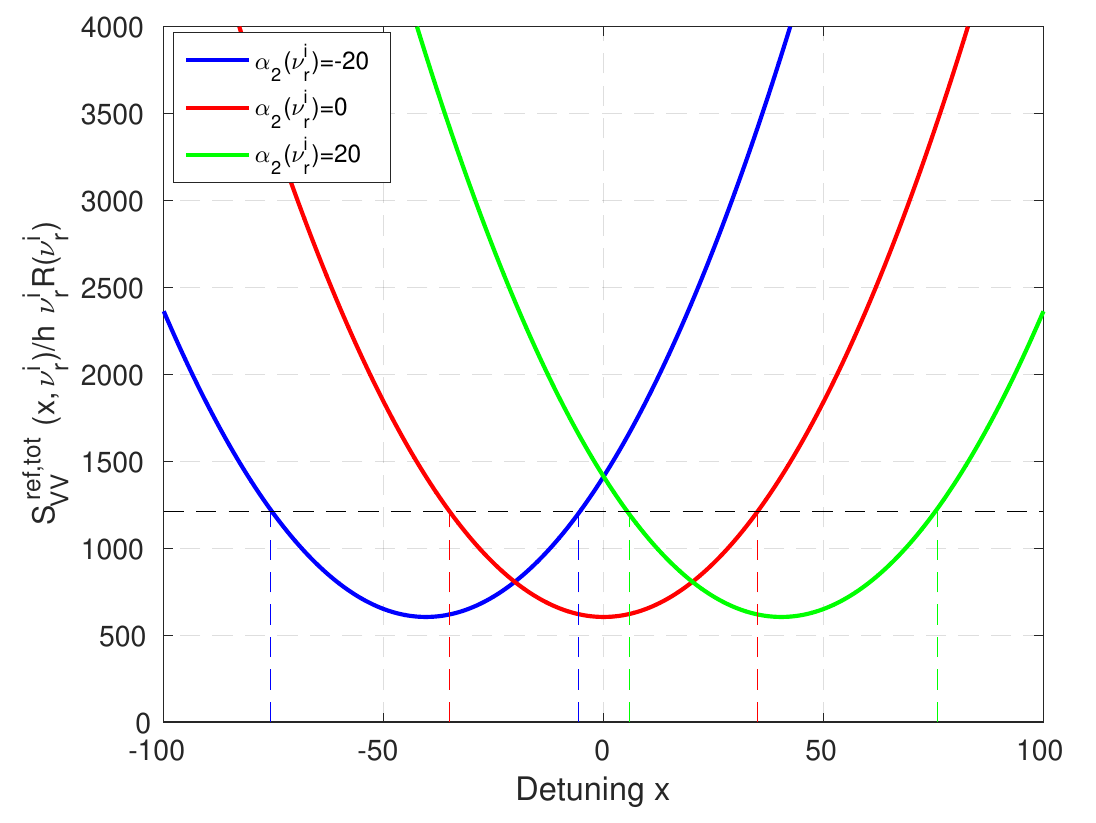}
\caption{Total noise of equation (\ref{eq:SVV_tot_Corr}) as a function of resonator detuning for $\alpha_{2}(\nu_{r}^{i})=-20,\ 0,\ +20$. The region of the total noise curve below the horizontal dashed line, and bounded by the colored vertical lines (same line positions as Fig. \ref{fig:SVV_diffalpha}) , represents the the sensitivity bandwidth.} \label{fig:SVV_tot_diffalpha}
\end{figure}

We plot each of these terms for $\eta(\nu_{r}^{i})=10$, $n(\nu_{r}^{i})=100$, $\alpha_{1}(\nu_{r}^{i})=\alpha_{1}^{\rm opt}(\nu_{r}^{i})$, as specified by equation (\ref{eq:alpha_opt}), for three different values of $\alpha_{2}(\nu_{r}^{i})$, $\alpha_{2}(\nu_{r}^{i})=-20,\ 0,\ +20$. See Figs. \ref{fig:SVV_diffalpha} and \ref{fig:SVV_tot_diffalpha}. Even though the backaction and correlated-noise terms have different values for the three situations, the minimum level of $S_{VV}^{\rm ref,tot}(x,\nu_{r}^{i})$ is the same ($\approx 604\ h\nu_{r}^{i}R(\nu_{r}^{i})$). The number of bandwidths over which the total noise is less than a factor of 2 above its minimum is indicated by points below the dashed horizontal line in Fig. \ref{fig:SVV_diffalpha}. This represents the sensitivity bandwidth and is the same for each of the three cases ($\approx 35$ resonator bandwidths). Indeed, Fig. \ref{fig:SVV_tot_diffalpha} shows that the total noise curves are simply translations of one another, as dictated by the sign and magnitude of the correlated noise.  

The signal-to-noise of the scan at a single dark-matter frequency of interest can be evaluated by approximating a continuum of resonator frequencies and using the optimal scan density function (\ref{eq:t_opt}) found in the main text, $\tau_{\rm opt}(\nu_{r})=\frac{T_{\rm tot}}{ln(\nu_{h}/\nu_{l})}$. Assuming fixed Q across resonance frequencies, we find, using methods similar to those already presented and the optimum readout defined by equations (\ref{eq:S_TN_min_PhiV})-(\ref{eq:alpha_opt}),
\begin{equation}
SNR^{2}(\nu_{\rm DM}^{0}) \approx 4\pi^{3}Q \frac{T_{\rm tot}}{ln(\nu_{h}/\nu_{l})} \bar{\mathcal{G}}(\nu_{\rm DM}^{0},\alpha_{1}^{\rm opt} (\nu_{\rm DM}^{0}),\eta(\nu_{\rm DM}^{0}))  \int_{\nu_{\rm DM}^{0}}^{\nu_{\rm DM}^{0} + \Delta \nu_{\rm DM}^{c}} d\nu \left( \frac{g_{\rm DM}^{2}V_{\rm PU} \frac{d\rho_{\rm DM}}{d\nu}}{h} \right)^{2}  \label{eq:SNR_PhiV_Gen_Simple} 
\end{equation} 
From this equation, sensitivity limits on $g_{\rm DM}$ and therefore, sensitivity limits on $\varepsilon$ and $g_{a\gamma \gamma}$ may be readily obtained. For a generic dense scan, we would replace $T_{\rm tot}/\ln(\nu_{h}/\nu_{l})$ with $\tau(\nu_{\rm DM}^{0})$ in the above equation.
\section{Comparison of Resonant and Broadband Searches with Flux-to-Voltage Amplifiers}
\label{sec:ResvBroad_SQUID}

\begin{figure}[htp] 
\includegraphics[width=\textwidth]{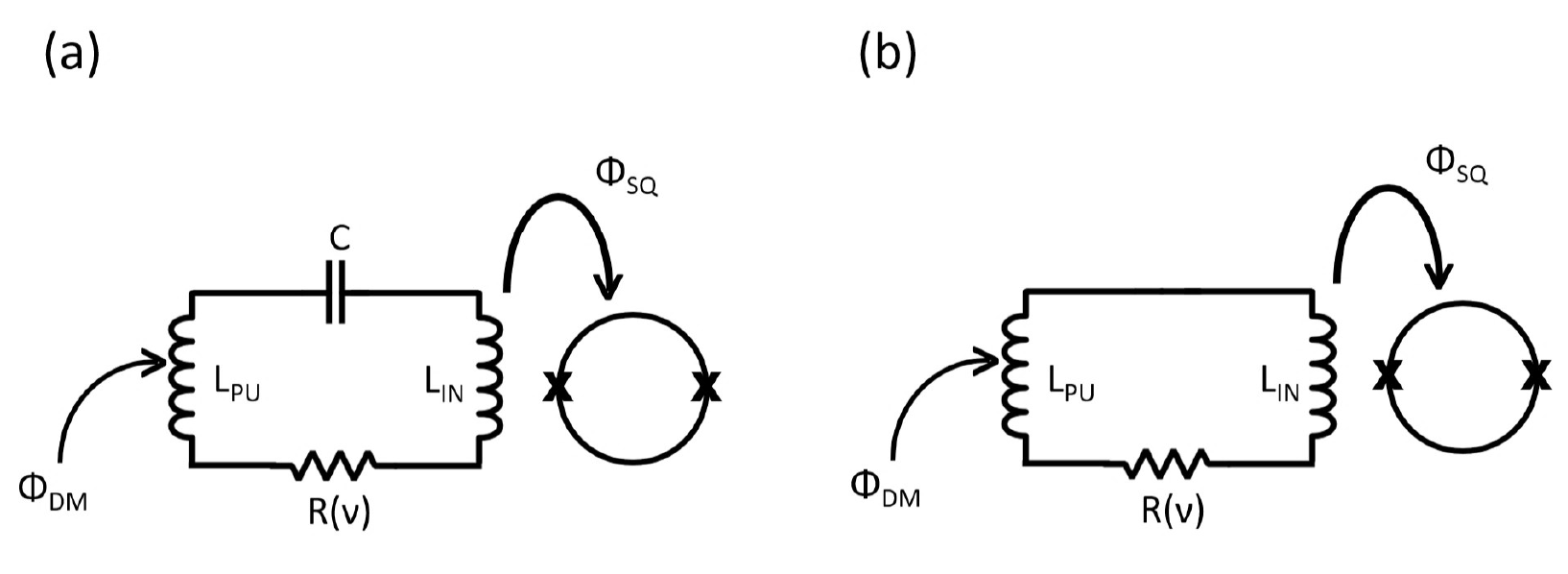}
\caption{Comparison of quasi-static resonant and reactive broadband searches. (a) Resonant dark matter search. (b) Broadband dark matter search. The oscillating dark-matter signal causes a flux in the inductor, which in turn causes currents to flow in the receiver. These currents are then read out as a flux in the flux-to-voltage amplifier, shown here schematically as a dc SQUID. For each of the two experiments, we assume that the pickup inductance $L_{\rm PU}$ is much larger than the input inductance $L_{\rm IN}$ in order to optimize sensitivity. For an apples-to-apples comparison, the inductor quality is assumed to be the same in the two experiments, but the inductance value is allowed to differ, e.g. the number of turns in the pickup coil is not necessarily the same.} \label{fig:ResvBroad}
\end{figure}

A broadband, non-resonant search in the quasi-static regime using flux-to-voltage amplifiers has been proposed as an alternative to a tunable resonant search. In ref. \cite{kahn2016broadband}, the sensitivity of a broadband search is compared to a tunable resonant search, with the conclusion that the broadband search is more sensitive at frequencies below 100 kHz. In this appendix, we develop an apples-to-apples comparison between broadband and resonant searches using our optimization framework. We show that, in contrast to the conclusion drawn in \cite{kahn2016broadband}, a tunable resonant search has better integrated scan sensitivity at any frequency at which an electromagnetic resonator can practically be constructed ($\gtrsim$100 Hz).

We first introduce the detection schemes. We then lay out the assumptions for performing an apples-to-apples comparison of an optimized single-pole tunable resonant search and an optimized broadband search. By analyzing the sensitivity of a fixed-frequency resonator, we show that, under any set of priors and in any search range, the tunable resonant search must be strictly superior to the broadband search. We then calculate the size of the advantage for example scans, which requires a treatment of amplifier noise and noise impedance in a broadband search.  We calculate the SNR of each experiment at each search frequency, and then determine the ratio of SNRs for two scans: a wide scan and a low-frequency scan. Motivated by these SNR calculations, we discuss the relationship of priors to optimized time allocation in a resonant search, which does not apply to a broadband search but which is required for an accurate comparison of the two techniques. We show that the optimized resonant search generically enables search times (integration time needed to reach a particular dark matter-photon coupling) a few orders of magnitude lower than an optimized broadband search.

Equivalent circuit diagrams are shown for both types of searches in Fig. \ref{fig:ResvBroad}. The two fundamental sources of noise in the detection schemes are thermal/zero-point noise from the resistor and amplifier noise. In the resonant search, the resonance frequency is scanned, and the sensitivity of the search in different frequency bands can be chosen by time allocation. At each scan step, the sensitivity bandwidth, which may be several resonator linewidths (e.g. see eq. \ref{eq:InvN_Spectrum_Q}), represents the ``useful" bandwidth for detection. Beyond this bandwidth, the SNR is degraded by the amplifier imprecision noise, and the resonator provides a negligible contribution to the overall scan sensitivity. In contrast, in a broadband search, there is no scanning and one uses the information over all bandwidth (up to $\sim$100 MHz, where parasitic capacitances not shown in Fig. \ref{fig:ResvBroad} degrade the coupling to the amplifier). Over most of this bandwidth, the SNR is greatly degraded by amplifier noise. But instead of discarding this information, it is integrated over the full time of the experiment. To some degree, the degradation of SNR from amplifier noise is mitigated by long integration times.

We now establish the assumptions under which we compare the optimized tunable resonator search to the optimized broadband search. We assume that the two experiments couple to the same amount of dark-matter signal energy, denoted $U_{\rm DM}$ in eq. (\ref{eq:UDM_def}). However, the two experiments may have different pickup inductances (e.g. different numbers of wiring turns in a lumped-element pickup coil).

The resonator is tuned by a variable lossless capacitor (e.g. a vacuum gap superconducting-parallel-plate capacitor for which the electrode overlap area or the separation distance may be changed), represented by $C$ in the figure. A broadband circuit can be realized by removing the capacitor and shorting the leads together. In this manner, we may assume that the loss in both circuits is the same; the loss is quantified by the inductor quality, which is the ratio of its reactance to the resistance $2 \pi \nu (L_{\rm PU} + L_{\rm IN})/R(\nu)$. Nevertheless, we emphasize that we do not require the values of the total inductance and resistance to be the same in the broadband and resonant searches. 
We assume that the tunable resonator is high Q, which implies $R(\nu_{\rm DM}^{0}) \ll 2\pi\nu_{\rm DM}^{0}(L_{\rm PU} + L_{\rm IN})$ at all dark matter frequencies in both searches.

We assume that both searches possess quantum-limited readout, with real-valued noise impedance. We assume that the total search time $T_{\rm tot}$ is long enough that we may dwell at each resonance frequency longer than the resonator ring-up/dark matter coherence time. We set the lower limit for all scans at $\nu_{l}=100$ Hz. Below $\sim$ 100 Hz, building a resonant circuit is challenging due to the physically large components required and resultant parasitic inductances and capacitances. We set the upper limit for all comparisons at $\nu_{h}=100$ MHz. We thus stay below high frequencies where uncontrolled impedances (for instance, stray inductances/capacitances) in the nominally physically-lumped-element input circuit degrade system performance in both the broadband and resonant experiment; for example, parasitic resonances associated with a dc SQUID flux-focusing washer create challenges for broadband amplification. 




Intuition suggests that the assertion that a broadband search has higher sensitivity below 100 kHz\cite{kahn2016broadband} may not be accurate. For a fixed resonance frequency, at frequencies high above resonance, the capacitor shorts out and the RLC circuit is approximately identical in filter characteristics to the broadband LR circuit. Near resonance, the current resulting from the thermal noise is resonantly enhanced. (Recall that the thermal noise can be represented as a voltage source in series with the resistance $R(\nu)$.) If the thermal noise dominates the amplifier noise, as is commonplace in SQUID readout of MHz resonant circuits \cite{myers2007calculated}, the SNR is not degraded by the readout. In a broadband search, the thermal noise is rolled off by the L/R pole of the input circuit and is generically lower than the amplifier noise, resulting in degraded sensitivity. Near resonance, one then expects the SNR of the resonant detector to be better than the broadband detector.
If the resonance frequency were not tuned at all, but fixed at a single frequency, one should then expect the SNR to be better than or equal to the broadband search at the resonance frequency and above. If we then scan the resonance frequency, one might expect better SNR than broadband at all frequencies at which a resonator can be built. As we will observe, a rigorous analysis is somewhat more complicated, but comes to the same conclusion.\footnote{A practical implementation of a resonant search might include a broadband flux transformer (an LR circuit), between the RLC circuit and the amplifier, which also couples to the dark matter signal; see, e.g. \cite{Chaudhuri:2014dla}, in which the resonator pickup inductor and the transformer couple to the same volume of dark-matter-induced electromagnetic energy. In this manner, the resonant search can possess sensitivity comparable to the broadband search at frequencies far below resonance (in addition to frequencies far above resonance), where the high impedance of the capacitor suppresses the current response of the RLC circuit to the dark-matter-induced magnetic flux and where the SQUID input is thus dominated by the excitation of the transformer.}

We may gain further insight from our Bode-Fano constraint in Appendix \ref{ssec:BodeFano_FV}. The optimized integrated sensitivity, as constrained by the Bode-Fano criterion, is achieved with a narrowband noise match (a top-hat spectrum when we consider $\zeta(Z_{T}(\nu))$ as a function of frequency), and the single-pole resonator is close to this limit. If the broadband LR search were better than a tunable resonant RLC search, then we would expect the Bode-Fano limit to point toward a wideband match, with $\zeta$ taking on nonzero values over a broad bandwidth of frequencies. However, this is not the case.

We are now ready to quantitatively compare the sensitivity of the optimized resonant and broadband searches. Our comparison consists of two steps. First, we consider the SNR formula for resonant and broadband searches, as given by (\ref{eq:SNR_1_QLamp_FV}). We use this formula to show that the optimized resonant search must be strictly superior to the optimized broadband search, at all frequencies at which a resonator can be built. This statement is independent of the priors, the search range, the inductor quality factor, the physical temperature, and the value of the noise impedance. Second, we evaluate precisely the advantage of the resonant search in example scans. This requires a discussion of amplifier noise and a particular model of noise impedance in the broadband search, to be discussed below. The SNR for the resonant search was considered in detail in Appendix \ref{ssec:FtoV_opt_QL} and is given for the optimized log-uniform search in eq. (\ref{eq:SNR_opt_total}) of Section \ref{ssec:opt_time}. We denote the square of this SNR as $SNR_{R}^{2}(\nu_{\rm DM}^{0})$, where the subscript `R' denotes a resonant search. It thus suffices to derive below only the optimized SNR for the broadband search.


From equations (\ref{eq:SNR_f_opt_PhiV}), (\ref{eq:SNR_1_QLamp_FV}), and (\ref{eq:NoiseNumber_FV}), the signal-to-noise ratio for the broadband search at search frequency $\nu_{\rm DM}^{0}$ is given by
\begin{align}
SNR_{B} & \left[ \nu_{\rm DM}^{0}, g_{\rm DM}, \frac{d\rho_{\rm DM}}{d\nu}(\nu,\nu_{\rm DM}^{0}), \tau \right]^{2} = (2\pi^{2})^{2} T_{\rm tot} \label{eq:SNR_Broad} \\
&\times \int_{\nu_{\rm DM}^{0}}^{\nu_{\rm DM}^{0} + \Delta\nu_{\rm DM}^{c}} d\nu\ \left( \frac{\nu L_{\rm PU}}{R(\nu)} \frac{U_{\rm DM}(\nu,\nu_{\rm DM}^{0}, g_{\rm DM}, \frac{d\rho_{\rm DM}}{d\nu}(\nu,\nu_{\rm DM}^{0}))}{h} \frac{1}{N_{B}^{FV}(\nu, Z_{B}(\nu), Z_{N}(\nu), n(\nu))} \right)^{2} \nonumber
\end{align}
where
\begin{equation}
Z_{B}(\nu)=R(\nu)+ 2\pi i \nu (L_{\rm PU} + L_{\rm IN})
\end{equation}
is the impedance seen by the amplifier. As before, we assume that the real part of the amplifier input impedance is negligible, and we absorb the imaginary part of the amplifier input impedance into the total inductance. We have substituted $T_{\rm tot}$ for $\tau$ because at each search frequency, we integrate for the total experiment time. The subscript `B' denotes that the SNR is for the broadband search. 
\begin{equation}\label{eq:NoiseNumber_Broad}
N_{B}^{FV}(\nu, Z_{B}(\nu), Z_{N}(\nu), n(\nu))= \frac{Re(Z_{B}(\nu))(n(\nu)+ \frac{1}{2}) + \frac{1}{4} \left( \frac{|Z_{B}(\nu)|^{2}}{Z_{N}(\nu)} + Z_{N}(\nu) \right)}{Re(Z_{B}(\nu))} 
\end{equation}
is the noise-equivalent number. 

Suppose that we are performing a broadband search between frequencies $\nu_{l}$ and $\nu_{h}$.  We insert a lossless series capacitor into the circuit such that it possesses resonant frequency $\nu_{l}$. We denote the impedance of this circuit as $Z_{R}(\nu,\nu_{l})$. We compare the sensitivity of the broadband search to a resonant search in which the resonance frequency is not tuned, but simply sits at frequency $\nu_{l}$ for the entire integration time. At frequencies $\nu \geq \nu_{l}$, $|Z_{B}(\nu)|^{2} > |Z_{R}(\nu,\nu_{l})|^{2}$ because the capacitance reduces the circuit reactance and therefore, the magnitude of the impedance. Since $Re(Z_{B}(\nu))=Re(Z_{R}(\nu,\nu_{l}))$ and the noise impedance is the same for both circuits, the noise-equivalent number of the resonant circuit, evaluated with (\ref{eq:NoiseNumber_FV}), is lower than that of the broadband circuit. Since the inductor quality and coupled energy are the same for the two searches, (\ref{eq:SNR_1_QLamp_FV}) and (\ref{eq:SNR_Broad}) show that the SNR of the resonant circuit is greater than that of the broadband circuit at all frequencies in the search range. The optimal resonant scan performs at least as well as this fixed-frequency-resonator scan strategy, so we conclude that the optimized resonant scan is better than the optimized broadband scan at all search frequencies. Note that this statement is independent of the particular model of real-valued noise impedance $Z_{N}(\nu)$, the inductor quality, and the physical temperature, as well as the search range and the set of priors.\footnote{One might argue, depending on the values of $L_{\rm PU}$ and $L_{\rm IN}$ in the broadband circuit and the value of $\nu_{l}$, that inserting a capacitor to produce the desired resonance is not feasible because the capacitance would be too large. However, our conclusions remain the same even if the resonator inductance is increased (above that in the broadband circuit) to reduce the capacitance. Suppose that the values of the total inductance and resistance in the resonant circuit are $N$ times larger than that in the broadband circuit. We may increase the noise impedance by a factor of $N$, e.g. by modifying amplifier parameters and/or by using flux transformers. Then, from (\ref{eq:NoiseNumber_FV}), the noise-equivalent number of the resonator is again found to be lower than the broadband circuit, resulting in higher SNR.}


It is informative to quantify the difference in sensitivity between the optimized broadband search and the optimized resonant search for example scans. We now establish the assumptions and methods for evaluating the difference, building on our resonator optimization for log-uniform priors in Appendix \ref{ssec:FtoV_opt_QL}.

A quantitative, apples-to-apples comparison of single-pole resonant and broadband sensitivities requires a careful treatment of the amplifier noise spectral densities and noise impedance. In Appendix \ref{sec:FtoV_opt}, we assumed that the amplifier noise spectral densities and noise impedance are, to good approximation, constant within a few sensitivity bandwidths of the resonance frequency. In a broadband search, practically speaking, the noise spectral densities and noise impedance of the flux-to-voltage amplifier (which is operated without retuning) vary significantly over the entire frequency range. In order to enable the sensitivity comparison, we must then make an assumption about the frequency-dependent noise impedance of the amplifier used in the broadband search. One natural choice for an amplifier in a broadband search is a dc SQUID. Because of correlations between imprecision noise and backaction noise in a dc SQUID, the noise impedance tends to have a complex value. We ignore this non-ideal property (stemming from Johnson noise in the resistive shunts across the Josephson junctions) for the purposes of this comparison and assume that the noise impedance of the amplifier is real. Furthermore, we make use of the fact that flux-to-voltage amplifiers tend to possess a noise impedance that increases linearly with frequency \cite{falferi1998back}.
A simple and natural model for this noise impedance is thus:
\begin{equation} \label{eq:Broad_ZN}
Z_{N}(\nu)=2\pi \nu L_{N},
\end{equation}
where $L_{N}$ is a characteristic inductance. Though we are assuming a specific model for the real-valued noise impedance, we stress that our above proof that the resonant search is superior to the broadband search does not depend on the model. We only need the model (\ref{eq:Broad_ZN}) to calculate the size of the advantage.

It is important to understand (\ref{eq:Broad_ZN}) in the context of the resonator optimization of Appendix \ref{ssec:FtoV_opt_QL} in order to ensure that we are establishing an apples-to-apples comparison of tunable-resonant and broadband sensitivity. In the resonator optimization, we assumed approximately constant noise impedance in the sensitivity bandwidth at each resonant scan step. However, the noise impedance of (\ref{eq:Broad_ZN}) is not constant with frequency, but rather, linear with frequency. We may reconcile the two sets of assumptions by observing that the sensitivity bandwidth is narrow (much smaller than the resonance frequency), and therefore, though the noise impedance (\ref{eq:Broad_ZN}) is frequency-dependent, we may substitute $\nu_{r}$ for $\nu$ for the purposes of sensitivity optimization and calculation. In other words, the noise impedance can be approximated as constant during one scan step in the resonant search. 

In the resonant search, after enough scan steps have been taken that the thermal occupation number changes significantly, the noise impedance of the amplifier is reoptimized. This optimization can be conducted, e.g., by tuning a variable transformer based on Josephson junctions, or by tuning the bias parameters of some flux-to-voltage amplifiers. At some frequency interval, perhaps a decade in frequency, the amplifier would need to be swapped with an amplifier with different coupling strength to enable a larger change in noise impedance. Optimizing the noise impedance is equivalent to optimizing $L_{N}$. 

For the broadband experiment, we also need to optimize $L_{N}$. In this case, we only choose a single value of $L_{N}$ because we do not scan and change circuit and readout parameters. Because we need to choose a single value for a wideband search, it would then seem that we would require a value function that weights the importance of different search frequencies. However, we will find that the value of $L_{N}$ that maximizes the SNR at each search frequency is independent of the frequency, and therefore, such a value function is not required. This observation suggests that an amplifier exhibiting a linear-in-frequency noise impedance scaling is an appropriate choice for a broadband search.


We evaluate analytically the optimized resonant and broadband scans--and the comparison of their SNRs--only for searches with lower limit $\nu_{l} \geq 1$ kHz. Below 1 kHz, some of the approximations in the resonator optimization of Appendix \ref{ssec:FtoV_opt_QL} begin to break down. However, we emphasize that the general results of our analysis apply down to the lowest frequency at which a resonator can be practically built,  $\nu_{l} \sim$ 100 Hz; as we have shown, for any search range and any set of priors, there exists a resonator scan strategy such that the resonant search gives better sensitivity than the broadband search at all frequencies in the search range. 

We consider two specific examples of scans. In the first scan, a wide band scan, we set the lower limit for comparison at $\nu_{l}=10$ kHz and the upper limit at $\nu_{h}=100$ MHz. The second scan we consider is a low-frequency scan, where $\nu_{l}=1$ kHz and $\nu_{h}=10$ kHz. The reason for considering two scans becomes clear when we discuss the role of priors in the sensitivity comparison. For the resonator scan time allocation, we assume the log-uniform search optimization represented by eq. (\ref{eq:t_opt}).

Using (\ref{eq:Broad_ZN}) and the assumption of low loss $R(\nu) \ll 2\pi \nu (L_{\rm PU} + L_{\rm IN})$, we may evaluate the noise-equivalent number of (\ref{eq:NoiseNumber_Broad}):
\begin{equation} \label{eq:NB_eval}
N_{B}^{FV}(\nu, Z_{B}(\nu), Z_{N}(\nu), n(\nu)) \approx n(\nu)+ \frac{1}{2} + \frac{1}{4} \frac{ 2\pi \nu \left( \frac{(L_{\rm PU}+L_{\rm IN})^{2}}{L_{N}} + L_{N} \right)}{R(\nu)}. 
\end{equation}
The noise equivalent number is minimized and the SNR is maximized when the characteristic inductance for the noise impedance is the sum of the pickup and input inductances: $L_{N}=L_{\rm PU} + L_{\rm IN}$. As promised, the optimal value of $L_{N}$ is independent of frequency. The optimal value of the noise equivalent number is then
\begin{equation}\label{eq:NoiseNumber_Broad_Opt}
N_{B}^{\rm opt}(\nu, Z_{B}(\nu), n(\nu))= n(\nu)+ \frac{1}{2} + \frac{1}{2} \frac{ 2\pi \nu ( L_{\rm PU}+L_{\rm IN} )}{R(\nu)}
\end{equation}
The first term on the right-hand side represents the thermal noise, the second the zero-point fluctuation noise, and the third the amplifier noise. We identify $2\pi \nu (L_{\rm PU} + L_{\rm IN})/R(\nu)$ as the inductor quality at frequency $\nu$. For the purposes of the sensitivity comparison, we set the quality factor to be a frequency-independent value $Q \gg 1$. Combining eqs. (\ref{eq:UDM_def}), (\ref{eq:SNR_Broad}), and (\ref{eq:NoiseNumber_Broad_Opt}), yields the SNR expression
\begin{equation}\label{eq:SNR_Broad_eval}
SNR_{B}^{2}(\nu_{\rm DM}^{0}) \approx \pi^{2} T_{\rm tot} \left( \frac{Q}{n(\nu_{\rm DM}^{0}) + \frac{1}{2}Q} \right)^{2} \int_{\nu_{\rm DM}^{0}}^{\nu_{\rm DM}^{0} + \Delta\nu_{\rm DM}^{c}} d\nu\ \left( \frac{g_{\rm DM}^{2} V_{\rm PU} \frac{d\rho_{\rm DM}}{d\nu}(\nu,\nu_{\rm DM}^{0})}{h} \right)^{2}
\end{equation}
where, as we did for the resonator, we have assumed $L_{\rm PU} \gg L_{\rm IN}$, so as to maximize SNR.

The ratio of the SNR-squared for broadband and resonant searches is, from (\ref{eq:SNR_opt_total}),
\begin{align}\label{eq:ResvBroad_Gamma}
\Gamma(\nu_{\rm DM}^{0}) &\equiv \frac{SNR_{R}^{2}(\nu_{\rm DM}^{0})}{SNR_{B}^{2}(\nu_{\rm DM}^{0})} \nonumber \\
&= \frac{4 \pi}{\rm ln(\nu_{h}/\nu_{l})} \frac{\xi^{\rm opt}(n(\nu_{\rm DM}^{0}))^{2}}{ (4\xi^{\rm opt}(n(\nu_{\rm DM}^{0}))n(\nu_{\rm DM}^{0}) + (1+\xi^{\rm opt}(n(\nu_{\rm DM}^{0})))^{2})^{3/2} } \frac{(n(\nu_{\rm DM}^{0}) + \frac{1}{2}Q)^{2}}{Q}
\end{align}
We now focus on particular limits of equation (\ref{eq:ResvBroad_Gamma}), which are relevant to the frequency range 1 kHz-100 MHz of our two example scans. Assume that both experiments are conducted at a temperature of 10 mK. Then, the high-thermal-occupation limit $n(\nu_{\rm DM}^{0}) \gg 1$ applies, and $\xi^{\rm opt}\approx 2n(\nu_{\rm DM}^{0})$, so we find
\begin{equation}
\Gamma(\nu_{\rm DM}^{0}) \approx \frac{2\pi}{\rm ln(\nu_{h}/\nu_{l})} \frac{1}{3\sqrt{3}} \frac{(n(\nu_{\rm DM}^{0})+ \frac{1}{2}Q)^{2}}{n(\nu_{\rm DM}^{0})Q}
\end{equation}
We calculated the optimal resonator sensitivity under the assumption that the sensitivity quality factor is much larger than unity, which is appropriate at 10 mK for internal Qs on the order of one million down to resonance frequencies of approximately $\sim$ 1 kHz. This condition implies that $Q \gg n(\nu_{\rm DM}^{0})$ (see eq. (\ref{eq:InvN_Spectrum_Q})) , so we find
\begin{equation}\label{eq:Gamma_parametric}
\Gamma(\nu_{\rm DM}^{0}) \approx \frac{\pi}{6\sqrt{3} \rm ln(\nu_{h}/\nu_{l})} \frac{Q}{n(\nu_{\rm DM}^{0})} \gg 1
\end{equation}
The above expression has the expected parametric dependence on quality factor and thermal occupation number. From the discussion of Fig. \ref{fig:PhiVcrit_v_opt} and eq. (\ref{eq:NoiseNumber_Broad_Opt}), the noise-equivalent number for an optimized resonator at frequency $\nu_{\rm DM}^{0}$ is lower than that for the broadband search by a factor of $\sim Q/n(\nu_{\rm DM}^{0})$. The noise number for the resonator is dominated by the thermal noise, varying as $\sim n$, while the noise number for the broadband search is dominated by the amplifier noise, varying as $\sim Q$. The broadband search spends more time integrating at a given frequency than the resonant search, with the ratio of integration times being proportional to the sensitivity quality factor $\sim Q/n(\nu_{\rm DM}^{0})$. The SNR is inversely proportional to the noise-equivalent number and varies as the square root of integration time. Thus, we find that the parametric dependence of $\Gamma$ on quality factor and thermal occupation number is $\sim (\sqrt{n(\nu_{\rm DM}^{0})/Q} \times Q/n(\nu_{\rm DM}^{0}))^{2} = Q/n(\nu_{\rm DM}^{0})$, which matches (\ref{eq:Gamma_parametric}). The numerical prefactor in the first fraction of (\ref{eq:Gamma_parametric}) is not far from unity--approximately 0.03 for the wide band scan ($\nu_{l}=10$ kHz and $\nu_{h}=100$ MHz) and 0.13 for the low-frequency scan ($\nu_{l}=1$ kHz and $\nu_{h}=10$ kHz). We thus find that the resonator sensitivity is superior to the sensitivity of the broadband search. 

\begin{figure}[htp]
\centering
\includegraphics[width=\textwidth]{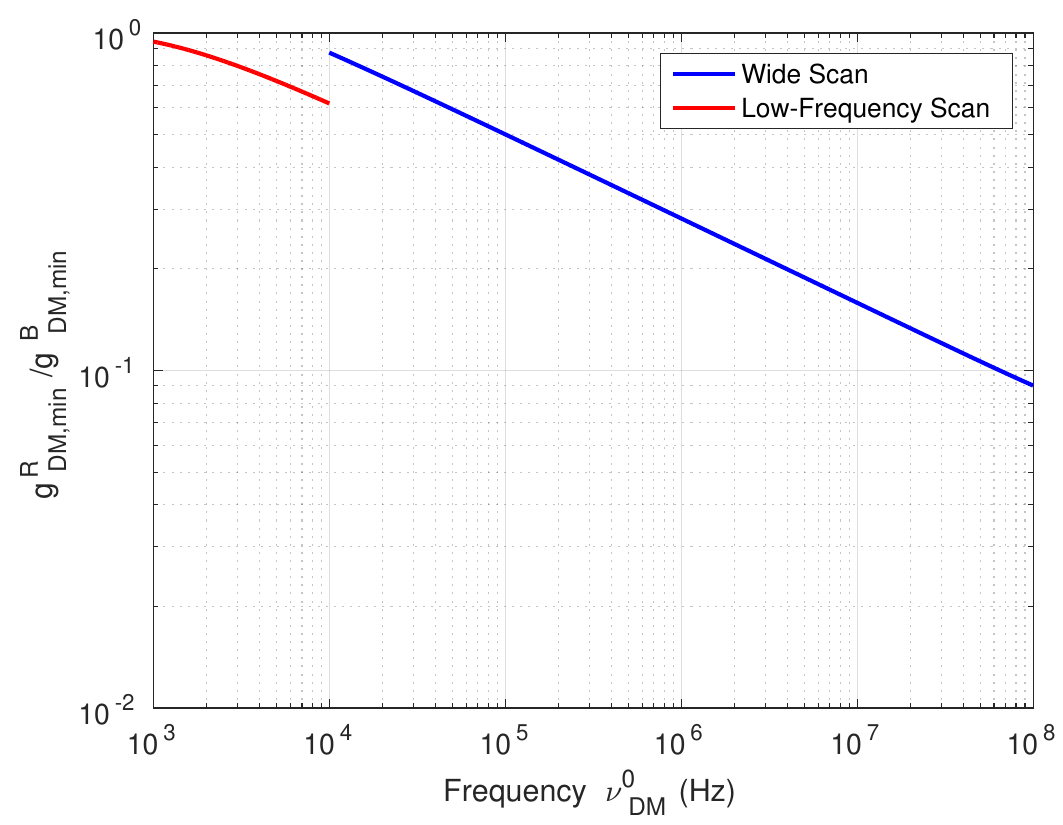}
\caption{Resonant vs broadband sensitivity. $\frac{g_{\rm DM,min}^{R}}{g_{\rm DM,min}^{B}}$ vs frequency for a quality factor of $10^{6}$ and temperature of 10 mK. The comparison for a wide scan between 10 kHz and 100 MHz is shown in blue, while that for a low-frequency scan between 1 kHz and 10 kHz is shown in red. The small curvature at the lowest frequencies in the red curve is the result of increasing error in the approximations used to evaluate resonant sensitivity.}  \label{fig:ResvBroad_gratio}
\end{figure}

To demonstrate the advantage of the resonant search more explicitly, we may consider the ratio of minimum couplings to which each search is sensitive,
\begin{equation}
\frac{g_{\rm DM, min}^{R}}{g_{\rm DM, min}^{B}}= \Gamma(\nu_{\rm DM}^{0})^{-1/4}
\end{equation}
A value less than unity implies that the resonant search is superior. For the wide band scan, the ratio is plotted, with the optimized log-uniform time allocation of eq. (\ref{eq:t_opt}), as the blue curve in Fig. \ref{fig:ResvBroad_gratio}. For the low-frequency scan, the ratio is plotted as the red curve. We have assumed a quality factor of one million and in both calculations have used the formula (\ref{eq:ResvBroad_Gamma}).

In contrast to the limits shown in \cite{kahn2016broadband}, combining the two scans in the plot demonstrates that the resonant search with Q of $10^{6}$ is considerably better than the broadband search at all frequencies above 1 kHz. The ratio is lower than one at all frequencies, dropping below 0.1 in coupling at the highest frequencies. This corresponds to a scan rate (reflecting the time to reach a particular coupling) that is more than four orders of magnitude higher. 

We have cut off the analysis at 1 kHz not because coils at lower frequencies cannot be made, but because the approximations used in the analysis break down.  At these low frequencies, the thermal occupation number is comparable to, or larger than, the resonator quality factor. This leads to an order-unity sensitivity quality factor, which manifests in Fig. \ref{fig:ResvBroad_gratio} as the curving of the red line at lower frequencies. At such low sensitivity quality factors, the integrals over frequency that lead to the optimized resonator sensitivity (\ref{eq:SNR_opt_total})--see in particular, Section \ref{sssec:res_match_opt} and eq. (\ref{eq:F_neu_res})--cannot be evaluated analytically in the manner stated. Nevertheless, we know from our demonstration below eq. (\ref{eq:NoiseNumber_Broad}) that the optimized tunable resonant search is strictly better than the optimized broadband search, as long as a resonator can practically be constructed (at frequencies $\gtrsim 100$ Hz).

If one were to extrapolate the sensitivity comparison implied by the wide scan to below 10 kHz, it would seem that a broadband approach may be fundamentally more suitable. However, this conclusion is not accurate because the time allocation for the scanned search was selected by assuming the prior of a log-uniform search over the 10 kHz-100 MHz search range. (Again, we refer the refer the reader to the proof below eq. (\ref{eq:NoiseNumber_Broad}).) Mathematically, this prior is represented by the natural logarithm in the denominator of eq. (\ref{eq:ResvBroad_Gamma}) and thus affects the calculated sensitivity curves. If there were reason to expect that dark matter is more likely at lower frequencies, a different prior would be used.


This observation begs the question of conducting two separate experiments in the two different frequency ranges, 1-10 kHz and 10 kHz-100 MHz. If the two experiments were conducted for the same integration time, might one expect that, by selecting a resonant experiment for one and a broadband experiment for the other, that the lower-frequency sensitivity might be improved? However, asking this question is assuming a prior that weights low frequency information as particularly interesting--in other words, as more valuable than high frequency information. Assuming that prior, we find that two resonant scans can always be designed to outperform a resonant scan and a broadband scan at all search frequencies. For example, one of the two resonant scans may integrate solely at low frequencies. We consider a second, low-frequency scan covering only the 1 kHz-10 kHz band, which yields the ratio of minimum detectable couplings shown in red. Again, the ratio is smaller than unity, indicating the superiority of the low-frequency resonant scan; the scan rate is a factor of few higher for resonant than for broadband in the 1-10 kHz range. The optimal choice for two experiments is thus two resonant scans, rather than a resonant scan and a broadband scan, at all frequencies above which a resonator can be built.

The data shown in Figure \ref{fig:ResvBroad_gratio} is plotted with $Q=10^6$. The advantage of a resonant search is diminished as the $Q$ is reduced. It may then seem that if an experiment only achieves a lower $Q \approx $50,000, a broadband search would have a fundamental advantage at lower frequencies. Again, one must be careful about extrapolating. Such an extrapolation is inaccurate here. Eqs. (\ref{eq:ResvBroad_Gamma}) and (\ref{eq:Gamma_parametric}) show that, as long as the Q is high enough that the approximations used are correct, the integrated scan sensitivity is better for a resonant search for any set of priors, including priors emphasizing lower frequencies. At $Q\approx $50,000, our approximations break down at $\approx$ 10-20 kHz (instead of 1 kHz), but we can still state that the advantage of a resonant search persists to lower frequencies and $Q$. This assertion is based on the original argument that a fixed-frequency resonant circuit  possesses higher sensitivity than a broadband circuit at all frequencies above resonance. 

In conclusion, given any set of priors, search range, inductor quality, physical temperature, and noise impedance model (for the quantum-limited flux-to-voltage amplifier), the optimized resonant search is fundamentally superior to the optimized broadband search over all frequencies at which a resonator may be built ($\gtrsim$ 100 Hz). The resonant search enables search times that are generically a few orders of magnitude lower than the broadband search.

As a final remark in this section, we consider some of the practical advantages for a resonant search. Both resonant and broadband searches are susceptible to false signals from electromagnetic interference. In a resonant search, an interfering signal at a given frequency is quickly detected, and after elimination of the source or modification of the experiment, confirmation that the spurious signal is gone can be quickly achieved. In contrast, in a broadband search, a weak spurious signal can only be detected (and its removal can only be confirmed) after a long integration time. Broadband searches are also more vulnerable to weak, intermittent transient signals, which can be eliminated in resonant experiments by repeated scans. Resonant scans also have more subtle advantages. Because of amplifier nonlinearities, in broadband scans, the many pickup lines that inevitably couple to the experiment at some level will produce a forest of weak intermodulation products at many points in the bandwidth of the experiment. These weak products will only be evident after a lengthy integration. A resonator filters the pickup lines outside of the sensitivity bandwidth before it is amplified, greatly reducing problems with intermodulation products. The rejection may be further enhanced by use of narrowband filters (e.g. those containing only the sensitivity bandwidth) in the readout lines following the first-stage amplifier. Such filters inherently cannot be used in a broadband experiment. Finally, if a weak candidate signal is identified in a broadband search, optimal follow-up will require a resonant measurement at the frequency of interest. A resonator can quickly integrate to a much higher signal-to-noise ratio than a broadband experiment at a given frequency in order to determine spatial, temporal, and directional properties and to measure the candidate signal power spectrum.

\newpage

\newpage
\bibliography{Bib_Opt}

\end{document}